\documentclass[12pt]{article}
\pdfoutput=1
\usepackage{yfonts}
\usepackage{color}
\usepackage{mhchem}
\usepackage{xcolor}
\usepackage{cite}
\usepackage{hyperref}
\hypersetup{colorlinks=true,linkcolor=red,anchorcolor=black,citecolor=green}
\usepackage[toc,page]{appendix}
\usepackage{amsfonts}
\usepackage{bbold}
\usepackage{textcomp}
\usepackage[DIV13]{typearea}
\usepackage{amsmath, amsthm, amssymb, mathtools,empheq,latexsym,dsfont}
\usepackage{bbm}
\usepackage{slashed, simplewick}
\usepackage[utf8]{inputenc}
\usepackage{graphicx,placeins}
\usepackage{makeidx}
\usepackage[font=small,labelfont=bf]{caption}
\usepackage{nicefrac}
\usepackage{subfigure}
\usepackage{array, bigdelim,multirow,multicol}
\usepackage[integrals]{wasysym}
\usepackage{fancybox}
\usepackage{bm}
\usepackage{float}
\usepackage{rotating}
\usepackage{colortbl}
\usepackage{booktabs}
\usepackage[top=2cm,textwidth=16.6cm,textheight=22.75cm]{geometry}
\usepackage{doi}
\usepackage{fancybox}
\usepackage[compat=1.0.0]{tikz-feynman}
\usepackage{tikz}
\usepackage{empheq}
\allowdisplaybreaks
\newcommand{\ignore}[1]{}
\usepackage{enumitem}

\usepackage{braket}

\graphicspath{{immagini/}}
\definecolor{Gray}{gray}{0.92}
\DeclareMathAlphabet{\mathscr}{OT1}{pzc}{m}{it}

\newcommand{\be}{\begin{equation}}
\newcommand{\ee}{\end{equation}}
\newcommand{\bea}{\begin{eqnarray}}
\newcommand{\eea}{\end{eqnarray}}

\newcommand{\taubar}{\overline{\tau}}

\makeatletter
\@addtoreset{equation}{section}
\makeatother

\begin{document}

\unitlength = 1mm
\setlength{\extrarowheight}{0.2 cm}
\thispagestyle{empty}
\bigskip
\vskip 1cm

\title{\Large \bf Non-holomorphic modular flavor symmetry and odd weight polyharmonic Maa{\ss} form \\[2mm] }

\date{}

\author{
Bu-Yao Qu$^{a}$\footnote{E-mail: {\tt
qubuyao@mail.ustc.edu.cn}},  \ Jun-Nan Lu$^{a}$\footnote{E-mail: {\tt junnanlu@ustc.edu.cn}},\
Gui-Jun Ding$^{a,b}$\footnote{E-mail: {\tt
dinggj@ustc.edu.cn}}
\\*[20pt]
\centerline{
\begin{minipage}{\linewidth}
\begin{center}
$^a${\it \small Department of Modern Physics,  and Anhui Center for fundamental sciences in theoretical physics,\\
University of Science and Technology of China, Hefei, Anhui 230026, China}\\[2mm]
$^b${\it \small College of Physics, Guizhou University, Guiyang 550025, China}
\end{center}
\end{minipage}}
\\[10mm]}
\maketitle
\thispagestyle{empty}

\centerline{\large\bf Abstract}

\begin{quote}
\indent

We extend the framework of non-holomorphic modular flavor symmetry to include the odd weight polyharmonic Maa{\ss} forms. The integer weight polyharmonic Maa{\ss} forms of level $N$ can be arranged into multipltets of the homogeneous finite modular group $\Gamma'_N$.  We propose to construct the integer weight, including weight one, non-holomorphic polyharmonic Maa{\ss} forms from the non-holomorphic Eisenstein series. The previous results of even weight polyharmonic Maa{\ss} forms are reproduced. We apply this formalism to address the flavor structure of the standard model. An example lepton model based on the modular group $\Gamma'_3\cong T'$ is constructed, where neutrino masses are generated via type-I seesaw mechanism with two right-handed neutrinos. This model can accommodate the experimental data for both normal and inverted neutrino mass orderings. We further extend this model to include quarks, so that the masses and mixing parameters of both quark and lepton sectors can be successfully described in terms of only thirteen real free parameters.  It is the modular invariant model with the smallest number of free parameters so far, only normal ordering neutrino mass is viable after including quarks, and the correlations among the input parameters and flavor observables are analyzed.

\end{quote}

\clearpage

\section{Introduction}

The origin of three families of quarks and leptons is a big challenge  of Standard Model (SM). We also don't know the fundamental principle governing the observed hierarchies among the masses of the three generations of up quarks, down quarks and charged leptons. The discovery of neutrino oscillations calls for new physics beyond the SM to dynamically generate the tiny neutrino masses and significant lepton flavor mixing. One of interesting approaches is to impose non-Abelian discrete symmetries for flavors at a high energy scale~\cite{King:2013eh,King:2017guk,Petcov:2017ggy,Feruglio:2019ybq,Xing:2020ijf,Ding:2023htn,Ding:2024ozt}.
In the models based on traditional discrete flavor symmetry, a number of scalar fields called flavons transforming nontrivially under the flavor symmetry group have to be introduced. In order to accommodate the large lepton mixing angles, the vacuum expectation values (VEVs) of flavon fields should be aligned along specific directions, and it is notoriously difficult to be realized dynamically. Generally shaping symmetries and extra dynamics responsible for the vacuum alignment are required, and the scalar potential has to be cleverly designed. As a consequence, the flavor symmetry breaking sector makes the model quite elaborate.

Recently the modular symmetry has been proposed to explain the flavor structure of SM~\cite{Feruglio:2017spp}.
In this framework, the supersymmetry is required to enforce that the Yukawa couplings are modular forms of level $N$, which are holomorphic functions of complex modulus $\tau$ and transform as irreducible representations of the finite modular groups $\Gamma_N$ or $\Gamma'_N$. Consequently  the vacuum expectation value of the modulus $\tau$ is the unique source of the symmetry breaking and the flavon fields are unnecessary so that one doesn't need to tackle the notorious vacuum alignment problem. Various aspects of modular symmetry has been comprehensively reviewed in~\cite{Ding:2023htn,Kobayashi:2023zzc}. The modular symmetry allows to construct quite predictive fermion mass models, all the lepton masses and mixing angles can be explained in terms of only four real couplings plus the modulus $\tau$ in the minimal modular invariant model~\cite{Ding:2022nzn,Ding:2023ydy}. The modular symmetry could connect  quark sector and lepton sector through the modulus $\tau$ as a portal, the flavor observables of quarks and leptons can be accommodated simultaneously by using fourteen real parameters including the real part and imaginary part of $\tau$ in the most predictive modular invariant model so far~\cite{Ding:2023ydy,Ding:2024pix}. It is remarkable that the modular symmetry could provide a solution to the strong CP problem without introducing axion~\cite{Feruglio:2023uof,Penedo:2024gtb,Feruglio:2024ytl,Feruglio:2025ajb}. Moreover, the complex modulus $\tau$ may serve as a bridge connecting particle physics and cosmology. It could drive the cosmic inflation~\cite{Gunji:2022xig,Abe:2023ylh,Ding:2024neh,King:2024ssx,Casas:2024jbw,Kallosh:2024ymt,Ding:2024euc,Aoki:2025wld}, it could also be responsible for the reheating of the Universe~\cite{Ding:2024euc} and generate the baryon asymmetry~\cite{Duch:2025abl}.

The framework of the non-holomorphic cousin of modular flavor symmetry was formulated in~\cite{Qu:2024rns}. In this approach, supersymmetry is unnecessary, and the Yukawa couplings are not necessarily constrained to be holomorphic. Instead, the harmonic condition takes the place of the holomorphicity assumption. As a result, the Yukawa couplings are required to be polyharmonic Maa{\ss} forms of level $N$ which can be decomposed into multiplets of the homogeneous finite modular group $\Gamma'_N$.
In this scenario, the generalized CP (gCP) symmetry can be consistently incorporated, as discussed in the supersymmetric cases~\cite{Novichkov:2019sqv}. The gCP acts on the complex modulus $\tau$ as $\tau \stackrel{{\rm CP}}{\longmapsto} - \tau^*$ up to modular transformation~\cite{Dent:2001cc,Dent:2001mn,Baur:2019kwi,Baur:2019iai,Novichkov:2019sqv}. In the basis where the representation matrices of $S$ and $T$ in all irreducible representations are unitary and symmetric, the gCP transformation would reduce to the traditional CP symmetry and the corresponding CP transformation is an identity matrix in flavor space. As a result, all the coupling constants are constrained to be real, which enhances the predictive power of the model.
So far only even weight polyharmonic Maa{\ss} forms are considered when constructing non-holomorphic modular invariant models. Some non-holomorphic modular invariant models for lepton masses and flavor mixing have been constructed at levels $N=2$~\cite{Okada:2025jjo}, $N=3$~\cite{Qu:2024rns,Kumar:2024uxn,Nomura:2024ghc,Nomura:2024atp,Nomura:2024ctl,Nomura:2024vus,Kobayashi:2025hnc,Loualidi:2025tgw}, $N=4$~\cite{Ding:2024inn} and $N=5$~\cite{Li:2024svh}.

In the paradigm of modular flavor symmetry, top-down constructions typically give rise to integer weight modular forms~\cite{Baur:2019kwi,Baur:2019iai,Nilles:2020nnc,Kikuchi:2020nxn}, the corresponding finite modular group is $\Gamma'_N$ which is the double cover of $\Gamma_N$. In the non-holomorphic cousin of modular flavor symmetry, even weight polyharmonic Maa{\ss} forms of level $N$ have been considered. The contribution of the general integral weight polyharmonic Maa{\ss} forms will be studied in the present work, and the odd weight especially weight $1$ polyharmonic Maa{\ss} forms will be investigated. The non-holomorphic polyharmonic Maa{\ss} forms of even weights were constructed by identifying the preimage of the holomorphic modular forms under the action of the differential operators $\xi_k$ and $D^{1-k}$ in previous work~\cite{Qu:2024rns}. However, it is challenging to generalize this approach for constructing polyharmonic Maa{\ss} forms with arbitrary integer weights. In particular for $k=1$, the condition of $D$ operator is trivially satisfied because of $D^{1-k}=1$, thus only the action of $\xi$ operator remains and it is insufficient to fix the expressions of polyharmonic Maa{\ss} forms. In the present work, we shall construct the integer weight polyharmonic Maa{\ss} forms from the non-holomorphic Eisenstein series, the even weight polyharmonic Maa{\ss} forms studied before are reproduced. The polyharmonic Maa{\ss} forms of level $N$ can be organized into irreducible multiplets of $\Gamma'_N$, their explicit expressions of $q$-expansion are presented for $N=3, 4, 5$.

Furthermore, we apply the integer weight polyharmonic Maa{\ss} forms to construct the modular invariant quark and lepton models based on the finite modular group $\Gamma'_3\cong T'$. The neutrino masses are assumed to be generated via the type-I seesaw mechanism. We are interested in the models with a small number of free parameters.
In the lepton sector, several models with gCP symmetry are shown to well describe the experimental data in terms of $7$ real parameters including $\text{Re}(\tau)$ and $\text{Im}(\tau)$. For the quark sector, at least $8$ free parameters are necessary to accommodate the hierarchical quark masses and mixing angles. Furthermore, we demonstrate that quark-lepton unification can be achieved by combining the lepton and quark sectors, which share a common value of the modulus $\tau$. In the resulting model, the twenty-two observables for the masses and mixing of quarks and leptons can be described by altogether only thirteen real parameters, which is the minimal number of real parameters in the modular invariant flavor models so far. Since number of parameters is smaller than the number of observables, the correlations between some observables are expected. We explore the correlations among flavor observables and input parameters by comprehensive numerical analysis.

This paper is organized as follows. In section~\ref{sec:frame}, we describe the framework of non-holomorphic modular symmetry. In section~\ref{sec:integer_weight}, we show how to construct the integer weight polyharmonic Maa{\ss} forms by using non-holomorphic Eisenstein series. In section~\ref{sec:models}, we investigate the non-holomorphic modular invariant models for quarks and leptons with the finite modular group $\Gamma'_3\cong T'$, and we present a predictive model which can explain the masses and mixing parameters of both quarks and leptons by thirteen real free parameters when gCP symmetry is imposed. It is the modular invariant model with the smallest number of free parameters to describe the flavor structure of quarks and leptons in the literature.
Finally, we draw the conclusion in section~\ref{sec:conclusion}. The non-holomorphic Eisenstein series are defined at each cup, and we give the cusps of $\Gamma(N)$ in Appendix~\ref{app:cusp} for $N=1, 2, \ldots, 7$. Appendix~\ref{app:math} contains the Dirichlet character and $L$-function which appear in the Fourier expansion coefficients of the non-holomorphic Eisenstein series. The finite modular group $\Gamma'_3\cong T'$, $\Gamma'_4\cong S'_4$ and $\Gamma'_5\cong A'_5$ and their irreducible representations are given in Appendix~\ref{app:Gamma3-MF3}. In Appendix~\ref{app:Polyharmonic_Maass_forms}, we present the integer weight polyharmonic Maa{\ss} forms of levels $N=3, 4, 5$ which are constructed from the non-holomorphic Eisenstein series.

\section{Framework\label{sec:frame}}

In string theory and certain low-energy field theories, modular symmetry plays a pivotal role in constraining the structure of physical models. Specifically, the modular group ${\rm SL}(2,\mathbb{Z})$ acts on fields in such a way that the theory remains invariant under this group's transformation. This is crucial in models where supersymmetry is either softly broken or absent altogether, leading to non-holomorphic transformations and the emergence of non-holomorphic modular functions.

The full modular group ${\rm SL}(2,\mathbb{Z})$ is the group of $2\times 2$ matrices with integer entries and unit determinant, i.e.
\begin{eqnarray}
{\rm SL}(2,\mathbb{Z}) =\left\{\begin{pmatrix}
a ~&~ b \\
c ~&~ d
\end{pmatrix}\Bigg| a, b, c, d\in\mathbb{Z},~~ad-bc=1\right\}\,,
\end{eqnarray}
which is usually denoted as $\Gamma$.
The group ${\rm SL}(2,\mathbb{Z})\equiv\Gamma$ can be generated by two matrices $S$ and $T$ with
\begin{eqnarray}
S=\begin{pmatrix}
0 ~&~ 1 \\
-1 ~&~ 0
\end{pmatrix}\,,~~~
T=\begin{pmatrix}
1 ~&~ 1 \\
0 ~&~ 1
\end{pmatrix}\,.
\end{eqnarray}
One can see $S^2=-\mathbb{1}$ and the modular generators $S$, $T$ satisfy the following relations
\begin{eqnarray}
S^4 = (ST)^3=\mathbb{1},\quad S^2T=TS^2\,.
\end{eqnarray}
The modular group acts on the upper half plane $\mathcal{H}=\{ \tau \in \mathbb{C} \mid {\rm Im}(\tau) >0 \}$ via the M{\"o}bius transformation
\begin{eqnarray}
\label{eq:mobius-trans}\gamma \tau = \dfrac{a \tau + b}{c \tau + d}\,,~~~\gamma=\begin{pmatrix}
a ~&~ b \\ c ~&~ d
\end{pmatrix} \in {\rm SL}(2,\mathbb{Z})\,,
\end{eqnarray}
which implies
\begin{equation}
S\tau=-\frac{1}{\tau}\,,~~~T\tau=\tau+1\,.
\end{equation}
The principal congruence subgroup $\Gamma(N)$ of level $N$ for any positive integer $N$ is defined as
\begin{eqnarray}
\Gamma(N) = \left\{ \begin{pmatrix}
a ~&~ b \\
c ~&~ d
\end{pmatrix} \in {\rm SL}(2, \mathbb{Z}) \Bigg| \begin{pmatrix}
a ~&~ b \\
c ~&~ d
\end{pmatrix} = \begin{pmatrix}
1 ~&~ 0 \\
0 ~&~ 1
\end{pmatrix}~~({\rm mod}\, N) \right\} \,,
\end{eqnarray}
which is the infinite normal subgroup of ${\rm SL}(2, \mathbb{Z})$ and $\Gamma(1)={\rm SL}(2, \mathbb{Z})$. One can see that $T^N$ is an element of $\Gamma(N)$. The quotient group $\Gamma'_N\equiv\Gamma / \Gamma(N)$ is the so called homogeneous finite modular group. It can be generated by $S$ and $T$ obeying the following relations
\begin{eqnarray}
S^2 = R\,,~~(ST)^3 = T^N = R^2 = \mathbb{1} \,,~~RT=TR  \,.
\end{eqnarray}
This relation is enough to fix the group $\Gamma'_N$ for $N\leq 5$, and additional relations are needed to render the group finite for $N\geq6$~\cite{deAdelhartToorop:2011re,Li:2021buv,Ding:2020msi}.

The modular invariance requires the Yukawa couplings to be some modular function $Y(\tau)$ of weight $k$ and level $N$, which transforms under modular symmetry as follows,
\begin{eqnarray}
\label{eq:modularity-def}Y(\gamma \tau) = (c\tau+d)^k Y(\tau)\,,~~\gamma \in \Gamma(N)
\end{eqnarray}
The absence of supersymmetry would release the holomorphic condition. Instead, the Yukawa couplings $Y(\tau)$ are supposed to fulfill the harmonic condition~\cite{Qu:2024rns}:
\begin{eqnarray}\label{eq:harmonoic-condition}
\label{eq:Delta-harmonic}\Delta_k Y(\tau)=0\,,
\end{eqnarray}
where $\Delta_k$ is the weight $k$ hyperbolic Laplacian operator
\begin{eqnarray}
\Delta_k = - y^2 \left( \dfrac{\partial^2}{\partial x^2} + \dfrac{\partial^2}{\partial y^2} \right) + iky \left( \dfrac{\partial}{\partial x} + i \dfrac{\partial}{\partial y} \right)=-4y^2\frac{\partial}{\partial\tau}\frac{\partial}{\partial\bar{\tau}}+2iky\frac{\partial}{\partial\bar{\tau}}\,.
\end{eqnarray}
Here $\tau\equiv x+iy$, $x$ and $y$ are real and imaginary parts of $\tau$ respectively. Moreover, the modular function should satisfy certain moderate growth condition. In this case, we require $Y(\tau)$ to be at most polynomial divergent, i.e.
\begin{eqnarray}
\label{eq:growth-cond}Y(x+iy)=\mathcal{O}(y^\alpha)
\end{eqnarray}
for some constant $\alpha$. Such functions are called polyharmonic Maa{\ss} forms which have played an important role in automorphic forms. They span a linear space of finite dimension.

From Eqs.~(\ref{eq:modularity-def}, \ref{eq:Delta-harmonic}, \ref{eq:growth-cond}), one can obtain the Fourier expansion of a polyharmonic Maa{\ss} form $Y(\tau)$ of weight $k$ and level $N$ as follow,
\begin{eqnarray}\label{eq:Fourier-exp}
Y(\tau) = \sum_{\substack{n\in \frac{1}{N}\mathbb{Z} \\ n \geq 0}} c^+(n) q^n + \sum_{\substack{n \in \frac{1}{N}\mathbb{Z}  \\ n\leq 0}} c^-(n) \beta_{1-k}(n, y) q^n
\end{eqnarray}
where $q\equiv e^{2\pi i\tau}$ and
\begin{eqnarray}
\beta_{1-k}(n,y) = \begin{cases}
\Gamma(1-k, -4\pi n y) \,,~~~ & n\neq 0 \\
y^{1-k} \,,~~~ & n = 0, k\neq 1 \\
\log y \,,~~~ & n = 0, k = 1
\end{cases}
\end{eqnarray}
Here $\Gamma(s,z)$ is the incomplete gamma function
\begin{eqnarray}
\Gamma(s,z) = \int_z^{+\infty} e^{-t} t^{s-1}\,.
\end{eqnarray}
Its analytical expression for integer $s$ is given in Appendix~\ref{app:Polyharmonic_Maass_forms}. It has been shown that the polyharmonic Maa{\ss} forms of integer weight $k$ and level $N$ can be arranged into different irreducible multiplets of $\Gamma'_N$ up to the automorphy factor $(c\tau + d)^{k}$~\cite{Qu:2024rns}. As a consequence, one can always find a basis such that the polyharmonic Maa{\ss} form multiplet $Y^{(k)}_{\bm{r}}(\tau)$ transforms in the irreducible unitary representation $\rho_{\bm{r}}$ of the finite modular group $\Gamma'_N$
\begin{eqnarray}
Y^{(k)}_{\bm{r}}(\gamma \tau) = (c\tau + d)^k \rho_{\bm{r}}(\gamma) Y^{(k)}_{\bm{r}}(\tau)\,,~~\gamma\in \Gamma\,.
\end{eqnarray}
Notice that $\rho_{\bm{r}}(\gamma)=1$ for $\gamma\in \Gamma(N)$, thus we have $\rho_{\bm{r}}(\gamma)=\rho_{\bm{r}}(\gamma\Gamma(N))$ and the finite set of matrices $\rho_{\bm{r}}(\gamma)$ form an irreducible representation of the quotient group $\Gamma'_N\equiv\Gamma / \Gamma(N)$.

The explicit form of the polyharmonic Maa{\ss} form multiplet $Y^{(k)}_{\bm{r}}(\tau)$ for $k\leq0$ can be determined from the known modular forms of level $N$ by using the differential operators $D^{1-k}$ and $\xi_k$~\cite{Qu:2024rns}. For any non-positive integer weight $k\leq 0$,
\begin{eqnarray}\label{eq:Bol-operator}
D^{1-k} = \left( \dfrac{1}{2\pi i} \dfrac{\partial}{\partial \tau} \right)^{1-k}\,.
\end{eqnarray}
The operator $D^{1-k}$ annihilates the non-holomorphic parts of the polyharmonic Maa{\ss} form, and it maps a weight $k$ polyharmonic Maa{\ss} form to a weight $2-k$ holomorphic modular form. This operator can be viewed as an iterated raising operator which raises the modular weight by $2$ each time. Another operator $\xi_k$ is given by
\begin{eqnarray}\label{eq:xi-operator}
\xi_k = 2i y^k \overline{\dfrac{\partial}{\partial \taubar}} \,.
\end{eqnarray}
which is well defined at any integer modular weight $k$. It turns out that $\xi_kY^{(k)}_{\bm{r}}(\tau)$ is a weight $2-k$ and level $N$ modular form multiplet in the representation $\bm{r}^{*}$. Obviously the action of $\xi_k$ on holomorphic modular form equals to zero. For the weight $k=1$, we have $D^{1-k}=1$ which is trivial, and $\xi_1Y^{(1)}_{\bm{r}}(\tau)$ is a weight $1$ and level $N$ modular form multiplet in the representation $\bm{r}^{*}$. However, it is insufficient to fix the expression of $Y^{(1)}_{\bm{r}}(\tau)$ from the action of $\xi_1Y^{(1)}_{\bm{r}}(\tau)$.

We shall briefly review the framework of the non-holomorphic modular invariant theory, in which the Yukawa couplings are generic integer weight polyharmonic Maa{\ss} forms. Considering the standard model gauge symmetry $SU(3)_C\times SU(2)_L \times U(1)_Y$ without low energy supersymmetry, the modular invariant Lagrangian for the Yukawa interactions can be written as
\begin{eqnarray}
\mathcal{L}^Y = Y^{(k_Y)}(\tau) \psi^c \psi H + \mathrm{h.c.} \,,
\end{eqnarray}
where $\psi$ and $\psi^c$ are the two-component spinors for matter fields which carry the integer modular weights $-k_{\psi}$ and $-k_{\psi^c}$ respectively. They transform under the modular group as
\begin{eqnarray}
\nonumber && \psi(x) \rightarrow (c\tau + d)^{-k_{\psi}} \rho_{\psi}(\gamma) \psi(x) \,,\\
&& \psi^c(x) \rightarrow (c\tau + d)^{-k_{\psi^c}} \rho_{\psi^c}(\gamma) \psi^c(x) \,,
\end{eqnarray}
where $\rho_{\psi}$ and $\rho_{\psi^c}$ are the irreducible representations of the finite modular group $\Gamma'_N$. Meanwhile, $H$ refers to the Higgs field whose modular weight is an integer $-k_H$.
The modular transformation of the Higgs field is given by
\begin{eqnarray}
H(x) \rightarrow (c\tau + d)^{-k_H} \rho_H(\gamma) H(x)\,,
\end{eqnarray}
where $\rho_H$ is a one-dimensional representation of $\Gamma'_N$, since there is only one Higgs field in SM. The modular invariance of $\mathcal{L}^Y$ requires the $Y^{(k_Y)}(\tau)$ to be a weight $k_Y$ polyharmonic Maa{\ss} form multiplet transforming in the representation $\rho_Y$ of $\Gamma'_N$ as follow
\begin{eqnarray}
Y^{(k_Y)}(\gamma \tau) = (c\tau + d)^{k_Y} \rho_Y(\gamma) Y^{(k_Y)}(\tau)\,.
\end{eqnarray}
The following condition should be fulfilled
\begin{eqnarray}
k_Y = k_{\psi} + k_{\psi^c} + k_{H}\,,~~~ \rho_{Y}\otimes\rho_{\psi^c}\otimes\rho_{\psi}\otimes\rho_H \ni \bm{1}\,,
\end{eqnarray}
where $\bm{1}$ refers to the trivial singlet representation of $\Gamma'_N$. We assume a minimal form of the kinetic terms of the matter fields, after modular symmetry breaking, it is of the following form,
\begin{eqnarray}
\mathcal{L}_K&=&\langle-i\tau+i\bar{\tau}\rangle^{-k_{\psi}}\,i\,\psi^{\dagger} \, \overline{\sigma}^{\mu}\partial_{\mu}\psi+\langle-i\tau+i\bar{\tau}\rangle^{-k_{\psi^c}}\,i\,\psi^{c\dagger} \, \overline{\sigma}^{\mu}\partial_{\mu}\psi^c\,.
\end{eqnarray}
 However, modular invariance allows for additional terms with
additional parameters in the kinetic terms in the bottom-up approach~\cite{Qu:2024rns}. As shown in the holomorphic modular flavor symmetry based on supersymmetry, these additional terms would reduce the predictive power of modular invariance and their impact on the mixing parameters is expected to be signiﬁcant~\cite{Chen:2019ewa}. The kinetic terms could be under control in the eclectic flavor groups
which are nontrivial products of modular and traditional flavor symmetries~\cite{Nilles:2020nnc,Nilles:2020kgo}, while the non-holomorphic version of eclectic flavor groups is still unclear.

The generalized CP (gCP) symmetry can be consistently combined with modular symmetry, and the modulus $\tau$ should transform under the action CP as~\cite{Dent:2001cc,Dent:2001mn,Baur:2019kwi,Baur:2019iai,Novichkov:2019sqv}:
\begin{eqnarray}
\tau \stackrel{\mathcal{CP}}{\longmapsto}- \tau^*\,.
\end{eqnarray}
The gCP symmetry acts on the field multiplet $\varphi$ as
\begin{eqnarray}
\varphi \stackrel{\mathcal{CP}}{\longmapsto} X_{\bm{r}} \varphi^*\,,
\end{eqnarray}
where the CP transformation matrix $X_{\bm{r}}$ is a unitary matrix acting on flavor space. The consistency condition between gCP symmetry and modular symmetry reads as follow~\cite{Novichkov:2019sqv}
\begin{eqnarray}
\label{eq:consistency-cond}X_{\bm{r}}\rho^{*}_{\bm{r}}(S)X^{-1}_{\bm{r}}=\rho^{-1}_{\bm{r}}(S),\qquad X_{\bm{r}}\rho^{*}_{\bm{r}}(T)X^{-1}_{\bm{r}}=\rho^{-1}_{\bm{r}}(T)\,.
\end{eqnarray}
Under the action of gCP, the polyharmonic Maa{\ss} forms transform in a similar way~\cite{Novichkov:2019sqv,Ding:2021iqp},
\begin{eqnarray}
Y^{(k)}_{\bm{r}}(\tau)\stackrel{\mathcal{CP}}{\longmapsto} Y^{(k)}_{\bm{r}}(-\tau^{*})=X_{\bm{r}}Y^{(k)*}_{\bm{r}}(\tau)\,.
\end{eqnarray}
Given the representation matrices $\rho_{\bm{r}}(S)$ and $\rho_{\bm{r}}(T)$, one can fix the CP transformation $X_{\bm{r}}$ from the consistency condition of Eq.~\eqref{eq:consistency-cond}. If working in the basis where both $S$ and $T$ are represented by symmetric and unitary matrices in all irreducible representations, one can find that Eq.~\eqref{eq:consistency-cond} is fulfilled by $X_{\bm{r}}=\mathbb{1}_{\bm{r}}$, and thus gCP symmetry reduces to the traditional CP. In such basis, all the Fourier coefficients $c^+(n)$ and $c^-(n)$ in Eq.~\eqref{eq:Fourier-exp} would be real. If all the Clebsch-Gordan coefficients of $\Gamma'_N$ are real as well in the symmetric basis, the gCP symmetry would constrain the coupling constants in the modular invariant Lagrangian to be real. As a consequence, the VEV of the complex modulus $\tau$ is the sole source breaking modular symmetry and gCP symmetry.

The even weight polyharmonic Maa{\ss} forms at levels $N=2, 3, 4, 5$ has been fixed by considering the differential operators $D^{1-k}$ and $\xi_k$~\cite{Qu:2024rns}, and they has been used to construct modular invariant flavor models for quarks and leptons in the framework of non-holomorphic modular flavor symmetry~\cite{Qu:2024rns,Okada:2025jjo,Kumar:2024uxn,Nomura:2024ghc,Nomura:2024atp,Nomura:2024ctl,Nomura:2024vus,Kobayashi:2025hnc,Loualidi:2025tgw,Ding:2024inn,Li:2024svh }. In this work, we shall study the odd weight polyharmonic Maa{\ss} form and its application to flavor puzzle.

\section{Integer weight polyharmonic Maa{\ss} forms\label{sec:integer_weight}}

In previous work, the differential operators $D^{1-k}$ and $\xi_k$ are used to determine the non-positive even integer weight $k$ polyharmonic Maa{\ss} form of level $N$~\cite{Qu:2024rns}. This approach can be easily extended to  negative odd integer weights~\footnote{Notice that there is no odd weight polyharmonic Maa{\ss} form for levels $N=1, 2$}. However, it fails at positive modular weights, due to the absence of the $D^{1-k}$ operator. For the weight $k\geq3$, there is no non-holomorphic polyharmonic Maa{\ss} form due to the absence of non-zero negative weight modular form. Consequently polyharmonic Maa{\ss} forms coincide with modular forms for $k\geq3$. At weight $k=2$, besides the holomorphic modular forms, the modified Eisenstein series $\widehat{E}_2(\tau)$ is the unique non-holomorphic polyharmonic Maa{\ss} form satisfying $\xi_2\widehat{E}_2(\tau)=\frac{3}{\pi}$ and $\Delta_2 \widehat{E}_2(\tau)=0$. The expression of $\widehat{E}_2(\tau)$ is
\begin{equation}
\label{eq:E2hat}\widehat{E}_2(\tau)=1-\frac{3}{\pi y}-24\sum^{\infty}_{n=1}\sigma_1(n)q^n=1-\frac{3}{\pi y}-24q-72q^{2}-96q^{3}-168q^{4}-144q^{5} -\ldots\,,
\end{equation}
where $\sigma_1(n)=\sum_{d|n}d$ is the sum of the divisors of $n$. At weight $k=1$, there exists non-holomorphic polyharmonic Maa{\ss} forms apart from the holomorphic ones, nevertheless they can not be determined by only considering the differential operators $D^{1-k}$ and $\xi_k$.

In this section we shall construct the polyharmonic Maa{\ss} forms in a uniform manner, and mainly focus on the weights $k\leq1$. The integer weight polyharmonic Maa{\ss} forms can be generated by the non-holomorphic Eisentein series~\cite{Lagarias2015poly}. For instance, the non-holomorphic Eisenstein series of weight $k$ for ${\rm SL}(2,\mathbb{Z})$ is defined as~\cite{cohen2017modular},
\begin{eqnarray}\label{eq:non_holo_Eisenstein_SL2Z}
E_k(\tau; s) = \sum_{\gcd(c,d)=1} \dfrac{y^s}{(c\tau + d)^k |c\tau + d|^{2s}}  \,,
\end{eqnarray}
where $\gcd(c, d)$ denotes the greatest common divisor of $c$ and $d$. This series can be expressed as a Fourier expansion, specially the $q$-expansion formula. Obviously one sees $E_k(\tau; s)$ is vanishing for odd $k$.
Similarly one can define non-holomorphic Eisenstein series for the principal congruence subgroup $\Gamma(N)$, actually they can be defined for each cusp of $\Gamma(N)$. A cusp is a representative of the equivalence class of the projective line over the rational numbers $\mathbb{Q}\cup \{i\infty\}$ under the action of $\Gamma(N)$. We can add an overline on the rational number or $i\infty$ as the representative of equivalence class under $\Gamma(N)$. Moreover explicitly, we can choose two coprime integers $A$ and $C$ to represent a cusp $\overline{A/C}$.
Two pairs of coprime numbers $(A,C)$ and $(A',C')$ describe the same cusp of $\Gamma(N)$ if and only if $(A,C)\equiv \pm (A',C')\,({\rm mod}\, N)$~\cite{cohen2017modular}. In other words, if and only if there exists $\varepsilon= \pm 1$ such that $A\equiv \varepsilon A'~~({\rm mod}~N)$ and $C\equiv \varepsilon C'~~({\rm mod}~N)$. We take $\overline{1/0}$ to represent the cusp $\overline{i\infty}$ and it is equivalent to the cusp $\overline{1/N}$ for level $N$. All the representatives of cusp of $\Gamma(N)$ would be denoted as $\mathcal{C}(N)$ in the following. The explicit expression of $\mathcal{C}(N)$ for $N\leq7$ is collected in Appendix~\ref{app:cusp}. The non-holomorphic Eisenstein series for each cusp of $\Gamma(N)$ are defined as follow~\cite{diamond2005first}, \begin{eqnarray}\label{eq:non_holo_Eisenstein}
E_k(N; \tau; s; \overline{A/C}) &=& \sum_{\substack{(c,d)\equiv (-C, A) \,({\rm mod}\, N) \\ \gcd(c,d)=1} }  \dfrac{y^s}{(c\tau + d)^k |c\tau + d|^{2s}} \,.
\end{eqnarray}
At the level $N=1$, there is only one cusp $\overline{i\infty}=\overline{1/0}$ for $\Gamma(1)={\rm SL}(2,\mathbb{Z})$. It can be seen that, in this case the level $1$ non-holomorphic Eisenstein series reduces to the series defined in Eq.~\eqref{eq:non_holo_Eisenstein_SL2Z}, i.e.
\begin{eqnarray}
E_k(1; \tau; s; \overline{1/0}) = E_k(\tau; s)\,.
\end{eqnarray}
The non-holomorphic Eisenstein series $E_k(N; \tau; s; \overline{A/C})$ depends slightly on the choice of the representative of the cusp up to an overall sign. For two equivalent descriptions $(A, C)$ and $(A',C')=\varepsilon (A,C)({\rm mod}\, N)$ of the same cusp, we have
\begin{equation}
E_k(N; \tau; s; \overline{A'/C'}) = \varepsilon^k E_k(N; \tau; s; \overline{A/C})\,.
\end{equation}
Under the action of a generic modular symmetry element $\gamma\in {\rm SL}(2,\mathbb{Z})$, one can chcek that the non-holomorphic Eisenstein series
$E_k(N; \tau; s; \overline{A/C})$ transforms as follow,
\begin{eqnarray}\label{eq:Eisenstein_trans}
E_k(N; \gamma \tau; s; \overline{A/C}) = (c\tau+d)^k E_k (N; \tau; s; \overline{\gamma^{-1}(A/C)} ) \,.
\end{eqnarray}
Hence $E_k(N; \tau; s; \overline{A/C})$ is invariant under $\Gamma(N)$ up to the automorphy factor $(c\tau+d)^k$, while it is mapped to another non-holomorphic Eisenstein series at the cusp $\overline{\gamma^{-1}(A/C)}$ under the quotient group $\Gamma/\Gamma(N)\cong \Gamma'_N$. As a result, all the non-holomorphic Eisenstein series of weight $k$ and level $N$ can be arranged into irreducible multiplets of the finite modular group $\Gamma'_N$.

The non-holomorphic Eisenstein series converges uniformly on any compact area of upper half plane $\mathcal{H}$ when $\text{Re}(s)>(2-k)/2$. It has an analytic continuation to the whole complex plane $s\in \mathbb{C}$. Furthermore, one can check that the non-holomorphic Eisenstein series of level $N$ is an eigenfunction of the Laplacian operator with eigenvalue $\lambda= - s(s+k-1)$~\cite{Lagarias2015poly}:
\begin{eqnarray}
\label{eq:Laplacian-cond}
\Delta_k E_k(N; \tau; s; \overline{A/C}) = - s(s+k-1) E_k(N; \tau; s; \overline{A/C})\,.
\end{eqnarray}
Hence the harmonic condition in Eq.~\eqref{eq:harmonoic-condition} is satisfied when $s=0$ or $s=1-k$. It is notable that the two conditions $s=0$ and $s=1-k$ coincide in the case of $k=1$. Thus both $E_k(N; \tau; 0; \overline{A/C})$ and $E_k(N; \tau; 1-k; \overline{A/C})$ are polyharmonic Maa{\ss} forms of level $N$. However, they are not linearly independent. For the lowest level $N=1$, $i\infty\equiv \overline{1/0}$ is the only cusp of the full modular group $\Gamma\cong \Gamma(1)$. The corresponding non-holomorphic Eisenstein series satisfies the following identity~\cite{andersen2017shifted}
\begin{eqnarray}
E_k(1; \tau; s; \overline{1/0})= \pi^{2s-1} \dfrac{\Gamma(1-s-\frac{k}{2}+ \frac{|k|}{2})}{\Gamma(s+\frac{k}{2}+ \frac{|k|}{2})} \dfrac{\zeta(2-k-2s)}{\zeta(k+2s)}  E_k(1; \tau; 1-k-s; \overline{1/0})\,.
\end{eqnarray}
Therefore the two polyharmonic Maa{\ss} forms $E_k(1; \tau; 0; \overline{1/0})$ and $E_k(1; \tau; 1-k; \overline{1/0})$ are proportional and they differ only by an overall constant. The situation becomes more complex for higher levels $N>1$. The Eisenstein series $E_k(N; \tau; 0; \overline{A/C})$ and $E_k(N; \tau; 1-k; \overline{A/C})$ at the same cusp $\overline{A/C}$ are not proportional to each other. In fact, $E_k(N; \tau; 0; \overline{A/C})$ can be expressed as a linear combination of the Eisenstein series $\left\{E_k(N; \tau; 1-k; \overline{A/C})\Big|\overline{A/C}\in \mathcal{C}(N)\right\}$, this has been explicitly checked at each weight $k$ for the concerned levels $N=3, 4, 5$. As a consequence, the Eisenstein series $\left\{E_k(N; \tau; 0; \overline{A/C})\Big|\overline{A/C}\in \mathcal{C}(N)\right\}$ and $\left\{E_k(N; \tau; 1-k; \overline{A/C})\Big|\overline{A/C}\in \mathcal{C}(N)\right\}$ span the same linear space. Without loss of generality, we shall be concerned with the non-holomorphic Eisenstein series $E_k(N; \tau; 1-k; \overline{A/C})$ which are exactly the polyharmonic Maa{\ss} forms for the weights $k\leq 0$, where $\overline{A/C}$ ranges over all cusps of $\Gamma(N)$, up to equivalence. This set of non-holomorphic Eisenstein series can be arranged into multiplets of $\Gamma'_N$. We find that the $q$-expansion of the even weight polyharmonic Maa{\ss} forms of level $N$ coincide with those in Ref.~\cite{Qu:2024rns}.

In the following, we proceed to consider the polyharmonic Maa{\ss} forms of weight $k=1$ which needs special treatment. By taking derivative with respect to $s$ on both sides of Eq.~\eqref{eq:Laplacian-cond}, we obtain
\begin{small}
\begin{eqnarray}
\Delta_k \dfrac{\partial}{\partial s} E_k(N; \tau; s; \overline{A/C})  &=& - (2s+k-1) E_k(N; \tau; s; \overline{A/C}) - s(s+k-1) \dfrac{\partial}{\partial s} E_k(N; \tau; s; \overline{A/C})\,.~~~~
\end{eqnarray}
\end{small}
One can see that the derivative $\left. \dfrac{\partial}{\partial s} E_k(N; \tau; s; \overline{A/C}) \right|_{s=0\;\text{or}\; s=1-k}$ satisfies the harmonic condition $\left. \Delta_k \dfrac{\partial}{\partial s} E_k(N; \tau; s; \overline{A/C})\right|_{s=0\;\text{or}\; s=1-k}=0$ only for $k=1$. Obviously the derivative $\left. \dfrac{\partial}{\partial s} E_k(N; \tau; s; \overline{A/C}) \right|_{s=0\;\text{or}\; s=1-k}$ also fulfills the modularity condition of Eq.~\eqref{eq:Eisenstein_trans}.
Hence non-holomorphic Eisenstein series $E_k(N; \tau; s; \overline{A/C})$ with unit weight $k=1$ is special, both $E_1(N; \tau; 0; \overline{A/C})$  and $\left. \dfrac{\partial}{\partial s} E_1(N; \tau; s; \overline{A/C}) \right|_{s=0}\equiv E^{(1)}_1(N; \tau; \overline{A/C})$ are polyharmonic Maa{\ss} forms of weight $k=1$ and level $N$~\cite{kudla1999derivative,yangsecond,yang2003taylor,duke2015harmonic}.
Moreover, we find that $E_1(N; \tau; 0; \overline{A/C})$ is a holomorphic
function of $\tau$, and it is related to $E^{(1)}_1(N; \tau; \overline{A/C})$ through the operator $\xi_1$, i.e.
\begin{eqnarray}
\xi_1 E^{(1)}_1(N; \tau; \overline{A/C}) = E_1(N; \tau; 0; \overline{A/C})\,.
\end{eqnarray}
Thus the Eisenstein series $\left\{E_1(N; \tau; 0; \overline{A/C})\Big|\overline{A/C}\in\mathcal{C}(N)\right\}$ span the linear space of holomorphic modular forms of weight $1$ at level $N$ while this basis is not linear independent. The first order derivative on $s$ of the non-holomorphic Eisenstein series $\left\{E_1^{(1)}(N; \tau; \overline{A/C})\Big|\overline{A/C}\in\mathcal{C}(N)\right\}$ span the full space of the weight 1 polyharmonic Maa{\ss} forms including the non-holomorphic ones. Notice that the Eisenstein series $E_k(N; \tau; 1-k; \overline{A/C})$ for $k<0$ converges uniformly on any compact area of upper half plane $\mathcal{H}$. However, we need to perform an analytic continuation for both $E_1(N; \tau; 0; \overline{A/C})$ and $E_1^{(1)}(N; \tau; \overline{A/C})$ to get their analytical expressions. At the weight $k=2$, the polyharmonic Maa{\ss} forms of level $N$ comprise the weight 2 holomorphic modular forms and the modified Eisenstein series $\widehat{E}_2(\tau)$ which is invariant under $\Gamma'_N$. At higher weights $k\geq3$, the polyharmonic Maa{\ss} forms coincides with the modular forms which are well known in the literature~\cite{Ding:2023htn}.

Although the non-holomorphic Eisenstein series is not intuitive and is difficult to calculate, it offers some advantages over differential operators. By using the Eisenstein series, we can directly determine the polyharmonic Maa{\ss} forms multiplets. This approach eliminates the need to distinguish whether a modular form can be lifted to a polyharmonic Maaß form, as required when using differential operators. We do not need to know explicit expression of the modular forms in advance. The analytic expression of the polyharmonic Maa{\ss} forms can be obtained at any negative weight, if needed, without calculating them order by order.

\subsection{Constructing integral weight polyharmonic Maa{\ss} forms of level $N=3$}

The polyharmonic Maa{\ss} forms coincide with the holomorphic modular forms for weight $k\geq 3$, and the polyharmonic Maa{\ss} forms of weight $k=2$ extend the modular form by including the modified Eisenstein series $\widehat{E}_2(\tau)$~\cite{Qu:2024rns}. All the holomorphic modular forms of weight $k\geq2$
can be constructed from the tensor products of weight one modular form. At level $N=3$, there are only two linearly independent weight $1$ modular forms which can be expressed in terms of Dedekind eta function $\eta(\tau)$, and they can be arranged into a doublet $\bm{\widehat{2}}$ of $\Gamma'_3\cong T'$ with the following $q$-expansion~\cite{Liu:2019khw,Lu:2019vgm,Ding:2022aoe},
\begin{eqnarray}\label{eq:wt1_Y2h}
Y_{\bm{\widehat{2}}}^{(1)} &=& \begin{pmatrix}
Y_1 \\ Y_2
\end{pmatrix} = \begin{pmatrix}
- 3 \sqrt{2} q^{1/3}\left( 1 + q + 2 q^2 + 2 q^4 + q^5 + 2 q^6 + q^8 + 2 q^9 \cdots \right)  \\
1 + 6 ( q + q^2 + q^4 + 2 q^7 + q^9 + q^{12} + 2 q^{13} )+\ldots
\end{pmatrix}\,.
\end{eqnarray}
In order to be self-contained,  we collect the level 3 polyharmonic Maa{\ss} forms of weight $k=2,3,4,5,6$ in Appendix~\ref{app:Polyharmonic_N_3}.

The non-trivial polyharmonic Maa{\ss} forms exists at weight $k\leq 1$. In the following we shall construct these non-trivial polyharmonic Maa{\ss} forms of level $N=3$ from the non-holomorphic Eisenstein series explicitly.
The representative of the cusps of $\Gamma(3)$ is given by $\{i\infty, 0, 1, 2 \}$. These cusps can be represented as the following comprime number pairs $\{\overline{1/0}, \overline{0/1}, \overline{1/1}, \overline{2/1}\}$, respectively. Therefore we have four different non-holomorphic Eisenstein series of level $3$ at each weight $k\leq0$, they are $E_k(\tau; s; i\infty)$, $E_k(\tau; s; 0)$, $E_k(\tau; s; 1)$ and $E_k(\tau; s; 2)$. Using Eq.~\eqref{eq:Eisenstein_trans}, one can obtain the modular transformations of the Eisenstein series under the generators $S$ and $T$ as follows,
\begin{eqnarray}
\nonumber E_k(S \tau; s; \overline{A/C}) &=& ( - \tau )^k E_k (\tau; s; \overline{(-C)/A} )\,,  \\
\label{eq:modular-trans-S-T}E_k(T \tau; s; \overline{A/C}) &=& E_k (\tau; s; \overline{(A-C)/C} )\,.
\end{eqnarray}
Thus these non-holomorphic Eisenstein series can be decomposed into irreducible multiplets of $\Gamma'_3\cong T'$. For simplicity, we drop the argument $N$ in the non-holomorphic Eisenstein series $E_k(N; \tau; s; \overline{A/C})$ herein and after. Considering the action of $S^2$, we have
\begin{eqnarray}
\nonumber E_k(S^2 \tau; s; \overline{A/C}) &=& (-1)^k E_k(\tau; s; \overline{-A/-C}) = E_k(\tau; s; \overline{A/C})=(-1)^k\rho(S^2)E_k(\tau; s; \overline{A/C})\,,
\end{eqnarray}
which implies $(-1)^k\rho(S^2)=1$. Hence the odd weight Eisenstein series can only be in the hatted representation of $\Gamma'_N$ with $\rho(S^2)=-1$ while the even weight Eisenstein series are in the unhatted representation of $\Gamma'_N$ with $\rho(S^2)=1$.
To be more explicit, under the action of the generator $T$, the non-holomorphic Eisenstein series of level $3$ for each cusp transform as
\begin{eqnarray}
\nonumber E_k(\tau; s; i\infty) &\stackrel{T}{\longmapsto}& E_k(\tau; s; i\infty) \,, \\
\nonumber E_k(\tau; s; 0) &\stackrel{T}{\longmapsto}& E_k(\tau; s; 2) \,, \\
\nonumber E_k(\tau; s; 1) &\stackrel{T}{\longmapsto}& E_k(\tau; s; 0)\,,  \\
E_k(\tau; s; 2) &\stackrel{T}{\longmapsto}& E_k(\tau; s; 1)\,.
\end{eqnarray}
Similarly, the modular transformation of the level 3 Eisenstein series under another generator $S$ are given by
\begin{eqnarray}
\nonumber  E_k(S\tau; s; i\infty) &\stackrel{S}{\longmapsto}& (-\tau)^{k} E_k(\tau; s; 0) \,, \\
\nonumber E_k(S\tau; s; 0) &\stackrel{S}{\longmapsto}& (-1)^k (-\tau)^{k} E_k(\tau; s; i\infty) \,, \\
\nonumber E_k(S\tau; s; 1) &\stackrel{S}{\longmapsto}& (-\tau)^{k} E_k(\tau; s; 2) \,, \\
E_k(S\tau; s; 2) &\stackrel{S}{\longmapsto}& (-1)^k (-\tau)^{k} E_k(\tau; s; 1)\,.
\end{eqnarray}
We observe that the even weight Eisenstein series of level 3 can be organzied into a singlet $\bm{1}$ and a triplet $\bm{3}$ representations $T'$. The harmonic condition can be satisfied when $s=1-k$, thus the weight $2k$ polyharmonic Maa{\ss} forms singlet $Y^{(2k)}_{\bm{1}}(\tau)$ and triplet $Y^{(2k)}_{\bm{3}}(\tau)=(Y^{(2k)}_{\bm{3},1}, Y^{(2k)}_{\bm{3},2}, Y^{(2k)}_{\bm{3},3})^T$ take the following form,
\begin{eqnarray}
\nonumber Y^{(2k)}_{\bm{1}}(\tau) &=& E_{2k}(\tau; 1-2k; i\infty) + E_{2k}(\tau; 1-2k; 0) + E_{2k}(\tau; 1-2k; 1) + E_{2k}(\tau; 1-2k; 2) \,, \\
\nonumber Y^{(2k)}_{\bm{3},1}(\tau) &=& E_{2k}(\tau; 1-2k; i\infty) - \dfrac{1}{3} \Big( E_{2k}(\tau; 1-2k; 0) + E_{2k}(\tau; 1-2k; 1) + E_{2k}(\tau; 1-2k; 2) \Big) \,, \\
\nonumber Y^{(2k)}_{\bm{3},2}(\tau) &=& \dfrac{2}{3} \Big( E_{2k}(\tau; 1-2k; 0) + \omega E_{2k}(\tau; 1-2k; 1) + \omega^2 E_{2k}(\tau; 1-2k; 2) \Big)\,, \\
\label{eq:even_dec} Y^{(2k)}_{\bm{3},3}(\tau) &=& \dfrac{2}{3} \Big( E_{2k}(\tau; 1-2k; 0) + \omega^2 E_{2k}(\tau; 1-2k; 1) + \omega E_{2k}(\tau; 1-2k; 2) \Big) \,.
\end{eqnarray}
Moreover, the weight $2k+1$  polyharmonic Maa{\ss} forms are linear combinations of the non-holomorphic Eisenstein series $E_{2k+1}(\tau; -2k; i\infty)$, $E_{2k+1}(\tau; -2k; 0)$, $E_{2k+1}(\tau; -2k; 1)$, $E_{2k+1}(\tau; -2k; 2)$ for $k<0$, they can be arranged into two doublets $Y^{(2k+1)}_{\bm{\widehat{2}}}(\tau)=(Y^{(2k+1)}_{\bm{\widehat{2}},1}, Y^{(2k+1)}_{\bm{\widehat{2}},2})^T$ and $Y^{(2k+1)}_{\bm{\widehat{2}''}}(\tau)=(Y^{(2k+1)}_{\bm{\widehat{2}''},1}, Y^{(2k+1)}_{\bm{\widehat{2}''},2})^T$ in the representation $\bm{\widehat{2}}$ and $\bm{\widehat{2}''}$ of $T'$ as follow,
\begin{eqnarray}
\nonumber Y^{(2k+1)}_{\bm{\widehat{2}},1}(\tau) &=& i\,\sqrt{\dfrac{2}{3}} \Big( E_{2k+1}(\tau; -2k; 0) + \omega E_{2k+1}(\tau; -2k; 1) + \omega^2 E_{2k+1}(\tau; -2k; 2) \Big) \,, \\
\nonumber Y^{(2k+1)}_{\bm{\widehat{2}},2}(\tau) &=& E_{2k+1}(\tau; -2k; i\infty) - \dfrac{i}{\sqrt{3}} \Big( E_{2k+1}(\tau; -2k; 0) + E_{2k+1}(\tau; -2k; 1) + E_{2k+1}(\tau; -2k; 2) \Big) \,, \\
\nonumber Y^{(2k+1)}_{\bm{\widehat{2}''},1}(\tau) &=& E_{2k+1}(\tau; -2k; i\infty) + \dfrac{i}{\sqrt{3}} \Big( E_{2k+1}(\tau; -2k; 0) + E_{2k+1}(\tau; -2k; 1) + E_{2k+1}(\tau; -2k; 2) \Big) \,, \\
\label{eq:odd_dec} Y^{(2k+1)}_{\bm{\widehat{2}''},2}(\tau) &=& i\,\sqrt{\dfrac{2}{3}}  \Big( E_{2k+1}(\tau; -2k; 0) + \omega^2 E_{2k+1}(\tau; -2k; 1) + \omega E_{2k+1}(\tau; -2k; 2) \Big) \,.
\end{eqnarray}
The expression of the $q$-expansion of the normalised negative even weight singlet polyharmonic Maa{\ss} form $Y^{(2k)}_{\bm{1}}(\tau)$ is given by
\begin{eqnarray}
\nonumber Y^{(2k)}_{\bm{1}}(\tau) &=& 1+(-1)^k 2^{-2k}\dfrac{\zeta(2-2k)}{\zeta(1-2k)}\dfrac{y^{1-2k}}{\pi} + \dfrac{1}{\zeta(1-2k)} \sum_{n=1}^{+\infty} \sigma_{2k-1}(n)q^n \\
&& + \dfrac{1}{(-2k)!\zeta(1-2k)} \sum_{n=1}^{+\infty} \sigma_{2k-1}(n)q^{-n} \Gamma(1-2k, 4\pi n y)  \,,
\end{eqnarray}
where $\sigma_k(n)= \sum_{d|n}d^k$ is the sum of the $k$-th power of the divisors of $n$ and $\zeta(s)$ is the Riemann zeta function.
Similarly, we can obtain the $q$-expansion of triplet polyharmonic Maa{\ss} form $Y^{(2k)}_{\bm{3}}(\tau)$ at level 3 in Eq.~\eqref{eq:even_dec} as follows
\begin{eqnarray}
\nonumber Y^{(2k)}_{\bm{3},1} &=& (-1)^k \dfrac{3 - 3^{1-2k}}{3^{2-2k}-1} \dfrac{2^{2k}\pi}{1-2k} \dfrac{\zeta(1-2k)}{\zeta(2-2k)} + \dfrac{y^{1-2k}}{1-2k}  \\
\nonumber && + \dfrac{(-1)^k\, 2^{2k}\, \pi}{(1-2k)(3-3^{2k-1})\zeta(2-2k)} \sum_{n=1}^{+\infty} \left[ (3+3^{2k})\sigma_{2k-1}(n)-4\sigma_{2k-1}(3n) \right] q^n  \\
\nonumber && + \dfrac{(-1)^k\, 2^{2k}\, \pi}{(1-2k)!(3-3^{2k-1})\zeta(2-2k)} \sum_{n=1}^{+\infty} \left[ (3+3^{2k})\sigma_{2k-1}(n)-4\sigma_{2k-1}(3n) \right] q^{-n}\Gamma(1-2k,4\pi n y)  \,, \\
\nonumber Y^{(2k)}_{\bm{3},2} &=& \dfrac{(-1)^k\, 2^{2k+1}\, \pi}{(1-2k)(3-3^{2k-1})\zeta(2-2k)} q^{1/3}\sum_{n=0}^{+\infty} \sigma_{2k-1}(3n+1) q^n  \\
\nonumber && + \dfrac{(-1)^k\, 2^{2k+1}\, \pi}{(1-2k)!(3-3^{2k-1})\zeta(2-2k)} q^{-2/3}\sum_{n=0}^{+\infty} \sigma_{2k-1}(3n+2) q^{-n}\Gamma\left( 1-2k,\frac{(12 n + 8)\pi y}{3} \right) \,, \\
\nonumber Y^{(2k)}_{\bm{3},3} &=& \dfrac{(-1)^k\, 2^{2k+1}\, \pi}{(1-2k)(3-3^{2k-1})\zeta(2-2k)} q^{2/3}\sum_{n=0}^{+\infty} \sigma_{2k-1}(3n+2) q^n  \\
\label{eq:N=3_HMF_even_qexp}&& + \dfrac{(-1)^k\, 2^{2k+1}\, \pi}{(1-2k)!(3-3^{2k-1})\zeta(2-2k)} q^{-1/3}\sum_{n=0}^{+\infty} \sigma_{2k-1}(3n+1) q^{-n}\Gamma\left( 1-2k,\frac{(12 n + 4)\pi y}{3} \right)\,.~~~~~~~~
\end{eqnarray}
Notice that the above expressions also apply to  weight zero poyharmonic Maa{\ss} forms of level 3 by taking the limit $k\rightarrow0$. The Laurent expansion of the Riemann zeta function $\zeta(s)$ near $s=1$ is
\begin{equation}
\label{eq:zeta_series}\zeta(s) = \frac{1}{s-1} + \gamma_E + \cdots\,,
\end{equation}
where $\gamma_E\simeq 0.577$ is the Euler-Mascheroni constant, thus we have $\displaystyle\lim_{k\rightarrow0}\zeta(1-2k)=\infty$ and the singlet polyharmonic Maa{\ss} form $Y^{(0)}_{\bm{1}}(\tau)$ of weight $0$ and level 3 is simply a constant with
\begin{equation}
Y^{(0)}_{\bm{1}} =1\,.
\end{equation}
For triplet polyharmonic Maa{\ss} form $Y^{(0)}_{\bm{3}}(\tau)$, because $\displaystyle\lim_{k\rightarrow0}(3-3^{1-2k})\zeta(1-2k) = - 3\log3$, we find that its explicit $q$-expansion is given by
\begin{eqnarray}
\nonumber Y^{(0)}_{\bm{3},1}(\tau) &=& - \dfrac{9\log 3}{4\pi} + y + \dfrac{9}{\pi} \sum_{n=1}^{+\infty} \Big(\sigma_{-1}(n) - \sigma_{-1}(3n) \Big) \left( q^n + \overline{q}^n \right) \\
\nonumber &=& - \dfrac{9\log 3}{4\pi} + y - \dfrac{3 }{\pi}\left( q + \dfrac{3 }{2}q^2+ \dfrac{ 1}{3}q^3 + \dfrac{7 }{4}q^4 + \dfrac{6 }{5}q^5 + \dfrac{ 1}{2}q^6+ \dfrac{8 }{7}q^7 + \cdots \right)    \\
\nonumber && - \dfrac{3 }{\pi}\left( \overline{q} + \dfrac{3}{2}\overline{q}^2+ \dfrac{1}{3}\overline{q}^3 + \dfrac{7}{4}\overline{q}^4 + \dfrac{6}{5}\overline{q}^5 + \dfrac{1}{2}\overline{q}^6+ \dfrac{8}{7}\overline{q}^7 + \cdots \right)  \,, \\
\nonumber Y^{(0)}_{\bm{3},2}(\tau) &=& \dfrac{9}{2\pi} \left( q^{1/3}\sum_{n=0}^{+\infty} \sigma_{-1}(3n+1) q^n + \overline{q}^{2/3}\sum_{n=0}^{+\infty} \sigma_{-1}(3n+2) \overline{q}^n \right)   \\
\nonumber &=& \dfrac{9}{2\pi} q^{1/3}\left( 1 + \dfrac{7}{4}q+ \dfrac{8}{7}q^2+ \dfrac{9 }{5}q^3+ \dfrac{14 }{13}q^4+ \dfrac{31 }{16}q^5+ \dfrac{20}{19}q^6+ \dfrac{18 }{11}q^7 + \cdots \right)  \\
\nonumber && + \dfrac{27 }{4 \pi}\overline{q}^{2/3}\left( 1 + \dfrac{4}{5}\overline{q}+ \dfrac{5}{4}\overline{q}^2+ \dfrac{8}{11}\overline{q}^3+ \dfrac{8}{7}\overline{q}^4+ \dfrac{12}{17}\overline{q}^5+ \dfrac{7}{5}\overline{q}^6+ \dfrac{16}{23}\overline{q}^7 + \cdots \right)  \,, \\
\nonumber Y^{(0)}_{\bm{3},3}(\tau) &=& \dfrac{9}{2\pi} \left( q^{2/3}\sum_{n=0}^{+\infty} \sigma_{-1}(3n+2) q^n + \overline{q}^{1/3}\sum_{n=0}^{+\infty} \sigma_{-1}(3n+1) \overline{q}^n \right)  \\
\nonumber &=& \dfrac{27 }{4 \pi}q^{2/3}\left( 1 + \dfrac{4}{5} q+ \dfrac{5 }{4}q^2+
\dfrac{8 }{11}q^3+ \dfrac{8 }{7}q^4+ \dfrac{12 }{17}q^5+ \dfrac{7}{5}q^6+ \dfrac{16 }{23}q^7 + \cdots \right) \\
&& + \dfrac{9 }{2 \pi}\overline{q}^{1/3}\left( 1 + \dfrac{7 }{4}\overline{q}+ \dfrac{8 }{7}\overline{q}^2+ \dfrac{9 }{5}\overline{q}^3+ \dfrac{14 }{13}\overline{q}^4+ \dfrac{31 }{16}\overline{q}^5+ \dfrac{20 }{19}\overline{q}^6+ \dfrac{18 }{11}\overline{q}^7 + \cdots \right)\,,
\end{eqnarray}
where $\overline{q} = e^{-2\pi i \taubar}$ represents the complex conjugate of $q$.

In a similar way, the $q$-expansion of the negative odd integer weight polyharmonic Maa{\ss} forms of level 3 in Eq.~\eqref{eq:odd_dec} is determined to be
\begin{eqnarray}
\nonumber Y^{(2k+1)}_{\bm{\widehat{2}},1} &=& -\dfrac{(-1)^k\, 2^{2k+1} \, \sqrt{2}\, \pi }{\sqrt{3}(-2k)\, L(1-2k,\chi_2)} q^{1/3} \sum_{n=0}^{+\infty} \widetilde{\sigma}^{\chi_2}_{2k}(3n+1) q^n  \\
\nonumber && + \dfrac{(-1)^k\, 2^{2k+1} \, \sqrt{2}\, \pi }{\sqrt{3}(-2k)!\, L(1-2k,\chi_2)} q^{-2/3}\sum_{n=0}^{+\infty} \widetilde{\sigma}^{\chi_2}_{2k}(3n+2) q^{-n}\Gamma\left( -2k,  \dfrac{(12n+8)\pi y}{3} \right) \,,  \\
\nonumber Y^{(2k+1)}_{\bm{\widehat{2}},2} &=& (-1)^k \dfrac{2^{2k+1}\,\pi L(-2k,\chi_2)}{\sqrt{3}(-2k)L(1-2k,\chi_2)} - \dfrac{y^{-2k}}{2k} + \dfrac{(-1)^k\, 2^{2k+1} \, \pi }{\sqrt{3}(-2k)\, L(1-2k,\chi_2)}\sum_{n=1}^{+\infty} \Big( \widehat{\sigma}^{\chi_2}_{2k}(n) + 3^{2k} \widetilde{\sigma}^{\chi_2}_{2k}(n) \Big) q^n  \\
\nonumber && + \dfrac{(-1)^k\, 2^{2k+1}\, \pi }{\sqrt{3}(-2k)!\, L(1-2k,\chi_2)} \sum_{n=1}^{+\infty} \Big( \widehat{\sigma}^{\chi_2}_{2k}(n) - 3^{2k} \widetilde{\sigma}^{\chi_2}_{2k}(n) \Big) q^{-n} \Gamma(-2k,4\pi n y)\,, \\
\nonumber Y^{(2k+1)}_{\bm{\widehat{2}''},1} &=& - (-1)^k \dfrac{2^{2k+1}\,\pi L(-2k,\chi_2)}{\sqrt{3}(-2k)L(1-2k,\chi_2)} - \dfrac{y^{-2k}}{2k} - \dfrac{(-1)^k\, 2^{2k+1} \, \pi }{\sqrt{3}(-2k)\, L(1-2k,\chi_2)}\sum_{n=1}^{+\infty} \Big( \widehat{\sigma}^{\chi_2}_{2k}(n) - 3^{2k} \widetilde{\sigma}^{\chi_2}_{2k}(n) \Big) q^n  \\
\nonumber && - \dfrac{(-1)^k\, 2^{2k+1}\, \pi }{\sqrt{3}(-2k)!\, L(1-2k,\chi_2)} \sum_{n=1}^{+\infty} \Big( \widehat{\sigma}^{\chi_2}_{2k}(n) + 3^{2k} \widetilde{\sigma}^{\chi_2}_{2k}(n) \Big) q^{-n} \Gamma(-2k,4\pi n y)\,,  \\
\nonumber Y^{(2k+1)}_{\bm{\widehat{2}''},2} &=& \dfrac{(-1)^k\, 2^{2k+1} \, \sqrt{2}\, \pi }{\sqrt{3}(-2k)\, L(1-2k,\chi_2)} q^{2/3} \sum_{n=0}^{+\infty} \widetilde{\sigma}^{\chi_2}_{2k}(3n+2) q^n  \\
\label{eq:N=3_HMF_odd_qexp} && - \dfrac{(-1)^k\, 2^{2k+1} \, \sqrt{2}\, \pi }{\sqrt{3}(-2k)!\, L(1-2k,\chi_2)} q^{-1/3}\sum_{n=0}^{+\infty} \widetilde{\sigma}^{\chi_2}_{2k}(3n+1) q^{-n}\Gamma\left( -2k, \dfrac{(12n+4)\pi y}{3} \right)\,,
\end{eqnarray}
where $\chi_2$ is Dirichlet character modulo $3$ with
\begin{eqnarray}
\chi_2(n) = \begin{cases}
1 \,, ~~ &n\equiv 1 \,~ ({\rm mod}\, 3) \\
-1 \,, ~~ &n\equiv 2 \,~ ({\rm mod}\, 3) \\
0 \,, ~~ &n\equiv 0 \,~ ({\rm mod}\, 3)
\end{cases}
\end{eqnarray}
and $L(s,\chi_2)=\sum_{n=1}^{+\infty} \chi_2(n)n^{-s}$ is the Dirichlet L-function associated with $\chi_2$, as explained in Appendix~\ref{app:math}. The notations $\widehat{\sigma}^\chi_k(n)$ and $\widetilde{\sigma}^\chi_k(n)$ refer to modified divisor sums involving some Dirichlet character $\chi$, and they are defined as
\begin{eqnarray}\label{eq:modified_divisorSum}
\widehat{\sigma}^{\chi}_k(n) = \sum_{d|n}\chi(d) d^k\,,~~~ \widetilde{\sigma}^{\chi}_k(n) = \sum_{d|n} \chi^* \left( \dfrac{n}{d} \right) d^k  \,,
\end{eqnarray}
where the sum runs over all the positive divisors $d$ of $n$.

As explained in section~\ref{sec:integer_weight}, the weight $1$ polyharmonic Maa{\ss} forms of level $N$ are spanned by the first derivative of the non-holomorphic Eisenstein series with respect to $s$ at $s=0$, i.e. $\left\{E_1^{(1)}(N; \tau; \overline{A/C})\Big|\overline{A/C}\in\mathcal{C}(N)\right\}$. The argument $N$ in $E_1^{(1)}(N; \tau; \overline{A/C})$ would be dropped for simplicity in the following. $E_1^{(1)}(N; \tau; \overline{A/C})$ and $E_k(N; \tau; s; \overline{A/C})$ transform in the same way under modular symmetry, since the modular transformations in Eq.~\eqref{eq:modular-trans-S-T} are kept intact under the derivative with respect to $s$. As a consequence, similar to Eq.~\eqref{eq:odd_dec}, the four polyharmonic Maa{\ss} forms $E_1^{(1)}(\tau; i\infty)$, $E_1^{(1)}(\tau; 0)$, $E_1^{(1)}(\tau; 1)$ and $E_1^{(1)}(\tau; 2)$ can be arranged into two doublets $Y^{(1)}_{\bm{\widehat{2}}}(\tau)=(Y^{(1)}_{\bm{\widehat{2}},1}, Y^{(1)}_{\bm{\widehat{2}},2})^T$
and $Y^{(1)}_{\bm{\widehat{2}''}}(\tau)=(Y^{(1)}_{\bm{\widehat{2}''},1}, Y^{(1)}_{\bm{\widehat{2}''},2})^T$. The multiplet $Y^{(1)}_{\bm{\widehat{2}}}(\tau)$ in the  representation $\bm{\widehat{2}}$ is identical with the weight 1 holomorphic modular form doublet in Eq.~\eqref{eq:wt1_Y2h}. On the other hand, another multiplet $Y^{(1)}_{\bm{\widehat{2}''}}(\tau)$ consists of non-holomorphic polyharmonic Maa{\ss} forms of weight 1 and level 3. Their explicit $q$-expansion is given by
\begin{eqnarray}
\nonumber Y^{(1)}_{\bm{\widehat{2}''},1} &=& a_0 - 6 \left( q \log 3 + 2 q^2 \log 2 + 2 q^3 \log 3 + q^4 \log 3 + 2 q^5 \log 5 + 2 q^6 \log 2 + 2 q^7 \log 3 + \cdots  \right) \\
\nonumber &&\hskip-0.2in  + \log y - 6 \left(\dfrac{\Gamma(0, 4\pi y)}{q} + \dfrac{ \Gamma(0, 12\pi y)}{q^3} + \dfrac{\Gamma(0, 16\pi y)}{q^4} + \dfrac{2 \Gamma(0, 28\pi y)}{q^7} + \dfrac{\Gamma(0, 36\pi y)}{q^9} + \cdots \right) \,, \\
\nonumber Y^{(1)}_{\bm{\widehat{2}''},2} &=& - 6 \sqrt{2}  q^{2/3} \left( \log 2 + q \log 5 + 2 q^2 \log 2 + q^3 \log 11 + 2 q^4 \log 2 + q^5 \log 17 + q^6 \log 5 + \cdots \right) \\
 &&\hskip-0.2in - 3 \sqrt{2}  q^{2/3} \left( \dfrac{\Gamma(0,4\pi y/3)}{q} + \dfrac{\Gamma(0,16\pi y/3)}{q^2} + \dfrac{2\Gamma(0,28\pi y/3)}{q^3} + \dfrac{2\Gamma(0,52\pi y/3)}{q^5}  + \cdots \right)\,,
\end{eqnarray}
where the constant $a_0$ is
\begin{eqnarray}
a_0 = \gamma_E + \log 3 + \log 4\pi - 6 \log \Gamma\left( \dfrac{1}{3} \right) + 6 \log \Gamma\left( \dfrac{2}{3} \right) \simeq 0.1132\,.
\end{eqnarray}
For convenience, we have normalized the above multiplets of polyharmonic Maa{\ss} forms by multiplying a real constant to set the Fourier  coefficient of $y^{1-k}$ being $1/(1-k)$. This normalization procedure doesn't change neither the modular structure nor the gCP transformation of the polyharmonic Maa{\ss} forms. As a consequence, acting the operator $\xi_k$ on the normalized multiplet of polyharmonic Maa{\ss} forms would give a modular form multiplet whose non-zero component is equal to 1 in the limit $y\rightarrow\infty$. Morevoer, we find that the even weight polyharmonic Maa{\ss} forms of level 3 in~\cite{Qu:2024rns} are reproduced exactly, and their explicit expressions are provided in Appendix~\ref{app:Polyharmonic_N_3} for the sake of self-containess.

\begin{table}[t!]
\centering
\begin{tabular}{|c|c|c|c|c|} \hline \hline
Observable & bf $\pm~1\sigma$ & $3\sigma$ range & Observable & bf $\pm~1\sigma$ \\ \hline
$\sin^2\theta^{\ell}_{12}$ $(\text{NO} \;\&\; \text{IO})$ & $0.308^{+0.012}_{-0.011}$ & $[0.275, 0.345]$  &$\theta_{12}^q$ & $0.2265 \pm 0.00113$ \\
$\sin^2\theta^{\ell}_{13}$(NO) & $0.02215^{+0.00056}_{-0.00058}$ & $[0.02030, 0.02388]$ &$\theta_{13}^q$ & $0.0035 \pm 0.0003$  \\
$\sin^2\theta^{\ell}_{13}$(IO) & $0.02231^{+0.00056}_{-0.00056}$ & $[0.02060, 0.02409]$ &$\theta_{23}^q$ & $0.0420 \pm 0.0013$  \\
$\sin^2\theta^{\ell}_{23}$(NO) & $0.470^{+0.017}_{-0.013}$ & $[0.435, 0.585]$ &$\delta_{CP}^q / {}^\circ $ & $69.6 \pm 3.3$ \\
$\sin^2\theta^{\ell}_{23}$(IO) & $0.550^{+0.012}_{-0.015}$ & $[0.440, 0.584]$ &$m_u/m_c$ & $0.00204 \pm 0.00086$ \\
$\delta^{\ell}_{CP}/\pi$(NO) & $1.18^{+0.14}_{-0.23}$ & $[0.69, 2.02]$ &$m_c/m_t$ & $0.00318 \pm 0.00046$ \\
$\delta^{\ell}_{CP}/\pi$(IO) & $1.52^{+0.12}_{-0.14}$ & $[1.12, 1.86]$ &$m_d/m_s$ & $0.0518 \pm 0.0272$ \\
$\Delta m_{21}^2 / 10^{-5}\text{eV}^2$ & $7.49^{+0.19}_{-0.19}$ & $[6.92, 8.05]$ &$m_s/m_b$ & $0.022\pm 0.005$  \\
$\Delta m_{31}^2 / 10^{-3}\text{eV}^2$(NO) & $2.513^{+0.021}_{-0.019}$ & $[2.451, 2.578]$ &$m_t/{\rm GeV}$ & $74_{-3.7}^{+4.0}$ \\
$\Delta m_{32}^2 / 10^{-3}\text{eV}^2$(IO) & $-2.484^{+0.020}_{-0.020}$ & $[-2.547, -2.421]$  & $m_b/{\rm GeV}$ & $1\pm 0.04$   \\ \hline
$m_e/m_\mu $ & $0.004737 $ & $-$ & \multirow{2}{*}{$m_e/{\rm MeV}$} & \multirow{2}{*}{$0.469652$} \\
$m_\mu/m_\tau$ & $0.05882$ & $-$ & &  \\
\hline \hline
\end{tabular}
\caption{\label{tab:lepton-data}The central values and the $1\sigma$ errors of the mass ratios and mixing angles and CP violation phases for both quarks and leptons. The charged lepton mass ratios are taken from~\cite{Xing:2007fb} where the uncertainties are very small. We set the uncertainties of the charged lepton mass ratios to be $0.1\%$ of their central value when scanning the parameter space of our models. We adopt the values of the lepton mixing parameters from NuFIT v6.0 with Super-Kamiokanda atmospheric data for normal ordering (NO) and inverted ordering (IO)~\cite{Esteban:2024eli}. The values of the quark mass ratios and mixing angels as well as CP violation phase are taken from ~\cite{Xing:2007fb,Joshipura:2011nn} }
\end{table}

\section{Promising non-holomorphic models for quarks and leptons with the modular group $\Gamma'_3\cong T'$ \label{sec:models}}

The integer weight (particularly odd weight) polyharmonic Maa{\ss} forms offer a novel avenue for understanding the flavor structure of SM from modular invariance. In the following, we apply the formalism outlined above to construct models of fermion masses and flavor mixing by using the integer weight polyharmonic Maa{\ss} forms of level $N=3$, which can be organized into multiplets of $\Gamma_{3}'\cong T'$, as listed in Appendix~\ref{app:Polyharmonic_N_3}. We are concerned with the predictive models with the minimal number of free real input parameters. Firstly we will give phenomenologically viable benchmark models for leptons and quarks individually, and then we present a benchmark model that can describe the flavor structures of quark and lepton simultaneously.

\subsection{\label{subsec:lepton_model-seesaw} Lepton sector }

In this section, we give a benchmark lepton model under the assumption of Majorana neutrinos, and the neutrino masses are generated via the type-I seesaw mechanism, and only two right-handed neutrinos $N^{c}$ are introduced. The three generations of the left-handed leptons $L$ are assumed to transform as a triplet of $T'$, while three right-handed charged leptons $E^{c}_{1,2,3}$ are $T'$ singlets. The two right-handed neutrinos $N^{c}$ are embedded into a $T'$ doublet, and we assume that the Higgs field $H$ is invariant under $T'$ with vanishing modular weight $k_{H}=0$. Thus the representation and modular-weight assignments of lepton fields under $T'$ are given by,
\begin{eqnarray}
\nonumber && L\sim\bm{3} \,,~~~E^c_{1}\sim\bm{1'}\,,~~~E^c_{2}\sim\bm{1'} \,,~~~E^c_{3}\sim\bm{1}\,,~~~N^c\sim\bm{\widehat{2}}\,,~~~H\sim\bm{1}\,,\\
&&k_{L}= -1 \,,~~ k_{E^c_{1}} = 1 \,,~~ k_{E^c_{2}} =3 \,,~~ k_{E^c_{3}} = 5 \,,~~ k_{N^c} = 2\,,~~k_{H}=0\,. \label{eq:lepton_content}
\end{eqnarray}
In this setting, the modular invariant Lagrangian for the lepton masses can be read off as follows,
\begin{eqnarray}
\nonumber -\mathcal{L}^{Y}_{l} &=& \alpha (E^c_1 L Y^{(0)}_{\bm{3}}H^*)_{\bm{1}}  + \beta (E^c_2 L Y^{(2)}_{\bm{3}}H^*)_{\bm{1}}  + \gamma (E^c_3 L Y^{(4)}_{\bm{3}}H^*)_{\bm{1}}  + \text{h.c.}\,, \\
-\mathcal{L}_\nu &=& g_1 (N^c L H Y^{(1)}_{\bm{\widehat{2}''}})_{\bm{1}}   + g_2 (N^c L H Y^{(1)}_{\bm{\widehat{2}}})_{\bm{1}}  + \dfrac{1}{2} \Lambda \left(N^c N^c Y^{(4)}_{\bm{3}}\right)_{\bm{1}} +\text{h.c.}~\,,
\end{eqnarray}
where the phases of coupling constants $\alpha$, $\beta$, and $\gamma$ can be absorbed by redefinition of the right-handed charged lepton fields $E^{c}_{1,2,3}$, $g_{1}$ and $\Lambda$ can be taken to be real and positive by redefining $L$ and $N^{c}$ respectively. The remaining coupling $g_2$ is generally complex, and it would be constrained to be real if gCP symmetry is imposed on the model. The charged lepton mass matrix $M_{e}$, the neutrino Dirac mass matrix $M_{D}$ and the heavy Majorana neutrino mass matrix $M_{N}$ take the following form:
\begin{eqnarray}
\nonumber M_e &=& \begin{pmatrix}
\alpha Y^{(0)}_{\bm{3},3} ~&~ \alpha Y^{(0)}_{\bm{3},2} ~&~ \alpha Y^{(0)}_{\bm{3},1} \\
\beta Y^{(2)}_{\bm{3},3} ~&~ \beta Y^{(2)}_{\bm{3},2} ~&~ \beta Y^{(2)}_{\bm{3},1} \\
\gamma Y^{(4)}_{\bm{3},1} ~&~ \gamma Y^{(4)}_{\bm{3},3} ~&~ \gamma Y^{(4)}_{\bm{3},2}
\end{pmatrix}v \,,  ~~~
M_N =\Lambda\begin{pmatrix}
 -Y^{(4)}_{\bm{3},2} ~&~ \frac{Y^{(4)}_{\bm{3},3}}{\sqrt{2}} \\
\frac{Y^{(4)}_{\bm{3},3}}{\sqrt{2}} ~&~ Y^{(4)}_{\bm{3},1}
\end{pmatrix}\,, \\
M_D &=& \begin{pmatrix}
g_1 Y^{(1)}_{\mathbf{\widehat{2}''},2} ~&~ \sqrt{2}g_2 Y^{(1)}_{\mathbf{\widehat{2}},1} ~&~ -\sqrt{2}g_1 Y^{(1)}_{\mathbf{\widehat{2}''},1}-g_2 Y^{(1)}_{\mathbf{\widehat{2}},2}  \\
g_1 Y^{(1)}_{\mathbf{\widehat{2}''},1}-\sqrt{2}g_2 Y^{(1)}_{\mathbf{\widehat{2}},2} ~&~ \sqrt{2}g_1 Y^{(1)}_{\mathbf{\widehat{2}''},2} ~&~ -g_2 Y^{(1)}_{\mathbf{\widehat{2}},1}
\end{pmatrix} v\,,
\end{eqnarray}
where $v\equiv \braket{H_{0}}\simeq174$ GeV is the VEV of the SM Higgs field. The effective light neutrino mass matrix can be obtained by seesaw formula,
\begin{equation}
  M_{\nu}=-M_{D}^{T}M_{N}^{-1}M_{D}\,.
\end{equation}
As a consequence, we find that this model describes all the lepton masses and mixing observables in terms of 8 real parameters: an overall scale factor $\alpha v$ and two real constants $\beta/\alpha$, $\gamma/\alpha$ in $M_{e}$, an overall scale factor $g_{1}^{2} v^{2}/\Lambda$ and a complex parameter $g_{2}/g_{1}$ in $M_{\nu}$, in addition to the complex modulus $\tau$. In order to quantitatively assess whether the lepton model can accommodate the experimental data, we perform a $\chi^{2}$ analysis for this model and the $\chi^{2}$ function is defined as
\begin{equation}\label{eq:chisq_lepton}
\chi^2_{\ell}=\sum^{7}_{i=1}\left(\frac{P_i(x)-\mu_i}{\sigma_i}\right)^2\,,
\end{equation}
where the vector $x$ contains the model parameters including $\beta/\alpha$, $\gamma/\alpha$, $g_{2}/g_{1}$, $g_{1}^{2} v^{2}/\Lambda$ and $\tau$, the overall scale $\alpha v$ is fixed by the electron mass $m_{e}$, $P_{i}(x)$ are the model predictions for the observables, $\mu_{i}$ and $\sigma_{i}$ denote the central values and $1\sigma$ standard deviations of the corresponding eight quantities:
\begin{equation}
\sin^{2}\theta^{\ell}_{12},~\sin^{2}\theta^{\ell}_{13},~\sin^{2}\theta^{\ell}_{23},~~\delta^{\ell}_{CP},~\Delta m_{21}^{2},~\Delta m_{31}^{2},~m_{e}/m_{\mu},~m_{\mu}/m_{\tau}.
\end{equation}
The experimental results of the charged lepton masses $m_{e,\mu,\tau}$ are taken from~\cite{Xing:2007fb}, while the neutrino mass squared differences $\Delta m_{21}^{2}$, $\Delta m_{31}^{2}$ and mixing angles $\theta^{\ell}_{12}$, $\theta^{\ell}_{13}$, $\theta^{\ell}_{23}$, $\delta^{\ell}_{CP}$ are taken from the latest NuFIT 6.0 with Super-Kamiokanda atmospheric data~\cite{Esteban:2024eli}-- see table~\ref{tab:lepton-data}. We employ the minimization package \textbf{TMinuit}~\cite{minuit} to determine the minima of the $\chi^{2}$ function. A successful description of the experimental data is achieved for both the normal ordering (NO) neutrino mass spectrum ($m_{1} < m_{2} < m_{3}$) and the inverted ordering (IO) neutrino mass spectrum ($m_{3} < m_{1} < m_{2}$). The best-fit values of the input parameters, along with the resulting lepton flavor observables at the best fit point for the NO and IO spectra, are listed in table~\ref{tab:lepton_obs_prediction}.
\begin{table}[t!]
\centering
\begin{tabular}{| c | c | c | c | c |} \hline \hline
\multirow{2}{*}{Inputs and Obs} & \multicolumn{2}{c|}{Without gCP} & \multicolumn{2}{c|}{With gCP} \\ \cline{2-5}
 & NO & IO & NO & IO  \\ \hline
$\rm{Re}(\tau)$ & $-0.1563$ &  $3.971\times 10^{-3}$ & $ -0.03777$  & $4.388\times 10^{-3}$ \\
$\rm{Im}(\tau)$ & $1.108$ &  $1.083$ & $1.090$  & $1.083$ \\
$\beta/\alpha$ & $4.896$ & $16.00$ &  $17.70$  & $16.00$ \\
$\gamma/\alpha$ & $3.267\times 10^{-3}$ &  $257.1$ & $284.6$  & $257.1$ \\
$|g_{2}/g_{1}|$ & $0.2946$ &  $0.08106$ & $0.1490$ & $-0.08106$ \\
$\rm{arg}(g_{2}/g_{1})$ & $1.888\,\pi$ &  $1.005\,\pi$ & $-$  & $-$ \\
$\alpha v /\rm{GeV}$ & $0.2916$ &  $5.158\times 10^{-3}$ & $4.696\times 10^{-3}$  & $5.156\times 10^{-3}$ \\
$\frac{g_{1}^{2}v^{2}}{\Lambda}/\rm{meV}$ & $89.75$ &  $603.9$ & $191.6$  & $603.8$ \\ \hline \hline
$\sin^2\theta^{\ell}_{12}$& $0.308$ & $0.308$ & $0.308$ & $0.308$   \\
$\sin^2\theta^{\ell}_{13}$& $0.02215$ & $0.02236$ & $0.02215$ & $0.02236$  \\
$\sin^2\theta^{\ell}_{23}$ & $0.470$ & $0.561$ & $0.459$ & $0.561$  \\
$\delta^{\ell}_{CP}$ & $1.15 \pi$ & $1.55 \pi$ & $1.09 \pi$ & $1.56 \pi$  \\
$\Delta m_{21}^2 / 10^{-5}\text{eV}^2$ & $7.49$ & $7.49$ & $7.48$ & $7.49$   \\
$\Delta m_{3l}^2 / 10^{-3}\text{eV}^2$ & $2.513$ & $-2.484$ & $2.513$ & $-2.484$  \\
$m_e/m_\mu $ & $0.004737$ & $0.004737$ & $0.004737$ & $0.004737$   \\
$m_\mu/m_\tau$ & $0.05882$ & $0.05882$ & $0.05882$ & $0.05882$  \\
$m_e/\rm{MeV}$ & $0.469652$ & $0.469652$ & $0.469652$ & $0.469652$ \\
  \hline
\hline
$\phi$ & $1.14\pi$ & $0.35\pi$ & $1.04\pi$ & $0.35\pi$  \\
$m_{1}/\text{meV}$ & $0$ & $49.08$ & $0$ & $49.08$   \\
$m_{2}/\text{meV}$ & $8.65$ & $49.84$ & $8.65$ & $49.84$   \\
$m_{3}/\text{meV}$ & $50.13$ & $0$ & $50.13$ & $0$   \\
$\sum_i m_{i}/\text{meV}$ & $58.78$ & $98.92$ & $58.78$ & $98.92$   \\
$m_{\beta\beta}/\text{meV}$ & $2.66$ & $42.34$ & $1.91$ & $42.34$   \\
  $m_{\beta}/\text{meV}$ & $8.84$ & $48.76$ & $8.84$ & $48.76$   \\ \hline
$\chi^{2}_{\ell}$ & $0.01$ & $0.91$ & $0.90$ & $0.92$ \\
\hline \hline
\end{tabular}
\caption{\label{tab:lepton_obs_prediction}The best fit results of lepton model in section~\ref{subsec:lepton_model-seesaw}. Note that $\Delta m_{3l}^{2}=\Delta m_{31}^{2}$ for NO and $\Delta m_{3l}^{2}=\Delta m_{32}^{2}$ for IO.}
\end{table}

We can find that the predictions are in agreement with experimental data at $1\sigma$ level. Besides the measured neutrino mixing angles and mass squared differences, we have also calculated the unknown neutrino observables, including the Majorana CP-violation phase $\phi$~\footnote{$\phi$ is the unique Majorana phase in case that the lightest neutrino is massless.} and the sum of light neutrino masses $\Sigma_{i} m_{i}$ as well as the effective Majorana neutrino mass $m_{\beta\beta}$ in neutrinoless double beta ($0\nu\beta\beta$) decay experiments and the kinematic mass $m_{\beta}$ in beta decay experiments. Note that $m_{\beta\beta}$ and $m_{\beta}$ depend on the light neutrino masses and mixing parameters:
\begin{eqnarray}
\nonumber m_{\beta\beta}&=&\left|m_1\cos^2\theta^{\ell}_{12}\cos^2\theta^{\ell}_{13} +m_2\sin^2\theta^{\ell}_{12}\cos^2\theta^{\ell}_{13} e^{i\alpha_{21}} +m_3\sin^2\theta^{\ell}_{13} e^{i(\alpha_{31}-2\delta^{\ell}_{CP})}\right|\,,\\
m_{\beta}&=&\sqrt{m^2_1\cos^2\theta^{\ell}_{12}\cos^2\theta^{\ell}_{13} +m^2_2\sin^2\theta^{\ell}_{12}\cos^2\theta^{\ell}_{13} +m^2_3\sin^2\theta^{\ell}_{13}}\,.
\end{eqnarray}
We adopt the convention $\alpha_{21}=\phi$, $\alpha_{31}=0$ for the Majorana phases if the mass of the lightest neutrino is zero. In the case of the NO spectrum, the predicted value $m_{\beta\beta} = 2.66~\text{meV}$ is too small to be detectable by both current and future $0\nu\beta\beta$ decay experiments. For instance, the most stringent bound on the effective neutrino mass from KamLAND-Zen collaboration is $m_{\beta\beta} < (28- 122)~\text{meV}$ at 90\% C.L.~\cite{KamLAND-Zen:2024eml}. The next generation tonne-scale $0\nu\beta\beta$ experiments, such as LEGEND-1000~\cite{LEGEND:2021bnm} and nEXO~\cite{nEXO:2021ujk}, are expected to improve sensitivities to $m_{\beta\beta} \sim (9 - 21)~\text{meV}$ and $m_{\beta\beta} \sim (4.7 - 20.3)~\text{meV}$, respectively, for ten years of running time. Meanwhile, the predicted value $m_{\beta\beta} = 42.34~\text{meV}$ for the IO spectrum falls within the reach of these forthcoming experiments.
The current experimental bound on the kinematical mass $m_{\beta}$, as reported by the KATRIN experiment, is $m_{\beta} < 0.45~\text{eV}$ at 90\% C.L.~\cite{Katrin:2024tvg}. Future sensitivities are anticipated to improve to $m_{\beta} < 0.2~\text{eV}$ for KATRIN~\cite{KATRIN:2021dfa} and $m_{\beta} < 0.04~\text{eV}$ for Project 8~\cite{Project8:2022wqh}. The predicted value $m_{\beta} = 8.80~\text{meV}$ for NO remains below these thresholds and thus undetectable, while the prediction $m_{\beta} = 48.76~\text{meV}$ for IO lies within the potential reach of future experiments. The predictions of sum of the neutrino masses $\Sigma_i m_{i}$ is $58.78~\rm{meV}$ ($98.92~\rm{meV}$) for NO (IO) neutrino spectrum, which is compatible with the Planck Collaboration result $\Sigma_i m_{i}\leq 120~\rm{meV}$ at $95\%$ CL~\cite{Planck:2018vyg}.

To enhance the predictive power of the model, we further reduce the number of input parameters by imposing the gCP symmetry~\cite{Novichkov:2019sqv}, which enforces all coupling constants to be real~\footnote{In this paper, we have adopted a basis in which the generators $S$ and $T$ are represented by symmetric matrices in all irreducible representations, and the Clebsch-Gordan coefficients are real, as shown in Appendix~\ref{app:group-MF-N3}. Consequently, the gCP symmetry reduces to the canonical CP transformation in flavor space~\cite{Novichkov:2019sqv}, i.e. $\varphi \stackrel{\mathcal{CP}}{\longmapsto} \varphi^*$ , and all coupling constants are constrained to be real.}. In this case, the lepton flavor sector is described by 7 real parameters: three real constants $\alpha$, $\beta$, and $\gamma$ for the charged lepton masses, two real constants $g_1^2/\Lambda$ and $g_2/g_1$, along with the real and imaginary parts of the modulus $\tau$, denoted as $\text{Re}\,\tau$ and $\text{Im}\,\tau$, respectively. These parameters collectively account for the 9 observable quantities in the lepton sector. We find that the model can successfully accommodate the experimental data on lepton masses and mixing angles even if gCP symmetry is included. The best-fit values of the input parameters, as well as the resulting lepton flavor observables for both NO and IO neutrino mass spectra, are summarized in table~\ref{tab:lepton_obs_prediction}. Similar to the scenario without gCP symmetry, the predictions $m_{\beta\beta} = 1.91~\text{meV}$ and $m_{\beta} = 8.84~\text{meV}$ for the NO spectrum are too small to be probed by current and upcoming experiments. However, the predictions $m_{\beta\beta} = 42.34~\text{meV}$ and $m_{\beta} = 48.76~\text{meV}$ for the IO spectrum fall within the sensitivity ranges of future experiments and could be tested.

\begin{figure}[hpt!]
\centering
\includegraphics[width=3.5in]{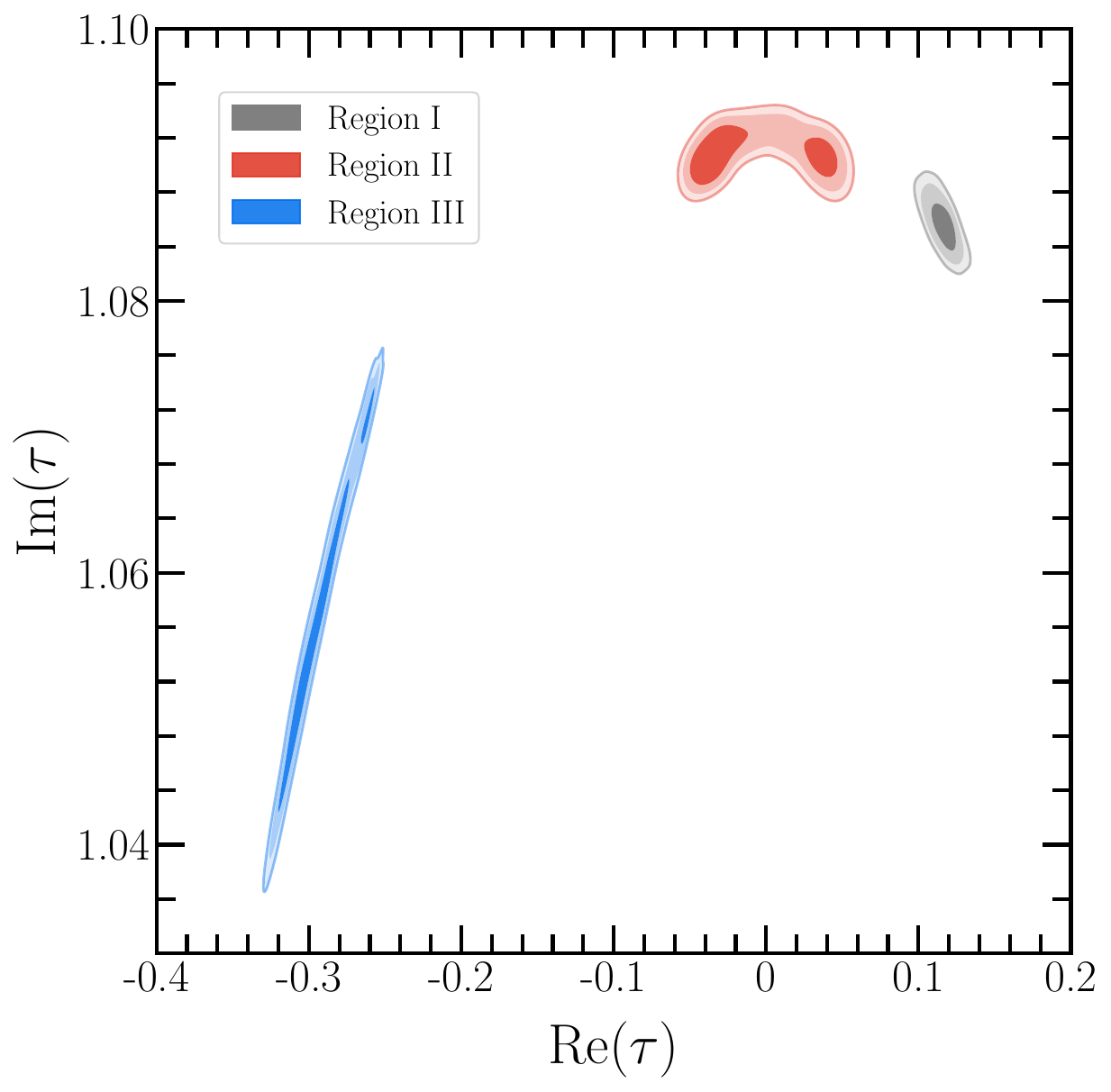}
\caption{Allowed regions in the $\tau$ plane for the lepton model with gCP in section~\ref{subsec:lepton_model-seesaw}. Different color shadings correspond to the $1\sigma$, $2\sigma$, and $3\sigma$ confidence levels.}
\label{fig:lepton_tau}
\end{figure}

To further investigate the phenomenological predictions of this model with gCP symmetry, we utilize the sampler \texttt{MultiNest}~\cite{Feroz:2007kg,Feroz:2008xx} to perform a comprehensive scan of the parameter space, requiring that all measured lepton observables fall within their experimentally preferred $3\sigma$ ranges.
In the case of NO neutrino masses, we find there are three separately allowed regions of $\tau$ as are shown in figure~\ref{fig:lepton_tau}. As an illustration, we focus on Region II which has overlap with the values of $\tau$ favored by the quark sector in section~\ref{sec:quark_model}. The allowed regions of lepton observables are summarized in table~\ref{tab:lepton_obs_prediction_ranges}. We can find that the atmospheric neutrino mixing angle $\sin^{2}\theta_{23}^{\ell}$ and the Dirac CP violation phase $\delta_{CP}^{\ell}$ are constrained in narrow regions, $\sin^{2}\theta_{23}^{\ell}$ is located in the first octant $[0.454,0.463]$ and second octant $[0.557,0.565]$ for NO and IO spectra respectively, and the lepton Dirac CP phase $\delta^{\ell}_{CP}\in[0.84\pi, 1.16\pi]$ for NO and $\delta^{\ell}_{CP}\in[1.54\pi, 1.58\pi]$ for IO. The sum of the neutrino masses $\sum_{i}m_{i}$ is predicted to lie in $[57.83~\text{meV},59.74~\text{meV}]$ for NO and $[97.59~\text{meV},100.24~\text{meV}]$ for IO which are compatible with the upper limit $\sum_{i}m_{i}<120~\text{meV}$ from the Planck collaboration~\cite{Planck:2018vyg}. The predicted regions of the effective neutrino masses $m_{\beta\beta}$ and $m_{\beta}$ for the IO spectrum fall within the sensitivity ranges of the next generation facilities, while they are far below the reach of future experiments in case of NO neutrino masses.

\begin{figure}[hptb!]
\centering
\includegraphics[width=6.5in]{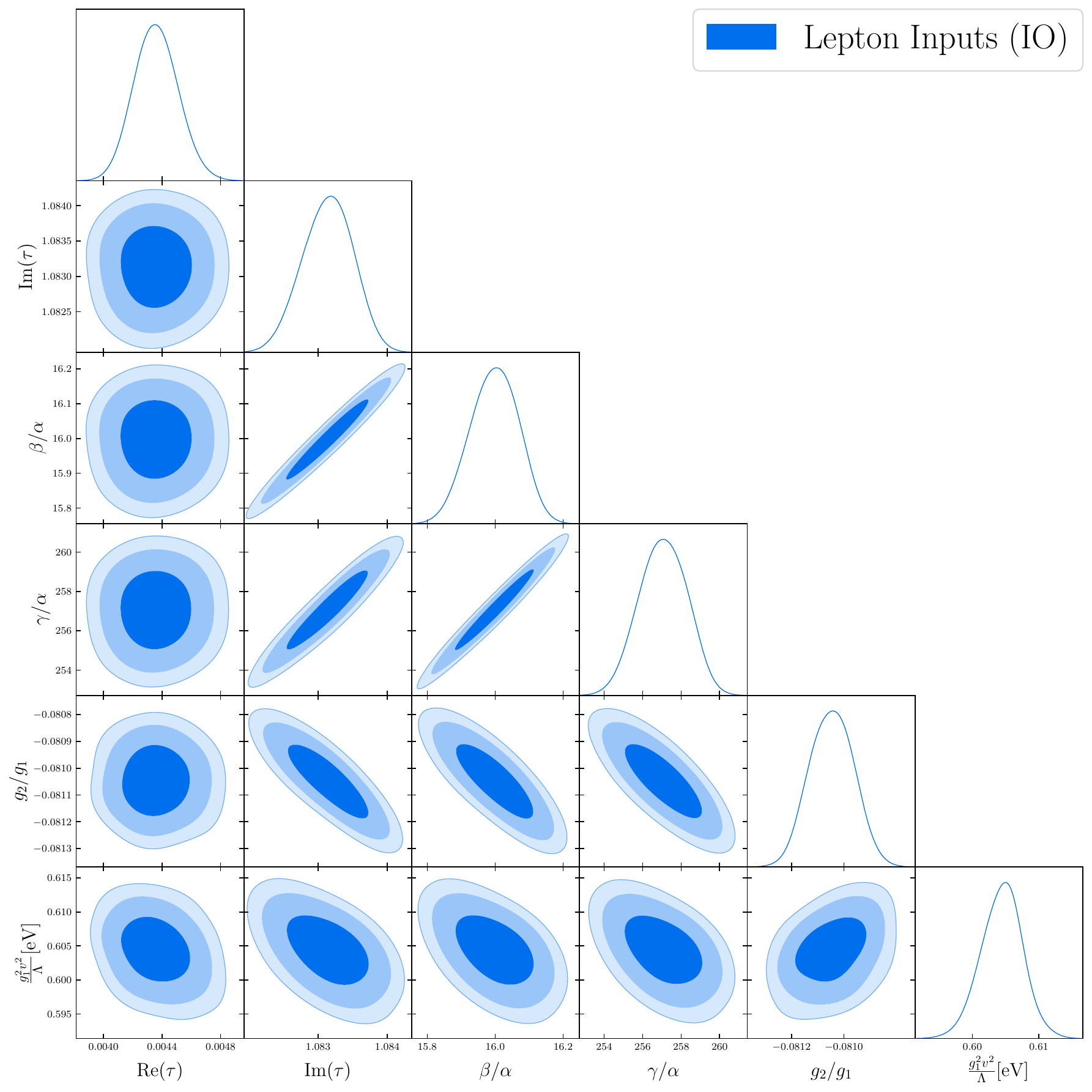}
\caption{Allowed regions for the lepton input parameters in case of IO neutrino masses spectrum. Different color shadings correspond to the $1\sigma$, $2\sigma$, and $3\sigma$ confidence levels. }
\label{fig:lepton_input_IO}
\end{figure}

\begin{figure}[hptb!]
\centering
\includegraphics[width=6.5in]{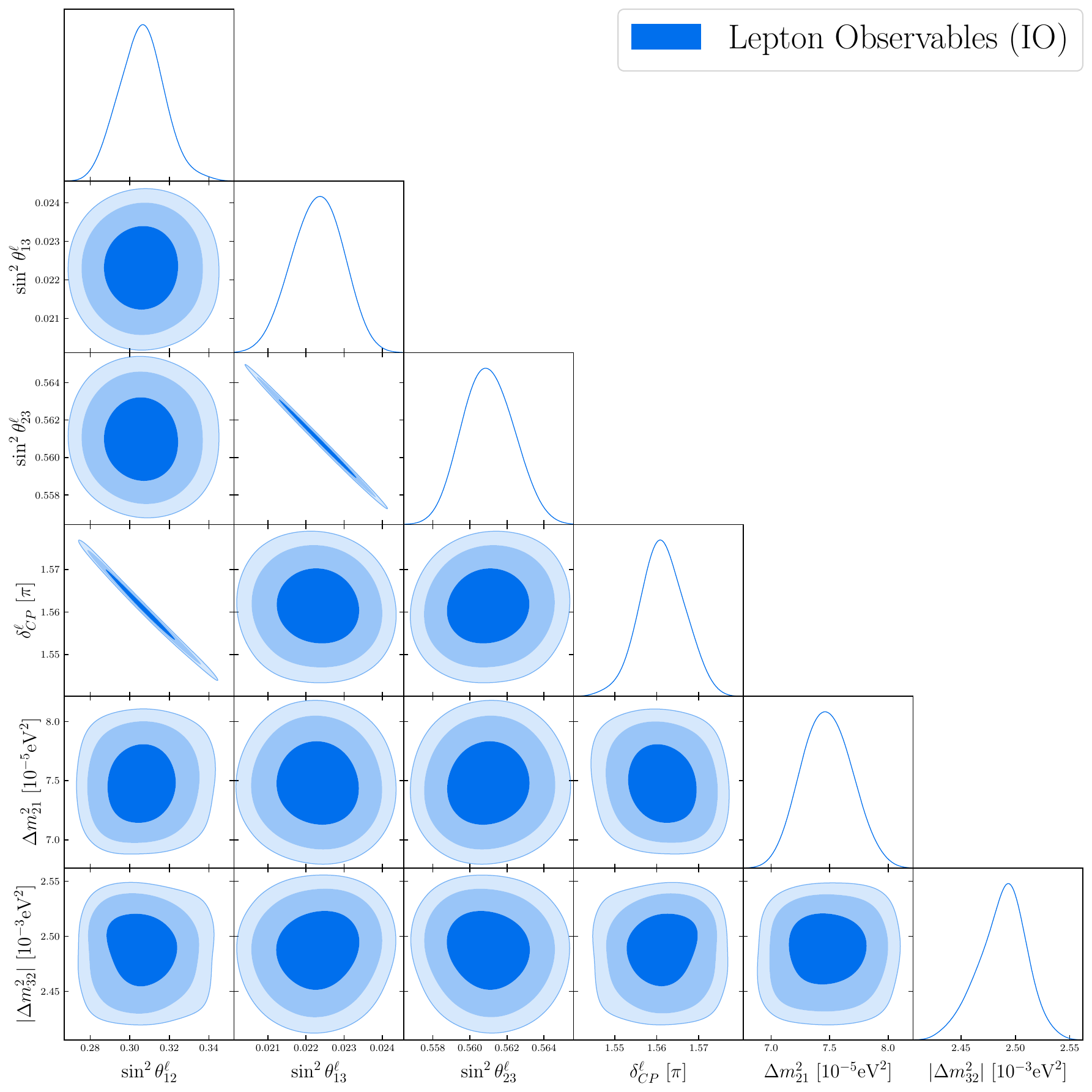}
\caption{Allowed regions for the lepton observables in case of IO neutrino masses spectrum. Different color shadings correspond to the $1\sigma$, $2\sigma$, and $3\sigma$ confidence levels. }
\label{fig:lepton_obs_IO}
\end{figure}

\begin{figure}[hptb!]
\centering
\includegraphics[width=6.5in]{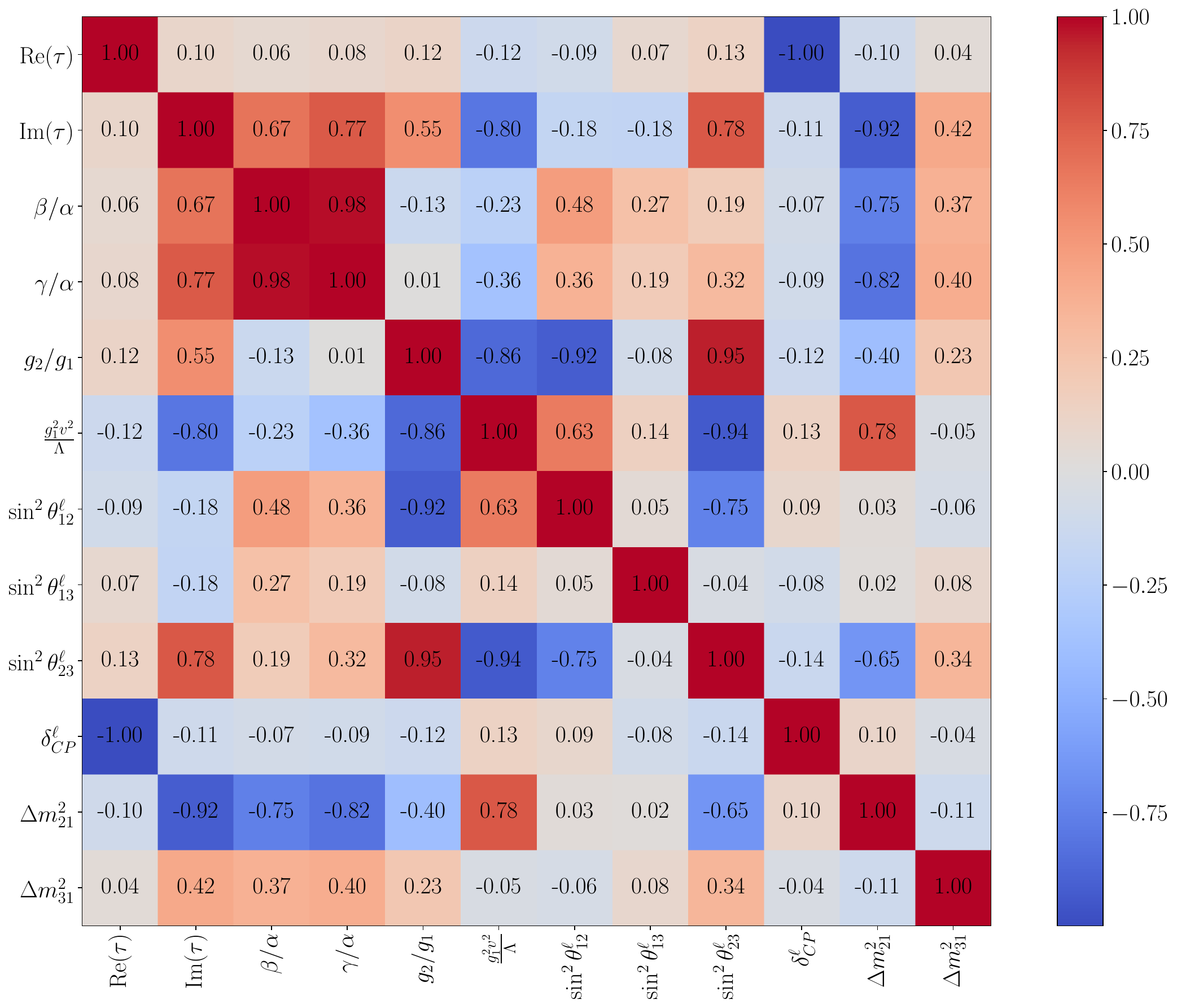}
\caption{Matrix of the correlations among the input parameters and lepton observables for the lepton model with gCP in section~\ref{subsec:lepton_model-seesaw}, and the neutrino mass spectrum is NO. }
\label{fig:lepton_correlation_matrix_NO}
\end{figure}

\begin{figure}[hptb!]
\centering
\includegraphics[width=6.5in]{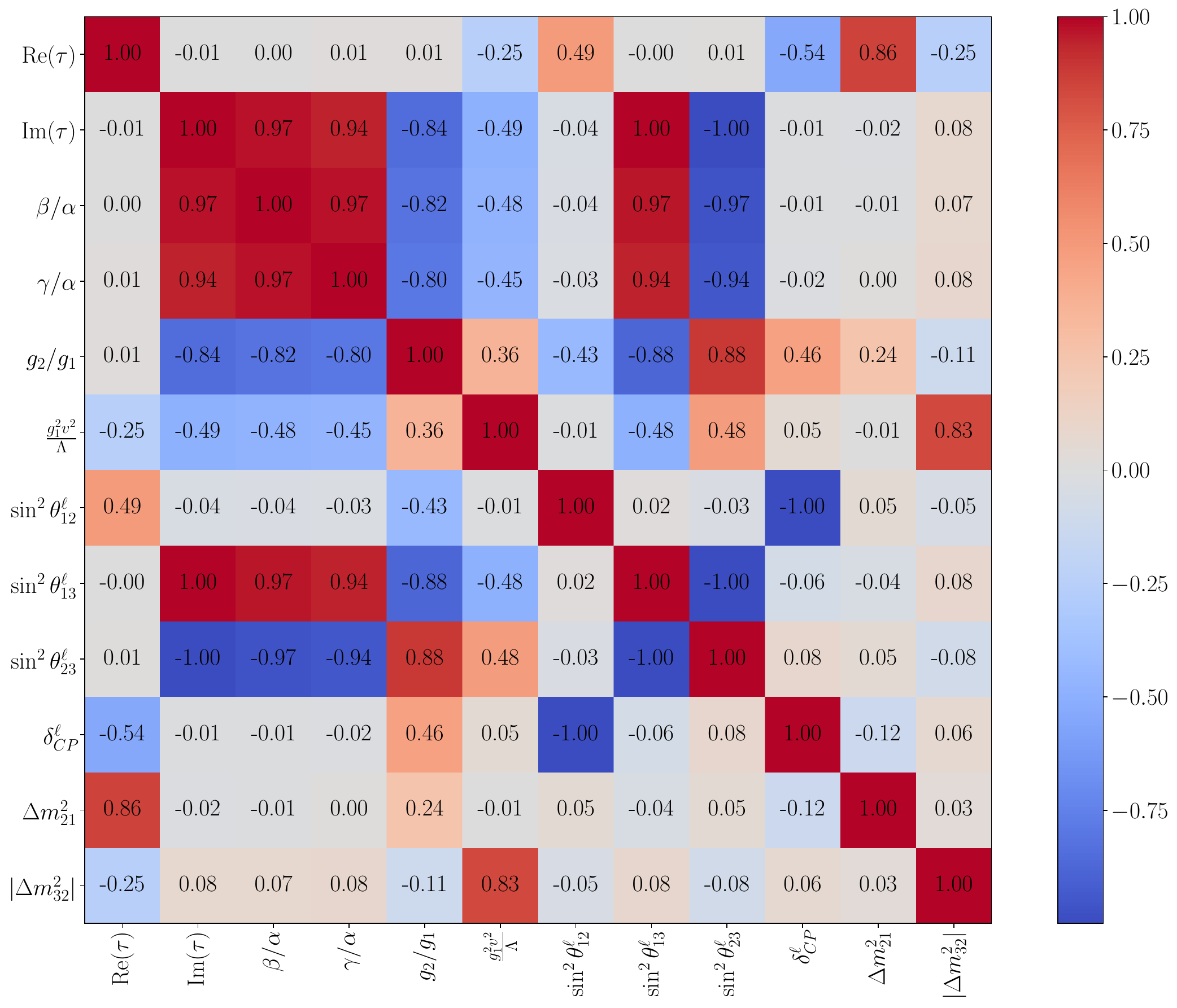}
\caption{Similar to figure~\ref{fig:lepton_correlation_matrix_NO}, but for IO neutrino mass spectrum. }
\label{fig:lepton_correlation_matrix_IO}
\end{figure}

Furthermore, we show correlations between the model free parameters in figure~\ref{fig:lepton_unified_input}. One can see that $g^2_{1}v^2/\Lambda$ is negatively correlated with both $g_{2}/g_{1}$ and $\rm{Im}(\tau)$, and there are strong positive correlations among $\beta/\alpha$, $\gamma/\alpha$ and $\rm{Im}(\tau)$. The correlations among the neutrino mixing observables are displayed in figure~\ref{fig:lepton_unified_obs}, the atmospheric neutrino mixing angle $\sin^{2}\theta_{23}^{\ell}$ is in negative correlation with the solar neutrino mixing angle $\sin^{2}\theta_{12}^{\ell}$ and the mass squares difference $\Delta m_{21}^{2}$.
For the case of the IO neutrino mass spectrum, the allowed regions for the lepton input parameters and the lepton observables are displayed in figure~\ref{fig:lepton_input_IO} and figure~\ref{fig:lepton_obs_IO}, respectively. From figure~\ref{fig:lepton_input_IO}, we see strong positive correlations among the parameters $\beta/\alpha$, $\gamma/\alpha$, and $\mathrm{Im}(\tau)$. In addition, negative correlations are found between $g_{2}/g_{1}$ and the parameters $\beta/\alpha$, $\gamma/\alpha$, and $\mathrm{Im}(\tau)$. Regarding the lepton mixing parameters shown in figure~\ref{fig:lepton_obs_IO}, strong negative correlations are observed between $\sin^{2}\theta_{12}^{\ell}$ and $\delta_{CP}^{\ell}$, as well as between $\sin^{2}\theta_{13}^{\ell}$ and $\sin^{2}\theta_{23}^{\ell}$. In order to quantitatively measure the strength of correlations, we have calculated the correlation matrix of the model parameters and the observables, and the results are shown in figure~\ref{fig:lepton_correlation_matrix_NO} and figure~\ref{fig:lepton_correlation_matrix_IO} for NO and IO spectrum respectively.

\begin{table}[t!]
\centering
\begin{tabular}{| c | c | c |} \hline \hline
Measured observables & Predicted regions (NO) & Predicted regions (IO) \\ \hline
$\sin^2\theta^{\ell}_{12}$& $[0.275,0.345]$ & $[0.275, 0.345]$   \\
$\sin^2\theta^{\ell}_{13}$& $[0.02030,0.02388]$ & $[0.02060,0.02409]$  \\
$\sin^2\theta^{\ell}_{23}$ & $[0.454,0.463]$ & $[0.557,0.565]$  \\
$\delta^{\ell}_{CP}/\pi$ & $[0.84,1.16]$ & $[1.54,1.58]$  \\
$\Delta m_{21}^2 / 10^{-5}\text{eV}^2$ & $[6.92,8.05]$ & $[6.92,8.05]$   \\
$\Delta m_{3l}^2 / 10^{-3}\text{eV}^2$ & $[2.451,2.578]$ & $[-2.547,-2.421]$  \\
$m_e/m_\mu $ & $[0.004723,0.004751]$ & $[0.004723,0.004751]$  \\
$m_\mu/m_\tau$ & $[0.05864,0.05900]$ & $[0.05864,0.05900]$  \\
$m_e/\rm{MeV}$ & $[0.469652,0.469652]$ & $[0.469652,0.469652]$  \\
  \hline
Predicted observables & Predicted regions (NO) & Predicted regions (IO) \\ \hline
$\phi/\pi$ & $[0.93,1.07]$ & $[0,2]$  \\
$m_{1}/\text{meV}$ & $0$ & $[48.38,49.78]$   \\
$m_{2}/\text{meV}$ & $[8.32,8.97]$ & $[49.20,50.47]$   \\
$m_{3}/\text{meV}$ & $[49.51,50.77]$ & $0$   \\
$\sum_i m_{i}/\text{meV}$ & $[57.83,59.74]$ & $[97.59,100.24]$   \\
$m_{\beta\beta}/\text{meV}$ & $[1.05,2.84]$ & $[41.68,42.98]$   \\
$m_{\beta}/\text{meV}$ & $[8.38,9.39]$ & $[48.04,49.47]$   \\
$J_{CP}$ & $[-0.0174,0.0173]$ & $[-0.0354,-0.0302]$   \\
  \hline \hline
\end{tabular}
\caption{\label{tab:lepton_obs_prediction_ranges}The allowed regions of lepton observables in the lepton model of section~\ref{subsec:lepton_model-seesaw} with gCP symmetry.}
\end{table}

\subsection{Quark sector\label{sec:quark_model}}

We turn to consider the representative models of quarks in this section. The minimal phenomenologically viable quark model is found to contain $8$ real input parameters. There are some quark models with $7$ parameters can be regarded as leading order approximation. As a consequence, we present one representative quark model with $7$ parameters and another viable quark model with $8$ parameters in the following.

\subsubsection{Quark model with $7$ input parameters\label{sec:quark_model_7para}}

Excluding the complex modulus $\tau$, the model would only depend on 5 real parameters so that the structure of the model is strongly constrained. Both left-handed and right-handed quarks have to be assigned the direct sum of a doublet and a singlet of the $T'$ modular group, either the up-type quark mass matrix or the down-type quark mass matrix would be block diagonal. In the following, we present a benchmark quark model involving $7$ input parameters and it can accommodate all the experimental data of quarks except that the predicted value of the mass ratio $m_s/m_b$ is too small. The representation and modular weight assignments of quark fields are given as follow,
\begin{eqnarray}
\nonumber&& u_{D}^c\sim \widehat{\bm{2}}\,,~~u_{3}^c\sim \bm{1'}\,,~~d_{D}^c\sim\widehat{\bm{2}}'\,,~~d_{3}^c\sim \bm{1}\,,~~Q_{D}\sim\widehat{\bm{2}}\,,~~Q_{3}\sim \bm{1'}\,,~~H\sim\bm{1}\,,\\
&& k_{Q_{D}}=-8-k_{u^c_{D}}=-5-k_{u^c_{3}}=-6-k_{d^c_{D}}=-3-k_{d^c_{3}}=k_{Q_{3}}-9\,,~~k_{H}=0\,.
\end{eqnarray}
Here $u^{c}_{D}\equiv \left( u_{1}^{c},u_{2}^{c}\right)^{T}$, $d^{c}_{D}\equiv \left( d_{1}^{c},d_{2}^{c}\right)^{T}$ and $Q_{D}\equiv \left( Q_{1},Q_{2}\right)^{T}$ denote the first two generations of quark fields, and the modular weight $k_{Q_{D}}$ is an arbitrary integer.
The modular invariant Lagrangian for the quark Yukawa coupling takes the following form,
\begin{eqnarray}
\nonumber -\mathcal{L}_u &=& \alpha_u \left( u^c_{D} Q_{D} Y^{(-8)}_{\bm{3}} H \right)_{\bm{1}} + \beta_u \left( u^c_3 Q_{3} Y^{(4)}_{\bm{1'}} H \right)_{\bm{1}}+ \text{h.c.}\,, \\
-\mathcal{L}_d &=& \alpha_d \left(d^c_{D} Q_{D} Y^{(-6)}_{\bm{3}} H^*\right)_{\bm{1}} + \beta_d \left( d^c_{3} Q_{D} Y^{(-3)}_{\widehat{\bm{2}}''} H^* \right)_{\bm{1}} + \gamma_d \left(d^c_{D} Q_{3} Y^{(3)}_{\widehat{\bm{2}}} H^* \right)_{\bm{1}}  + \text{h.c.}\,,~~
\end{eqnarray}
where the phases of the couplings $\alpha_{u}$, $\beta_{u}$, $\alpha_{d}$, $\beta_{d}$ and $\gamma_{d}$ can be absorbed into the quark fields. The up-type and down-type quark mass matrices read as
\begin{eqnarray}
M_u &=& \begin{pmatrix}
-\alpha_u Y^{(-8)}_{\bm{3},2} ~& \alpha_u \frac{Y^{(-8)}_{\bm{3},3}}{\sqrt{2}} ~& 0 \\
\alpha_u \frac{Y^{(-8)}_{\bm{3},3}}{\sqrt{2}} ~& \alpha_u Y^{(-8)}_{\bm{3},1} ~& 0  \\
0 ~& 0 ~& \beta_u Y^{(4)}_{\bm{1'}}
\end{pmatrix} v\,,~~
M_d = \begin{pmatrix}
-\alpha_d Y^{(-6)}_{\bm{3},1} ~& \alpha_d \frac{Y^{(-6)}_{\bm{3},2}}{\sqrt{2}} ~& -\gamma_{d}Y^{(3)}_{\widehat{\bm{2}},2} \\
\alpha_d \frac{Y^{(-6)}_{\bm{3},2}}{\sqrt{2}} ~& \alpha_d Y^{(-6)}_{\bm{3},3} ~& \gamma_{d}Y^{(3)}_{\widehat{\bm{2}},1}  \\
\beta_{d}Y^{(-3)}_{\widehat{\bm{2}}'',2} ~& -\beta_{d}Y^{(-3)}_{\widehat{\bm{2}}'',1} ~& 0
\end{pmatrix} v\,,~~~~~~
\end{eqnarray}
To assess whether this quark model can accommodate the experimental data of quark masses and mixing parameters. We perform a $\chi^{2}$ analysis of this model and the $\chi^{2}$ function is defined as
\begin{equation}\label{eq:chisq_quark}
\chi^2_{q}=\sum^{8}_{i=1}\left(\frac{P_i(x)-\mu_i}{\sigma_i}\right)^2\,,
\end{equation}
where the vector $x$ contains the model parameters $\beta_u/\alpha_u$, $\beta_d/\alpha_d$, $\gamma_d/\alpha_d$, $\text{Re}(\tau)$ and $\text{Im}(\tau)$, $P_{i}(x)$ are the model predictions for the observables, $\mu_{i}$ and $\sigma_{i}$ denote the central values and standard deviations of the corresponding eight dimensionless quantities:
\begin{equation}
  \theta^{q}_{12},~\theta^{q}_{13},~\theta^{q}_{23},~\delta^{q}_{CP},~m_{u}/m_{c},~m_{c}/m_{t},~m_{d}/m_{s},~m_{s}/m_{b}.
\end{equation}
The central values and the uncertainties of the quark masses and mixings parameters are taken from~\cite{Xing:2007fb}, they are listed in table~\ref{tab:lepton-data}. Analogous to the lepton sector, we use the central values of $m_{t}$ and $m_{b}$ to determine the overall scales $\alpha_u v$ and $\alpha_d v$ of the quark mass matrices $M_{u}$ and $M_{d}$ respectively. By performing $\chi^{2}$ analysis, the best fit values of the input parameters and corresponding values of the quark observables are presented in table~\ref{tab:quark_obs_prediction}. We find that the largest contribution to $\chi^2$ comes from $m_{s}/m_{b}$ which is outside the corresponding 3 times of $1\sigma$ range, and all others are compatible with the experimental data. Consequently this model can be regarded as a good leading order approximation.

\begin{table}[t!]
\centering
\begin{tabular}{| c | c | c |} \hline \hline
Inputs & 7 parameters & 8 parameters \\ \hline
$\rm{Re}(\tau)$ & $-0.4957$ &  $5.226\times 10^{-3}$ \\
$\rm{Im}(\tau)$ & $0.8749$ &  $1.101$ \\
$\beta_{u}/\alpha_{u}$ & $7.129$ & $6.304\times 10^{-3}$ \\
$\gamma_{u}/\alpha_{u}$ & $-$ & $497.0$ \\
$\beta_{d}/\alpha_{d}$ & $7.676\times 10^{-4}$ & $29.83$ \\
$\gamma_{d}/\alpha_{d}$ & $0.5647$ & $0.1003$ \\
$\alpha_{u} v /\rm{GeV}$ & $4.178$ &  $0.3823$ \\
$\alpha_{d} v /\rm{GeV}$ & $0.6646$ &  $0.0965$ \\
\hline \hline
\end{tabular}
~~~
\begin{tabular}{| c | c | c |} \hline \hline
Observables & 7 parameters & 8 parameters \\ \hline
$\theta^{q}_{12}$& $0.2256$ & $0.2257$ \\
$\theta^{q}_{13}$& $0.0038$ & $0.0037$ \\
$\theta^{q}_{23}$& $0.0417$ & $0.0398$ \\
$\delta^{q}_{CP}$& $60.18$ & $73.32$ \\
$m_u/m_c $ & $0.00277$ & $0.00204$  \\
$m_c/m_t $ & $0.00318$ & $0.00318$  \\
$m_d/m_s $ & $0.0676$ & $0.0950$  \\
$m_s/m_b $ & $0.0023$ & $0.0187$  \\ \hline \hline
$\chi^{2}_{q}$ & $26.03$ & $8.11$ \\
  \hline \hline
\end{tabular}
\caption{\label{tab:quark_obs_prediction}The best fit results of the benchmark quark models.}
\end{table}

\subsubsection{Quark model with $8$ input parameters\label{sec:quark_model_8para}}

To fully account for the quark data in table~\ref{tab:lepton-data}, we find that the quark model requires at least 8 input parameters within the framework of the modular group $T'$. We will present a viable model in which the Yukawa couplings are level 3 polyharmonic Maa{\ss} forms of non-positive weights. The left-handed quark fields are assumed to transform as a triplet under $T'$, while all right-handed quark fields are taken to be singlets of $T'$. The representations and modular weight assignments of the quark fields under the modular group are given as follows,
\begin{eqnarray}\nonumber
&& u^c\sim\bm{1}\,,~~c^c\sim\bm{1'}\,,~~t^c \sim\bm{1''}\,,~~d^c \sim\bm{1}\,,~~ s^c \sim\bm{1''}\,,~~b^c\sim\bm{1'} \,,~~Q\sim\bm{3}\,,~~H\sim\bm{1}\,,  \\
&& k_{Q}=-2-k_{u^c}=-k_{c^c}=-k_{t^c} = -4-k_{d^c} = -4-k_{s^c} = -k_{b^c}\,,~~k_{H}=0\,. \label{eq:quark_content}
\end{eqnarray}
Here $Q\equiv \left( Q_{1},Q_{2},Q_{3}\right)^{T}$ denotes the three generations of the left-handed quark doublets, and $k_{Q}$ is an arbitrary integer. Then one can read off the modular invariant Lagrangian for the quark masses
\begin{eqnarray}
-\mathcal{L}_u &=& \alpha_u \left( u^c_1 Q Y^{(-2)}_{\bm{3}} H \right)_{\bm{1}} + \beta_u \left( u^c_2 Q Y^{(0)}_{\bm{3}} H \right)_{\bm{1}} + \gamma_u \left( u^c_3 Q Y^{(0)}_{\bm{3}} H \right)_{\bm{1}}  + \text{h.c.}\,, \\
-\mathcal{L}_d &=& \alpha_d \left( d^c_1 Q Y^{(-4)}_{\bm{3}} H^* \right)_{\bm{1}} + \beta_d \left( d^c_2 Q Y^{(-4)}_{\bm{3}} H^* \right)_{\bm{1}} + \gamma_d \left( d^c_3 Q Y^{(0)}_{\bm{3}} H^* \right)_{\bm{1}}  + \text{h.c.}\,,
\end{eqnarray}
where all coupling constants can be taken to be real by using the freedom of the right-handed quark fields redefinition. The corresponding up- and down-type quark mass matrices are given by
\begin{eqnarray}
M_u &=& \begin{pmatrix}
\alpha_u Y^{(-2)}_{\bm{3},1} ~&~ \alpha_u Y^{(-2)}_{\bm{3},3} ~&~ \alpha_u Y^{(-2)}_{\bm{3},2} \\
\beta_u Y^{(0)}_{\bm{3},3} ~&~ \beta_u Y^{(0)}_{\bm{3},2} ~&~ \beta_u Y^{(0)}_{\bm{3},1}  \\
\gamma_u Y^{(0)}_{\bm{3},2} ~&~ \gamma_u Y^{(0)}_{\bm{3},1} ~&~ \gamma_u Y^{(0)}_{\bm{3},3}
\end{pmatrix} v\,,~~
M_d = \begin{pmatrix}
\alpha_d Y^{(-4)}_{\bm{3},1} ~&~ \alpha_d Y^{(-4)}_{\bm{3},3} ~&~ \alpha_d Y^{(-4)}_{\bm{3},2} \\
\beta_d Y^{(-4)}_{\bm{3},2} ~&~ \beta_d Y^{(-4)}_{\bm{3},1} ~&~ \beta_d Y^{(-4)}_{\bm{3},3}  \\
\gamma_d Y^{(0)}_{\bm{3},3} ~&~ \gamma_d Y^{(0)}_{\bm{3},2} ~&~ \gamma_d Y^{(0)}_{\bm{3},1}
\end{pmatrix} v\,.~~~~~~
\end{eqnarray}
The best fit values of the input parameters and the quark observables are given in table~\ref{tab:quark_obs_prediction}. We can find that all of the quark observables are predicted to lie in the experimentally preferred $3\sigma$ ranges given in table~\ref{tab:lepton-data}.

\begin{figure}[hptb!]
\centering
\includegraphics[width=6.5in]{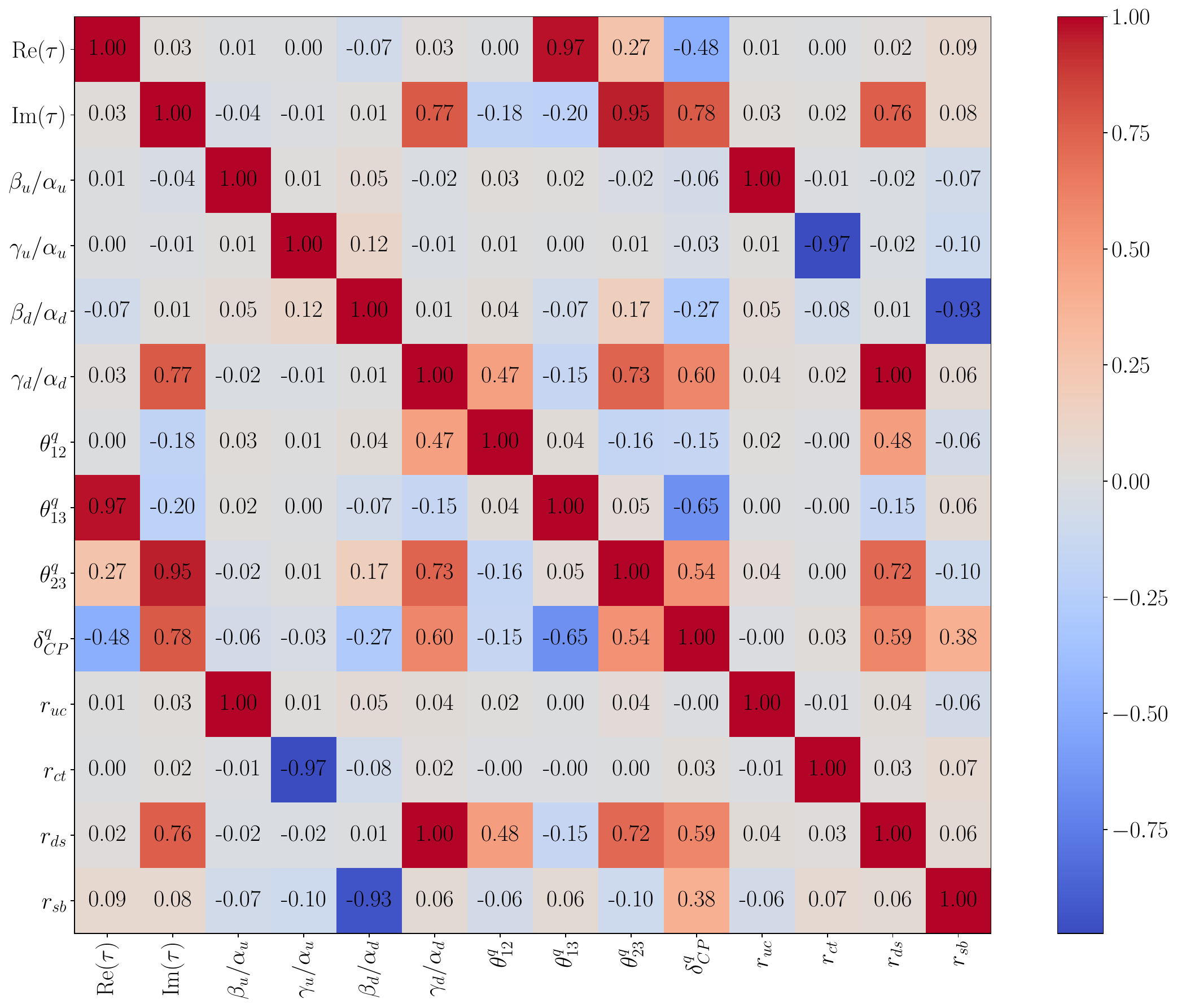}
\caption{Matrix of correlations among the free parameters and observables for the quark model in section~\ref{sec:quark_model_8para}. }
\label{fig:quark_correlation_matrix}
\end{figure}

By scanning the parameter space of this model, we plot the allowed regions for input parameters and quark observables in figure~\ref{fig:quark_unified_input} and figure~\ref{fig:quark_unified_obs}, respectively. It is notable that the modulus $\tau$ is highly constrained within a narrow range close to the imaginary axis, and $\gamma_{d}/\alpha_{d}$ exhibits a positive correlation with $\mathrm{Im}(\tau)$. From figure~\ref{fig:quark_unified_obs}, we observe that the mass ratio $r_{ds}\equiv m_d/m_s$ correlates positively with $\theta_{12}^{q}$, $\theta_{23}^{q}$, and $\delta_{CP}^{q}$, while $\theta_{13}^{q}$ and $\delta_{CP}^{q}$ are negatively correlated. The correlation matrix in figure~\ref{fig:quark_correlation_matrix} further quantitatively shows the strong dependence of quark flavor observables on the free parameters of the model. Specifically, $r_{uc}\equiv m_{u}/m_c$ and $r_{ds}$ correlates positively with $\beta_{u}/\alpha_{u}$ and $\gamma_{d}/\alpha_{d}$ respectively, whereas $r_{ct}\equiv m_c/m_t$ and $r_{sb}\equiv m_s/m_b$ exhibit strong negative correlations with $\gamma_{u}/\alpha_{u}$ and $\beta_{d}/\alpha_{d}$ respectively. These results indicate that quark mass ratios are primarily governed by the Yukawa coupling structure in this model.

\subsection{\label{sec:unified_model}Unified model of leptons and quarks}

In the previous sections, we examined phenomenologically viable models for the lepton and quark sectors individually. We now explore whether the $T'$ modular symmetry can simultaneously account for the flavor structure of both quarks and leptons.
For the lepton sector, we adopt the lepton model of section~\ref{subsec:lepton_model-seesaw}, while the quark sector follows the framework of section~\ref{sec:quark_model_8para}. The complete modular transformations of the SM quark and lepton fields are provided in Eq.~\eqref{eq:lepton_content} and Eq.~\eqref{eq:quark_content}. Thus there are 8 (7) real parameters in the lepton sector and 8 parameters in the quark sector, in the absence (presence) of gCP. Since both sectors share the same complex modulus $\tau$, the total number of real free parameters is $8(7)+8-2=14 (13)$ without (with) gCP. It is the minimal modular invariant model with the smallest number of free parameters at present to address the flavor structure of quarks and leptons. Since the 22 quark and lepton observables including masses and mixing angles and CP violation phases are described by only 14 (13) real parameters when gCP isn't (is) imposed, the model is highly predictive and strong correlation between flavor observables are expected.

We adopt a joint $\chi^{2}$ analysis of this unified model and the total $\chi^{2}$ function is the sum of $\chi^2_{\ell}$ and $\chi^2_q$,
\begin{equation}
\chi^2_{tot}=\chi^{2}_{\ell}+\chi^{2}_{q}\,,
\end{equation}
where $\chi^{2}_{\ell}$ and $\chi^{2}_{q}$ are in Eq.~\eqref{eq:chisq_lepton} and Eq.~\eqref{eq:chisq_quark}, respectively. A joint $\chi^{2}$ analysis of lepton and quark sectors reveals that only the NO neutrino mass spectrum is compatible with experimental data. The best-fit values of the input parameters and the corresponding predictions for the flavor observables, with and without gCP, are summarized in table~\ref{tab:combine_obs_prediction}. The minimal value of $\chi^{2}_{tot}$ is 22.65 with gCP, compared to 13.33 without gCP. All flavor observables are within their experimental $3\sigma$ ranges in both scenarios. When gCP symmetry is considered, the 19 measured observables (9 in the lepton sector~\footnote{The leptonic Dirac CP violation phase $\delta^{\ell}_{CP}$ has been included in the fit, although there are still large uncertainties and its exact value is yet unknown with high precision. However, the forthcoming neutrino oscillation experiments such as DUNE~\cite{DUNE:2020ypp}  and T2HK~\cite{Hyper-Kamiokande:2018ofw} are expected to measure this phase more precisely.} and 10 in the quark sector) are described by only 13 real input parameters, highlighting the strong predictive power of the unified model.

\begin{figure}[hptb!]
\centering
\includegraphics[width=3.5in]{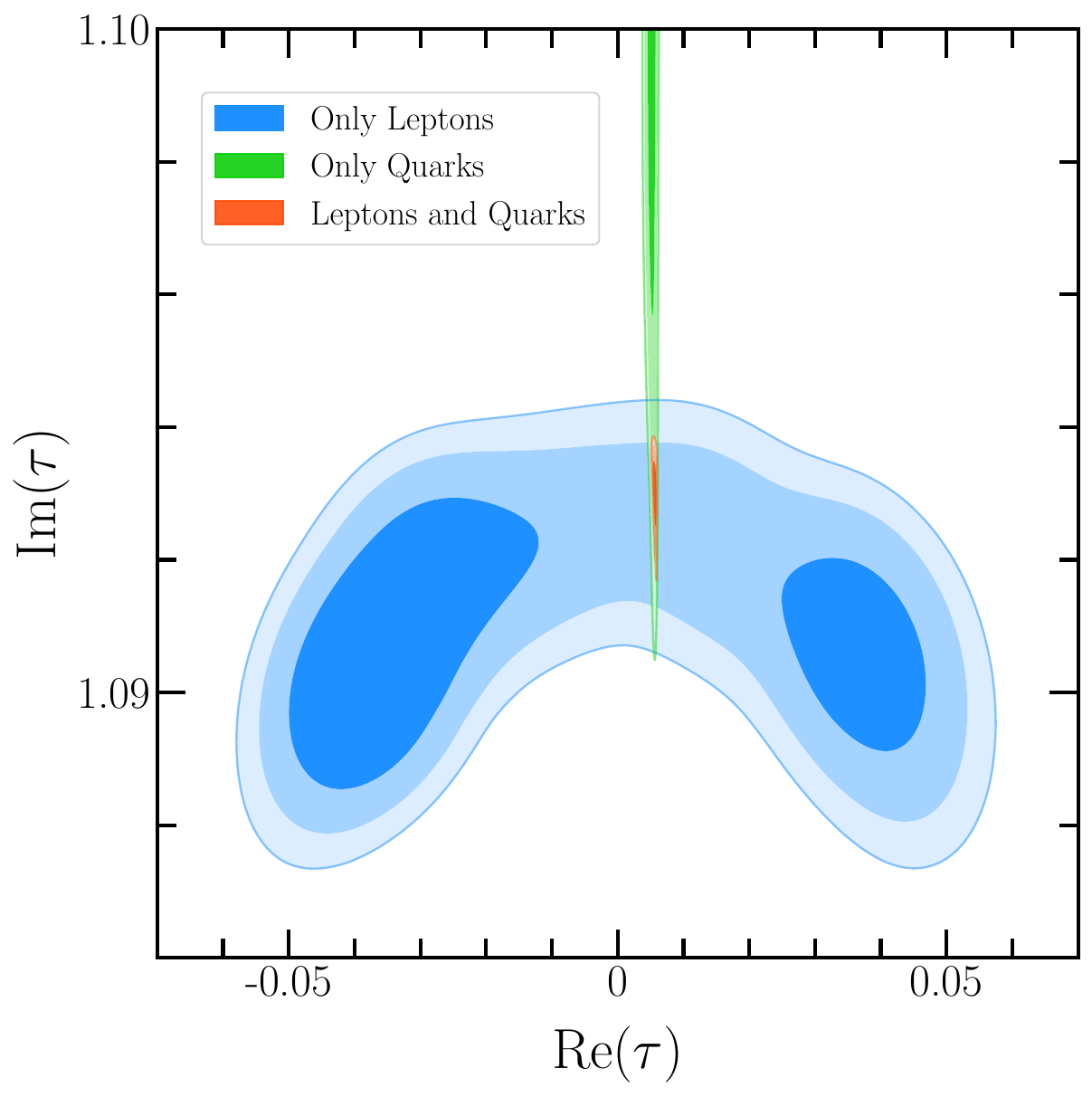}
\caption{Allowed regions in the $\tau$ plane from the lepton-only (blue), quark-only (green),
and combined (red) analyses. Different color shadings correspond to the $1\sigma$, $2\sigma$, and $3\sigma$ confidence levels.}
\label{fig:tau}
\end{figure}

We now investigate the parameter space of the model with gCP to identify potential correlations between input parameters and fermion masses and mixings. In section~\ref{subsec:lepton_model-seesaw} and section~\ref{sec:quark_model_8para}, we have independently studied the lepton and quark sectors, requiring all observables lie in their experimental $3\sigma$ ranges. In the unified model, the quark and lepton sectors share the same modulus \(\tau\). By plotting the allowed regions of \(\tau\) for each sector, we confirm an overlapping region in the \(\tau\) plane, as shown in figure~\ref{fig:tau}, which is consistent with the best-fit result. The final
viable region of $\tau$ is located at the intersection of lepton-allowed and quark allowed regions.

\begin{figure}[hptb!]
\centering
\includegraphics[width=6.5in]{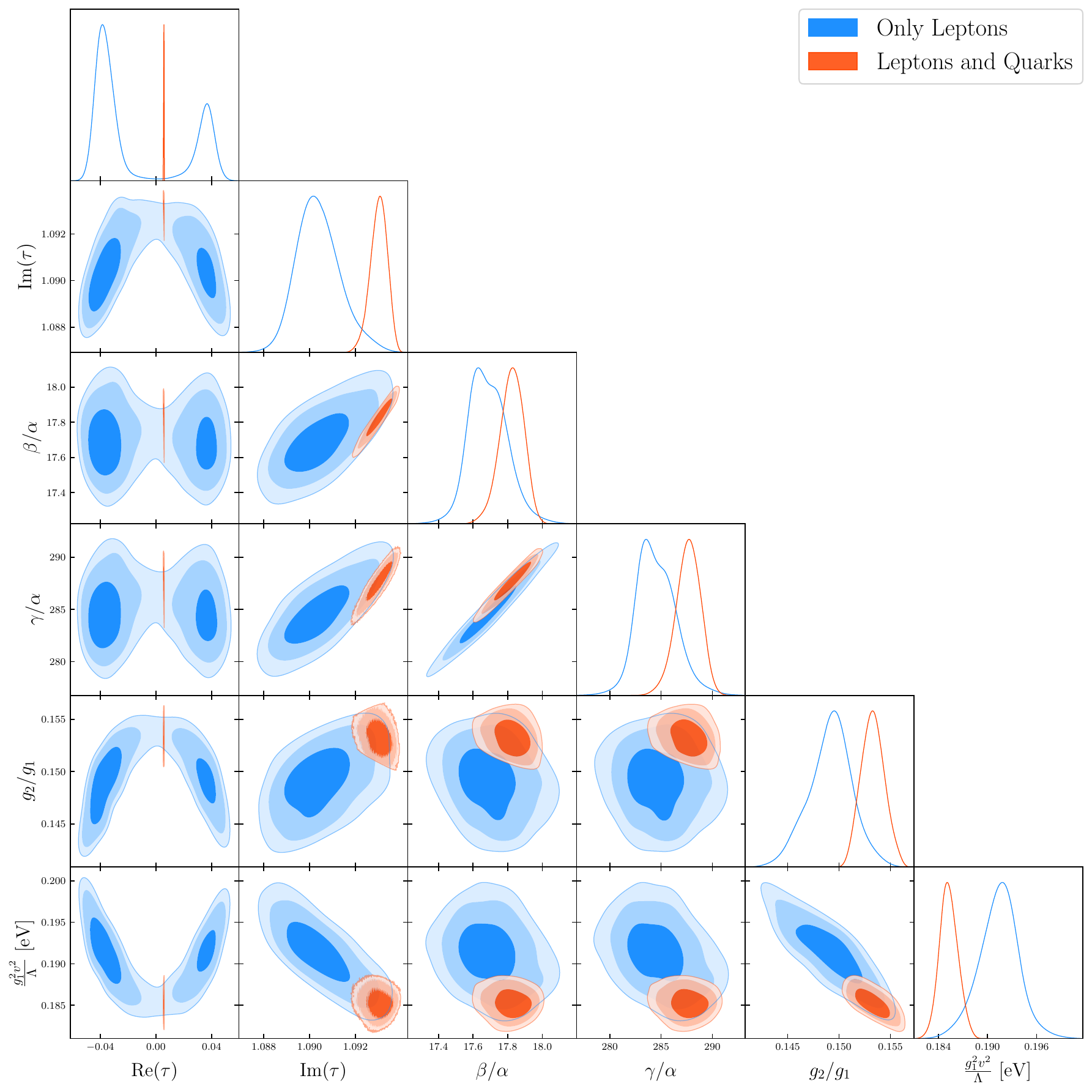}
\caption{Allowed regions for the lepton input parameters from the lepton-only (blue) and combined (red) analyses. Different color shadings correspond to the $1\sigma$, $2\sigma$, and $3\sigma$ confidence levels. }
\label{fig:lepton_unified_input}
\end{figure}

\begin{figure}[hptb!]
\centering
\includegraphics[width=6.5in]{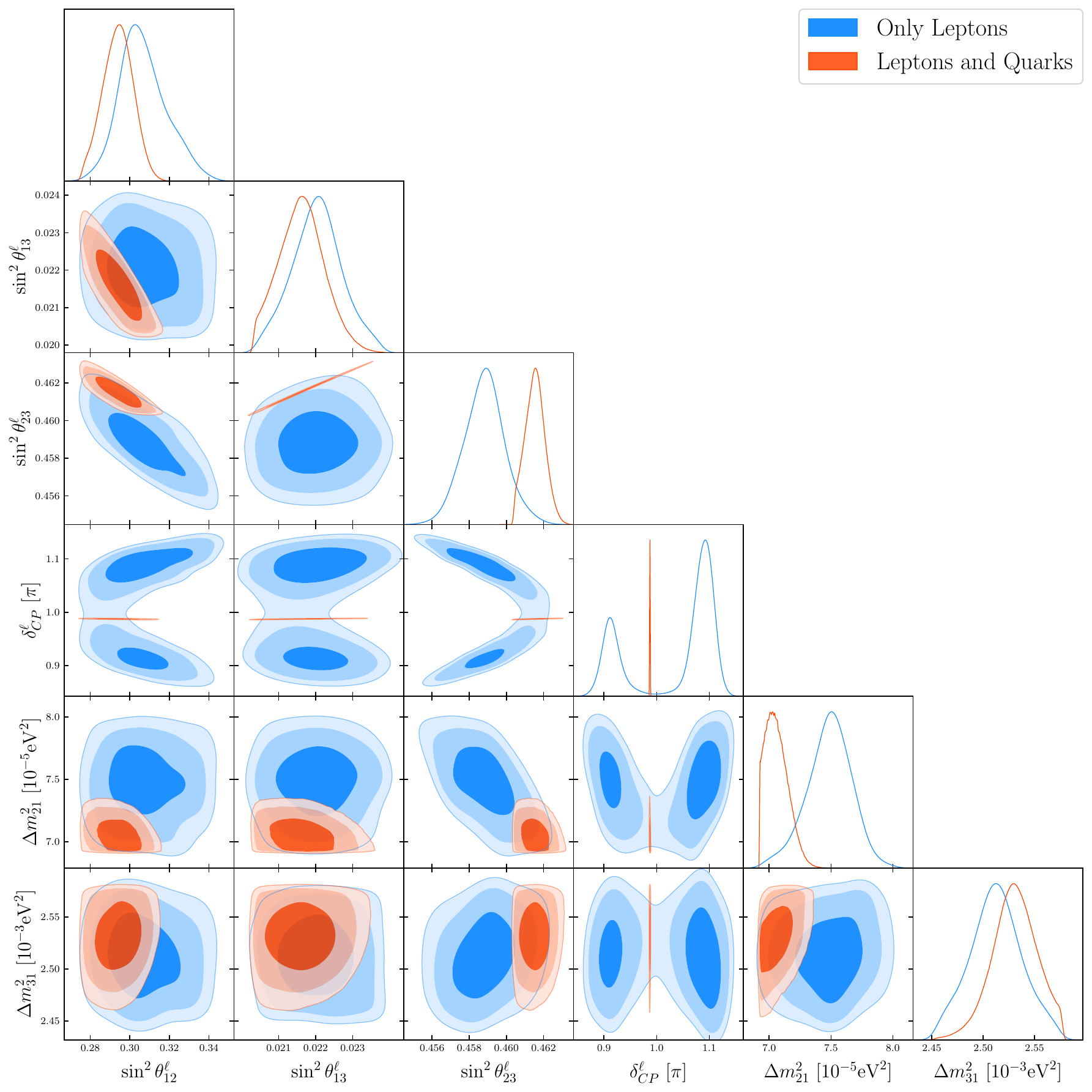}
\caption{Allowed regions for the lepton observables from the lepton-only (blue) and combined (red) analyses. Different color shadings correspond to the $1\sigma$, $2\sigma$, and $3\sigma$ confidence levels. }
\label{fig:lepton_unified_obs}
\end{figure}

In comparison with the lepton-only analysis, the allowed regions of input parameters in the lepton sector are significantly more constrained in the combined analysis, as can be seen from figure~\ref{fig:lepton_unified_input}. This effect is particularly evident in the one-dimensional projections, where the most notable shift occurs for \(\mathrm{Re}(\tau)\). The underlying reason is that imposing experimental constraints from both quark and lepton sectors greatly reduces the viable \(\tau\) region, and the quark data favor \(\tau\) in a narrow band near the imaginary axis, as shown in figure~\ref{fig:tau}.  A similar trend is observed in the predictions of flavor observables. As can be seen from figure~\ref{fig:lepton_unified_obs}, the red regions for lepton observables from the joint analysis are significantly smaller than the blue regions from the lepton-only analysis. In particular, the Dirac CP violation phase \(\delta_{CP}^{\ell}\) is tightly constrained around \(\pi\). This restriction arises from the smallness of \(\mathrm{Re}(\tau)\) in the allowed region of $\tau$ shown in figure~\ref{fig:tau}. In the presence of gCP, where all coupling constants are real, \(\mathrm{Re}(\tau)\) serves as the sole source of CP-violating phases. Furthermore, we observe a decrease in fit quality for \(\sin^{2}\theta_{12}^{\ell}\), \(\sin^{2}\theta_{23}^{\ell}\), and \(\Delta m_{21}^{2}\). The one-dimensional projections reveal noticeable shifts in the preferred values of the measured lepton observables, except for \(\sin^{2}\theta_{13}^{\ell}\) and \(\Delta m_{31}^{2}\). Moreover, a strong positive correlation is found between \(\sin^{2}\theta_{13}^{\ell}\) and \(\sin^{2}\theta_{23}^{\ell}\), they are almost linearly dependent in the combined analysis. Critical tests of the model’s viability will come from future high-precision measurements of \(\sin^2 \theta^{\ell}_{23}\) and \(\delta^{\ell}_{CP}\) at T2HK~\cite{Hyper-Kamiokande:2018ofw} and DUNE~\cite{DUNE:2020ypp}, as well as at the proposed ESS\(\nu\)SB experiment~\cite{Alekou:2022emd}.

\begin{figure}[hptb!]
\centering
\includegraphics[width=6.5in]{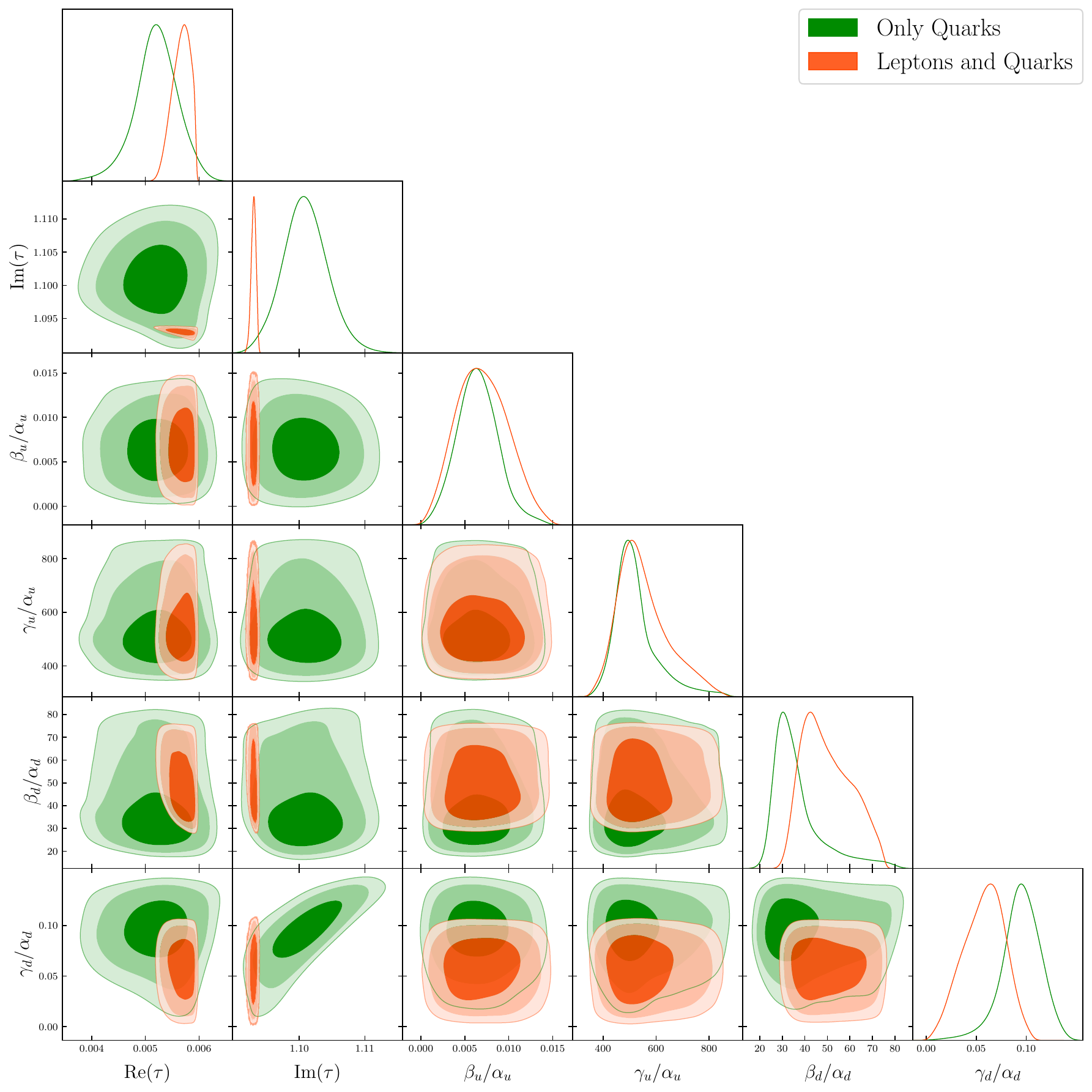}
\caption{Allowed regions of the free parameters from the quark-only (green) and combined (red) analyses for the quark model in section~\ref{sec:quark_model_8para}. Different color shadings correspond to the $1\sigma$, $2\sigma$, and $3\sigma$ confidence levels.}
\label{fig:quark_unified_input}
\end{figure}

\begin{figure}[hptb!]
\centering
\includegraphics[width=6.5in]{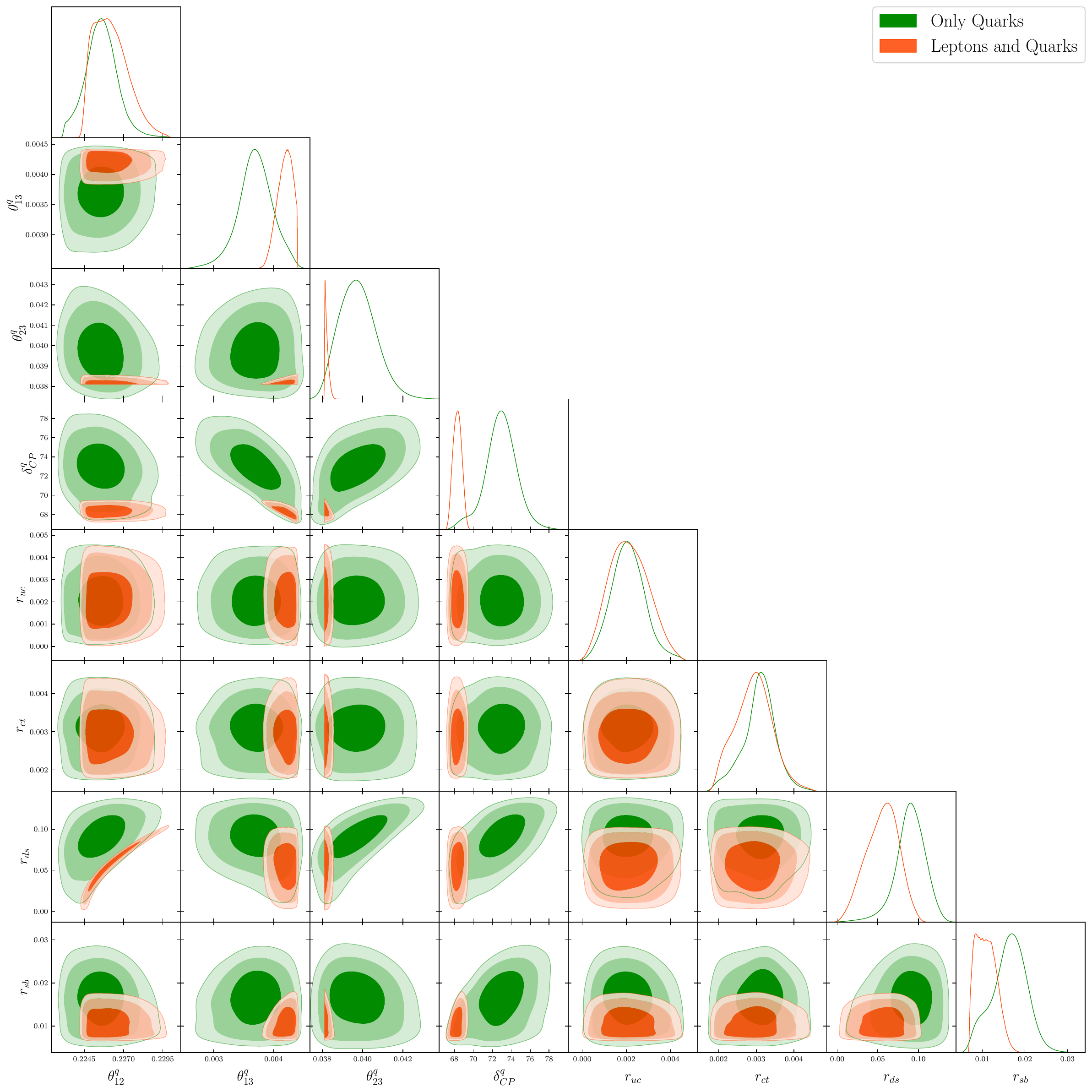}
\caption{Allowed regions of the quark observables from the quark-only (green) and combined (red) analyses for the quark model in section~\ref{sec:quark_model_8para}, where $r_{uc}=m_u/m_c$, $r_{ct}=m_c/m_t$, $r_{ds}=m_d/m_s$ and $r_{sb}=m_s/m_b$ denote the ratios of quark masses. Different color shadings correspond to the $1\sigma$, $2\sigma$, and $3\sigma$ confidence levels. }
\label{fig:quark_unified_obs}
\end{figure}

Similarly, as regards the input parameters in the quark sector, the red region for \(\tau\) in the combined analysis is considerably smaller than the green regions from the quark-only analysis, as illustrated in figure~\ref{fig:quark_unified_input}. Clear shifts in the preferred values are observed in the one-dimensional projections, except for \(\beta_{u}/\alpha_{u}\) and \(\gamma_{u}/\alpha_{u}\) which remain largely unchanged. Likewise we can see from figure~\ref{fig:quark_unified_obs} that the allowed regions of quark observables in the combined analysis get reduced with respect to the quark-only analysis. Particularly the regions of \(\theta_{13}^{q}\), \(\theta_{23}^{q}\), and \(\delta_{CP}^{q}\) become much smaller, due to the stricter constraint on \(\tau\). Additionally, we observe a strong correlation between \(\theta_{12}^{q}\) and \(r_{ds}\) in the combined analysis, which is absent in the quark-only scenario. Since \(\theta_{13}^{q}\) is strongly correlated with \(\mathrm{Re}(\tau)\), while \(\theta_{23}^{q}\) and $\delta_{CP}^{q}$ are positively correlated with $\mathrm{Im}(\tau)$, the one-dimensional projections of these parameters exhibit clear shifts in comparison with the quark-only case.

\begin{figure}[hptb!]
\centering
\includegraphics[width=6.5in]{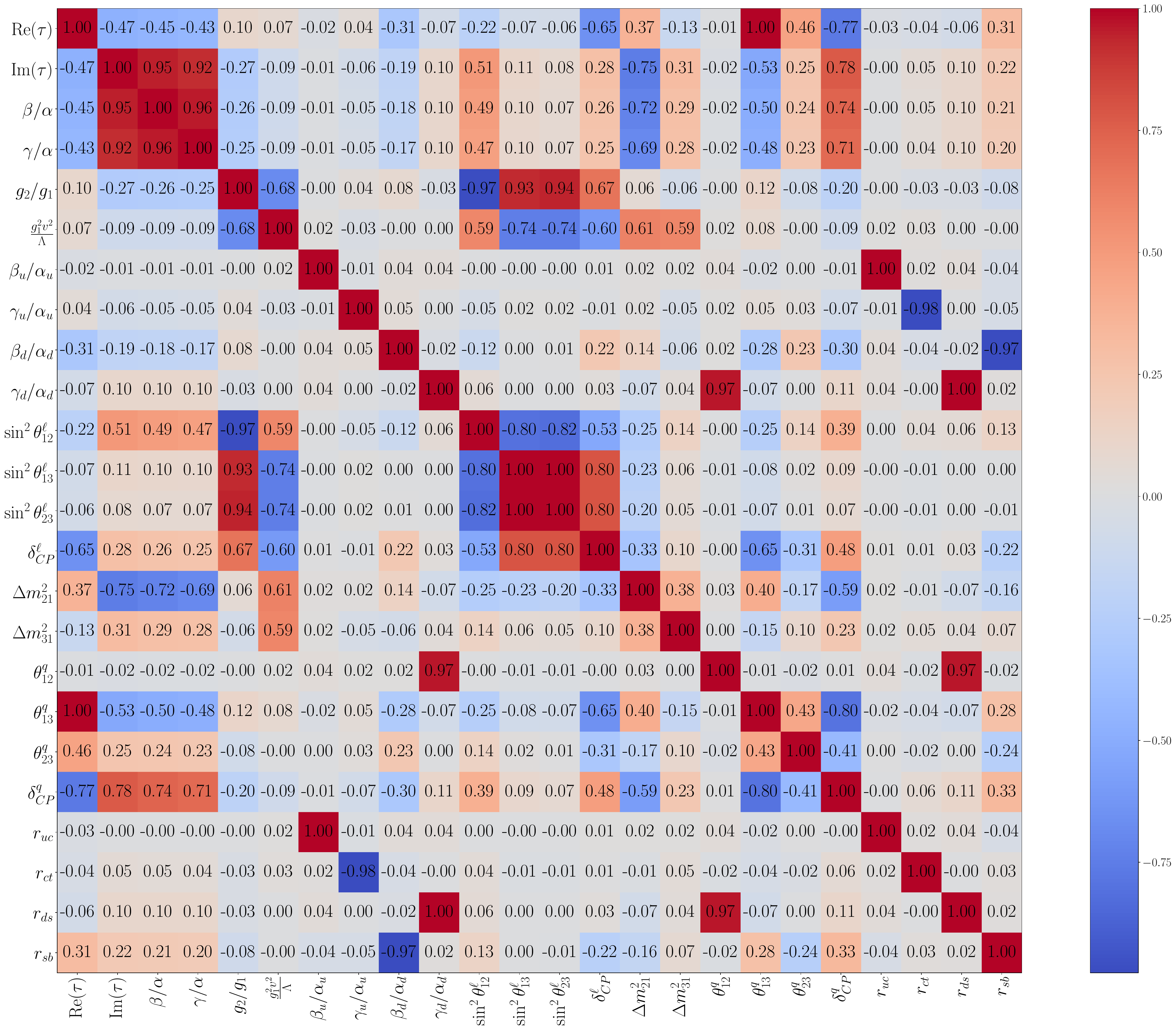}
\caption{Matrix of the correlations among the free parameters and flavor observables in the jonit analysis of the quark and lepton sectors. }
\label{fig:combined_correlation_matrix_input_obs}
\end{figure}

Because of the nontrivial interplay between the quark and lepton sectors, the separate (lepton-only and quark-only) and combined fittings give different correlations among model parameters and observables. We show the correlation matrix predicted by combined analysis in figure~\ref{fig:combined_correlation_matrix_input_obs}. For input parameters, we identify strong positive correlations between \(\mathrm{Im}(\tau)\), \(\beta/\alpha\), and \(\gamma/\alpha\). Additionally, moderate negative correlations are observed between \(\mathrm{Re}(\tau)\) and \(\mathrm{Im}(\tau)\), as well as between \(g_{2}/g_{1}\) and \(g^2_{1}v^2/\Lambda\). However, no clear correlations are found among the coupling constants in the quark sector, nor between the lepton and quark sector coupling constants.

As regards the flavor observables, we identify strong positive correlations in the lepton sector between \(\sin^{2}\theta_{13}^{\ell}\), \(\sin^{2}\theta_{23}^{\ell}\), and \(\delta_{CP}^{\ell}\), while \(\sin^{2}\theta_{12}^{\ell}\) exhibits a negative correlation with them. In the quark sector, the Cabibbo angle \(\theta_{12}^{q}\) is strongly positively correlated with the mass ratio \(r_{ds}\equiv m_d/m_s\), and a negative correlation  between \(\theta_{13}^{q}\) and \(\delta_{CP}^{q}\) is observed, as is evident in figure~\ref{fig:quark_unified_obs}. It is notable that correlations between lepton and quark observables also emerge. The Dirac CP violation phase \(\delta_{CP}^{\ell}\) has a moderate positive correlation with \(\delta_{CP}^{q}\) and a negative correlation with \(\theta_{13}^{q}\). Moreover, \(\Delta m_{21}^{2}\) is negatively correlated with \(\delta_{CP}^{q}\). These nontrivial correlations indicate potential interplay between the flavor structures of quarks and leptons.

We proceed to discuss the relations between the input parameters and flavor observables, we find that \(\text{Re}(\tau)\) exhibits a negative correlation with \(\delta_{CP}^{\ell}\) and \(\delta_{CP}^{q}\), while showing a strong positive correlation with \(\theta_{13}^{q}\). The observables \(\sin^{2}\theta_{12}^{\ell}\) and \(\delta_{CP}^{q}\) positively correlate with the input parameters \(\text{Im}(\tau)\), \(\beta/\alpha\), and \(\gamma/\alpha\),
nevertheless \(\Delta m_{21}^{2}\) and \(\theta_{13}^{q}\) are negatively correlated with \(\text{Im}(\tau)\), \(\beta/\alpha\), and \(\gamma/\alpha\). The coupling constants \(g_{2}/g_{1}\) and \(g^2_{1}v^2/\Lambda\) in neutrino masses also exhibit correlations with lepton observables. Specifically, \(g_{2}/g_{1}\) is positively correlated with \(\sin^{2}\theta_{13}^{\ell}\), \(\sin^{2}\theta_{23}^{\ell}\), and \(\delta_{CP}^{\ell}\), while negatively correlated with \(\sin^{2}\theta_{12}^{\ell}\). Likewise \(g^2_{1}v^2/\Lambda\) shows positive correlations with \(\sin^{2}\theta_{12}^{\ell}\), \(\Delta m_{21}^{2}\), and \(\Delta m_{31}^{2}\), but negative correlations with \(\sin^{2}\theta_{13}^{\ell}\), \(\sin^{2}\theta_{23}^{\ell}\), and \(\delta_{CP}^{\ell}\). Similar to the quark-only analysis, strong dependencies between quark mass ratios and Yukawa coupling constants persist. In particular, \(\theta_{12}^{q}\) exhibits a strong positive correlation with \(\gamma_{d}/\alpha_{d}\).

In summary, the results of the combined analysis are not identical to those from the separate lepton-only and quark-only analyses. This discrepancy arises from the interplay between lepton and quark observables in the joint fit. The combined analysis reveals an intricate structure of correlations between input parameters and flavor observables. The common modulus $\tau$ sharged by the quark and lepton sectors is the portal linking the lepton observables with quark observables, in particular the CP violation phases $\delta^{\ell}_{CP}$ and $\delta^q_{CP}$, although   the correlations between the lepton and quark sector couplings are somewhat weak in the present model.

\begin{table}[t!]
\centering
\begin{tabular}{| c | c | c |} \hline \hline
Inputs  & Without gCP & With gCP  \\ \hline
$\rm{Re}(\tau)$ & $0.005315$ & $0.005056$\\ $\rm{Im}(\tau)$ & $1.104$ & $1.093$\\ $\beta/\alpha$ & $5.084$ & $17.77$\\ $\gamma/\alpha$ & $0.004373$ & $286.8$\\ $|g_{2}/g_{1}|$ & $0.1647$ & $0.1535$\\ $\rm{arg}(g_{2}/g_{1})$ & $5.799$ & $0$\\ $\alpha v /\rm{GeV}$ & $0.2789$ & $0.004679$\\ $\frac{g_{1}^{2}v^{2}}{\Lambda}/\rm{meV}$ & $164.7$ & $185.3$\\ $\beta_{u}/\alpha_{u}$ & $0.006165$ & $0.006747$\\ $\gamma_{u}/\alpha_{u}$ & $497.3$ & $496.2$\\ $\beta_{d}/\alpha_{d}$ & $30.11$ & $29.12$\\ $\gamma_{d}/\alpha_{d}$ & $0.1131$ & $0.06111$\\ $\alpha_{u} v /\rm{GeV}$ & $0.3806$ & $0.3869$\\ $\alpha_{d} v /\rm{GeV}$ & $0.09457$ & $0.1018$\\
\hline \hline
\end{tabular}~~
\begin{tabular}{| c | c | c |} \hline \hline
Observables  & Without gCP & With gCP  \\ \hline
$\sin^2\theta^{\ell}_{12}$ & $0.308$ & $0.291$\\ $\sin^2\theta^{\ell}_{13}$ & $0.02205$ & $0.02158$\\ $\sin^2\theta^{\ell}_{23}$ & $0.443$ & $0.462$\\ $\delta^{\ell}_{CP}/\pi$ & $1.09$ & $0.99$\\ $\Delta m_{21}^2 / 10^{-5}\text{eV}^2$ & $7.57$ & $7.11$\\ $\Delta m_{31}^2 / 10^{-3}\text{eV}^2$ & $2.51$ & $2.53$\\ $m_e/m_\mu $ & $0.004737$ & $0.004737$\\ $m_\mu/m_\tau$ & $0.05882$ & $0.05882$\\ $\phi/\pi$ & $1.10$ & $0.99$\\ $m_{1}/\text{meV}$ & $0$ & $0$\\ $m_{2}/\text{meV}$ & $8.70$ & $8.43$\\ $m_{3}/\text{meV}$ & $50.11$ & $50.26$\\ $\sum m_{i}/\text{meV}$ & $58.81$ & $58.69$\\ $m_{\beta\beta}/\text{meV}$ & $2.13$ & $1.32$\\ $m_{\beta}/\text{meV}$ & $8.84$ & $8.65$\\ $\theta^{q}_{12}$ & $0.2256$ & $0.2263$\\ $\theta^{q}_{13}$ & $0.0037$ & $0.0037$\\ $\theta^{q}_{23}$ & $0.0406$ & $0.0374$\\ $\delta^{q}_{CP}/^{\circ}$ & $74.33$ & $70.52$\\ $m_u/m_c$ & $0.00204$ & $0.00204$\\ $m_c/m_t $ & $0.00318$ & $0.00318$\\ $m_d/m_s $ & $0.1066$ & $0.0588$\\ $m_s/m_b $ & $0.019$ & $0.018$\\ \hline $\chi^{2}_{\ell}$ & $4.57$ & $8.86$\\ $\chi^{2}_{q}$ & $8.77$ & $13.79$\\ $\chi^{2}_{tot}$ & $13.33$ & $22.65$\\
\hline \hline
\end{tabular}
\caption{\label{tab:combine_obs_prediction}The best fit results of the quark-lepton unified model. Left panel: the input parameters. Right panel: the predictions of the flavor observables. }
\end{table}

\section{Conclusion\label{sec:conclusion}}

The non-holomorphic modular flavor symmetry proposed in~\cite{Qu:2024rns} is an interesting extension of the paradigm of modular invariance, the Yukawa couplings are polyharmonic Maa{\ss} forms of level $N$ in this proposal. The even weight polyharmonic Maa{\ss} forms of level $N$ have been constructed from the known modular forms by using the differential operators $\xi_k$ and $D^{1-k}$~\cite{Qu:2024rns}. In the present work, we have extended the framework of the non-holomorphic modular invariance to include odd weight polyharmonic Maa{\ss} forms. However, the differential operators $\xi_k$ and $D^{1-k}$ are not sufficient to fix all the integer weight polyharmonic Maa{\ss} forms, especially the weight one polyharmonic Maa{\ss} forms. We propose a universal approach to construct the integer weight polyharmonic Maa{ss} forms of level $N$. We find that the non-holomorphic Eisenstein series can span the linear space of polyharmonic Maa{\ss} forms. We explicitly compute the $q$-expansion of these non-holomorphic Eisenstein series, allowing them to be evaluated numerically. In this approach, we don't need to determine which modular forms can be lifted to polyharmonic Maa{\ss} forms, as required when using differential operators.  Moreover, we find that the negative weight polyharmonic Maa{\ss} forms can only be decomposed into certain representations of $\Gamma'_N$. At level $N=3$, the negative even weight polyharmonic Maa{\ss} forms can be arranged into representations $\bm{1}$ and $\bm{3}$ of $T'$ while the negative odd weight ones decompose into the doublets $\bm{\widehat{2}}$ and $\bm{\widehat{2}''}$ of $T'$.
In the same fashion, we present the results for $N=4$ and $N=5$.
At level $4$, the negative even weight polyharmonic Maa{\ss} forms can be arranged into the representations $\bm{1}$, $\bm{2}$ and $\bm{3}$ of $S'_4$. The negative odd weight polyharmonic Maa{\ss} forms can be arranged into two triplets $\bm{\widehat{3}}$ and $\bm{\widehat{3}'}$ of $S'_4$.
At level $5$, the negative even weight polyharmonic Maa{\ss} forms can be arranged into the representations $\bm{1}$, $\bm{3}$, $\bm{3'}$ and $\bm{5}$ of $A'_5$. The negative odd weight polyharmonic Maa{\ss} forms can be arranged into two linearly independent sextets $\bm{\widehat{6}}$ of $A'_5$.

The non-holomorphic modular flavor symmetry with integer weight polyharmonic Maa{\ss} forms may open new avenues of understanding the flavor structure of SM. To illustrate this, we have built typical lepton and quark models based on the modular group \(\Gamma'_3 \cong T'\). The lepton model is presented in section~\ref{subsec:lepton_model-seesaw}, the light neutrino masses are generated by the type-I seesaw mechanism with two right-handed neutrinos transforming as a doublet $\hat{\bm{2}}$ of $T'$. This model features 8 (7) free real parameters in the absence (presence) of gCP symmetry. The experimental data of leptons can be successfully accommodated for both NO and IO neutrino mass spectrum, no matter whether gCP symmetry is included or not. The best fit values and the allowed $3\sigma$ regions for the input parameters and lepton observables are presented in tables~\ref{tab:lepton_obs_prediction} and~\ref{tab:lepton_obs_prediction_ranges}, respectively. For the quark sector, we propose two benchmark models in section~\ref{sec:quark_model}. The first quark model, with 7 real parameters, accurately predicts all quark masses and mixings except for \(m_s/m_b\) (see section~\ref{sec:quark_model_7para}). The second quark model, detailed in section~\ref{sec:quark_model_8para}, fully accommodates the experimental data of quark masses and CKM mixing matrix by using 8 real parameters, as summarized in table~\ref{tab:quark_obs_prediction}. Furthermore, we can unify the lepton model with the viable 8-parameter quark model to provide a comprehensive description of both leptons and quarks simultaneously, as shown in table~\ref{tab:combine_obs_prediction}.
The allowed regions of free parameters and flavor observables get reduced significantly with respect to the lepton only and quark-only analyses, and neutrino mass spectrum can only be NO. In this combined model, the lepton and quark sectors share a common modulus \(\tau\), leading to a total of 14 (13) parameters in the absence (presence) of gCP symmetry. Notably, in the presence of gCP, this quark-lepton unified model involves only 13 real parameters and it is the most minimal yet phenomenologically viable fermion models with modular invariance up to now. Since the number of free parameters is much small the number of observables, there are strong correlations among free input parameters and flavor observables.

In the present work, we have studied the finite modular group $\Gamma'_3\cong T'$, it is interesting to explore the application of holomorphic modular flavor symmetry to flavor puzzle of SM with the other finite modular group such as $\Gamma'_4$, $\Gamma'_5$ and $\Gamma'_6\cong S_3\times T'$ etc. The benchmark models in section~\ref{sec:models} is the based on the SM. The quarks and leptons are unified into very few irreducible representations of the gauge group in Grand Unified Theories (GUTs), so that the quark and lepton mass matrices are related. We expect that implementing the non-holomorphic modular symmetry in the context of GUTs could lead to more restrictive and more predictive theory of flavor.

We have been concerned with the bottom-up approach to the non-holomorphic modular flavor symmetry in this paper. However, the connection with top-down approach is still unclear, although there are some evidences that the polyharmonic Maa{\ss} forms could appear as coefficients in the low energy expansion of string theory compactification~\cite{Green:2008uj,Green:2010wi,DHoker:2015gmr} and low-energy supersymmetry may not be required for modular invariance~\cite{Cremades:2004wa,Almumin:2021fbk}. The meeting of bottom-up and top-down approaches can guide us toward more realistic and elegant flavor models.

\section*{Acknowledgements}

GJD and BYQ are supported by the National Natural Science Foundation of China under Grant No.~12375104 and Guizhou Provincial Major Scientific and Technological Program XKBF(2025)010. JNL is supported by the Grants No. NSFC-12147110 and the China Post-doctoral Science Foundation under Grant No. 2021M70.

\clearpage

\begin{appendix}

\section{\label{app:cusp} Cusps of $\Gamma(N)$}

In this section, we describe the cusp of the principal congruence subgroup $\Gamma(N)$. The cusps of $\Gamma(N)$ are the set of $\Gamma(N)$-equivalence classes of $\mathbb{Q}\cup \{i\infty\}$, where $\mathbb{Q}$ denoted the set of rational numbers. We use two coprime number $(A,C)$ to describe a cusp $\overline{A/C}$ which is a subset of $\mathbb{Q}\cup \{i\infty\}$ and its elements are related to the rational number $A/C$ by $\Gamma(N)$ transformations. We let $\mathcal{C}(N)$ denote a system of representatives of the cusps of $\Gamma(N)$. The action of a modular transformation $\gamma=\begin{pmatrix}
a & b\\
c & d
\end{pmatrix}$ on $\mathbb{Q}\cup \{i\infty\}$ comes from the action of ${\rm SL}(2,\mathbb{Z})$ on a two-dimensional column vector.
Let $s=A/C$ be an element of $\mathbb{Q}\cup \{i\infty\}$ with $\gcd(A, C) = 1$, one has
\begin{eqnarray}
\gamma \begin{pmatrix}
A \\ C
\end{pmatrix} =
\begin{pmatrix}
a ~&~ b \\
c ~&~ d
\end{pmatrix}
\begin{pmatrix}
A \\ C
\end{pmatrix}
= \begin{pmatrix}
a A + b C \\ c A + d C
\end{pmatrix}\equiv \begin{pmatrix}
A' \\ C'
\end{pmatrix} \,.
\end{eqnarray}
We can see that $A'$ and $C'$ are also coprime so that $s'= A'/C'$ is another element of $\mathbb{Q}\cup \{i\infty\}$. Hence the modular symmetry acts on the cusps via the linear fraction transformation Eq.~\eqref{eq:mobius-trans} with $\gamma(A/C)=A'/C'$,
thus we have $\gamma(i\infty)=a/c$.

The number of cusps of $\Gamma(N)$ reads as~\cite{cohen2017modular}
\begin{eqnarray}
|\mathcal{C}(N)| = \begin{cases}
1\,,~~~& N = 1 \,,\\
3 \,,~~~ & N = 2  \,,\\
\dfrac{1}{2}N^2 \prod_{p|N} \left( 1 - \dfrac{1}{p^2} \right) \,,~~~ & N > 2 \,,
\end{cases}
\end{eqnarray}
where the product is over the prime divisors $p$ of $N$.
To be more explicit, we list the cusps of $\Gamma(N)$ for the level $N\leq7$ in the following:
\begin{eqnarray}
\nonumber \mathcal{C}(1) &=& \{\overline{1/0} \} \,,~~~ \mathcal{C}(2) = \{ \overline{1/0}, \overline{0/1}, \overline{1/1} \} \,, \\
\nonumber \mathcal{C}(3) &=& \{ \overline{1/0}, \overline{0/1}, \overline{1/1}, \overline{2/1} \} \,,  \\
\nonumber \mathcal{C}(4) &=& \{ \overline{1/0}, \overline{0/1}, \overline{1/1}, \overline{2/1}, \overline{3/1}, \overline{1/2} \} \,,  \\
\nonumber \mathcal{C}(5) &=& \{ \overline{1/0}, \overline{0/1}, \overline{1/1}, \overline{2/1}, \overline{3/1}, \overline{4/1}, \overline{1/2}, \overline{3/2}, \overline{5/2}, \overline{7/2}, \overline{9/2}, \overline{2/5} \} \,,  \\
\nonumber \mathcal{C}(6) &=& \{ \overline{1/0}, \overline{0/1}, \overline{1/1}, \overline{2/1}, \overline{3/1}, \overline{4/1}, \overline{5/1}, \overline{1/2}, \overline{3/2}, \overline{5/2}, \overline{1/3}, \overline{2/3} \} \,,  \\
\nonumber \mathcal{C}(7) &=& \{ \overline{1/0}, \overline{0/1}, \overline{1/1}, \overline{2/1}, \overline{3/1}, \overline{4/1}, \overline{5/1}, \overline{6/1}, \overline{1/2}, \overline{3/2}, \overline{5/2}, \overline{7/2}, \\
&& ~~ \overline{9/2}, \overline{11/2}, \overline{13/2}, \overline{1/3}, \overline{2/3}, \overline{4/3}, \overline{5/3}, \overline{7/3}, \overline{10/3}, \overline{13/3}, \overline{2/7}, \overline{3/7} \}  \,. \label{eq:cusps-N}
\end{eqnarray}
Note that $\overline{1/0}$ stands for the cusp $i\infty$.

\section{\label{app:math}Dirichlet character and $L$-function}

A Dirichlet character $\chi$ of modulus $N$
is a function on the set of integers that satisfies the following conditions~\cite{apostol1998introduction}:
\begin{eqnarray}
\nonumber &&\chi(a+N) = \chi(a)\,,~~~~\text{for all }a\in\mathbb{Z} \,, \\
\nonumber &&\chi(a) = 0  \,,~~~~ \text{if and only if }\gcd(a,N)>1 \,, \\
&&\chi(ab) = \chi(a) \chi(b) \,,~~~~ \text{for }a,b \in \mathbb{Z} \,,\label{eq:Dirichlet-cha-def}
\end{eqnarray}
where $\gcd(a,N)$ denotes the greatest common divisor of $a$ and $N$. In other words, a Dirichlet character of modulus $N$ is an arithmetic function that is not identically equal to zero, and that is totally multiplicative and periodic with the period $N$. There could be more than one Dirichlet character $\chi$ of modulus $N$ satisfying Eq.~\eqref{eq:Dirichlet-cha-def}, and the  number of distinct Dirichlet characters modulo $N$ equals the Euler's totient function $\varphi(N) =  N \prod_{p|N} (1-1/p)$, which is the number of positive integers not exceeding $N$ which are relatively prime to $N$~\cite{apostol1998introduction}. The simplest character modulo $N$ is the principal character $\chi_1$ which is defined as
\begin{eqnarray}
\chi_1(a) = \begin{cases}
1 \,,~~& \text{if } \gcd(N,a) = 1 \\
0\,,~~& \text{if } \gcd(N,a)\neq 1\,.
\end{cases}
\end{eqnarray}
When $N=1$ or $N=2$, then the only Dirichlet character is the principal character $\chi_1$. For $N=3$, there are two Dirichlet characters,
\begin{eqnarray}
\chi_1(n) = \begin{cases}
1 \,, ~~ &n\equiv \pm 1 \,~ ({\rm mod}\, 3) \\
0 \,, ~~ &n\equiv 0 \,~ ({\rm mod}\, 3)
\end{cases}\,,~~~
\chi_2(n) = \begin{cases}
1 \,, ~~ &n\equiv 1 \,~ ({\rm mod}\, 3) \\
-1 \,, ~~ &n\equiv 2 \,~ ({\rm mod}\, 3) \\
0 \,, ~~ &n\equiv 0 \,~ ({\rm mod}\, 3)
\end{cases}
\end{eqnarray}
There are also two Dirichlet characters for $N=4$ as follow,
\begin{eqnarray}
\chi_1(n) = \begin{cases}
1 \,, ~~ &n\equiv \pm 1 \,~ ({\rm mod}\, 4) \\
0 \,, ~~ &\text{other cases}
\end{cases}\,,~~~
\chi_2(n) = \begin{cases}
1 \,, ~~ &n\equiv 1 \,~ ({\rm mod}\, 4) \\
-1 \,, ~~ &n\equiv 3 \,~ ({\rm mod}\, 4) \\
0 \,, ~~ &\text{other cases}
\end{cases}
\end{eqnarray}
There are four possible Dirichlet characters for $N=5$,
\begin{eqnarray}
\chi_1(n) &=& \begin{cases}
0 \,, ~~ & n\equiv 0 \,~ ({\rm mod}\, 5) \\
1 \,, ~~ & \text{other cases}
\end{cases}\,,~~~
\chi_3(n) = \begin{cases}
0 \,, ~~ & n\equiv 0 \,~ ({\rm mod}\, 5) \\
1 \,, ~~ & n\equiv \pm 1 \,~ ({\rm mod}\, 5) \\
-1 \,, ~~ & n\equiv \pm 2 \,~ ({\rm mod}\, 5)
\end{cases}   \\
\chi_2(n) &=& \begin{cases}
0 \,, ~~ & n\equiv 0 \,~ ({\rm mod}\, 5) \\
1 \,, ~~ & n\equiv 1 \,~ ({\rm mod}\, 5) \\
i \,, ~~ & n\equiv 2 \,~ ({\rm mod}\, 5) \\
-i \,, ~~ & n\equiv 3 \,~ ({\rm mod}\, 5) \\
-1 \,, ~~ & n\equiv 4 \,~ ({\rm mod}\, 5)
\end{cases}\,,~~~
\chi_4(n) = \begin{cases}
0 \,, ~~ & n\equiv 0 \,~ ({\rm mod}\, 5) \\
1 \,, ~~ & n\equiv 1 \,~ ({\rm mod}\, 5) \\
-i \,, ~~ & n\equiv 2 \,~ ({\rm mod}\, 5) \\
i \,, ~~ & n\equiv 3 \,~ ({\rm mod}\, 5) \\
-1 \,, ~~ & n\equiv 4 \,~ ({\rm mod}\, 5)
\end{cases}
\end{eqnarray}
Similar to the Riemann zeta function $\zeta(s)$, the Dirichlet $L$ function is defined as~\cite{apostol1998introduction,cohen2017modular}
\begin{eqnarray}
L(s,\chi) = \sum_{n=1}^{+\infty} \dfrac{\chi(n)}{n^s}\,.
\end{eqnarray}
As a special case, the $L$-function of the principal character $\chi_1$ of modulus $N$ can be expressed in terms of the Riemann zeta function
\begin{eqnarray}
L(s,\chi_1) = \zeta(s) \prod_{p|N} (1-p^{-s}) \,,
\end{eqnarray}
where $p$ is the prime factor of $N$. The Dirichlet $L$-function can be conveniently calculated by the \texttt{Wolfram Mathematica} function \texttt{DirichletL[$N$,$i$, $s$]} which gives $L(s,\chi_i)$ for the Dirichlet character $\chi_i$ of modulus $N$.

\section{\label{app:Gamma3-MF3} Finite modular groups $\Gamma'_N$ }

\subsection{\label{app:group-MF-N3} $N=3$}

The homogeneous finite modular group $\Gamma'_3\cong T'$ is the double covering of the inhomogeneous finite modular group $\Gamma_3\cong A_4$ which is the symmetry group a regular tetrahedron. It can be generated by three generators $S$, $T$ and $R$ satisfying the following multiplication rules~\cite{Liu:2019khw,Kobayashi:2022moq}
\begin{equation}
S^{2}=R,~~ (ST)^{3}=T^{3}=R^{2}=1,~~RT = TR\,.
\end{equation}
Obviously the generator $R$ commutes with all elements of the group. Notice that $A_4$ is not a subgroup of $T'$, yet it is isomorphic to the quotient group $T'/Z^R_2$, where $Z^R_2$ refers to the cyclic group generated by $R$. The $T'$ group has three singlets representations $\bm{1}$, $\bm{1'}$ and $\bm{1''}$, three doublet representations $\bm{\widehat{2}}$, $\bm{\widehat{2}'}$ and $\bm{\widehat{2}''}$, and a triplet representation $\bm{3}$. The explicit form of the generators $S$, $T$ and $R$ in each of the irreducible representations are given by~\cite{Ding:2022aoe}:
\begin{eqnarray}
\label{eq:irr-Tp} \begin{array}{cccc}
\bm{1:} & S=1\,, ~&~ T=1\,,~&~ S=1\,, \\
\bm{1'}:& S=1\,, ~&~ T=\omega\,,~&~ S=1\,, \\
\bm{1''}: & S=1\,, ~&~ T=\omega^{2}\,,~&~ S=1 \,,\\
\bm{\widehat{2}:} ~&~ S=-\frac{i}{\sqrt{3}}
\begin{pmatrix}
1  ~& \sqrt{2}  \\
\sqrt{2}  ~&  -1
\end{pmatrix}\,, ~&~ T=\left(\begin{array}{cc}
\omega & 0 \\
0 & 1
\end{array}\right) \,, ~&~ R=-\left(\begin{array}{cc}
1 & 0 \\
0 & 1
\end{array}\right)\,,\\
\bm{\widehat{2}'}: ~&~ S=-\frac{i}{\sqrt{3}}
\begin{pmatrix}
1  ~& \sqrt{2}  \\
\sqrt{2}  ~&  -1
\end{pmatrix}\,, ~&~ T=\left(\begin{array}{cc}
\omega^{2} & 0 \\
0 & \omega
\end{array}\right)\,, ~&~ R=-\left(\begin{array}{cc}
1 & 0 \\
0 & 1
\end{array}\right) \,,\\
\bm{\widehat{2}''}: ~&~ S=-\frac{i}{\sqrt{3}}
\begin{pmatrix}
1  ~& \sqrt{2}  \\
\sqrt{2}  ~&  -1
\end{pmatrix}\,, ~&~ T=\left(\begin{array}{cc}
1 & 0 \\
0 & \omega^{2}
\end{array}\right)\,, ~&~ R=-\left(\begin{array}{cc}
1 & 0 \\
0 & 1
\end{array}\right) \,,\\
\bm{3}: ~&~ S=\frac{1}{3}\left(\begin{array}{ccc}
-1&2&2\\
2&-1&2\\
2&2&-1
\end{array}\right)\,,
~&~ T=\left(
\begin{array}{ccc}
1~&0~&0\\
0~&\omega~&0\\
0~&0~&\omega^2
\end{array}\right)\,, ~&~ R=\left(\begin{array}{ccc}
1 & 0 & 0 \\
0 & 1 & 0 \\
0 & 0 & 1
\end{array}\right) \,,
\end{array}
\end{eqnarray}
with $\omega=e^{i2\pi/3}=-1/2+i\sqrt{3}/2$. The two groups $T'$ and $A_4$ are represented by the same set of matrices in the irreducible representations $\bm{1}$, $\bm{1'}$, $\bm{1''}$ and $\bm{3}$. All the three doublet representations $\bm{\widehat{2}}$, $\bm{\widehat{2}'}$, $\bm{\widehat{2}''}$ are faithful representations of $T'$, $\bm{\widehat{2}}$ and $\bm{\widehat{2}''}$ are complex conjugated to each other, while $\bm{\widehat{2}'}$ and its complex conjugation are equivalent. The tensor products between different irreducible representations of $T'$ are given by
\begin{eqnarray}
\nonumber&&\bm{1}\otimes\bm{1} = \bm{1'}\otimes \bm{1''}=\bm{1}\,,~~\bm{1}\otimes\bm{1'} = \bm{1''}\otimes \bm{1''}=\bm{1'}\,,~~\bm{1}\otimes\bm{1''} = \bm{1'}\otimes \bm{1'}=\bm{1''}\,, \\
\nonumber&&\bm{1}\otimes\bm{\widehat{2}} = \bm{1'}\otimes \bm{\widehat{2}''}= \bm{1''}\otimes \bm{\widehat{2}'}=\bm{\widehat{2}}\,,~~\bm{1}\otimes\bm{\widehat{2}'} = \bm{1'}\otimes \bm{\widehat{2}}= \bm{1''}\otimes \bm{\widehat{2}''}=\bm{\widehat{2}'}\,,\\
\nonumber&&\bm{1}\otimes\bm{\widehat{2}''} = \bm{1'}\otimes \bm{\widehat{2}'}= \bm{1''}\otimes \bm{\widehat{2}}=\bm{\widehat{2}''}\,,~~\bm{1}\otimes\bm{3}=\bm{1'}\otimes\bm{3}=\bm{1''}\otimes\bm{3}=\bm{3}
\,, \\
\nonumber&&\bm{\widehat{2}}\otimes\bm{\widehat{2}}=\bm{\widehat{2}'}\otimes \bm{\widehat{2}''}=\bm{1'}\oplus\bm{3}\,,~~\bm{\widehat{2}}\otimes\bm{\widehat{2}'}=\bm{\widehat{2}''}\otimes\bm{\widehat{2}''}=\bm{1''}\oplus\bm{3}\,,~~\bm{\widehat{2}}\otimes\bm{\widehat{2}''}=\bm{\widehat{2}'}\otimes\bm{\widehat{2}'}=\bm{1}\oplus \bm{3}\,,\\
&&\bm{\widehat{2}}\otimes\bm{3}=\bm{\widehat{2}'}\otimes\bm{3}=\bm{\widehat{2}''}\otimes\bm{3}=\bm{\widehat{2}}\oplus\bm{\widehat{2}'}\oplus\bm{\widehat{2}''}\,,~~\bm{3} \otimes \bm{3} =  \bm{1} \oplus \bm{1'} \oplus \bm{1''}\oplus\bm{3}_S \oplus \bm{3}_A\,,
\end{eqnarray}
where $\bm{3}_S$ and $\bm{3}_A$ stand for the symmetric and antisymmetric triplet contractions respectively. The corresponding Clebsch-Gordon (CG) coefficients are listed in table~\ref{tab:Tprime_CG}.

\begin{table}[hptb!]
\centering \resizebox{1.0\textwidth}{!}{
\begin{tabular}{|c|c|c|c|c|c|c|c|c|c|c|c|} \hline
\hline
\multicolumn{4}{|c}{$\bm{1}\otimes\bm{\widehat{2}} = \bm{1'}\otimes \bm{\widehat{2}''}= \bm{1''}\otimes \bm{\widehat{2}'}=\bm{\widehat{2}}$} & \multicolumn{4}{|c}{$\bm{1}\otimes\bm{\widehat{2}'} = \bm{1'}\otimes \bm{\widehat{2}}= \bm{1''}\otimes \bm{\widehat{2}''}=\bm{\widehat{2}'}$} & \multicolumn{4}{|c|}{$\bm{1}\otimes\bm{\widehat{2}''} = \bm{1'}\otimes \bm{\widehat{2}'}= \bm{1''}\otimes \bm{\widehat{2}}=\bm{\widehat{2}''}$} \\ \hline
\multicolumn{4}{|c}{$\bm{\widehat{2}}:
~\alpha\left(\begin{array}{c}\beta_1\\
\beta_2 \\ \end{array}\right)$} &
\multicolumn{4}{|c}{$\bm{\widehat{2}}':
~\alpha\left(\begin{array}{c}\beta_1\\
\beta_2 \\ \end{array}\right)$} &
\multicolumn{4}{|c|}{$\bm{\widehat{2}}'':
~\alpha\left(\begin{array}{c}\beta_1\\
\beta_2 \\ \end{array}\right)$} \\ \hline
\multicolumn{4}{|c}{$\bm{1}\otimes\bm{3} = \bm{3}$} & \multicolumn{4}{|c}{$\bm{1}'\otimes\bm{3} = \bm{3}$}&\multicolumn{4}{|c|}{$\bm{1}''\otimes\bm{3} = \bm{3}$}  \\ \hline
\multicolumn{4}{|c}{$\bm{3}:~ \alpha \left(\begin{array}{c}\beta_1\\
\beta_2 \\
\beta_3 \end{array}\right)$} &
\multicolumn{4}{|c}{$\bm{3}:~ \alpha \left(\begin{array}{c}\beta_3\\
\beta_1 \\
\beta_2 \end{array}\right)$} &
\multicolumn{4}{|c|}{$\bm{3}:~ \alpha \left(\begin{array}{c}\beta_2\\
\beta_3 \\
\beta_1 \end{array}\right)$}  \\ \hline
\multicolumn{4}{|c}{$\bm{\widehat{2}}\otimes\bm{\widehat{2}}=\bm{\widehat{2}}'\otimes \bm{\widehat{2}}''=\bm{1}'\oplus\bm{3}$} & \multicolumn{4}{|c}{$\bm{\widehat{2}}\otimes\bm{\widehat{2}}'=\bm{\widehat{2}}''\otimes\bm{\widehat{2}}''
=\bm{1}''\oplus\bm{3}$} & \multicolumn{4}{|c|}{$\bm{\widehat{2}}\otimes\bm{\widehat{2}}''=\bm{\widehat{2}}'\otimes\bm{\widehat{2}}'=\bm{1}\oplus \bm{3}$}  \\ \hline
\multicolumn{4}{|c}{$\begin{array}{l}
\bm{1}':~\alpha_1\beta_2-\alpha_2\beta_1 \\
\bm{3}:~\left(\begin{array}{c}\alpha_2\beta_2 \\
\frac{1}{\sqrt{2}}(\alpha_1\beta_2+\alpha_2\beta_1)  \\
-\alpha_1\beta_1 \end{array}\right)
\end{array}$ } &
\multicolumn{4}{|c}{  $\begin{array}{l}
\bm{1}'':~\alpha_1\beta_2-\alpha_2\beta_1\\
\bm{3}:~
 \left(\begin{array}{c} -\alpha_1\beta_1\\
\alpha_2\beta_2 \\
 \frac{1}{\sqrt{2}}(\alpha_1\beta_2+\alpha_2\beta_1) \end{array}\right)
\end{array}$ } &
\multicolumn{4}{|c|}{  $\begin{array}{l}
\bm{1}:~\alpha_1\beta_2-\alpha_2\beta_1 \\
\bm{3}:~
\left(\begin{array}{c}  \frac{1}{\sqrt{2}}(\alpha_1\beta_2+\alpha_2\beta_1)\\
 -\alpha_1\beta_1\\
\alpha_2\beta_2  \end{array}\right)
\end{array}$ } \\ \hline
\multicolumn{4}{|c}{$\bm{\widehat{2}}\otimes\bm{3}=\bm{\widehat{2}}\oplus\bm{\widehat{2}}'\oplus\bm{\widehat{2}}''$} & \multicolumn{4}{|c}{$\bm{\widehat{2}}'\otimes\bm{3}=\bm{\widehat{2}}\oplus\bm{\widehat{2}}'\oplus\bm{\widehat{2}}''$} & \multicolumn{4}{|c|}{$\bm{\widehat{2}}''\otimes\bm{3}=\bm{\widehat{2}}\oplus\bm{\widehat{2}}'\oplus\bm{\widehat{2}}''$}  \\ \hline
\multicolumn{4}{|c}{  $\begin{array}{l}
\bm{\widehat{2}}:~
 \left(\begin{array}{c}\alpha_1\beta_1+\sqrt{2}\alpha_2\beta_2 \\
 -\alpha_2\beta_1+\sqrt{2}\alpha_1\beta_3 \end{array}\right)  \\ [0.1in]
\bm{\widehat{2}}':~
\left(\begin{array}{c}\alpha_1\beta_2+\sqrt{2}\alpha_2\beta_3 \\
-\alpha_2\beta_2+\sqrt{2}\alpha_1\beta_1  \end{array}\right) \\ [0.1in]
\bm{\widehat{2}}'':~
\left(\begin{array}{c}\alpha_1\beta_3+\sqrt{2}\alpha_2\beta_1 \\
-\alpha_2\beta_3+\sqrt{2}\alpha_1\beta_2  \end{array}\right) \\ [0.1in]
\end{array}$ } &
\multicolumn{4}{|c}{  $\begin{array}{l}
\bm{\widehat{2}}:~
\left(\begin{array}{c}\alpha_1\beta_3+\sqrt{2}\alpha_2\beta_1 \\
-\alpha_2\beta_3+\sqrt{2}\alpha_1\beta_2  \end{array}\right)  \\ [0.1in]
\bm{\widehat{2}}':~
 \left(\begin{array}{c}\alpha_1\beta_1+\sqrt{2}\alpha_2\beta_2 \\
 -\alpha_2\beta_1+\sqrt{2}\alpha_1\beta_3 \end{array}\right) \\ [0.1in]
\bm{\widehat{2}}'':~
\left(\begin{array}{c}\alpha_1\beta_2+\sqrt{2}\alpha_2\beta_3 \\
-\alpha_2\beta_2+\sqrt{2}\alpha_1\beta_1  \end{array}\right) \\ [0.1in]
\end{array}$ } &
\multicolumn{4}{|c|}{  $\begin{array}{l}
\bm{\widehat{2}}:~
\left(\begin{array}{c}\alpha_1\beta_2+\sqrt{2}\alpha_2\beta_3 \\
-\alpha_2\beta_2+\sqrt{2}\alpha_1\beta_1  \end{array}\right)  \\ [0.1in]
\bm{\widehat{2}}':~
\left(\begin{array}{c}\alpha_1\beta_3+\sqrt{2}\alpha_2\beta_1 \\
-\alpha_2\beta_3+\sqrt{2}\alpha_1\beta_2  \end{array}\right) \\ [0.1in]
\bm{\widehat{2}}'':~
 \left(\begin{array}{c}\alpha_1\beta_1+\sqrt{2}\alpha_2\beta_2 \\
 -\alpha_2\beta_1+\sqrt{2}\alpha_1\beta_3 \end{array}\right) \\ [0.1in]
\end{array}$ } \\ \hline
\multicolumn{12}{|c|}{$\bm{3}\otimes\bm{3}=\bm{3}_S\oplus\bm{3}_A\oplus\bm{1}\oplus\bm{1}'\oplus\bm{1}''$}  \\ \hline
\multicolumn{12}{|c|}{  $\begin{array}{l}
\bm{3}_S:~
 \left(\begin{array}{c} 2\alpha_1\beta_1 -\alpha_2\beta_3 -\alpha_3\beta_2 \\
 2\alpha_3\beta_3 -\alpha_1\beta_2 -\alpha_2\beta_1  \\
 2\alpha_2\beta_2 -\alpha_1\beta_3 -\alpha_3\beta_1 \end{array}\right) \\ [0.1in]
\bm{3}_A:~
 \left(\begin{array}{c}\alpha_2\beta_3 -\alpha_3\beta_2 \\
\alpha_1\beta_2 -\alpha_2\beta_1  \\
\alpha_3\beta_1 -\alpha_1\beta_3 \end{array}\right) \\ [0.1in]
\bm{1} :~\alpha_1\beta_1 +\alpha_2\beta_3 +\alpha_3\beta_2 \\ [0.1in]
\bm{1}' :~\alpha_3\beta_3 +\alpha_1\beta_2 +\alpha_2\beta_1 \\ [0.1in]
\bm{1}'' :~\alpha_2\beta_2 +\alpha_1\beta_3 +\alpha_3\beta_1 \\ [0.1in]
\end{array}$ } \\
\hline
\end{tabular}}
\caption{\label{tab:Tprime_CG} Tensor products and the corresponding CG coefficients for the $T'$ group. Here $\alpha_i$ and $\beta_i$ denote the elements of the first and second representations respectively in the tensor product. }
\end{table}

\subsection{\label{app:group-MF-N4} $N=4$}

The homogeneous finite modular group $\Gamma'_4\cong S'_4$ is the double covering of $\Gamma_4\cong S_4$ which is the symmetry group of a cube. It has three generators $S$, $T$ and $R$ obeying the following relations:
\begin{equation}
S^2=R,~~(ST)^3=T^4=R^2=1,~~ TR=RT\,.
\end{equation}
The $S_4$ group is isomorphic to $S'_4/Z^R_2$, and it can be reproduced by setting $R=1$. The group $S'_4$ has four singlet representations $\bm{1}$, $\bm{1^{\prime}}$, $\bm{\widehat{1}}$ and $\bm{\widehat{1}^{\prime}}$, two doublet representations $\bm{2}$ and $\bm{\widehat{2}}$, and four triplet representations $\bm{3}$, $\bm{3^{\prime}}$, $\bm{\widehat{3}}$ and $\bm{\widehat{3}^{\prime}}$. The representation matrices of the generators $S$, $T$ and $R$ are given by
\begin{eqnarray}
\begin{array}{cccc}
\bm{1}: & S=1, ~&~ T=1\,, ~&~ R=1\,, \\
\bm{1^{\prime}}: ~&~ S=-1, ~&~ T=-1\,,~&~ R=1\,, \\
\bm{\widehat{1}}: ~&~ S=i, ~&~ T=-i \,,~&~ R=-1\,,\\
\bm{\widehat{1}^{\prime}}: ~&~ S=-i, ~&~ T=i\,,~&~ R=-1\,,\\
\bm{2}: ~&~ S=\dfrac{1}{2}\begin{pmatrix}
  -1 ~&\sqrt{3} \\
\sqrt{3} ~& 1 \\
\end{pmatrix}\,, ~&~ T=\begin{pmatrix}
 1 ~& 0 \\
 0 ~& -1 \\
\end{pmatrix}\,, ~&~ R=\begin{pmatrix}
 1 ~& 0 \\
 0 ~& 1 \\
\end{pmatrix}\,, \\
\bm{\widehat{2}}: ~&~ S=\dfrac{i}{2}\begin{pmatrix}
 -1 ~&\sqrt{3} \\
\sqrt{3} ~& 1 \\
\end{pmatrix}\,, ~&~ T=-i\begin{pmatrix}
 1 ~& 0 \\
 0 ~& -1 \\
\end{pmatrix}\,, ~&~ R=-\begin{pmatrix}
 1 ~& 0 \\
 0 ~& 1 \\
\end{pmatrix}\,, \\
\bm{3}: ~&~ S=\dfrac{1}{2}\begin{pmatrix}
 0 ~&\sqrt{2} ~&\sqrt{2} \\
\sqrt{2} ~& -1 ~& 1 \\
\sqrt{2} ~& 1 ~& -1
\end{pmatrix}\,, ~&~ T=\begin{pmatrix}
 1 ~& 0 ~& 0 \\
 0 ~& i ~& 0 \\
 0 ~& 0 ~& -i
\end{pmatrix}\,, ~&~ R=\begin{pmatrix}
 1 ~& 0 ~& 0 \\
 0 ~& 1 ~& 0 \\
 0 ~& 0 ~& 1
\end{pmatrix}\,,\\
\bm{3^{\prime}}: ~&~ S=-\dfrac{1}{2}\begin{pmatrix}
 0 ~&\sqrt{2} ~&\sqrt{2} \\
\sqrt{2} ~& -1 ~& 1 \\
\sqrt{2} ~& 1 ~& -1
\end{pmatrix}\,, ~&~ T=-\begin{pmatrix}
 1 ~& 0 ~& 0 \\
 0 ~& i ~& 0 \\
 0 ~& 0 ~& -i
\end{pmatrix}\,, ~&~ R=\begin{pmatrix}
 1 ~& 0 ~& 0 \\
 0 ~& 1 ~& 0 \\
 0 ~& 0 ~& 1
\end{pmatrix}\,, \\
\bm{\widehat{3}}: ~&~ S=\dfrac{i}{2}\begin{pmatrix}
 0 ~&\sqrt{2} ~&\sqrt{2} \\
\sqrt{2} ~& -1 ~& 1 \\
\sqrt{2} ~& 1 ~& -1
\end{pmatrix}\,, ~&~ T=-i\begin{pmatrix}
 1 ~& 0 ~& 0 \\
 0 ~& i ~& 0 \\
 0 ~& 0 ~& -i
\end{pmatrix}\,,~&~ R=-\begin{pmatrix}
 1 ~& 0 ~& 0 \\
 0 ~& 1 ~& 0 \\
 0 ~& 0 ~& 1
\end{pmatrix}\,, \\
\bm{\widehat{3}^{\prime}}: ~&~ S=-\dfrac{i}{2}\begin{pmatrix}
 0 ~&\sqrt{2} ~&\sqrt{2} \\
\sqrt{2} ~& -1 ~& 1 \\
\sqrt{2} ~& 1 ~& -1
\end{pmatrix}\,, ~&~ T=i\begin{pmatrix}
 1 ~& 0 ~& 0 \\
 0 ~& i ~& 0 \\
 0 ~& 0 ~& -i
\end{pmatrix}\,,~&~ R=-\begin{pmatrix}
 1 ~& 0 ~& 0 \\
 0 ~& 1 ~& 0 \\
 0 ~& 0 ~& 1
\end{pmatrix}\,.
\end{array}
\end{eqnarray}
For simplicity, the CG coefficients of the $S'_4$ group are not listed here and they can be found in Appendix C of Ref.~\cite{Ding:2023htn}.

\subsection{\label{app:group-MF-N5} $N=5$}

The homogeneous finite modular group $\Gamma'_5$ which is the double covering of $\Gamma_5\cong A_5$ which is the symmetry group of the regular icosahedron. It can be generated by three generators $S$, $T$ and $R$ satisfying the following multiplication rules,
\begin{equation}
S^2=R,~~ T^5=(ST)^3=R^2=1,~~ RT=TR\,.
\end{equation}
The group $A_5$ is not a subgroup of $A'_5$, and it is isomorphic to $A'_5/Z^R_2$. The group $A'_5$ has 120 elements which is twice as many elements as $A_5$. Besides the irreducible representations $\bm{1}$, $\bm{3}$, $\bm{3'}$, $\bm{4}$, $\bm{5}$ in common with these of $A_5$, the $A'_5$ group has four spinor representations $\bm{\widehat{2}}$, $\bm{\widehat{2}'}$, $\bm{\widehat{4}'}$ and $\bm{\widehat{6}}$. The representation matrices of the generators $S$ and $T$ are collected in table~\ref{tab:A5prime_rep}.

\begin{table}[hptb!]
\centering \centering \resizebox{0.90\textwidth}{!}{
\begin{tabular}{|c|c|c|c|}
\hline\hline
Reps & $S$ & $T$ & $R$ \\
\hline
$\bm{1}$ & $1$ & $1$ & $1$\\
\hline
$\bm{\widehat{2}}$ & $i\sqrt{\frac{1}{\sqrt{5}\phi}}\left(\begin{array}{cc}
 \phi  & 1 \\
 1 & -\phi
\end{array}\right)$ & $\left(\begin{array}{cc}
 \omega_5^2 & 0 \\
 0 & \omega_5^3
\end{array}\right)$ & $-\left(\begin{array}{cc}
 1 & 0 \\
 0 & 1
\end{array}\right)$\\
\hline
$\bm{\widehat{2}'}$ & $i\sqrt{\frac{1}{\sqrt{5}\phi}}\left(\begin{array}{cc}
 1 & \phi  \\
 \phi & -1
\end{array}\right)$ & $\left(\begin{array}{cc}
 \omega_5 & 0 \\
 0 & \omega_5^4
\end{array}\right)$ & $-\left(\begin{array}{cc}
 1 & 0 \\
 0 & 1
\end{array}\right)$\\
\hline
$\bm{3}$ & $\frac{1}{\sqrt{5}}\left(\begin{array}{ccc}
 1 & -\sqrt{2} & -\sqrt{2} \\
 -\sqrt{2} & -\phi & \frac{1}{\phi} \\
 -\sqrt{2} & \frac{1}{\phi} & -\phi
\end{array}\right)$ & $\left(\begin{array}{ccc}
 1 & 0 & 0 \\
 0 & \omega_5 & 0 \\
 0 & 0 & \omega_5^4
\end{array}\right)$ & $\left(\begin{array}{ccc}
 1 & 0 & 0 \\
 0 & 1 & 0 \\
 0 & 0 & 1
\end{array}\right)$\\
\hline
$\bm{3'}$ & $\frac{1}{\sqrt{5}}\left(\begin{array}{ccc}
 -1 & \sqrt{2} & \sqrt{2} \\
 \sqrt{2} & -\frac{1}{\phi} & \phi \\
 \sqrt{2} & \phi & -\frac{1}{\phi}
\end{array}\right)$ & $\left(\begin{array}{ccc}
 1 & 0 & 0 \\
 0 & \omega_5^2 & 0 \\
 0 & 0 & \omega_5^3
\end{array}\right)$ & $\left(\begin{array}{ccc}
 1 & 0 & 0 \\
 0 & 1 & 0 \\
 0 & 0 & 1
\end{array}\right)$\\
\hline
$\bm{4}$ & $\frac{1}{\sqrt{5}}\left(\begin{array}{cccc}
 1 & \frac{1}{\phi} & \phi & -1 \\
 \frac{1}{\phi} & -1 & 1 & \phi \\
 \phi & 1 & -1 & \frac{1}{\phi} \\
 -1 & \phi & \frac{1}{\phi} & 1
\end{array}\right)$ & $\left(\begin{array}{cccc}
 \omega_5 & 0 & 0 & 0 \\
 0 & \omega_5^2 & 0 & 0 \\
 0 & 0 & \omega_5^3 & 0 \\
 0 & 0 & 0 & \omega_5^4
\end{array}\right)$ & $\left(\begin{array}{cccc}
 1 & 0 & 0 & 0 \\
 0 & 1 & 0 & 0 \\
 0 & 0 & 1 & 0 \\
 0 & 0 & 0 & 1
\end{array}\right)$ \\
\hline
$\bm{\widehat{4}'}$ & $i\sqrt{\frac{1}{5\sqrt{5}\phi}}\left(\begin{array}{cccc}
 -\phi^2 & \sqrt{3}\phi & -\sqrt{3} & -\frac{1}{\phi} \\
 \sqrt{3}\phi & \frac{1}{\phi} & -\phi^2 & -\sqrt{3} \\
 -\sqrt{3} & -\phi^2 & -\frac{1}{\phi} & -\sqrt{3}\phi \\
 -\frac{1}{\phi} & -\sqrt{3} & -\sqrt{3}\phi & \phi^2
\end{array}\right)$ & $\left(\begin{array}{cccc}
 \omega_5 & 0 & 0 & 0 \\
 0 & \omega_5^2 & 0 & 0 \\
 0 & 0 & \omega_5^3 & 0 \\
 0 & 0 & 0 & \omega_5^4
\end{array}\right)$ & $-\left(\begin{array}{cccc}
 1 & 0 & 0 & 0 \\
 0 & 1 & 0 & 0 \\
 0 & 0 & 1 & 0 \\
 0 & 0 & 0 & 1
\end{array}\right)$\\
\hline
$\bm{5}$ & $\frac{1}{5}\left(\begin{array}{ccccc}
 -1 & \sqrt{6} & \sqrt{6} & \sqrt{6} & \sqrt{6} \\
 \sqrt{6} & \frac{1}{\phi^2} & -2\phi & \frac{2}{\phi} & \phi^2 \\
 \sqrt{6} & -2\phi & \phi^2 & \frac{1}{\phi^2} & \frac{2}{\phi} \\
 \sqrt{6} & \frac{2}{\phi} & \frac{1}{\phi^2} & \phi^2 & -2\phi \\
 \sqrt{6} & \phi^2 & \frac{2}{\phi} & -2\phi & \frac{1}{\phi^2}
\end{array}\right)$ & $\left(\begin{array}{ccccc}
 1 & 0 & 0 & 0 & 0 \\
 0 & \omega_5 & 0 & 0 & 0 \\
 0 & 0 & \omega_5^2 & 0 & 0 \\
 0 & 0 & 0 & \omega_5^3 & 0 \\
 0 & 0 & 0 & 0 & \omega_5^4
\end{array}\right)$ & $\left(\begin{array}{ccccc}
 1 & 0 & 0 & 0 & 0 \\
 0 & 1 & 0 & 0 & 0 \\
 0 & 0 & 1 & 0 & 0 \\
 0 & 0 & 0 & 1 & 0 \\
 0 & 0 & 0 & 0 & 1
\end{array}\right)$\\
\hline
$\bm{\widehat{6}}$ & $i\sqrt{\frac{1}{5\sqrt{5}\phi}}\left(\begin{array}{cccccc}
 -1 & \phi & \frac{1}{\phi} & \sqrt{2}\phi & \sqrt{2} & \phi^2 \\
 \phi & 1 & \phi^2 & \sqrt{2} & -\sqrt{2}\phi & -\frac{1}{\phi} \\
 \frac{1}{\phi} & \phi^2 & 1 & -\sqrt{2} & \sqrt{2}\phi & -\phi \\
 \sqrt{2}\phi & \sqrt{2} & -\sqrt{2} & -\phi & -1 & \sqrt{2}\phi \\
 \sqrt{2} & -\sqrt{2}\phi & \sqrt{2}\phi & -1 & \phi & \sqrt{2} \\
 \phi^2 & -\frac{1}{\phi} & -\phi & \sqrt{2}\phi & \sqrt{2} & -1
\end{array}\right)$ & $\left(\begin{array}{cccccc}
 1 & 0 & 0 & 0 & 0 & 0 \\
 0 & 1 & 0 & 0 & 0 & 0 \\
 0 & 0 & \omega_5 & 0 & 0 & 0 \\
 0 & 0 & 0 & \omega_5^2 & 0 & 0 \\
 0 & 0 & 0 & 0 & \omega_5^3 & 0 \\
 0 & 0 & 0 & 0 & 0 & \omega_5^4
\end{array}\right)$ & $-\left(\begin{array}{cccccc}
 1 & 0 & 0 & 0 & 0 & 0 \\
 0 & 1 & 0 & 0 & 0 & 0 \\
 0 & 0 & 1 & 0 & 0 & 0 \\
 0 & 0 & 0 & 1 & 0 & 0 \\
 0 & 0 & 0 & 0 & 1 & 0 \\
 0 & 0 & 0 & 0 & 0 & 1
\end{array}\right)$ \\
\hline \hline
\end{tabular}}
\caption{\label{tab:A5prime_rep} The representation matrices of the generators $S$, $T$ and $R$ of $\Gamma'_5\cong A'_5$. Here $\phi=(1+\sqrt{5})/2$ and $\omega_5=e^{2\pi i/5}$.}
\end{table}

The generator $R$ is represented by the identity matrix $\mathds{1}$ in the hatless representations $\bm{1}$, $\bm{3}$, $\bm{3'}$, $\bm{4}$ and $\bm{5}$, and it is $-\mathds{1}$ in the hatted representations $\bm{\widehat{2}}$, $\bm{\widehat{2}'}$, $\bm{\widehat{4}'}$ and $\bm{\widehat{6}}$. Note that the hatted representations are novel and specific to $A'_5$. The CG coefficients of the $A'_5$ group are too lengthy to be shown here, their explicit expressions can be found in the Appendix C of Ref.~\cite{Ding:2023htn}.

\section{\label{app:Polyharmonic_Maass_forms}Constructing integer weight polyharmonic Maa{\ss} forms from non-holomorphic Eisenstein series  }

In the following we report the $q$-expansion of the polyharmonic Maa{\ss} forms which are constructed from the non-holomorphic Eisenstein series for level $N=3, 4, 5$. All these expressions in \texttt{Wolfram Mathematica} format can be downloaded from the website~\cite{Qu:2025PHMFsup}. In this approach, we need to evaluate the Fourier expansion of non-holomorphic Eisenstein series $E_k(N; \tau; s; \overline{A/C})$ defined in Eq.~\eqref{eq:non_holo_Eisenstein}. After lengthy and tedious algebraic calculations, we find that be seen that $E_k(N; \tau; s; \overline{A/C})$ can be expanded into the form of Eq.~\eqref{eq:Fourier-exp} when $s=1-k$. The coefficients are expressed in terms of Riemann zeta function and Dirichlet $L$-function.

The incomplete gamma function $\Gamma(s, x)$ appears in the following $q$-expansion of the non-holomorphic polyharmonic Maa{\ss} forms. For different low integer values of $s$ of interest for our analysis, $\Gamma(s, x)$ is given by simple analytical expressions:
\begin{eqnarray}
\nonumber \Gamma(1,x)&=&e^{-x}\,, \\
\nonumber \Gamma(2,x)&=&(x+1)\,e^{-x}\,, \\
\nonumber \Gamma(3,x)&=&(x^2+2x+2)\,e^{-x}\,, \\
\nonumber \Gamma(4,x)&=&(x^3+3x^2+6x+6)\,e^{-x}\,, \\
\Gamma(5,x)&=&(x^4+4x^3+12x^2+24x+24)\,e^{-x}\,.
\end{eqnarray}
In general the incomplete gamma function can be written as
\begin{eqnarray}
\Gamma(s,x) = (s-1)! \, e^{-x}\, \sum_{k=0}^{s-1} \dfrac{x^k}{k!}
\end{eqnarray}
for any positive integer $s$. The incomplete gamma function $\Gamma(0,x)$ is involved in the weight one polyharmonic Maa{\ss} forms. Unfortunately, it is not an elementary function. One can use the continued fraction formula to efficiently evaluate the incomplete gamma function as follow~\cite{cuyt2008handbook},
\begin{eqnarray}
\Gamma(s,x) = \dfrac{x^s \,e^{-x}}{1+x-s + \dfrac{s-1}{3+x-s + \dfrac{2(s-2)}{5+x-s + \dfrac{3(s-3)}{7+x-s + \cdots}}}}
\end{eqnarray}
for $s=0$.

\subsection{$N=3$\label{app:Polyharmonic_N_3}}

In the following, we shall present the polyharmonic Maa{\ss} forms of level $N=3$ which are constructed from the non-holomorphic Eisenstein series  $E_k(N; \tau; s; \overline{A/C})$ for $s=1-k$, the procedure is explained in section~\ref{sec:integer_weight}. In the following we will omit the level $N$, and denote $E_k(N; \tau; s; \overline{A/C})$ as $E_k(\tau; s; \overline{A/C})$ for simplicity. The principal congruence subgroup $\Gamma(3)$ has four cusps $\{0,1,2,i\infty\}=\{\overline{0/1},\overline{1/1},\overline{2/1},\overline{1/0}\}$, thus there are four types of weight $k$ non-holomorphic Eisenstein series $E_k(\tau; s; 0)$, $E_k(\tau; s; 1)$, $E_k(\tau; s; 2)$ and $E_k(\tau; s; i\infty)$ at level $N=3$. As shown in Eq.~\eqref{eq:Eisenstein_trans}, the modular symmetry relates the non-holomorphic Eisenstein series at different cusps. Under the action of the modular generators $S$ and $T$, we have
\begin{eqnarray}
\nonumber  E_k(S\tau; s; i\infty) &=& (-\tau)^{k} E_k(\tau; s; 0)\,, \\
\nonumber E_k(S\tau; s; 0) &=& (-1)^k (-\tau)^{k} E_k(\tau; s; i\infty)\,, \\
\nonumber E_k(S\tau; s; 1) &=& (-\tau)^{k} E_k(\tau; s; 2)\,,  \\
\nonumber E_k(S\tau; s; 2) &=& (-1)^k (-\tau)^{k} E_k(\tau; s; 1) \,, \\
\nonumber E_k(T\tau; s; i\infty) &=& E_k(\tau; s; i\infty) \,, \\
\nonumber E_k(T\tau; s; 0) &=& E_k(\tau; s; 2)\,,  \\
\nonumber E_k(T\tau; s; 1) &=& E_k(\tau; s; 0) \,, \\
E_k(T\tau; s; 2) &=& E_k(\tau; s; 1)\,.
\end{eqnarray}
Therefore the even weight Eisenstein series of level $N=3$ can be arranged into a singlet $Y^{(2k)}_{\bm{1}}(\tau)$ and a triplet $Y^{(2k)}_{\bm{3}}(\tau)=(Y^{(2k)}_{\bm{3}, 1}, Y^{(2k)}_{\bm{3}, 2}, Y^{(2k)}_{\bm{3}, 3})^T$ of $\Gamma'_3\cong T'$, while the odd weight Eisenstein series can be organized into two doublets $Y^{(2k+1)}_{\bm{\widehat{2}}}(\tau)=(Y^{(2k+1)}_{\bm{\widehat{2}}, 1}, Y^{(2k+1)}_{\bm{\widehat{2}}, 2})^T$ and $Y^{(2k+1)}_{\bm{\widehat{2}''}}(\tau)=(Y^{(2k+1)}_{\bm{\widehat{2}''}, 1}, Y^{(2k+1)}_{\bm{\widehat{2}''}, 2})^T$ for the modular weight $k\leq0$. Each component of these modular multiplets can be expressed as
\begin{eqnarray}
\nonumber Y^{(2k)}_{\bm{1}}(\tau) &=& E_{2k}(\tau; s; i\infty) + E_{2k}(\tau; s; 0) + E_{2k}(\tau; s; 1) + E_{2k}(\tau; s; 2)\,,  \\
\nonumber Y^{(2k)}_{\bm{3},1}(\tau) &=& E_{2k}(\tau; s; i\infty) - \dfrac{1}{3} \Big( E_{2k}(\tau; s; 0) + E_{2k}(\tau; s; 1) + E_{2k}(\tau; s; 2) \Big)\,, \\
\nonumber Y^{(2k)}_{\bm{3},2}(\tau) &=& \dfrac{2}{3} \Big( E_{2k}(\tau; s; 0) + \omega E_{2k}(\tau; s; 1) + \omega^2 E_{2k}(\tau; s; 2) \Big)\,, \\
\label{eq:eis_dec_even_N3} Y^{(2k)}_{\bm{3},3}(\tau) &=& \dfrac{2}{3} \Big( E_{2k}(\tau; s; 0) + \omega^2 E_{2k}(\tau; s; 1) + \omega E_{2k}(\tau; s; 2) \Big) \,,
\end{eqnarray}
and
\begin{eqnarray}
\nonumber Y^{(2k+1)}_{\bm{\widehat{2}},1}(\tau) &=& i\,\sqrt{\dfrac{2}{3}} \Big( E_{2k+1}(\tau; s; 0) + \omega E_{2k+1}(\tau; s; 1) + \omega^2 E_{2k+1}(\tau; s; 2) \Big)\,,  \\
\nonumber Y^{(2k+1)}_{\bm{\widehat{2}},2}(\tau) &=& E_{2k+1}(\tau; s; i\infty) - \dfrac{i}{\sqrt{3}} \Big( E_{2k+1}(\tau; s; 0) + E_{2k+1}(\tau; s; 1) + E_{2k+1}(\tau;s; 2) \Big)\,, \\
\nonumber Y^{(2k+1)}_{\bm{\widehat{2}''},1}(\tau) &=& E_{2k+1}(\tau; s; i\infty) + \dfrac{i}{\sqrt{3}} \Big( E_{2k+1}(\tau; s; 0) + E_{2k+1}(\tau; s; 1) + E_{2k+1}(\tau; s; 2) \Big)\,, \\
\label{eq:eis_dec_odd_N3} Y^{(2k+1)}_{\bm{\widehat{2}''},2}(\tau) &=& i\,\sqrt{\dfrac{2}{3}}  \Big( E_{2k+1}(\tau; s; 0) + \omega^2 E_{2k+1}(\tau; s; 1) + \omega E_{2k+1}(\tau; s; 2) \Big)\,.
\end{eqnarray}
The Fourier expansion of the negative even weight $2k$ Eisenstein series $E_{2k}(\tau; 1-2k; \overline{A/C})$ for $k\leq 0$ are given by
\begin{eqnarray}
\nonumber E_{2k}(\tau; 1-2k; i\infty) &=& y^{1-2k} + (-1)^k \dfrac{2^{1+2k}\cdot 3^{2k-1}}{1-3^{2k-2}} \dfrac{\zeta(1-2k)}{\zeta(2-2k)} \dfrac{\pi}{3} \\
 \nonumber && \hspace{-3cm}+ \dfrac{(-1)^{k} 2^{2k}}{(1 - 3^{2k-2})\zeta(2-2k)} \dfrac{\pi}{3} \sum_{n=1}^{+\infty} q^{n/3} \sum_{\substack{m|n,\, m\in\mathbb{Z} \\ \frac{n}{m}\equiv 0\,({\rm mod}\, 3)}} \left| \dfrac{n}{m} \right|^{2k-1} \, \omega^{2m}  \\
 \nonumber &&\hspace{-3cm}+ \dfrac{(-1)^{k} 2^{2k}}{(-2k)!(1 - 3^{2k-2})\zeta(2-2k)} \dfrac{\pi}{3} \sum_{n=-1}^{-\infty} \Gamma\left(1-2k, - \dfrac{4\pi n y}{3} \right) q^{n/3} \sum_{\substack{m|n,\, m\in\mathbb{Z} \\ \frac{n}{m}\equiv 0\,({\rm mod}\, 3)}} \left| \dfrac{n}{m} \right|^{2k-1} \, \omega^{2m} \,, \\
\nonumber E_{2k}(\tau; 1-2k; 0) &=& (-1)^k 2^{2k} \dfrac{1-3^{2k-1}}{1-3^{2k-2}} \dfrac{\zeta(1-2k)}{\zeta(2-2k)} \dfrac{\pi}{3} \\
\nonumber &&\hspace{-3cm}  + \dfrac{(-1)^{k} 2^{2k}}{(1 - 3^{2k-2})\zeta(2-2k)} \dfrac{\pi}{3} \sum_{n=1}^{+\infty} q^{n/3} \sum_{\substack{m|n,\, m\in\mathbb{Z} \\ \frac{n}{m}\equiv 1\,({\rm mod}\, 3)}} \left| \dfrac{n}{m} \right|^{2k-1}   \\
\nonumber &&\hspace{-3cm}+ \dfrac{(-1)^{k} 2^{2k}}{(-2k)!(1 - 3^{2k-2})\zeta(2-2k)} \dfrac{\pi}{3} \sum_{n=-1}^{-\infty} \Gamma\left(1-2k, - \dfrac{4\pi n y}{3} \right) q^{n/3} \sum_{\substack{m|n,\, m\in\mathbb{Z} \\ \frac{n}{m}\equiv 1\,({\rm mod}\, 3)}} \left| \dfrac{n}{m} \right|^{2k-1} \,,  \\
\nonumber E_{2k}(\tau; 1-2k; 1) &=& (-1)^k 2^{2k} \dfrac{1-3^{2k-1}}{1-3^{2k-2}} \dfrac{\zeta(1-2k)}{\zeta(2-2k)} \dfrac{\pi}{3} \\
\nonumber && \hspace{-3cm} + \dfrac{(-1)^{k} 2^{2k}}{(1 - 3^{2k-2})\zeta(2-2k)} \dfrac{\pi}{3} \sum_{n=1}^{+\infty} q^{n/3} \sum_{\substack{m|n,\, m\in\mathbb{Z} \\ \frac{n}{m}\equiv 1\,({\rm mod}\, 3)}} \left| \dfrac{n}{m} \right|^{2k-1} \, \omega^{2m}  \\
\nonumber &&\hspace{-3cm}  + \dfrac{(-1)^{k} 2^{2k}}{(-2k)!(1 - 3^{2k-2})\zeta(2-2k)} \dfrac{\pi}{3} \sum_{n=-1}^{-\infty} \Gamma\left(1-2k, - \dfrac{4\pi n y}{3} \right) q^{n/3} \sum_{\substack{m|n,\, m\in\mathbb{Z} \\ \frac{n}{m}\equiv 1\,({\rm mod}\, 3)}} \left| \dfrac{n}{m} \right|^{2k-1} \, \omega^{2m} \,, \\
\nonumber E_{2k}(\tau; 1-2k; 2) &=&  (-1)^k 2^{2k} \dfrac{1-3^{2k-1}}{1-3^{2k-2}} \dfrac{\zeta(1-2k)}{\zeta(2-2k)} \dfrac{\pi}{3} \\
\nonumber &&\hspace{-3cm}  + \dfrac{(-1)^{k} 2^{2k}}{(1 - 3^{2k-2})\zeta(2-2k)} \dfrac{\pi}{3} \sum_{n=1}^{+\infty} q^{n/3} \sum_{\substack{m|n,\, m\in\mathbb{Z} \\ \frac{n}{m}\equiv 1\,({\rm mod}\, 3)}} \left| \dfrac{n}{m} \right|^{2k-1} \, \omega^{m}  \\
&&\hspace{-3cm}+ \dfrac{(-1)^{k} 2^{2k}}{(-2k)!(1 - 3^{2k-2})\zeta(2-2k)} \dfrac{\pi}{3} \sum_{n=-1}^{-\infty} \Gamma\left(1-2k, - \dfrac{4\pi n y}{3} \right) q^{n/3} \sum_{\substack{m|n,\, m\in\mathbb{Z} \\ \frac{n}{m}\equiv 1\,({\rm mod}\, 3)}} \left| \dfrac{n}{m} \right|^{2k-1} \, \omega^{m}\,,~~~\quad~~
\end{eqnarray}
where $\omega = e^{2\pi i /3}=-1/2+i\sqrt{3}/2$. Plugging the above expressions of the Eisenstein series into Eq.~\eqref{eq:eis_dec_even_N3}, we can obtain the Fourier expansion of the even weight polyharmonic Maa{\ss} forms of level $N=3$ in Eq.~\eqref{eq:N=3_HMF_even_qexp} up to an overall normalization factor.

Analogously the Fourier expansion of the negative odd weight $2k+1$ Eisenstein series $E_{2k+1}(\tau; -2k; \overline{A/C})$ for $k<0$ are given by
\begin{small}
\begin{eqnarray}
\nonumber E_{2k+1}(\tau; -2k; i\infty) &=& y^{-2k} - \dfrac{(-1)^{k} 2^{2k+1}}{L(1-2k,\chi_2)} \dfrac{i\pi}{3} \sum_{n=1}^{+\infty} q^{n/3} \sum_{\substack{m|n,\, m\in\mathbb{Z} \\ \frac{n}{m}\equiv 0\,({\rm mod}\, 3)}} {\rm sign}(m) \left| \dfrac{n}{m} \right|^{2k} \omega^{2m}   \\
\nonumber &&\hspace{-2.8cm} - \dfrac{(-1)^{k} 2^{2k+1}}{(-2k-1)! L(1-2k,\chi_2)} \dfrac{i\pi}{3} \sum_{n=-1}^{-\infty} \Gamma(-2k,-4\pi n y/3) q^{n/3} \sum_{\substack{m|n,\, m\in\mathbb{Z} \\ \frac{n}{m}\equiv 0\,({\rm mod}\, 3)}} {\rm sign}(m) \left| \dfrac{n}{m} \right|^{2k} \omega^{2m}\,,   \\
\nonumber E_{2k+1}(\tau; -2k; 0) &=& (-1)^k  2^{2k+1} \dfrac{L(-2k,\chi_2)}{L(1-2k,\chi_2)}\dfrac{i\pi}{3} + \dfrac{(-1)^{k} 2^{2k+1}}{L(1-2k,\chi_2)} \dfrac{i\pi}{3} \sum_{n=1}^{+\infty} q^{n/3} \sum_{\substack{m|n,\, m\in\mathbb{Z} \\ \frac{n}{m}\equiv 1\,({\rm mod}\, 3)}} {\rm sign}(m) \left| m \right|^{2k}    \\
\nonumber &&\hspace{-2.8cm} - \dfrac{(-1)^{k} 2^{2k+1}}{(-2k-1)! L(1-2k,\chi_2)} \dfrac{i\pi}{3} \sum_{n=-1}^{-\infty} \Gamma(-2k,-4\pi n y/3) q^{n/3} \sum_{\substack{m|n,\, m\in\mathbb{Z} \\ \frac{n}{m}\equiv 1\,({\rm mod}\, 3)}} {\rm sign}(m) \left| \dfrac{n}{m} \right|^{2k} \,,  \\
\nonumber E_{2k+1}(\tau; -2k; 1) &=& (-1)^k  2^{2k+1} \dfrac{L(-2k,\chi_2)}{L(1-2k,\chi_2)}\dfrac{i\pi}{3} + \dfrac{(-1)^{k} 2^{2k+1}}{L(1-2k,\chi_2)} \dfrac{i\pi}{3} \sum_{n=1}^{+\infty} q^{n/3} \sum_{\substack{m|n,\, m\in\mathbb{Z} \\ \frac{n}{m}\equiv 1\,({\rm mod}\, 3)}} {\rm sign}(m) \left| m \right|^{2k} \omega^{2m} \,,   \\
\nonumber &&\hspace{-2.8cm} - \dfrac{(-1)^{k} 2^{2k+1}}{(-2k-1)! L(1-2k,\chi_2)} \dfrac{i\pi}{3} \sum_{n=-1}^{-\infty} \Gamma(-2k,-4\pi n y/3) q^{n/3} \sum_{\substack{m|n,\, m\in\mathbb{Z} \\ \frac{n}{m}\equiv 1\,({\rm mod}\, 3)}} {\rm sign}(m) \left| \dfrac{n}{m} \right|^{2k} \omega^{2m} \,,  \\
\nonumber E_{2k+1}(\tau; -2k; 2) &=& (-1)^k  2^{2k+1} \dfrac{L(-2k,\chi_2)}{L(1-2k,\chi_2)}\dfrac{i\pi}{3} + \dfrac{(-1)^{k} 2^{2k+1}}{L(1-2k,\chi_2)} \dfrac{i\pi}{3} \sum_{n=1}^{+\infty} q^{n/3} \sum_{\substack{m|n,\, m\in\mathbb{Z} \\ \frac{n}{m}\equiv 1\,({\rm mod}\, 3)}} {\rm sign}(m) \left| \dfrac{n}{m} \right|^{2k} \omega^{m}   \\
&&\hspace{-2.8cm} - \dfrac{(-1)^{k} 2^{2k+1}}{(-2k-1)! L(1-2k,\chi_2)} \dfrac{i\pi}{3} \sum_{n=-1}^{-\infty} \Gamma(-2k,-4\pi n y/3) q^{n/3} \sum_{\substack{m|n,\, m\in\mathbb{Z} \\ \frac{n}{m}\equiv 1\,({\rm mod}\, 3)}} {\rm sign}(m) \left| \dfrac{n}{m} \right|^{2k} \omega^{m}\,.
\end{eqnarray}
\end{small}
Inserting the above odd weight Eisenstein series into Eq.~\eqref{eq:eis_dec_odd_N3} with $s=-2k$, we can obtain the polyharmonic Maa{\ss} form multiplets given in Eq.~\eqref{eq:N=3_HMF_odd_qexp} after proper normalization.

Then we proceed to study the weight $1$ Eisenstein series whose $q$-expansion is given by
\begin{eqnarray}
\nonumber  E_1(\tau; 0; i\infty)  &=& 1 + 6 \sum_{n=1}^{+\infty} \sum_{d|n\,, d>0}\chi_2(d) q^n\,,  \\
\nonumber E_1(\tau; 0; 0) &=& \dfrac{i}{\sqrt{3}} + 2\sqrt{3} i \sum_{n=1}^{+\infty} \sum_{d|n\,,d>0} \chi_2(d) q^n + 2\sqrt{3}i\, q^{1/3} \sum_{n=0}^{+\infty} \sum_{d|(3n+1)\,,d>0}\chi_2(d)q^n \,, \\
\nonumber E_1(\tau; 0; 1) &=& \dfrac{i}{\sqrt{3}} + 2\sqrt{3} i \sum_{n=1}^{+\infty} \sum_{d|n\,,d>0} \chi_2(d) q^n + 2\sqrt{3}i\, \omega^2 q^{1/3} \sum_{n=0}^{+\infty} \sum_{d|(3n+1)\,,d>0}\chi_2(d)q^n \,, \\
\label{eq:Eisen-series-N3}E_1(\tau; 0; 2) &=& \dfrac{i}{\sqrt{3}} + 2\sqrt{3} i \sum_{n=1}^{+\infty} \sum_{d|n\,,d>0} \chi_2(d) q^n + 2\sqrt{3}i\, \omega q^{1/3} \sum_{n=0}^{+\infty} \sum_{d|(3n+1)\,,d>0}\chi_2(d)q^n\,.~~~~~
\end{eqnarray}
The above Eisenstein series can be arranged into weight one polyharmonic Maa{\ss} forms in $\bm{\widehat{2}}$ and $\bm{\widehat{2}''}$ representations, as shown in Eq.~\eqref{eq:eis_dec_odd_N3}. The doublet $Y_{\bm{\widehat{2}}}^{(1)}(\tau)=(Y_{\bm{\widehat{2}},1}^{(1)}(\tau), Y_{\bm{\widehat{2}},2}^{(1)}(\tau))^{T}$ takes the following form,
\begin{eqnarray}
\nonumber Y_{\bm{\widehat{2}},1}^{(1)}(\tau) &=& \dfrac{i}{\sqrt{6}}  \left[E_1(\tau; 0; 0) + \omega E_1(\tau; 0; 1) + \omega^2 E_1(\tau; 0; 2) \right] \\
\nonumber &=& - 3\sqrt{2}  q^{1/3} \sum_{n=0}^{+\infty} \sum_{d|(3n+1)\,,d>0}\chi_2(d)q^n  \\
\nonumber &=& - 3 \sqrt{2} q^{1/3}\left( 1 + q + 2 q^2 + 2 q^4 + q^5 + 2 q^6 + q^8 + 2 q^9 \cdots \right) \,, \\
\nonumber Y_{\bm{\widehat{2}},2}^{(1)}(\tau) &=& \dfrac{1}{2} \left[ E_1(\tau; 0; i\infty) - \dfrac{i}{\sqrt{3}} \left( E_1(\tau; 0; 0) + E_1(\tau; 0; 1) + E_1(\tau; 0; 2) \right) \right] \\
\nonumber &=& 1 + 6 \sum_{n=1}^{+\infty} \sum_{m|n,m>0} \chi_2(m) q^n  \\
\label{eq:Eisenstein_holo_wt1_N3} &=& 1 + 6(q+q^3 + q^4 + 2 q^7 + q^9 + q^{12} + 2 q^{13} +\ldots)\,,
\end{eqnarray}
which coincide with the expression of the weight 1 and level 3 modular forms in Ref.~\cite{Liu:2019khw}. However, another Maa{\ss} forms doublet $Y_{\bm{\widehat{2}''}}^{(1)}(\tau)$ are found to be vanishing, i.e.
\begin{eqnarray}
\nonumber&& E_1(\tau; 0; i\infty) + \dfrac{i}{\sqrt{3}} \left(E_1(\tau; 0; 0) + E_1(\tau; 0; 1) + E_1(\tau; 0; 2)\right) =0\,, \\
&&E_1(\tau; 0; 0) + \omega^2 E_1(\tau; 0; 1) + \omega E_1(\tau; 0; 2)=0\,.
\end{eqnarray}
Therefore the four Eisenstein series of weight one in Eq.~\eqref{eq:Eisen-series-N3} are not linearly independent.

As explained in section~\ref{sec:integer_weight}, the weight 1 polyharmonic Maa{\ss} forms are a bit special, and one needs to consider the first order derivative $E^{(1)}_1(N; \tau; \overline{A/C})\equiv\left. \dfrac{\partial}{\partial s} E_1(N; \tau; s; \overline{A/C}) \right|_{s=0}$. The argument $N$ would be dropped for simplicity, and their $q$-expansion is found to be
\begin{eqnarray}
\nonumber E_1^{(1)}(\tau; i\infty) &=&  \log y + \gamma_E + \log 12\pi - 6 \log \Gamma\left( \dfrac{1}{3} \right) + 6 \log \Gamma\left( \dfrac{2}{3} \right) \\
\nonumber &&\hskip-0.7in - 6 \sum_{n=1}^{+\infty} \sum_{d|n,\ d>0} \chi_2(d) \Gamma\left( 0, 4\pi n y \right) q^{-n}
+ 6 \sum_{n=1}^{+\infty}   \sum_{d|n} \chi_2(d) \left( 2\log d - \log 3n \right) q^n \\
\nonumber &&\hskip-0.7in + \left( - \gamma_E - \log 12\pi + 6 \log \Gamma\left( \dfrac{1}{3} \right) - 6 \log \Gamma\left( \dfrac{2}{3} \right) \right)   E_1(\tau; 0; i\infty)\,, \\
\nonumber E_1^{(1)}(\tau; 0) &=& \dfrac{i}{\sqrt{3}} \left( - \log y - \gamma_E - \log 12\pi + 6 \log \Gamma\left( \dfrac{1}{3} \right) - 6 \log \Gamma\left( \dfrac{2}{3} \right) \right)  \\
\nonumber &&\hskip-0.7in + 2\sqrt{3}i \sum_{n=1}^{+\infty} \sum_{d|n}^{d>0} \chi_2(d) q^{-n}\Gamma(0,4\pi n y) + 2\sqrt{3} i q^{-1/3}\sum_{n=0}^{+\infty} \sum_{d|(3n+1)} \chi_2(d) q^{-n}\Gamma\left( 0, 4\pi\left(n+\dfrac{1}{3}\right) y \right) \\
\nonumber &&\hskip-0.7in + 2\sqrt{3} i \sum_{n=1}^{+\infty} \sum_{d|n} \chi_2(d) \left( \log 3n - 2\log d \right)q^n  - 4\sqrt{3} i q^{2/3}\sum_{n=0}^{+\infty} q^n \sum_{d|(3n+2)} \chi_2(d) \log d  \\
\nonumber &&\hskip-0.7in + 2\sqrt{3} i q^{1/3}\sum_{n=0}^{+\infty} \sum_{d|(3n+1)} \chi_2(d) \left( \log (3n+1) - 2\log d \right)q^n  \\
\nonumber &&\hskip-0.7in + \left( - \gamma_E - \log 12\pi + 6 \log \Gamma\left( \dfrac{1}{3} \right) - 6 \log \Gamma\left( \dfrac{2}{3} \right) \right)   E_1(\tau; 0; 0) \,, \\
\nonumber E_1^{(1)}(\tau; 1) &=& \dfrac{i}{\sqrt{3}} \left( - \log y - \gamma_E - \log 12\pi + 6 \log \Gamma\left( \dfrac{1}{3} \right) - 6 \log \Gamma\left( \dfrac{2}{3} \right) \right)  \\
\nonumber &&\hskip-0.7in + 2\sqrt{3}i \sum_{n=1}^{+\infty} \sum_{d|n}^{d>0} \chi_2(d) q^{-n}\Gamma(0,4\pi n y) + 2\sqrt{3} i \omega q^{-1/3}\sum_{n=0}^{+\infty} \sum_{d|(3n+1)}\chi_2(d) q^{-n}\Gamma\left( 0, 4\pi\left(n+\dfrac{1}{3}\right) y \right) \\
\nonumber &&\hskip-0.7in + 2\sqrt{3} i \sum_{n=1}^{+\infty} \sum_{d|n} \chi_2(d) \left( \log 3n - 2\log d \right)q^n  - 4\sqrt{3} i \omega q^{2/3}\sum_{n=0}^{+\infty} q^n \sum_{d|(3n+2)} \chi_2(d) \log d  \\
\nonumber &&\hskip-0.7in + 2\sqrt{3} i \omega^2 q^{1/3}\sum_{n=0}^{+\infty} \sum_{d|(3n+1)} \chi_2(d) \left( \log (3n+1) - 2\log d \right)q^n  \\
\nonumber &&\hskip-0.7in + \left( - \gamma_E - \log 12\pi + 6 \log \Gamma\left( \dfrac{1}{3} \right) - 6 \log \Gamma\left( \dfrac{2}{3} \right) \right)   E_1(\tau; 0; 1)\,, \\
\nonumber E_1^{(1)}(\tau; 2) &=& \dfrac{i}{\sqrt{3}} \left( - \log y - \gamma_E - \log 12\pi + 6 \log \Gamma\left( \dfrac{1}{3} \right) - 6 \log \Gamma\left( \dfrac{2}{3} \right) \right)  \\
\nonumber &&\hskip-0.7in + 2\sqrt{3}i \sum_{n=1}^{+\infty} \sum_{d|n}^{d>0} \chi_2(d) q^{-n}\Gamma(0,4\pi n y) + 2\sqrt{3} i  \omega^2 q^{-1/3}\sum_{n=0}^{+\infty} \sum_{d|(3n+1)} \chi_2(d) q^{-n}\Gamma\left( 0, 4\pi\left(n+\dfrac{1}{3}\right) y \right) \\
\nonumber &&\hskip-0.7in + 2\sqrt{3} i \sum_{n=1}^{+\infty} \sum_{d|n} \chi_2(d) \left( \log 3n - 2\log d \right)q^n  - 4\sqrt{3} i \omega^2 q^{2/3}\sum_{n=0}^{+\infty} q^n \sum_{d|(3n+2)} \chi_2(d) \log d  \\
\nonumber &&\hskip-0.7in + 2\sqrt{3} i \omega q^{1/3}\sum_{n=0}^{+\infty} \sum_{d|(3n+1)} \chi_2(d) \left( \log (3n+1) - 2\log d \right)q^n  \\
&&\hskip-0.7in + \left( - \gamma_E - \log 12\pi + 6 \log \Gamma\left( \dfrac{1}{3} \right) - 6 \log \Gamma\left( \dfrac{2}{3} \right) \right)   E_1(\tau; 0; 2)\,.
\end{eqnarray}
With the decomposition rules given in Eq.~\eqref{eq:eis_dec_odd_N3}, we find the
non-holomorphic polyharmonic Maa{\ss} forms doublet $Y^{(1)}_{\bm{\widehat{2}''}}(\tau)$ which is a linear combination of $E^{(1)}_1(\tau,\overline{A/C})$,
\begin{eqnarray}
\nonumber Y^{(1)}_{\bm{\widehat{2}''},1} &=& \gamma_E + \log 3 + \log 4\pi - 6 \log \Gamma\left( \dfrac{1}{3} \right) + 6 \log \Gamma\left( \dfrac{2}{3} \right)\\
\nonumber &&\hskip-0.3in - 6 \left( q \log 3 + 2 q^2 \log 2 + 2 q^3 \log 3 + q^4 \log 3 + 2 q^5 \log 5 + 2 q^6 \log 2 + 2 q^7 \log 3 + \cdots  \right) \\
\nonumber && \hskip-0.3in + \log y - 6 \left(\dfrac{\Gamma(0, 4\pi y)}{q} + \dfrac{ \Gamma(0, 12\pi y)}{q^3} + \dfrac{\Gamma(0, 16\pi y)}{q^4} + \dfrac{2 \Gamma(0, 28\pi y)}{q^7} + \dfrac{\Gamma(0, 36\pi y)}{q^9} + \cdots \right) \,, \\
\nonumber Y^{(1)}_{\bm{\widehat{2}''},2} &=& - 6 \sqrt{2}  q^{2/3} \left( \log 2 + q \log 5 + 2 q^2 \log 2 + q^3 \log 11 + 2 q^4 \log 2 + q^5 \log 17 + q^6 \log 5 + \cdots \right) \\
&&\hskip-0.3in - 3 \sqrt{2}  q^{2/3} \left( \Gamma(0,4\pi y/3) + \dfrac{\Gamma(0,16\pi y/3)}{q} + \dfrac{2\Gamma(0,2\pi y/3)}{q^2} + \dfrac{2\Gamma(0,52\pi y/3)}{q^4} + \cdots \right)\,. \label{eq:Y2hatpp-N3}
\end{eqnarray}
Moreover, another doublet in the representation $\bm{\widehat{2}}$ constructed from $E^{(1)}_1(\tau,\overline{A/C})$ is proportional to the holomorphic modular form multiplet $Y^{(1)}_{\bm{\widehat{2}}}$ in Eq.~\eqref{eq:Eisenstein_holo_wt1_N3}.

In the following we present the explicit $q$-expansions of the polyharmonic Maa{\ss} form multiplets from weight $-4$ to weight $6$ at level $N=3$.

\begin{itemize}[labelindent=-0.8em,leftmargin=0.3em]

\item{$k_Y = -4$}

The weight $k=-4$ polyharmonic Maa{\ss} forms of level $3$ can be arranged into a singlet $\bm{1}$ and a triplet $\bm{3}$ of $T'$. Actually, for each negative even weight at level $3$, the polyharmonic Maa{\ss} forms are non-holomorphic and they can be organized into two multiplets transforming as $\bm{1}$ and $\bm{3}$ of $T'$ up to the automorphy factor.
\begin{eqnarray}
\nonumber Y_{\bm{1}}^{(-4)}&=& \dfrac{y^5}{5} + \dfrac{63\Gamma(5,4\pi y)}{128\pi^5 q} + \dfrac{2079\Gamma(5,8\pi y)}{4096\pi^5 q^2} + \dfrac{427\Gamma(5,12\pi y)}{864\pi^5 q^3} + \dfrac{66591\Gamma(5,16\pi y)}{131072 \pi^5 q^4} + \cdots   \\
\nonumber &&+\dfrac{\pi}{80}\dfrac{\zeta(5)}{\zeta(6)} + \dfrac{189 q}{16\pi^5} + \dfrac{6237 q^2}{512 \pi^5} + \dfrac{427 q^3}{36 \pi^5} + \dfrac{199773 q^4}{16384 \pi^5} + \cdots \,,\\
\nonumber Y_{\bm{3},1}^{(-4)} &=& \dfrac{y^5}{5} - \dfrac{ 549 }{3328 \pi^5} \left( \dfrac{\Gamma(5, 4 \pi y)}{q}+ \dfrac{33 \Gamma(5, 8 \pi y)}{32 q^2}+ \dfrac{14641 \Gamma(5, 12 \pi y)}{14823 q^3}+ \dfrac{1057 \Gamma(5, 16 \pi y)}{1024 q^4} + \cdots \right) \\
\nonumber&&-\dfrac{3\pi}{728}\dfrac{\zeta(5)}{\zeta(6)} - \dfrac{1647 }{416 \pi^5} \left( q + \dfrac{33 q^2}{32}+ \dfrac{14641 q^3}{14823}+ \dfrac{1057 q^4}{1024}+ \dfrac{3126 q^5}{3125} + \cdots \right)\,, \\
\nonumber Y_{\bm{3},2}^{(-4)} &=& \dfrac{72171 q^{1/3}}{212992 \pi^5} \left( \dfrac{\Gamma(5, 8 \pi y/3)}{q}+ \dfrac{33344 \Gamma(5, 20 \pi y/3)}{34375 q^2}+ \dfrac{1025 \Gamma(5, 32 \pi y/3)}{1024 q^3} + \cdots  \right) \\
\nonumber&& + \dfrac{6561 q^{1/3}}{832 \pi^5} \left( 1 + \dfrac{1057 q}{1024}+ \dfrac{16808 q^2}{16807}+ \dfrac{51579 q^3}{50000}+ \dfrac{371294 q^4}{371293} + \cdots \right)\,, \\
\nonumber Y_{\bm{3},3}^{(-4)} &=& \dfrac{2187 q^{2/3}}{6656 \pi^5} \left( \dfrac{\Gamma(5, 4 \pi y/3)}{q}+ \dfrac{1057 \Gamma(5, 16 \pi y/3)}{1024 q^2}+ \dfrac{16808 \Gamma(5, 28 \pi y/3)}{16807 q^3} + \cdots  \right) \\
&& +\dfrac{216513 q^{2/3}}{26624 \pi^5} \left( 1 + \dfrac{33344 q}{34375}+ \dfrac{1025 q^2}{1024}+ \dfrac{1717888 q^3}{1771561}+ \dfrac{16808 q^4}{16807} + \cdots \right)\,.
\end{eqnarray}

\item{$k_Y = -3$}

Then negative odd weight polyharmonic Maa{\ss} forms of level $3$ can be arranged into two doublets $\bm{\widehat{2}}$ and $\bm{\widehat{2}''}$ of $T'$.
\begin{eqnarray}
\nonumber Y_{\bm{\widehat{2}},1}^{(-3)} &=& - \dfrac{3645 q^{1/3}}{2048 \sqrt{2} \pi^4}\left( \dfrac{\Gamma(4, 8 \pi y/3)}{ q}+ \dfrac{3328 \Gamma(4, 20 \pi y/3)}{3125 q^2}+ \dfrac{257 \Gamma(4, 32 \pi y/3)}{256 q^3} + \cdots  \right)   \\
\nonumber &&\hskip-0.4in - \dfrac{729 q^{1/3}}{64 \sqrt{2} \pi^4}\left( 1 + \dfrac{241 q}{256}+ \dfrac{2402 q^2}{2401}+ \dfrac{117 q^3}{125}+ \dfrac{28562 q^4}{28561}+ \dfrac{61681 q^5}{65536} + \cdots \right)\,,  \\
\nonumber Y_{\bm{\widehat{2}},2}^{(-3)} &=& \dfrac{y^4}{4} + \dfrac{15 }{16 \pi^4}\left( \dfrac{\Gamma(4, 4 \pi y)}{ q}+ \dfrac{123 \Gamma(4, 8 \pi y)}{128 q^2}+ \dfrac{82 \Gamma(4, 12 \pi y)}{81 q^3}+ \dfrac{241 \Gamma(4, 16 \pi y)}{256 q^4} + \cdots  \right)  \\
\nonumber && \hskip-0.4in + \dfrac{\pi}{32\sqrt{3}} \dfrac{L(4,\chi_2)}{L(5,\chi_2)} + \dfrac{369 }{64 \pi^4}\left( q + \dfrac{75 q^2}{82}+ \dfrac{3281 q^3}{3321}+ \dfrac{241 q^4}{256}+ \dfrac{4992 q^5}{5125}+ \dfrac{25 q^6}{27} + \cdots \right)\,,\\
\nonumber Y_{\bm{\widehat{2}''},1}^{(-3)} &=& \dfrac{y^4}{4} - \dfrac{123 }{128 \pi^4}\left( \dfrac{\Gamma(4, 4 \pi y)}{ q}+ \dfrac{75 \Gamma(4, 8 \pi y)}{82 q^2}+ \dfrac{3281 \Gamma(4, 12 \pi y)}{3321 q^3}+ \dfrac{241 \Gamma(4, 16 \pi y)}{256 q^4} + \cdots  \right)   \\
\nonumber && \hskip-0.4in - \dfrac{\pi}{32\sqrt{3}} \dfrac{L(4,\chi_2)}{L(5,\chi_2)} - \dfrac{45 }{8 \pi^4}\left( q + \dfrac{123 q^2}{128}+ \dfrac{82 q^3}{81}+ \dfrac{241 q^4}{256}+ \dfrac{3198 q^5}{3125}+ \dfrac{3281 q^6}{3456} + \cdots \right) \,,  \\
\nonumber Y_{\bm{\widehat{2}''},2}^{(-3)} &=& - \dfrac{243 q^{2/3}}{128 \sqrt{2} \pi^4}\left( \dfrac{\Gamma(4, 4 \pi y/3)}{ q}+ \dfrac{241 \Gamma(4, 16 \pi y/3)}{256 q^2}+ \dfrac{2402 \Gamma(4, 28 \pi y/3)}{2401 q^3} + \cdots  \right)   \\
&& \hskip-0.4in - \dfrac{10935 q^{2/3}}{1024 \sqrt{2} \pi^4}\left( 1 + \dfrac{3328 q}{3125}+ \dfrac{257 q^2}{256}+ \dfrac{15616 q^3}{14641}+ \dfrac{2402 q^4}{2401}+ \dfrac{89088 q^5}{83521}+ \dfrac{3133 q^6}{3125} + \cdots \right)\,.~~~~~~\quad
\end{eqnarray}

\item{$k_Y = -2$}

Similar to the case of $k_Y=-4$, one can organize the weight $k_Y=-2$ polyharmonic Maa{\ss} forms of levle $N=3$ into $T'$ multiplets in the representations $\bm{1}$ and $\bm{3}$.
\begin{eqnarray}
\nonumber Y_{\bm{1}}^{(-2)}&=& \dfrac{y^3}{3} - \dfrac{15\Gamma(3,4\pi y)}{4\pi^3 q} - \dfrac{135\Gamma(3,8\pi y)}{32\pi^3 q^2} - \dfrac{35\Gamma(3,12\pi y)}{9\pi^3 q^3} + \cdots \\
\nonumber&&-\dfrac{\pi}{12}\dfrac{\zeta(3)}{\zeta(4)} - \dfrac{15 q}{2\pi^3} - \dfrac{135 q^2}{16\pi^3} - \dfrac{70 q^3}{9\pi^3} - \dfrac{1095 q^4}{128\pi^3} - \dfrac{189 q^5}{25\pi^3} - \dfrac{35 q^6}{4\pi^3} + \cdots\,,\\
\nonumber Y_{\bm{3},1}^{(-2)}&=& \dfrac{y^3}{3} + \dfrac{21\Gamma(3,4\pi y)}{16\pi^3 q} + \dfrac{189\Gamma(3,8\pi y)}{128\pi^3 q^2} + \dfrac{169\Gamma(3,12\pi y)}{144\pi^3 q^3} + \dfrac{1533 \Gamma(3,16\pi y)}{1024\pi^3 q^4} + \cdots \\
\nonumber&&+\dfrac{\pi}{40}\dfrac{\zeta(3)}{\zeta(4)} + \dfrac{21 q}{8\pi^3} + \dfrac{189 q^2}{64\pi^3} + \dfrac{169 q^3}{72\pi^3} + \dfrac{1533 q^4}{512\pi^3} + \dfrac{1323 q^5}{500\pi^3} + \dfrac{169 q^6}{64\pi^3} + \cdots\,, \\
\nonumber Y_{\bm{3},2}^{(-2)}&=&-\dfrac{729 q^{1/3}}{16\pi^3}\left( \dfrac{\Gamma(3,8\pi y/3)}{16 q} + \dfrac{7\Gamma(3,20\pi y/3)}{125 q^2} + \dfrac{65\Gamma(3,32\pi y/3)}{1024 q^3} + \dfrac{74 \Gamma(3,44\pi y/3)}{1331 q^4} + \cdots \right)  \\
\nonumber&&-\dfrac{81q^{1/3}}{16\pi^3}\left( 1 + \dfrac{73 q}{64} + \dfrac{344 q^2}{343} + \dfrac{567 q^3}{500} + \dfrac{20198 q^4}{2197} + \dfrac{4681 q^5}{4096} + \cdots \right) \,, \\
\nonumber Y_{\bm{3},3 }^{(-2)}&=&-\dfrac{81 q^{2/3}}{32\pi^3} \left( \dfrac{\Gamma(3,4\pi y/3)}{q} + \dfrac{73 \Gamma(3,16\pi y/3)}{64 q^2} + \dfrac{344 \Gamma(3,28\pi y/3)}{343 q^3} + \dfrac{567 \Gamma(3,40\pi y/3)}{500 q^4} + \cdots \right)  \\
&&-\dfrac{729 q^{2/3}}{8\pi^3}\left( \dfrac{1}{16} + \dfrac{7 q}{125} + \dfrac{65 q^2}{1024} + \dfrac{74 q^3}{1331} + \cdots \right)\,.
\end{eqnarray}

\item{$k_Y = -1$}

Similar to the case of $k_Y=-3$, one can organize the weight $k=-1$ polyharmonic Maa{\ss} forms to $T'$ doublets $\bm{\widehat{2}}$ and $\bm{\widehat{2}''}$.
\begin{eqnarray}
\nonumber Y_{\bm{\widehat{2}},1}^{(-1)} &=& \dfrac{243 q^{1/3}}{32 \sqrt{2} \pi^2}\left( \dfrac{\Gamma(2, 8 \pi y/3)}{ q}+ \dfrac{32 \Gamma(2, 20 \pi y/3)}{25 q^2}+ \dfrac{17 \Gamma(2, 32 \pi y/3)}{16 q^3}+ \dfrac{160 \Gamma(2, 44 \pi y/3)}{121 q^4} + \cdots  \right)   \\
\nonumber && + \dfrac{81 q^{1/3}}{8 \sqrt{2} \pi^2}\left( 1 + \dfrac{13 q}{16}+ \dfrac{50 q^2}{49}+ \dfrac{18 q^3}{25}+ \dfrac{170 q^4}{169}+ \dfrac{205 q^5}{256}+ \dfrac{362 q^6}{361} + \cdots \right)\,,  \\
\nonumber Y_{\bm{\widehat{2}},2}^{(-1)} &=& \dfrac{y^2}{2} - \dfrac{9 }{2 \pi^2}\left( \dfrac{\Gamma(2, 4 \pi y)}{ q}+ \dfrac{15 \Gamma(2, 8 \pi y)}{16 q^2}+ \dfrac{10 \Gamma(2, 12 \pi y)}{9 q^3}+ \dfrac{13 \Gamma(2, 16 \pi y)}{16 q^4} + \cdots  \right)  \\
\nonumber && - \dfrac{\pi}{4\sqrt{3}} \dfrac{L(2,\chi_2)}{L(3,\chi_2)} - \dfrac{45 }{8 \pi^2}\left( q + \dfrac{3 q^2}{5}+ \dfrac{41 q^3}{45}+ \dfrac{13 q^4}{16}+ \dfrac{96 q^5}{125}+ \dfrac{2 q^6}{3} + \cdots \right)\,,\\
\nonumber Y_{\bm{\widehat{2}''},1}^{(-1)} &=& \dfrac{y^2}{2} + \dfrac{45 }{8 \pi^2}\left( \dfrac{\Gamma(2, 4 \pi y)}{ q}+ \dfrac{3 \Gamma(2, 8 \pi y)}{5 q^2}+ \dfrac{41 \Gamma(2, 12 \pi y)}{45 q^3}+ \dfrac{13 \Gamma(2, 16 \pi y)}{16 q^4} + \cdots  \right)   \\
\nonumber && + \dfrac{\pi}{4\sqrt{3}} \dfrac{L(2,\chi_2)}{L(3,\chi_2)} + \dfrac{9 }{2 \pi^2}\left( q + \dfrac{15 q^2}{16}+ \dfrac{10 q^3}{9}+ \dfrac{13 q^4}{16}+ \dfrac{6 q^5}{5}+ \dfrac{41 q^6}{48} + \cdots \right) \,, \\
\nonumber Y_{\bm{\widehat{2}''},2}^{(-1)} &=& \dfrac{81 q^{2/3}}{8 \sqrt{2} \pi^2}\left( \dfrac{\Gamma(2, 4 \pi y/3)}{ q}+ \dfrac{13 \Gamma(2, 16 \pi y/3)}{16 q^2}+ \dfrac{50 \Gamma(2, 28 \pi y/3)}{49 q^3}+ \dfrac{18 \Gamma(2, 40 \pi y/3)}{25 q^4} + \cdots  \right)   \\
&& + \dfrac{243 q^{2/3}}{32 \sqrt{2} \pi^2}\left( 1 + \dfrac{32 q}{25}+ \dfrac{17 q^2}{16}+ \dfrac{160 q^3}{121}+ \dfrac{50 q^4}{49}+ \dfrac{384 q^5}{289} + \cdots \right) \,.
\end{eqnarray}

\item{$k_Y = 0$}

At weight $k_Y=0$, besides the constant which can be chosen to be 1, the polyharmonic Maa{\ss} forms of level $N=3$ comprise a non-holomorphic triplet $Y_{\bm{3}}^{(0)}(\tau)$ which are given by
\begin{eqnarray}
\nonumber Y_{\bm{1}}^{(0)} &=& 1 \,, \\
\nonumber Y_{\bm{3},1}^{(0)}&=&y-\dfrac{3\,e^{-4\pi y}}{\pi q} - \dfrac{9\,e^{-8\pi y}}{2\pi q^2} - \dfrac{e^{-12\pi y}}{\pi q^3} - \dfrac{21\,e^{-16\pi y}}{4\pi q^4} - \dfrac{18\,e^{-20\pi y}}{5\pi q^5} - \dfrac{3\,e^{-24\pi y}}{2\pi q^6} + \cdots \\
\nonumber&& - \dfrac{9\log 3}{4\pi} - \dfrac{3q}{\pi} - \dfrac{9q^2}{2\pi} - \dfrac{q^3}{\pi} - \dfrac{21q^4}{4\pi} - \dfrac{18q^5}{5\pi} - \dfrac{3q^6}{2\pi} + \cdots \,, \\
\nonumber Y_{\bm{3},2}^{(0)}&=&\dfrac{27 q^{1/3}e^{\pi y /3}}{\pi} \left( \dfrac{e^{-3\pi y}}{4 q} +\dfrac{e^{-7\pi y}}{5 q^2} + \dfrac{5\,e^{-11\pi y}}{16 q^3} + \dfrac{2\,e^{-15\pi y}}{11 q^4} + \dfrac{2\,e^{-19\pi y}}{7 q^5} + \dfrac{4\,e^{-23\pi y}}{17 q^6} + \cdots \right) \\
\nonumber&&+\dfrac{9q^{1/3}}{2\pi}\left( 1 + \dfrac{7 q}{4} + \dfrac{8 q^2}{7} + \dfrac{9 q^3}{5} + \dfrac{14 q^4}{13} + \dfrac{31 q^5}{16} + \dfrac{20 q^6}{19} + \cdots \right)\,, \\
\nonumber Y_{\bm{3},3}^{(0)}&=&\dfrac{9\,q^{2/3}e^{2\pi y /3}}{2\pi}\left( \dfrac{e^{-2\pi y}}{q} + \dfrac{7\,e^{-6\pi y}}{4 q^2} + \dfrac{8\,e^{-10\pi y}}{7 q^3} + \dfrac{9\,e^{-14\pi y}}{5 q^4} + \dfrac{14\,e^{-18\pi y}}{13 q^5} + \dfrac{31\,e^{-22\pi y}}{16 q^6} + \cdots \right) \\
&&+\dfrac{27 q^{2/3}}{\pi}\left( \dfrac{1}{4} + \dfrac{q}{5}  + \dfrac{5 q^2}{16} + \dfrac{2 q^3}{11} + \dfrac{2 q^4}{7} + \dfrac{9 q^5}{17} + \dfrac{21 q^6}{20} + \cdots \right)\,.
\end{eqnarray}

\item{$k_Y = 1$}

The weight $1$ polyharmonic Maa{\ss} forms can be arranged into $\bm{\widehat{2}}$ and $\bm{\widehat{2}''}$ of $T'$. The $\bm{\widehat{2}''}$ multiplet is non-holomorphic modular function, it can be constructed from the derivative of the weight $1$ non-holomorphic Eisenstein series. As shown in Eq.~\eqref{eq:Y2hatpp-N3}, their Fourier expansion is given by
\begin{eqnarray}
\nonumber Y^{(1)}_{\bm{\widehat{2}''},1} &=& a_0 - 6 \left( q \log 3 + 2 q^2 \log 2 + 2 q^3 \log 3 + q^4 \log 3 + 2 q^5 \log 5 + 2 q^6 \log 2 + 2 q^7 \log 3 + \cdots  \right) \\
\nonumber &&\hskip-0.4in  + \log y - 6 \left(\dfrac{\Gamma(0, 4\pi y)}{q} + \dfrac{ \Gamma(0, 12\pi y)}{q^3} + \dfrac{\Gamma(0, 16\pi y)}{q^4} + \dfrac{2 \Gamma(0, 28\pi y)}{q^7} + \dfrac{\Gamma(0, 36\pi y)}{q^9} + \cdots \right)\,,  \\
\nonumber Y^{(1)}_{\bm{\widehat{2}''},2} &=& - 6 \sqrt{2}  q^{2/3} \left( \log 2 + q \log 5 + 2 q^2 \log 2 + q^3 \log 11 + 2 q^4 \log 2 + q^5 \log 17 + q^6 \log 5 + \cdots \right) \\
&&\hskip-0.4in  - 3 \sqrt{2}  q^{2/3} \left(  \dfrac{\Gamma(0,4\pi y/3)}{q} + \dfrac{\Gamma(0,16\pi y/3)}{q^2} + \dfrac{2\Gamma(0,28\pi y/3)}{q^3} + \dfrac{2\Gamma(0,52\pi y/3)}{q^5}  + \cdots \right)\,,
\end{eqnarray}
where the constant term $a_0$ is
\begin{eqnarray}
a_0 = \gamma_E + \log 3 + \log 4\pi - 6 \log \Gamma\left( \dfrac{1}{3} \right) + 6 \log \Gamma\left( \dfrac{2}{3} \right) \simeq 0.1132\,.
\end{eqnarray}
The weight 1 polyharmonic Maa{\ss} form in the doublet representation $\bm{\widehat{2}}$ is the holomorphic modular form which has been obtained from the Eisenstein series in Eq.~\eqref{eq:Eisenstein_holo_wt1_N3}. They have the following $q$-expansion,
\begin{eqnarray}
Y_{\bm{\widehat{2}}}^{(1)} &=& \begin{pmatrix}
Y_1 \\ Y_2
\end{pmatrix} = \begin{pmatrix}
- 3 \sqrt{2} q^{1/3}\left( 1 + q + 2 q^2 + 2 q^4 + q^5 + 2 q^6 + q^8 + 2 q^9 \cdots \right)  \\
1 + 6 ( q + q^2 + q^4 + 2 q^7 + q^9 + q^{12} + 2 q^{13} )\,.
\end{pmatrix}
\end{eqnarray}
The higher weight modular forms of level $N=3$ can be constructed form tensor product of lower weight ones.

\item{$k_Y = 2$}

The weight $2$ polyharmonic Maa{\ss} forms are composed of the modified Eisenstein series $\widehat{E}_2(\tau)$ and modular forms constructed from the tensor product of $Y_{\bm{\widehat{2}}}^{(1)}$.
\begin{eqnarray}
\nonumber Y^{(2)}_{\bm{1}} &=& \widehat{E}_2(\tau)\,,  \\
Y^{(2)}_{\bm{3}} &=& \left(Y^{(1)}_{\bm{\widehat{2}}}Y^{(1)}_{\bm{\widehat{2}}}\right)_{\bm{3}}=\left(Y^2_2,\quad\sqrt{2}Y_1Y_2,\quad -Y^2_1 \right)^T\,,
\label{eq:modfvec1-N3}
\end{eqnarray}
where the expression of $\widehat{E}_2(\tau)$ has been given in Eq.~\eqref{eq:E2hat}.

\item{$k_Y = 3$}

There are no non-holomorphic polyharmonic Maa{\ss} forms for weight $k_Y\geq 3$. All polyharmonic Maa{\ss} forms with weight $k_Y\geq 3$ are exactly modular forms.
\begin{eqnarray}
\nonumber Y^{(3)}_{\bm{\widehat{2}}}&=&\left(Y^{(1)}_{\bm{\widehat{2}}}Y^{(2)}_{\bm{3}}\right)_{\bm{\widehat{2}}}= \left(3Y_1Y^2_2 ,\quad -\sqrt{2}Y^3_1-Y^3_2 \right)^T,\\
Y^{(3)}_{\bm{\widehat{2}}''}&=&\left(Y^{(1)}_{\bm{\widehat{2}}}Y^{(2)}_{\bm{3}}\right)_{\bm{\widehat{2}}''}=\left(-Y^3_1+\sqrt{2}Y^3_2,\quad 3Y_2Y^2_1 \right)^T \,.
\end{eqnarray}

\item{$k_Y = 4$}

At weight $k_Y=4$, we find five independent modular forms which can be arranged into two singlets $\bm{1}$ and $\bm{1'}$ and a triplet of $T'$. The singlet modular form $Y^{(4)}_{\bm{1'}}(\tau)$ is a cusp form which cannot be related to the polyharmonic Maa{\ss} forms through the differential operators $\xi_{-2}$ and $D^{3}$.
\begin{eqnarray}
\nonumber \hskip-0.12in Y^{(4)}_{\bm{3}}&=&\left(Y^{(1)}_{\bm{\widehat{2}}}Y^{(3)}_{\bm{\widehat{2}}}\right)_{\bm{3}} = \left(-\sqrt{2}Y^3_1Y_2-Y^4_2,~ -Y^4_1+\sqrt{2}Y_1Y^3_2, ~ -3Y^2_1Y^2_2 \right)^T,\\
\nonumber \hskip-0.12in Y^{(4)}_{\bm{1'}}&=&\left(Y^{(1)}_{\bm{\widehat{2}}}Y^{(3)}_{\bm{\widehat{2}}}\right)_{\bm{1'}}=-\sqrt{2}Y^4_1-4Y_1Y^3_2 \,, \\
\hskip-0.12in Y^{(4)}_{\bm{1}}&=&\left(Y^{(1)}_{\bm{\widehat{2}}}Y^{(3)}_{\bm{\widehat{2}}''}\right)_{\bm{1}}=4Y^3_1Y_2-\sqrt{2}Y^4_2 \,.
\end{eqnarray}

\item{$k_Y = 5$}

Similarly, the linearly independent weight 5 modular forms can be constructed from the tensor products of weight 1 and weight 4 modular forms as follows,
\begin{eqnarray}
\nonumber Y^{(5)}_{\bm{\widehat{2}}}&=&\left(Y^{(1)}_{\bm{\widehat{2}}}Y^{(4)}_{\bm{3}}\right)_{\bm{\widehat{2}}}= \left[-2\sqrt{2}Y_{1}^{3}Y_{2}+Y_{2}^{4}\right]\left(Y_{1},\quad Y_{2}\right)^T,\\
\nonumber Y^{(5)}_{\bm{\widehat{2}}'}&=&\left(Y^{(1)}_{\bm{\widehat{2}}}Y^{(4)}_{\bm{3}}\right)_{\bm{\widehat{2}}'}=\left[-Y_{1}^{4}-2\sqrt{2}Y_{1}Y_{2}^{3}\right]\left(Y_{1},\quad Y_{2}\right)^T,\\
Y^{(5)}_{\bm{\widehat{2}}''}&=&\left(Y^{(1)}_{\bm{\widehat{2}}}Y^{(4)}_{\bm{3}}\right)_{\bm{\widehat{2}}''}=\left(-5Y^3_1Y^2_2-\sqrt{2}Y^5_2,~ -\sqrt{2}Y^5_1+5Y^2_1Y^3_2 \right)^T \,.
\end{eqnarray}

\item{$k_Y = 6$}

Finally, the linearly independent weight 6 modular forms of level 3 can be decomposed into one singlet $\bm{1}$ and two triplets $\bm{3}$ under $T'$,
\begin{eqnarray}
\nonumber Y^{(6)}_{\bm{3}I}&=& \left(Y^{(1)}_{\bm{\widehat{2}}}Y^{(5)}_{\bm{\widehat{2}}}\right)_{\bm{3}} = \left[-2\sqrt{2}Y_{1}^{3}Y_{2}+Y_{2}^{4}\right] \left( Y_{2}^{2},\quad\sqrt{2}Y_{1}Y_{2},\quad -Y_{1}^{2}\right)^{T},\\
\nonumber Y^{(6)}_{\bm{3}II}&=&\left(Y^{(1)}_{\bm{\widehat{2}}}Y^{(5)}_{\bm{\widehat{2}}'}\right)_{\bm{3}}=\left[-Y_{1}^{4}-2\sqrt{2}Y_{1}Y_{2}^{3}\right] \left( -Y_{1}^{2},\quad Y_{2}^{2},\quad\sqrt{2}Y_{1}Y_{2}\right)^{T},\\
Y^{(6)}_{\bm{1}}&=&\left(Y^{(1)}_{\bm{\widehat{2}}}Y^{(5)}_{\bm{\widehat{2}}''}\right)_{\bm{1}}=\sqrt{2}Y^6_2-\sqrt{2}Y^6_1+10Y^3_1Y^3_2 \,.
\label{eq:modfvec2}
\end{eqnarray}

\end{itemize}

\subsection{$N=4$\label{app:Polyharmonic_N_4}}

As shown in Eq.~\eqref{eq:cusps-N}, the cusps of $\Gamma(4)$ are $\{0,1,2,3,1/2,i\infty\}= \{\overline{0/1}, \overline{1/1}, \overline{2/1}, \overline{3/1}, \overline{1/2}, \overline{1/0}\}$. Consequently there are six weight $k$ non-holomorphic Eisenstein series including $E_k(\tau; s; 0)$, $E_k(\tau; s; 1)$, $E_k(\tau; s; 2)$, $E_k(\tau; s; 3)$, $E_k(\tau; s; 1/2)$, $E_k(\tau; s; i\infty)$
at level $N=4$. From Eq.~\eqref{eq:Eisenstein_trans}, we know that the modular generators $S$ and $T$ act on these non-holomorphic Eisenstein series as follow,
\begin{eqnarray}
\nonumber E_k(S\tau; s; i\infty) &=& (-\tau)^{k} E_k(\tau; s; 0) \,, \\
\nonumber E_k(S\tau; s; 0) &=& (-1)^k (-\tau)^{k} E_k(\tau; s; i\infty)\,,  \\
\nonumber E_k(S\tau; s; 1) &=& (-\tau)^{k} E_k(\tau; s; 3) \,, \\
\nonumber E_k(S\tau; s; 2) &=& (-1)^k (-\tau)^{k} E_k(\tau; s; 1/2)\,,  \\
\nonumber E_k(S\tau; s; 3) &=& (-1)^k (-\tau)^{k} E_k(\tau; s; 1)\,,  \\
 E_k(S\tau; s; 1/2) &=& (-\tau)^{k} E_k(\tau; s; 2) \,,
\end{eqnarray}
and
\begin{eqnarray}
\nonumber E_k(T\tau; s; i\infty) &=& E_k(\tau; s; i\infty)\,,  \\
\nonumber E_k(T\tau; s; 0) &=& E_k(\tau; s; 3) \,, \\
\nonumber E_k(T\tau; s; 1) &=& E_k(\tau; s; 0) \,, \\
\nonumber E_k(T\tau; s; 2) &=& E_k(\tau; s; 1) \,, \\
\nonumber E_k(T\tau; s; 3) &=& E_k(\tau; s; 2) \,, \\
E_k(T\tau; s; 1/2) &=& (-1)^k E_k(\tau; s; 1/2)\,.
\end{eqnarray}
The linear combinations of even weight Eisenstein series can be arranged into a singlet $Y^{(2k)}_{\bm{1}}(\tau)$,
a doublet $Y^{(2k)}_{\bm{2}}(\tau)=(Y^{(2k)}_{\bm{2}, 1}, Y^{(2k)}_{\bm{2}, 2})^T$ and a triplet $Y^{(2k)}_{\bm{3}}(\tau)=(Y^{(2k)}_{\bm{3}, 1}, Y^{(2k)}_{\bm{3}, 2}, Y^{(2k)}_{\bm{3}, 3})^T$ of $\Gamma'_4\cong S'_4$.
Each component of these modular multiplets are given by
\begin{eqnarray}
\nonumber Y^{(2k)}_{\bm{1}} &=& E_{2k}(\tau; s; i\infty) + E_{2k}(\tau; s; 1/2) + E_{2k}(\tau; s; 0) + E_{2k}(\tau; s; 1) + E_{2k}(\tau; s; 2) + E_{2k}(\tau; s; 3) \,, \\
\nonumber Y^{(2k)}_{\bm{2},1} &=& E_{2k}(\tau; s; i\infty) + E_{2k}(\tau; s; 1/2) - \dfrac{1}{2} \Big( E_{2k}(\tau; s; 0) + E_{2k}(\tau; s; 1) + E_{2k}(\tau; s; 2) + E_{2k}(\tau; s; 3) \Big)\,, \\
\nonumber Y^{(2k)}_{\bm{2},2} &=& \dfrac{\sqrt{3}}{2}\Big( E_{2k}(\tau; s; 0) - E_{2k}(\tau; s; 1) + E_{2k}(\tau; s; 2) - E_{2k}(\tau; s; 3) \Big)\,,\\
\nonumber Y^{(2k)}_{\bm{3},1} &=& E_{2k}(\tau; s; i\infty) - E_{2k}(\tau; s; 1/2) \,,\\
\nonumber Y^{(2k)}_{\bm{3},2} &=& \dfrac{\sqrt{2}}{2} \Big( E_{2k}(\tau; s; 0) + i E_{2k}(\tau; s; 1) - E_{2k}(\tau; s; 2) - i E_{2k}(\tau; s; 3) \Big) \,, \\
\label{eq:N=4_Y1-Y2-Y3} Y^{(2k)}_{\bm{3},3} &=& \dfrac{\sqrt{2}}{2} \Big( E_{2k}(\tau; s; 0) - i E_{2k}(\tau; s; 1) - E_{2k}(\tau; s; 2) + i E_{2k}(\tau; s; 3) \Big) \,.
\end{eqnarray}
On the other hand, the odd weight Eisenstein series of level $N=4$ can be arranged into two $S'_4$ triplets $Y^{(2k+1)}_{\bm{\widehat{3}}}=(Y^{(2k+1)}_{\bm{\widehat{3}},1}, Y^{(2k+1)}_{\bm{\widehat{3}}, 2}, Y^{(2k+1)}_{\bm{\widehat{3}}, 3})^T$ and $Y^{(2k+1)}_{\bm{\widehat{3}'}}=(Y^{(2k+1)}_{\bm{\widehat{3}'},1}, Y^{(2k+1)}_{\bm{\widehat{3}'}, 2}, Y^{(2k+1)}_{\bm{\widehat{3}'}, 3})^T$ with
\begin{small}
\begin{eqnarray}
\nonumber Y^{(2k+1)}_{\bm{\widehat{3}},1} &=& - \dfrac{i}{\sqrt{2}} \Big( E_{2k+1}(\tau; s; 0) - i E_{2k+1}(\tau; s; 1) - E_{2k+1}(\tau; s; 2) + i E_{2k+1}(\tau; s; 3) \Big) \,,  \\
\nonumber Y^{(2k+1)}_{\bm{\widehat{3}},2} &=& E_{2k+1}(\tau; s; i\infty) + \dfrac{i}{2} \left( E_{2k+1}(\tau; s; 0) + E_{2k+1}(\tau; s; 1) + E_{2k+1}(\tau; s; 2) + E_{2k+1}(\tau; s; 3) \right)\,,  \\
\nonumber Y^{(2k+1)}_{\bm{\widehat{3}},3} &=& - E_{2k+1}(\tau; s; 1/2) - \dfrac{i}{2} \left( E_{2k+1}(\tau; s; 0) - E_{2k+1}(\tau; s; 1) + E_{2k+1}(\tau; s; 2) - E_{2k+1}(\tau; s; 3) \right)\,,\\
\nonumber Y^{(2k+1)}_{\bm{\widehat{3}'},1} &=& \dfrac{i}{\sqrt{2}} \Big( E_{2k+1}(\tau; s; 0) + i E_{2k+1}(\tau; s; 1) - E_{2k+1}(\tau; s; 2) - i E_{2k+1}(\tau; s; 3) \Big)\,,  \\
\nonumber Y^{(2k+1)}_{\bm{\widehat{3}'},2} &=& - E_{2k+1}(\tau; s; 1/2) + \dfrac{i}{2} \Big( E_{2k+1}(\tau; s; 0) - E_{2k+1}(\tau; s; 1) + E_{2k+1}(\tau; s; 2) - E_{2k+1}(\tau; s; 3) \Big)\,,  \\
\label{eq:N=4_Y3h_Y3hp} Y^{(2k+1)}_{\bm{\widehat{3}'},3} &=& E_{2k+1}(\tau; s; i\infty) - \dfrac{i}{2} \Big( E_{2k+1}(\tau; s; 0) + E_{2k+1}(\tau; s; 1) + E_{2k+1}(\tau; s; 2) + E_{2k+1}(\tau; s; 3) \Big)\,.~~~~~\quad
\end{eqnarray}
\end{small}
As explained in section~\ref{sec:integer_weight}, the non-holomorphic
Eisenstein series $E_k(N; \tau; s; \overline{A/C})$ satisfies the harmonic condition for $s=1-k$, and they span the linear space of the polyharmonic Maa{\ss} forms at level $N$ for $k\leq1$. In the following, we report the Fourier expansion of $E_k(N; \tau; 1-k; \overline{A/C})$ for $N=4$ which will be dropped for simplicity. The non-holomorphic Eisenstein series of weight $2k$ for $k\leq 0$ are given by
\begin{eqnarray}
E_{2k}(\tau; 1-2k; i\infty) &=& y^{1-2k} + (-1)^k \dfrac{2^{6k-1}}{1-2^{2k-2}} \dfrac{\zeta(1-2k)}{\zeta(2-2k)} \dfrac{\pi}{4} \nonumber \\
\nonumber &&\hspace{-2.8cm} + \dfrac{(-1)^k 2^{2k}}{(1-2^{2k-2})\zeta(2-2k)} \dfrac{\pi}{4} \sum_{n=1}^{+\infty} q^{n/4} \sum_{\substack{m|n,\, m\in\mathbb{Z} \\ \frac{n}{m}\equiv 0\,({\rm mod}\, 4)}} \left| \dfrac{n}{m} \right|^{2k-1} \, (-i)^{m}  \\
\nonumber &&\hspace{-2.8cm} + \dfrac{(-1)^k 2^{2k}}{(-2k)!(1-2^{2k-2})\zeta(2-2k)} \dfrac{\pi}{4} \sum_{n=-1}^{-\infty} q^{n/4} \Gamma(1-2k,-\pi n y) \sum_{\substack{m|n,\, m\in\mathbb{Z} \\ \frac{n}{m}\equiv 0\,({\rm mod}\, 4)}} \left| \dfrac{n}{m} \right|^{2k-1} \, (-i)^{m} \,, \\
E_{2k}(\tau; 1-2k; 0) &=& (-1)^k \dfrac{2^{2k}(1-2^{2k-1})}{1-2^{2k-2}} \dfrac{\zeta(1-2k)}{\zeta(2-2k)} \dfrac{\pi}{4} \nonumber \\
\nonumber &&\hspace{-2.8cm} + \dfrac{(-1)^k 2^{2k}}{(1-2^{2k-2})\zeta(2-2k)} \dfrac{\pi}{4} \sum_{n=1}^{+\infty} q^{n/4} \sum_{\substack{m|n,\, m\in\mathbb{Z} \\ \frac{n}{m}\equiv 1\,({\rm mod}\, 4)}} \left| \dfrac{n}{m} \right|^{2k-1} \nonumber   \\
\nonumber && \hspace{-2.8cm}+ \dfrac{(-1)^k 2^{2k}}{(-2k)!(1-2^{2k-2})\zeta(2-2k)} \dfrac{\pi}{4} \sum_{n=-1}^{-\infty} q^{n/4} \Gamma(1-2k,-\pi n y) \sum_{\substack{m|n,\, m\in\mathbb{Z} \\ \frac{n}{m}\equiv 1\,({\rm mod}\, 4)}} \left| \dfrac{n}{m} \right|^{2k-1} \,,  \\
E_{2k}(\tau; 1-2k; 1) &=& (-1)^k \dfrac{2^{2k}(1-2^{2k-1})}{1-2^{2k-2}} \dfrac{\zeta(1-2k)}{\zeta(2-2k)} \dfrac{\pi}{4} \nonumber \\
\nonumber &&\hspace{-2.8cm} + \dfrac{(-1)^k 2^{2k}}{(1-2^{2k-2})\zeta(2-2k)} \dfrac{\pi}{4} \sum_{n=1}^{+\infty} q^{n/4} \sum_{\substack{m|n,\, m\in\mathbb{Z} \\ \frac{n}{m}\equiv 1\,({\rm mod}\, 4)}} \left| \dfrac{n}{m} \right|^{2k-1} \, (-i)^{m}  \\
\nonumber &&\hspace{-2.8cm} + \dfrac{(-1)^k 2^{2k}}{(-2k)!(1-2^{2k-2})\zeta(2-2k)} \dfrac{\pi}{4} \sum_{n=-1}^{-\infty} q^{n/4} \Gamma(1-2k,-\pi n y) \sum_{\substack{m|n,\, m\in\mathbb{Z} \\ \frac{n}{m}\equiv 1\,({\rm mod}\, 4)}} \left| \dfrac{n}{m} \right|^{2k-1} \, (-i)^{m}  \,, \\
E_{2k}(\tau; 1-2k; 2) &=& (-1)^k \dfrac{2^{2k}(1-2^{2k-1})}{1-2^{2k-2}} \dfrac{\zeta(1-2k)}{\zeta(2-2k)} \dfrac{\pi}{4} \nonumber \\
\nonumber &&\hspace{-2.8cm} + \dfrac{(-1)^k 2^{2k}}{(1-2^{2k-2})\zeta(2-2k)} \dfrac{\pi}{4} \sum_{n=1}^{+\infty} q^{n/4} \sum_{\substack{m|n,\, m\in\mathbb{Z} \\ \frac{n}{m}\equiv 1\,({\rm mod}\, 4)}} \left| \dfrac{n}{m} \right|^{2k-1} \, (-1)^m  \\
\nonumber && \hspace{-2.8cm} + \dfrac{(-1)^k 2^{2k}}{(-2k)!(1-2^{2k-2})\zeta(2-2k)} \dfrac{\pi}{4} \sum_{n=-1}^{-\infty} q^{n/4} \Gamma(1-2k,-\pi n y) \sum_{\substack{m|n,\, m\in\mathbb{Z} \\ \frac{n}{m}\equiv 1\,({\rm mod}\, 4)}} \left| \dfrac{n}{m} \right|^{2k-1} \, (-1)^m  \\
E_{2k}(\tau; 1-2k; 3) &=& (-1)^k \dfrac{2^{2k}(1-2^{2k-1})}{1-2^{2k-2}} \dfrac{\zeta(1-2k)}{\zeta(2-2k)} \dfrac{\pi}{4} \nonumber \\
\nonumber &&\hspace{-2.8cm} + \dfrac{(-1)^k 2^{2k}}{(1-2^{2k-2})\zeta(2-2k)} \dfrac{\pi}{4} \sum_{n=1}^{+\infty} q^{n/4} \sum_{\substack{m|n,\, m\in\mathbb{Z} \\ \frac{n}{m}\equiv 1\,({\rm mod}\, 4)}} \left| \dfrac{n}{m} \right|^{2k-1} \, i^{m}  \\
\nonumber &&\hspace{-2.8cm} + \dfrac{(-1)^k 2^{2k}}{(-2k)!(1-2^{2k-2})\zeta(2-2k)} \dfrac{\pi}{4} \sum_{n=-1}^{-\infty} q^{n/4} \Gamma(1-2k,-\pi n y) \sum_{\substack{m|n,\, m\in\mathbb{Z} \\ \frac{n}{m}\equiv 1\,({\rm mod}\, 4)}} \left| \dfrac{n}{m} \right|^{2k-1} \, i^{m}  \\
E_{2k}(\tau; 1-2k; 1/2) &=& (-1)^k \dfrac{2^{2k}(1-2^{2k-1})}{1-2^{2k-2}} \dfrac{\zeta(1-2k)}{\zeta(2-2k)} \dfrac{\pi}{4}\nonumber \\
\nonumber &&\hspace{-2.8cm} + \dfrac{(-1)^k 2^{2k}}{(1-2^{2k-2})\zeta(2-2k)} \dfrac{\pi}{4} \sum_{n=1}^{+\infty} q^{n/4} \sum_{\substack{m|n,\, m\in\mathbb{Z} \\ \frac{n}{m}\equiv 2\,({\rm mod}\, 4)}} \left| \dfrac{n}{m} \right|^{2k-1} \, (-i)^{m}  \\
 && \hspace{-2.8cm} + \dfrac{(-1)^k 2^{2k}}{(-2k)!(1-2^{2k-2})\zeta(2-2k)} \dfrac{\pi}{4} \sum_{n=-1}^{-\infty} q^{n/4} \Gamma(1-2k,-\pi n y) \sum_{\substack{m|n,\, m\in\mathbb{Z} \\ \frac{n}{m}\equiv 2\,({\rm mod}\, 4)}} \left| \dfrac{n}{m} \right|^{2k-1} \, (-i)^{m}\,.~~~~~\quad
\end{eqnarray}
Plugging the above expressions into Eq.~\eqref{eq:N=4_Y1-Y2-Y3} with $s=1-2k$, we can obtain the explicit expressions of the even weight polyharmonic Maa{\ss} form multiplets of level $N=4$. Up to an overall normalization constant which can not be fixed in the bottom-up approach, the results
\begin{eqnarray}
\nonumber Y_{\bm{1}}^{(2k)}(\tau) &=& (-1)^k \dfrac{2^{2k}\, \pi}{1-2k}\dfrac{\zeta(1-2k)}{\zeta(2-2k)} + \dfrac{y^{1-2k}}{1-2k}  + \dfrac{(-1)^k\, 2^{2k}\,\pi}{(1-2k)\zeta(2-2k)} \sum_{n=1}^{+\infty} \sigma_{2k-1}(n) q^n \\
\nonumber&&\hskip-0.4in + \dfrac{(-1)^k\, 2^{2k}\,\pi}{(1-2k)!\zeta(2-2k)} \sum_{n=1}^{+\infty} \sigma_{2k-1}(n) q^{-n}\Gamma(1-2k,4\pi n y)\,,\\
\nonumber Y_{\bm{2},1}^{(2k)}(\tau) &=& (-1)^k \dfrac{2^{2k-1}(2^{2k}-1)\,\pi}{(1-2k)(1-2^{2k-2})} \dfrac{\zeta(1-2k)}{\zeta(2-2k)} + \dfrac{y^{1-2k}}{1-2k}  \\
\nonumber && \hskip-0.4in + \dfrac{(-1)^k\, 2^{2k-1}\, \pi}{(1-2k)(1-2^{2k-2})\zeta(2-2k)} \sum_{n=1}^{+\infty} \Big( (2+2^{2k})\sigma_{2k-1}(n) - 3\sigma_{2k-1}(2n) \Big) q^n  \\
\nonumber &&\hskip-0.4in + \dfrac{(-1)^k\, 2^{2k-1}\, \pi}{(1-2k)!(1-2^{2k-2})\zeta(2-2k)} \sum_{n=1}^{+\infty} \Big( (2+2^{2k})\sigma_{2k-1}(n) - 3\sigma_{2k-1}(2n) \Big) q^{-n} \Gamma(1-2k, 4\pi n y) \,, \\
\nonumber Y_{\bm{2},2}^{(2k)}(\tau) &=& \dfrac{(-1)^k\, 2^{2k-1} \sqrt{3}\, \pi}{(1-2k)(1-2^{2k-2})\zeta(2-2k)} \sum_{n=0}^{+\infty} \sigma_{2k-1}(2n+1) q^{(2n+1)/2}  \\
\nonumber &&\hskip-0.4in +\dfrac{(-1)^k\, 2^{2k-1} \sqrt{3}\, \pi}{(1-2k)!(1-2^{2k-2})\zeta(2-2k)} \sum_{n=0}^{+\infty} \sigma_{2k-1}(2n+1) q^{-(2n+1)/2} \Gamma(1-2k,2\pi(2n+1) y) \,,\\
\nonumber Y_{\bm{3},1}^{(2k)}(\tau) &=& (-1)^k \dfrac{2^{2k}(2^{2k}-1)\,\pi}{(1-2k)(2^{2-2k}-1)} \dfrac{\zeta(1-2k)}{\zeta(2-2k)} + \dfrac{y^{1-2k}}{1-2k}  \\
\nonumber &&\hskip-0.4in + \dfrac{(-1)^k \,\pi}{(1-2k)(2^{2-2k}-1) \zeta(2-2k)} \times \left( \sum_{n=0}^{+\infty} 2^{2k} \sigma_{2k-1}(n) q^{2n+1}  \right. \\
\nonumber && \hskip-0.4in \left. +\sum_{n=1}^{+\infty} \Big( (4 + 2^{2k} + 2^{4k})\sigma_{2k-1}(n) - (4 + 2^{2k+1}) \sigma_{2k-1}(2n) \Big) q^{2n} \right) \\
\nonumber &&\hskip-0.4in +  \dfrac{(-1)^k \,\pi}{(1-2k)!(2^{2-2k}-1) \zeta(2-2k)} \times  \left( \sum_{n=0}^{+\infty} 2^{2k} \sigma_{2k-1}(n) q^{-(2n+1)}  \Gamma(1-2k, 4\pi(2n+1)y) \right.  \\
\nonumber &&\hskip-0.4in \left. +\sum_{n=1}^{+\infty} \Big( (4 + 2^{2k} + 2^{4k})\sigma_{2k-1}(n) - (4 + 2^{2k+1}) \sigma_{2k-1}(2n) \Big)  q^{-2n}  \Gamma(1-2k, 8\pi n y) \right)\,, \\
\nonumber Y_{\bm{3},2}^{(2k)}(\tau) &=& \dfrac{(-1)^k \, 2\sqrt{2}\,\pi}{(1-2k)(2^{2-2k}-1) \zeta(2-2k)} \sum_{n=0}^{+\infty} \sigma_{2k-1}(4n+1) q^{(4n+1)/4} \\
\nonumber &&\hskip-0.4in + \dfrac{(-1)^k \, 2\sqrt{2}\,\pi}{(1-2k)!(2^{2-2k}-1) \zeta(2-2k)} \sum_{n=0}^{+\infty} \sigma_{2k-1}(4n+3) q^{-(4n+3)/4} \Gamma(1-2k,(4n+3)\pi y) \,, \\
\nonumber Y_{\bm{3},2}^{(2k)}(\tau) &=& \dfrac{(-1)^k \, 2\sqrt{2}\,\pi}{(1-2k)(2^{2-2k}-1) \zeta(2-2k)} \sum_{n=0}^{+\infty} \sigma_{2k-1}(4n+3) q^{(4n+3)/4} \\
 &&\hskip-0.4in + \dfrac{(-1)^k \, 2\sqrt{2}\,\pi}{(1-2k)!(2^{2-2k}-1) \zeta(2-2k)} \sum_{n=0}^{+\infty} \sigma_{2k-1}(4n+1) q^{-(4n+1)/4} \Gamma(1-2k,(4n+1)\pi y)\,.
\end{eqnarray}
Similar to the cases of $N=3$, the weight zero  polyharmonic Maa{\ss} form can be obtain from the above expressions by taking the limit $k\rightarrow 0$ and using the Laurent expansion of Riemann zeta function in Eq.~\eqref{eq:zeta_series}.

Analogously the Fourier expansion of the negative odd weight $2k+1$ Eisenstein series for $k<0$ are given by
\begin{small}
\begin{eqnarray}
\nonumber E_{2k+1}(\tau; -2k; i\infty) &=& y^{-2k} - \dfrac{(-1)^k 2^{2k+1}}{L(1-2k,\chi_2)} \dfrac{i\pi}{4} \sum_{n=1}^{+\infty} q^{n/4} \sum_{\substack{m|n,\, m\in\mathbb{Z} \\ \frac{n}{m}\equiv 0\,({\rm mod}\, 4)}} {\rm sign}(m) \left| \dfrac{n}{m} \right|^{2k} \, (-i)^{m} \\
\nonumber &&\hspace{-2.8cm} - \dfrac{(-1)^k 2^{2k+1}}{(-2k-1)! L(1-2k,\chi_2)} \dfrac{i\pi}{4}   \sum_{n=-1}^{-\infty} q^{n/4} \Gamma(-2k,-\pi n y) \sum_{\substack{m|n,\, m\in\mathbb{Z} \\ \frac{n}{m}\equiv 0\,({\rm mod}\, 4)}} {\rm sign}(m) \left| \dfrac{n}{m} \right|^{2k} \, (-i)^{m} \,, \\
\nonumber E_{2k+1}(\tau; -2k; 0) &=& -(-1)^k 2^{2k+1} \dfrac{L(-2k,\chi_2)}{L(1-2k,\chi_2)} \dfrac{i\pi}{4} + \dfrac{(-1)^k 2^{2k+1}}{L(1-2k,\chi_2)} \dfrac{i\pi}{4} \sum_{n=1}^{+\infty} q^{n/4} \sum_{\substack{m|n,\, m\in\mathbb{Z} \\ \frac{n}{m}\equiv 1\,({\rm mod}\, 4)}} {\rm sign}(m) \left| \dfrac{n}{m} \right|^{2k}  \\
\nonumber &&\hspace{-2.8cm} - \dfrac{(-1)^k 2^{2k+1}}{(-2k-1)! L(1-2k,\chi_2)} \dfrac{i\pi}{4}  \sum_{n=-1}^{-\infty} q^{n/4} \Gamma(-2k,-\pi n y) \sum_{\substack{m|n,\, m\in\mathbb{Z} \\ \frac{n}{m}\equiv 1\,({\rm mod}\, 4)}} {\rm sign}(m) \left| \dfrac{n}{m} \right|^{2k}  \,,\\
\nonumber E_{2k+1}(\tau; -2k; 1) &=& -(-1)^k 2^{2k+1} \dfrac{L(-2k,\chi_2)}{L(1-2k,\chi_2)} \dfrac{i\pi}{4} + \dfrac{(-1)^k 2^{2k+1}}{L(1-2k,\chi_2)} \dfrac{i\pi}{4} \sum_{n=1}^{+\infty} q^{n/4} \sum_{\substack{m|n,\, m\in\mathbb{Z} \\ \frac{n}{m}\equiv 1\,({\rm mod}\, 4)}} {\rm sign}(m) \left| \dfrac{n}{m} \right|^{2k} \, (-i)^{m} \\
\nonumber &&\hspace{-2.8cm} - \dfrac{(-1)^k 2^{2k+1}}{(-2k-1)! L(1-2k,\chi_2)} \dfrac{i\pi}{4}  \sum_{n=-1}^{-\infty} q^{n/4} \Gamma(-2k,-\pi n y) \sum_{\substack{m|n,\, m\in\mathbb{Z} \\ \frac{n}{m}\equiv 1\,({\rm mod}\, 4)}} {\rm sign}(m) \left| \dfrac{n}{m} \right|^{2k} \, (-i)^{m}\,,  \\
\nonumber E_{2k+1}(\tau; -2k; 2) &=& -(-1)^k 2^{2k+1} \dfrac{L(-2k,\chi_2)}{L(1-2k,\chi_2)} \dfrac{i\pi}{4} + \dfrac{(-1)^k 2^{2k+1}}{L(1-2k,\chi_2)} \dfrac{i\pi}{4} \sum_{n=1}^{+\infty} q^{n/4} \sum_{\substack{m|n,\, m\in\mathbb{Z} \\ \frac{n}{m}\equiv 1\,({\rm mod}\, 4)}} {\rm sign}(m) \left| \dfrac{n}{m} \right|^{2k} \, (-1)^{m} \\
\nonumber &&\hspace{-2.8cm} - \dfrac{(-1)^k 2^{2k+1}}{(-2k-1)! L(1-2k,\chi_2)} \dfrac{i\pi}{4}  \sum_{n=-1}^{-\infty} q^{n/4} \Gamma(-2k,-\pi n y) \sum_{\substack{m|n,\, m\in\mathbb{Z} \\ \frac{n}{m}\equiv 1\,({\rm mod}\, 4)}} {\rm sign}(m) \left| \dfrac{n}{m} \right|^{2k} \, (-1)^{m} \,, \\
\nonumber E_{2k+1}(\tau; -2k; 3) &=& -(-1)^k 2^{2k+1} \dfrac{L(-2k,\chi_2)}{L(1-2k,\chi_2)} \dfrac{i\pi}{4} + \dfrac{(-1)^k 2^{2k+1}}{L(1-2k,\chi_2)} \dfrac{i\pi}{4} \sum_{n=1}^{+\infty} q^{n/4} \sum_{\substack{m|n,\, m\in\mathbb{Z} \\ \frac{n}{m}\equiv 1\,({\rm mod}\, 4)}} {\rm sign}(m) \left| \dfrac{n}{m} \right|^{2k} \, i^{m} \\
\nonumber &&\hspace{-2.8cm} - \dfrac{(-1)^k 2^{2k+1}}{(-2k-1)! L(1-2k,\chi_2)} \dfrac{i\pi}{4}  \sum_{n=-1}^{-\infty} q^{n/4} \Gamma(-2k,-\pi n y) \sum_{\substack{m|n,\, m\in\mathbb{Z} \\ \frac{n}{m}\equiv 1\,({\rm mod}\, 4)}} {\rm sign}(m) \left| \dfrac{n}{m} \right|^{2k} \, i^{m} \,, \\
\nonumber E_{2k+1}(\tau; -2k; 1/2) &=&  -\dfrac{(-1)^k 2^{2k+1}}{L(1-2k,\chi_2)} \dfrac{i\pi}{4} \sum_{n=1}^{+\infty} q^{n/4} \sum_{\substack{m|n,\, m\in\mathbb{Z} \\ \frac{n}{m}\equiv 2\,({\rm mod}\, 4)}} {\rm sign}(m) \left| \dfrac{n}{m} \right|^{2k} \, (-i)^{m} \\
&&\hspace{-2.8cm} - \dfrac{(-1)^k 2^{2k+1}}{(-2k-1)! L(1-2k,\chi_2)} \dfrac{i\pi}{4}  \sum_{n=-1}^{-\infty} q^{n/4} \Gamma(-2k,-\pi n y) \sum_{\substack{m|n,\, m\in\mathbb{Z} \\ \frac{n}{m}\equiv 2\,({\rm mod}\, 4)}} {\rm sign}(m) \left| \dfrac{n}{m} \right|^{2k} \, (-i)^{m}\,. \label{eq:Einstein-2k+1-N4}
\end{eqnarray}
\end{small}
Plugging the above expressions into Eq.~\eqref{eq:N=4_Y3h_Y3hp} with $s=-2k$, we can obtain the $q$-expansion of the negative odd weight polyharmonic Maa{\ss} form multiplets of level $N=4$ up to an overall normalization constant,
\begin{eqnarray}
\nonumber Y_{\bm{\widehat{3}},1}^{(2k+1)}(\tau)&=& - \dfrac{(-1)^k\, 2^{2k}\sqrt{2}\, \pi}{(-2k)L(1-2k,\chi_2)} \sum_{n=0}^{+\infty} \widetilde{\sigma}^{\chi_2}_{2k}(4n+3) q^{(4n+3)/4}  \\
\nonumber &&\hskip-0.6in + \dfrac{(-1)^k\, 2^{2k}\sqrt{2}\, \pi }{(-2k)!L(1-2k,\chi_2)} \sum_{n=0}^{+\infty} \widetilde{\sigma}^{\chi_2}_{2k}(4n+1) q^{-(4n+1)/4} \Gamma(-2k,(4n+1)\pi y) \,, \\
\nonumber Y_{\bm{\widehat{3}},2}^{(2k+1)}(\tau) &=& -(-1)^k \dfrac{2^{2k}\,\pi}{(-2k)} \dfrac{L(-2k,\chi_2)}{L(1-2k,\chi_2)} -\dfrac{y^{-2k}}{2k} \\
\nonumber &&\hskip-0.6in - \dfrac{(-1)^k\, 2^{2k}\, \pi}{(-2k)L(1-2k,\chi_2)}\sum_{n=1}^{+\infty} \Big( \widehat{\sigma}^{\chi_2}_{2k}(n) - 4^{2k} \widetilde{\sigma}^{\chi_2}_{2k}(n) \Big) q^n  \\
\nonumber &&\hskip-0.6in - \dfrac{(-1)^k\, 2^{2k}\, \pi}{(-2k)!L(1-2k,\chi_2)}\sum_{n=1}^{+\infty} \Big( \widehat{\sigma}^{\chi_2}_{2k}(n) + 4^{2k} \widetilde{\sigma}^{\chi_2}_{2k}(n) \Big) q^{-n} \Gamma(-2k,4\pi n y) \,, \\
\nonumber Y_{\bm{\widehat{3}},3}^{(2k+1)}(\tau) &=& \dfrac{(-1)^k\, 2^{2k}\, \pi}{(-2k)L(1-2k,\chi_2)}\sum_{n=0}^{+\infty} \Big( \widehat{\sigma}^{\chi_2}_{2k}(2n+1) - 2^{2k} \widetilde{\sigma}^{\chi_2}_{2k}(2n+1) \Big) q^{(2n+1)/2}  \\
\nonumber &&\hskip-0.6in + \dfrac{(-1)^k\, 2^{2k}\, \pi}{(-2k)!L(1-2k,\chi_2)}\sum_{n=0}^{+\infty} \Big( \widehat{\sigma}^{\chi_2}_{2k}(2n+1) + 2^{2k} \widetilde{\sigma}^{\chi_2}_{2k}(2n+1) \Big) q^{-(2n+1)/2} \Gamma(-2k,(4n+2)\pi y)\,, \\
\nonumber Y_{\bm{\widehat{3}'},1}^{(2k+1)}(\tau) &=& - \dfrac{(-1)^k\, 2^{2k}\sqrt{2}\, \pi}{(-2k)L(1-2k,\chi_2)} \sum_{n=0}^{+\infty} \widetilde{\sigma}^{\chi_2}_{2k}(4n+1) q^{(4n+1)/4}  \\
\nonumber &&\hskip-0.6in + \dfrac{(-1)^k\, 2^{2k}\sqrt{2}\, \pi }{(-2k)!L(1-2k,\chi_2)} \sum_{n=0}^{+\infty} \widetilde{\sigma}^{\chi_2}_{2k}(4n+3) q^{-(4n+3)/4} \Gamma(-2k,(4n+3)\pi y) \,, \\
\nonumber Y_{\bm{\widehat{3}'},2}^{(2k+1)}(\tau) &=& - \dfrac{(-1)^k\, 2^{2k}\, \pi}{(-2k)L(1-2k,\chi_2)}\sum_{n=0}^{+\infty} \Big( \widehat{\sigma}^{\chi_2}_{2k}(2n+1) + 2^{2k} \widetilde{\sigma}^{\chi_2}_{2k}(2n+1) \Big) q^{(2n+1)/2}  \\
\nonumber &&\hskip-0.6in - \dfrac{(-1)^k\, 2^{2k}\, \pi}{(-2k)!L(1-2k,\chi_2)}\sum_{n=0}^{+\infty} \Big( \widehat{\sigma}^{\chi_2}_{2k}(2n+1) - 2^{2k} \widetilde{\sigma}^{\chi_2}_{2k}(2n+1) \Big) q^{-(2n+1)/2} \Gamma(-2k,(4n+2)\pi y)\,, \\
\nonumber Y_{\bm{\widehat{3}'},3}^{(2k+1)}(\tau) &=& (-1)^k \dfrac{2^{2k}\,\pi}{(-2k)} \dfrac{L(-2k,\chi_2)}{L(1-2k,\chi_2)} -\dfrac{y^{-2k}}{2k}  \\
\nonumber && \hskip-0.6in + \dfrac{(-1)^k\, 2^{2k}\, \pi}{(-2k)L(1-2k,\chi_2)}\sum_{n=1}^{+\infty} \Big( \widehat{\sigma}^{\chi_2}_{2k}(n) + 4^{2k} \widetilde{\sigma}^{\chi_2}_{2k}(n) \Big) q^n  \\
&&\hskip-0.6in + \dfrac{(-1)^k\, 2^{2k}\, \pi}{(-2k)!L(1-2k,\chi_2)}\sum_{n=1}^{+\infty} \Big( \widehat{\sigma}^{\chi_2}_{2k}(n) - 4^{2k} \widetilde{\sigma}^{\chi_2}_{2k}(n) \Big) q^{-n} \Gamma(-2k,4\pi n y) \,,
\end{eqnarray}
where $\widehat{\sigma}^{\chi_2}_{2k}$ and $\widetilde{\sigma}^{\chi_2}_{2k}$ denote the modified divisor sums associated with $\chi_2$, as defined in Eq.~{\eqref{eq:modified_divisorSum}}.

In what follows, we discuss the weight $1$ non-holomorphic Eisenstein series $E_1(N; \tau; 0; \overline{A/C})$ and the first order derivative $E^{(1)}_1(N; \tau; \overline{A/C})$ at level $N=4$, and the argument $N$ will be omitted by default. The $q$-expansion of $E_1(N; \tau; 0; \overline{A/C})$ at different cusps are of the following form,
\begin{eqnarray}
\nonumber E_1(\tau; 0; i\infty) &=& 1 + 4 \sum_{n=1}^{+\infty} \sum_{d|n} \chi_2(d) q^n \\
\nonumber E_1(\tau; 0; 0) &=& \dfrac{i}{2} + 2i\sum_{n=1}^{+\infty} \sum_{d|n} \chi_2(d) q^n + 2i q^{1/2} \sum_{n=0}^{+\infty} \sum_{d|(2n+1)} \chi_2(d) q^n + 2i q^{1/4} \sum_{n=0}^{+\infty} \sum_{d|(4n+1)} \chi_2(d) q^n \\
\nonumber E_1(\tau; 0; 1) &=& \dfrac{i}{2} + 2i\sum_{n=1}^{+\infty} \sum_{d|n} \chi_2(d) q^n - 2i q^{1/2} \sum_{n=0}^{+\infty} \sum_{d|(2n+1)} \chi_2(d) q^n + 2 q^{1/4} \sum_{n=0}^{+\infty} \sum_{d|(4n+1)} \chi_2(d) q^n \\
\nonumber E_1(\tau; 0; 2) &=& \dfrac{i}{2} + 2i\sum_{n=1}^{+\infty} \sum_{d|n} \chi_2(d) q^n + 2i q^{1/2} \sum_{n=0}^{+\infty} \sum_{d|(2n+1)} \chi_2(d) q^n - 2i q^{1/4} \sum_{n=0}^{+\infty} \sum_{d|(4n+1)} \chi_2(d) q^n \\
\nonumber E_1(\tau; 0; 3) &=& \dfrac{i}{2} + 2i\sum_{n=1}^{+\infty} \sum_{d|n} \chi_2(d) q^n - 2i q^{1/2} \sum_{n=0}^{+\infty} \sum_{d|(2n+1)} \chi_2(d) q^n - 2 q^{1/4} \sum_{n=0}^{+\infty} \sum_{d|(4n+1)} \chi_2(d) q^n \\
E_1(\tau; 0; 1/2) &=&  4 q^{1/2} \sum_{n=0}^{+\infty} \sum_{d|(2n+1)} \chi_2(d) q^n\,,
\end{eqnarray}
which can be arranged into two $S'_4$ triplets $Y^{(1)}_{\bm{\widehat{3}}}(\tau)=(Y^{(1)}_{\bm{\widehat{3}},1}, Y^{(1)}_{\bm{\widehat{3}}, 2}, Y^{(1)}_{\bm{\widehat{3}}, 3})^T$ and $Y^{(1)}_{\bm{\widehat{3}'}}(\tau)=(Y^{(1)}_{\bm{\widehat{3}'},1}, Y^{(1)}_{\bm{\widehat{3}'}, 2}, Y^{(1)}_{\bm{\widehat{3}'}, 3})^T$, as shown in Eq.~\eqref{eq:N=4_Y3h_Y3hp}. It is notable that $Y^{(1)}_{\bm{\widehat{3}'}}(\tau)$ is the holomorphic modular forms of weight one and level 4,
\begin{eqnarray}
\nonumber Y^{(1)}_{\bm{\widehat{3}'},1}(\tau) \nonumber&=& \dfrac{i}{2\sqrt{2}} \left(E_1(\tau; 0; 0) + i E_1(\tau; 0; 1) - E_1(\tau; 0; 2) - i E_1(\tau; 0; 3) \right) \\
\nonumber&=& - 2\sqrt{2} q^{1/4} \sum_{n=0}^{+\infty} \sum_{d|(4n+1)} \chi_2(d) q^n \\
\nonumber &=& - 2\sqrt{2} q^{1/4} \left( 1 + 2 q + q^2 + 2 q^3 + 2 q^4 + 3 q^6 + 2 q^7 + \cdots \right)\,,  \\
\nonumber Y^{(1)}_{\bm{\widehat{3}'}, 2}(\tau) &=& - \dfrac{1}{2} E_1(\tau; 0; 1/2) + \dfrac{i}{4} \left( E_1(\tau; 0; 0) - E_1(\tau; 0; 1) + E_1(\tau; 0; 2) - E_1(\tau; 0; 3) \right)   \\
\nonumber &=& - 4 q^{1/2} \sum_{n=0}^{+\infty} \sum_{d|(2n+1)} \chi_2(d) q^n \\
\nonumber &=& - 4 q^{1/2} \left( 1 + 2 q^2 + q^4 + 2 q^6 + 2 q^8 + 3 q^{12} + 2 q^{14} + \cdots \right) \,, \\
\nonumber Y^{(1)}_{\bm{\widehat{3}'},3}(\tau) &=& \dfrac{1}{2} E_1(\tau; 0; i\infty) - \dfrac{i}{4} \left( E_1(\tau; 0; 0) + E_1(\tau; 0; 1) + E_1(\tau; 0; 2) + E_1(\tau; 0; 3) \right)   \\
\nonumber &=& 1 + 4 \sum_{n=1}^{+\infty} \sum_{d|n} \chi_2(d) q^n  \\
\label{eq:w1l4-3hatp}&=& 1 + 4 q + 4 q^2 + 4 q^4 + 8 q^5 + 4 q^8 + 4 q^9 + \cdots \,.
\end{eqnarray}
The above expressions are identical with weight 1 modular forms of level 4 which are constructed from the theta constants~\cite{Novichkov:2020eep,Liu:2020akv,Ding:2022nzn}. However, another triplet $Y^{(1)}_{\bm{\widehat{3}}}(\tau)$ is exactly vanishing, i.e.
\begin{eqnarray}
\nonumber E_1(\tau; 0; 0) - i E_1(\tau; 0; 1) - E_1(\tau; 0; 2) + i E_1(\tau; 0; 3) &=& 0 \,, \\
\nonumber E_1(\tau; 0; i\infty) + \dfrac{i}{2} \left( E_1(\tau; 0; 0) + E_1(\tau; 0; 1) + E_1(\tau; 0; 2) + E_1(\tau; 0; 3) \right)  &=&0 \,, \\
-E_1(\tau; 0; 1/2) - \dfrac{i}{2} \left( E_1(\tau; 0; 0) - E_1(\tau; 0; 1) + E_1(\tau; 0; 2) - E_1(\tau; 0; 3) \right) &=& 0\,.
\end{eqnarray}
The first order derivative of the Eisenstein series $E^{(1)}_1(\tau; \overline{A/C})$ gives rise to all the polyharmonic Maa{\ss} form of weight 1, and its Fourier expansion takes the following form,
\begin{eqnarray}
\nonumber E_1^{(1)}(\tau; \overline{A/C}) &=& \left[-\delta\left( \dfrac{C}{4} \right) \chi^{*}_2(3A) + \dfrac{i}{2} \chi^{*}_2(C)   \right] \left[\log y + \gamma_E + \log 16\pi - 4 \log \Gamma\left( \dfrac{1}{4} \right) + 4\log  \Gamma\left( \dfrac{3}{4} \right)  \right]  \\
\nonumber && + 2i \sum_{n=1}^{+\infty} \sum_{\substack{m|n\,, m\in\mathbb{Z} \\ \frac{n}{m}\equiv C\, ({\rm mod}\, 4)}} {\rm sign}(m) i^{3Am} \left( 2\log |m| - \log n \right) q^{n/4}  \\
\nonumber && -  2i \sum_{n=-1}^{-\infty} \sum_{\substack{m|n\,, m\in\mathbb{Z} \\ \frac{n}{m}\equiv C\, ({\rm mod}\, 4)}} {\rm sign}(m) i^{3Am} \Gamma(0,-\pi ny) q^{n/4}  \\
&& + \left[-\gamma_E - \log 16\pi + 4 \log \Gamma\left( \dfrac{1}{4} \right) - 4\log  \Gamma\left( \dfrac{3}{4} \right) \right] E_1(\tau; 0; \overline{A/C})\,.
\end{eqnarray}
where $\delta(x)=1$ if $x\in \mathbb{Z}$ and $\delta(x)=0$ if $x\not\in \mathbb{Z}$.
The modular function $E_1^{(1)}(\tau; \overline{A/C})$ at the six cusps of $\Gamma(4)$ can be  arranged into two triplets $Y_{\bm{\widehat{3}'}}^{(1)}(\tau)$ and $Y^{(1)}_{\bm{\widehat{3}}}(\tau)$ transforming as $\bm{\widehat{3}'}$ and $\bm{\widehat{3}}$ respectively under $S'_4$. Moreover, $Y_{\bm{\widehat{3}'}}^{(1)}(\tau)$ is the  weight one and level 4 holomorphic modular form given in Eq.~\eqref{eq:w1l4-3hatp},  and $Y^{(1)}_{\bm{\widehat{3}}}(\tau)$ is the non-holomorphic  polyharmonic Maa{\ss} form. With the above general results, we give the explicit $q$-expansion of the level $N=4$ polyharmonic Maa{\ss} form multiplets from weight $-4$ to weight $6$ in the following.

\begin{itemize}[labelindent=-0.8em,leftmargin=0.3em]
\item{$k_Y=-4$}

We can organize the weight $k=-4$ polyharmonic Maa{\ss} forms of level $4$ into a singlet $Y_{\bm{1}}^{(-4)}(\tau)$, a doublet $Y_{\bm{2}}^{(-4)}(\tau)=(Y_{\bm{2},1}^{(-4)}, Y_{\bm{2},2}^{(-4)})^{T}$ and a triplet $Y_{\bm{3}}^{(-4)}(\tau)=(Y_{\bm{3},1}^{(-4)}, Y_{\bm{3},2}^{(-4)}, Y_{\bm{3},3}^{(-4)})^T$ of $S'_4$ with
\begin{eqnarray}
\nonumber Y_{\bm{1}}^{(-4)}&=& \dfrac{y^5}{5} + \dfrac{63\Gamma(5,4\pi y)}{128\pi^5 q} + \dfrac{2079\Gamma(5,8\pi y)}{4096\pi^5 q^2} + \dfrac{427\Gamma(5,12\pi y)}{864\pi^5 q^3} + \dfrac{66591\Gamma(5,16\pi y)}{131072 \pi^5 q^4} + \cdots   \\
\nonumber&&\hskip-0.4in +\dfrac{\pi}{80}\dfrac{\zeta(5)}{\zeta(6)} + \dfrac{189 q}{16\pi^5} + \dfrac{6237 q^2}{512 \pi^5} + \dfrac{427 q^3}{36 \pi^5} + \dfrac{199773 q^4}{16384 \pi^5} + \cdots \,,\\
\nonumber Y_{\bm{2},1}^{(-4)} &=& \dfrac{y^5}{5} - \dfrac{ 33 \Gamma(5, 4 \pi y)}{128 \pi^5 q}- \dfrac{ 993 \Gamma(5, 8 \pi y)}{4096 \pi^5 q^2}- \dfrac{ 671 \Gamma(5, 12 \pi y)}{2592 \pi^5 q^3}- \dfrac{ 31713 \Gamma(5, 16 \pi y)}{131072 \pi^5 q^4} + \cdots  \\
\nonumber &&\hskip-0.4in - \dfrac{\pi}{168}\dfrac{\zeta(5)}{\zeta(6)} - \dfrac{99 q}{16 \pi^5}- \dfrac{2979 q^2}{512 \pi^5}- \dfrac{671 q^3}{108 \pi^5}- \dfrac{95139 q^4}{16384 \pi^5}- \dfrac{154737 q^5}{25000 \pi^5} + \cdots \,, \\
\nonumber Y_{\bm{2},2}^{(-4)} &=& \dfrac{\sqrt{3}q^{1/2}}{4\pi^5} \left( \dfrac{\Gamma(5, 2 \pi y)}{q}+ \dfrac{244 \Gamma(5, 6 \pi y)}{243 q^2}+ \dfrac{3126 \Gamma(5, 10 \pi y)}{3125 q^3}+ \dfrac{16808 \Gamma(5, 14 \pi y)}{16807 q^4} + \cdots  \right) \\
\nonumber&&\hskip-0.4in + \dfrac{6\sqrt{3}q^{1/2}}{\pi^5} \left( 1 + \dfrac{244 q}{243}+ \dfrac{3126 q^2}{3125}+ \dfrac{16808 q^3}{16807}+ \dfrac{59293 q^4}{59049}+ \dfrac{161052 q^5}{161051} + \cdots \right)\,, \\
\nonumber Y_{\bm{3},1}^{(-4)}&=& \dfrac{y^5}{5} + \dfrac{\Gamma(5, 4 \pi y)}{128 \pi^5 q}- \dfrac{ 33 \Gamma(5, 8 \pi y)}{4096 \pi^5 q^2}+ \dfrac{61 \Gamma(5, 12 \pi y)}{7776 \pi^5 q^3}- \dfrac{ 993 \Gamma(5, 16 \pi y)}{131072 \pi^5 q^4}+ \dfrac{1563 \Gamma(5, 20 \pi y)}{200000 \pi^5 q^5} + \cdots  \\
\nonumber &&\hskip-0.4in - \dfrac{\pi}{5376}\dfrac{\zeta(5)}{\zeta(6)}  + \dfrac{3 q}{16 \pi^5}- \dfrac{99 q^2}{512 \pi^5}+ \dfrac{61 q^3}{324 \pi^5}- \dfrac{2979 q^4}{16384 \pi^5}+ \dfrac{4689 q^5}{25000 \pi^5} + \cdots \,, \\
\nonumber Y_{\bm{3},2}^{(-4)}&=& \dfrac{61\sqrt{2}q^{1/4}}{243\pi^5} \left( \dfrac{\Gamma(5, 3 \pi y)}{q}+ \dfrac{1021086 \Gamma(5, 7 \pi y)}{1025227 q^2}+ \dfrac{9783909 \Gamma(5, 11 \pi y)}{9824111 q^3}+ \dfrac{3126 \Gamma(5, 15 \pi y)}{3125 q^4} + \cdots  \right)  \\
\nonumber &&\hskip-0.4in + \dfrac{6\sqrt{2}q^{1/4}}{\pi^5} \left( 1 + \dfrac{3126 q}{3125}+ \dfrac{59293 q^2}{59049}+ \dfrac{371294 q^3}{371293}+ \dfrac{1419858 q^4}{1419857}+ \dfrac{4101152 q^5}{4084101} + \cdots \right) \,, \\
\nonumber Y_{\bm{3},3}^{(-4)}&=& \dfrac{q^{3/4}}{2\sqrt{2}\pi^5} \left( \dfrac{\Gamma(5, \pi y)}{q}+ \dfrac{3126 \Gamma(5, 5 \pi y)}{3125 q^2}+ \dfrac{59293 \Gamma(5, 9 \pi y)}{59049 q^3}+ \dfrac{371294 \Gamma(5, 13 \pi y)}{371293 q^4} + \cdots  \right) \\
&&\hskip-0.4in +\dfrac{488\sqrt{2}q^{3/4}}{81\pi^5} \left( 1 + \dfrac{1021086 q}{1025227}+ \dfrac{9783909 q^2}{9824111}+ \dfrac{3126 q^3}{3125}+ \dfrac{150423075 q^4}{151042039} + \cdots \right)\,.
\end{eqnarray}

\item{$k_Y=-3$}

The weight $k_Y=-3$ polyharmonic Maa{\ss} forms of level $4$ can be arranged into two a triplets $\bm{\widehat{3}}$ and $\bm{\widehat{3}'}$ of $S'_4$.
\begin{eqnarray}
\nonumber Y_{\bm{\widehat{3}},1}^{(-3)} &=& \dfrac{4\sqrt{2} q^{3/4}}{5 \pi^4}\left( \dfrac{\Gamma(4, \pi y)}{ q}+ \dfrac{626 \Gamma(4, 5 \pi y)}{625 q^2}+ \dfrac{6481 \Gamma(4, 9 \pi y)}{6561 q^3}+ \dfrac{28562 \Gamma(4, 13 \pi y)}{28561 q^4} + \cdots  \right)  \\
\nonumber &&\hskip-0.4in + \dfrac{128 \sqrt{2} q^{3/4}}{27 \pi^4}\left( 1 + \dfrac{2430 q}{2401}+ \dfrac{14823 q^2}{14641}+ \dfrac{626 q^3}{625}+ \dfrac{131949 q^4}{130321}+ \dfrac{283338 q^5}{279841} + \cdots \right) \,, \\
\nonumber Y_{\bm{\widehat{3}},2}^{(-3)} &=& \dfrac{y^4}{4} - \dfrac{257 }{320 \pi^4}\left( \dfrac{\Gamma(4, 4 \pi y)}{ q}+ \dfrac{4097 \Gamma(4, 8 \pi y)}{4112 q^2}+ \dfrac{6800 \Gamma(4, 12 \pi y)}{6939 q^3}+ \dfrac{65537 \Gamma(4, 16 \pi y)}{65792 q^4} + \cdots  \right)  \\
\nonumber &&\hskip-0.4in - \dfrac{\pi}{64} \dfrac{L(4,\chi_2)}{L(5,\chi_2)} - \dfrac{153 }{32 \pi^4}\left( q + \dfrac{273 q^2}{272}+ \dfrac{4112 q^3}{4131}+ \dfrac{257 q^4}{256}+ \dfrac{626 q^5}{625}+ \dfrac{241 q^6}{243} + \cdots \right) \,, \\
\nonumber Y_{\bm{\widehat{3}},3}^{(-3)} &=& \dfrac{17 q^{1/2}}{20 \pi^4}\left( \dfrac{\Gamma(4, 2 \pi y)}{ q}+ \dfrac{400 \Gamma(4, 6 \pi y)}{459 q^2}+ \dfrac{626 \Gamma(4, 10 \pi y)}{625 q^3}+ \dfrac{36000 \Gamma(4, 14 \pi y)}{40817 q^4} + \cdots  \right)  \\
\nonumber&&\hskip-0.4in + \dfrac{9 q^{1/2}}{2 \pi^4}\left( 1 + \dfrac{272 q}{243}+ \dfrac{626 q^2}{625}+ \dfrac{2720 q^3}{2401}+ \dfrac{6481 q^4}{6561}+ \dfrac{16592 q^5}{14641}+ \dfrac{28562 q^6}{28561} + \cdots \right)\,,\\
\nonumber Y_{\bm{\widehat{3}'},1}^{(-3)} &=& - \dfrac{64 \sqrt{2} q^{1/4}}{81 \pi^4}\left( \dfrac{\Gamma(4, 3 \pi y)}{ q}+ \dfrac{2430 \Gamma(4, 7 \pi y)}{2401 q^2}+ \dfrac{14823 \Gamma(4, 11 \pi y)}{14641 q^3}+ \dfrac{626 \Gamma(4, 15 \pi y)}{625 q^4} + \cdots  \right)  \\
\nonumber &&\hskip-0.4in - \dfrac{24 \sqrt{2} q^{1/4}}{5 \pi^4}\left( 1 + \dfrac{626 q}{625}+ \dfrac{6481 q^2}{6561}+ \dfrac{28562 q^3}{28561}+ \dfrac{83522 q^4}{83521}+ \dfrac{64000 q^5}{64827}+ \dfrac{391251 q^6}{390625} + \cdots \right)\,,  \\
\nonumber Y_{\bm{\widehat{3}'},2}^{(-3)} &=& - \dfrac{3 q^{1/2}}{4 \pi^4}\left( \dfrac{\Gamma(4, 2 \pi y)}{ q}+ \dfrac{272 \Gamma(4, 6 \pi y)}{243 q^2}+ \dfrac{626 \Gamma(4, 10 \pi y)}{625 q^3}+ \dfrac{2720 \Gamma(4, 14 \pi y)}{2401 q^4} + \cdots  \right)  \\
\nonumber &&\hskip-0.4in - \dfrac{51 q^{1/2}}{10 \pi^4}\left( 1 + \dfrac{400 q}{459}+ \dfrac{626 q^2}{625}+ \dfrac{36000 q^3}{40817}+ \dfrac{6481 q^4}{6561}+ \dfrac{219600 q^5}{248897}+ \dfrac{28562 q^6}{28561} + \cdots \right) \,,  \\
\nonumber Y_{\bm{\widehat{3}'},3}^{(-3)} &=& \dfrac{y^4}{4} + \dfrac{51 }{64 \pi^4}\left( \dfrac{\Gamma(4, 4 \pi y)}{ q}+ \dfrac{273 \Gamma(4, 8 \pi y)}{272 q^2}+ \dfrac{4112 \Gamma(4, 12 \pi y)}{4131 q^3}+ \dfrac{257 \Gamma(4, 16 \pi y)}{256 q^4} + \cdots  \right)  \\
&&\hskip-0.4in + \dfrac{\pi}{64} \dfrac{L(4,\chi_2)}{L(5,\chi_2)} + \dfrac{771 }{160 \pi^4}\left(q + \dfrac{4097 q^2}{4112}+ \dfrac{6800 q^3}{6939}+ \dfrac{65537 q^4}{65792}+ \dfrac{626 q^5}{625}+ \dfrac{2275 q^6}{2313} + \cdots \right) \,.
\end{eqnarray}

\item{$k_Y=-2$}

The linear combination of the weight $k_Y=-2$ polyharmonic Maa{\ss} forms of level $4$ can gives a singlet $\bm{1}$, a doublet $\bm{2}$ and a triplet $\bm{3}$ of $S'_4$.
\begin{eqnarray}
\nonumber Y_{\bm{1}}^{(-2)}&=& \dfrac{y^3}{3} - \dfrac{15\Gamma(3,4\pi y)}{4\pi^3 q} - \dfrac{135\Gamma(3,8\pi y)}{32\pi^3 q^2} - \dfrac{35\Gamma(3,12\pi y)}{9\pi^3 q^3} + \cdots \\
\nonumber &&\hskip-0.4in -\dfrac{\pi}{12}\dfrac{\zeta(3)}{\zeta(4)} - \dfrac{15 q}{2\pi^3} - \dfrac{135 q^2}{16\pi^3} - \dfrac{70 q^3}{9\pi^3} - \dfrac{1095 q^4}{128\pi^3} - \dfrac{189 q^5}{25\pi^3} - \dfrac{35 q^6}{4\pi^3} + \cdots \,, \\
\nonumber Y_{\bm{2},1}^{(-2)}&=& \dfrac{y^3}{3} + \dfrac{9 \Gamma(3, 4 \pi y)}{4 \pi^3 q}+ \dfrac{57 \Gamma(3, 8 \pi y)}{32 \pi^3 q^2}+ \dfrac{7 \Gamma(3, 12 \pi y)}{3 \pi^3 q^3}+ \dfrac{441 \Gamma(3, 16 \pi y)}{256 \pi^3 q^4} + \cdots  \\
\nonumber &&\hskip-0.4in +\dfrac{\pi}{30} \dfrac{\zeta(3)}{\zeta(4)} + \dfrac{9 q}{2 \pi^3}+ \dfrac{57 q^2}{16 \pi^3}+ \dfrac{14 q^3}{3 \pi^3}+ \dfrac{441 q^4}{128 \pi^3}+ \dfrac{567 q^5}{125 \pi^3} + \cdots \,, \\
\nonumber Y_{\bm{2},2}^{(-2)}&=& -\dfrac{2\sqrt{3}q^{1/2}}{\pi^3}\left( \dfrac{\Gamma(3, 2 \pi y)}{q}+ \dfrac{28 \Gamma(3, 6 \pi y)}{27 q^2}+ \dfrac{126 \Gamma(3, 10 \pi y)}{125 q^3}+ \dfrac{344 \Gamma(3, 14 \pi y)}{343 q^4} + \cdots  \right) \\
\nonumber&&\hskip-0.4in -\dfrac{4\sqrt{3}q^{1/2}}{\pi^3} \left( 1 + \dfrac{28 q}{27}+ \dfrac{126 q^2}{125}+ \dfrac{344 q^3}{343}+ \dfrac{757 q^4}{729}+ \dfrac{1332 q^5}{1331} + \cdots \right) \,, \\
\nonumber Y_{\bm{3},1}^{(-2)}&=& \dfrac{y^3}{3} - \dfrac{ \Gamma(3, 4 \pi y)}{4 \pi^3 q}+ \dfrac{9 \Gamma(3, 8 \pi y)}{32 \pi^3 q^2}- \dfrac{ 7 \Gamma(3, 12 \pi y)}{27 \pi^3 q^3}+ \dfrac{57 \Gamma(3, 16 \pi y)}{256 \pi^3 q^4}- \dfrac{ 63 \Gamma(3, 20 \pi y)}{250 \pi^3 q^5} + \cdots  \\
\nonumber &&\hskip-0.4in + \dfrac{\pi}{240}\dfrac{\zeta(3)}{\zeta(4)} - \dfrac{q}{2 \pi^3}+ \dfrac{9 q^2}{16 \pi^3}- \dfrac{14 q^3}{27 \pi^3}+ \dfrac{57 q^4}{128 \pi^3}- \dfrac{63 q^5}{125 \pi^3} + \cdots \,, \\
\nonumber Y_{\bm{3},2}^{(-2)}&=&-\dfrac{56\sqrt{2} q^{1/4}}{27\pi^3} \left( \dfrac{\Gamma(3, 3 \pi y)}{q}+ \dfrac{2322 \Gamma(3, 7 \pi y)}{2401 q^2}+ \dfrac{8991 \Gamma(3, 11 \pi y)}{9317 q^3}+ \dfrac{126 \Gamma(3, 15 \pi y)}{125 q^4} + \cdots  \right) \\
\nonumber &&\hskip-0.4in - \dfrac{4\sqrt{2} q^{1/4}}{\pi^3} \left( 1 + \dfrac{126 q}{125}+ \dfrac{757 q^2}{729}+ \dfrac{2198 q^3}{2197}+ \dfrac{4914 q^4}{4913}+ \dfrac{1376 q^5}{1323} + \cdots \right) \,, \\
\nonumber Y_{\bm{3},3}^{(-2)}&=& -\dfrac{2\sqrt{2} q^{3/4}}{\pi^3} \left( \dfrac{\Gamma(3, \pi y)}{q}+ \dfrac{126 \Gamma(3, 5 \pi y)}{125 q^2}+ \dfrac{757 \Gamma(3, 9 \pi y)}{729 q^3}+ \dfrac{2198 \Gamma(3, 13 \pi y)}{2197 q^4} + \cdots  \right)  \\
&&\hskip-0.4in -\dfrac{112\sqrt{2} q^{3/4}}{27\pi^3} \left( 1 + \dfrac{2322 q}{2401}+ \dfrac{8991 q^2}{9317}+ \dfrac{126 q^3}{125}+ \dfrac{6615 q^4}{6859}+ \dfrac{82134 q^5}{85169} + \cdots \right)\,.
\end{eqnarray}

\item{$k_Y=-1$}

There are two triplets $Y_{\bm{\widehat{3}}}^{(-1)}(\tau)$ and $Y_{\bm{\widehat{3}'}}^{(-1)}(\tau)$ of the weight $k_Y=-1$ polyharmonic Maa{\ss} forms at level $4$ as follow,
\begin{eqnarray}
\nonumber Y_{\bm{\widehat{3}},1}^{(-1)} &=& - \dfrac{4 \sqrt{2} q^{3/4}}{\pi^2}\left( \dfrac{\Gamma(2, \pi y)}{ \
q}+ \dfrac{26 \Gamma(2, 5 \pi y)}{25 q^2}+ \dfrac{73 \Gamma(2, 9 \pi y)}{81 q^3}+ \dfrac{170 \Gamma(2, 13 \pi y)}{169 q^4} + \cdots  \right) \\
\nonumber && \hskip-0.4in - \dfrac{32 \sqrt{2} q^{3/4}}{9 \pi^2}\left( 1 + \dfrac{54 q}{49}+ \dfrac{135 q^2}{121}+ \dfrac{26 q^3}{25}+ \dfrac{405 q^4}{361}+ \dfrac{594 q^5}{529} + \cdots \right) \,, \\
\nonumber Y_{\bm{\widehat{3}},2}^{(-1)} &=& \dfrac{y^2}{2} + \dfrac{17 }{4 \pi^2} \left( \dfrac{\Gamma(2, 4 \pi y)}{ q}+ \dfrac{65 \Gamma(2, 8 \pi y)}{68 q^2}+ \dfrac{40 \Gamma(2, 12 \pi y)}{51 q^3}+ \dfrac{257 \Gamma(2, 16 \pi y)}{272 q^4} + \cdots  \right)   \\
\nonumber &&\hskip-0.4in +\dfrac{\pi}{8} \dfrac{L(2,\chi_2)}{L(3,\chi_2)} + \dfrac{15 }{4 \pi^2}\left( q + \dfrac{21 q^2}{20}+ \dfrac{136 q^3}{135}+ \dfrac{17 q^4}{16}+ \dfrac{26 q^5}{25}+ \dfrac{26 q^6}{27}+ \dfrac{272 q^7}{245} + \cdots \right) \,, \\
\nonumber Y_{\bm{\widehat{3}},3}^{(-1)} &=& - \dfrac{5 q^{1/2}}{\pi^2}\left( \dfrac{\Gamma(2, 2 \pi y)}{ q}+ \dfrac{8 \Gamma(2, 6 \pi y)}{15 q^2}+ \dfrac{26 \Gamma(2, 10 \pi y)}{25 q^3}+ \dfrac{144 \Gamma(2, 14 \pi y)}{245 q^4} + \cdots  \right)   \\
\nonumber&&\hskip-0.4in - \dfrac{3 q^{1/2}}{\pi^2}\left( 1 + \dfrac{40 q}{27}+ \dfrac{26 q^2}{25}+ \dfrac{80 q^3}{49}+ \dfrac{73 q^4}{81}+ \dfrac{200 q^5}{121}+ \dfrac{170 q^6}{169} + \cdots \right)\,, \\
\nonumber Y_{\bm{\widehat{3}'},1}^{(-1)} &=& \dfrac{32 \sqrt{2} q^{1/4}}{9 \pi^2}\left( \dfrac{\Gamma(2, 3 \pi y)}{ q}+ \dfrac{54 \Gamma(2, 7 \pi y)}{49 q^2}+ \dfrac{135 \Gamma(2, 11 \pi y)}{121 q^3}+ \dfrac{26 \Gamma(2, 15 \pi y)}{25 q^4} + \cdots  \right)   \\
\nonumber &&\hskip-0.4in + \dfrac{4 \sqrt{2} q^{1/4}}{\pi^2}\left( 1 + \dfrac{26 q}{25}+ \dfrac{73 q^2}{81}+ \dfrac{170 q^3}{169}+ \dfrac{290 q^4}{289}+ \dfrac{128 q^5}{147} + \cdots \right) \,,  \\
\nonumber Y_{\bm{\widehat{3}'},2}^{(-1)} &=& \dfrac{3 q^{1/2}}{\pi^2}\left( \dfrac{\Gamma(2, 2 \pi y)}{ q}+ \dfrac{40 \Gamma(2, 6 \pi y)}{27 q^2}+ \dfrac{26 \Gamma(2, 10 \pi y)}{25 q^3}+ \dfrac{80 \Gamma(2, 14 \pi y)}{49 q^4} + \cdots  \right)  \\
\nonumber &&\hskip-0.4in + \dfrac{5 q^{1/2}}{\pi^2}\left( 1 + \dfrac{8 q}{15}+ \dfrac{26 q^2}{25}+ \dfrac{144 q^3}{245}+ \dfrac{73 q^4}{81}+ \dfrac{72 q^5}{121}+ \dfrac{170 q^6}{169} + \cdots \right) \,,  \\
\nonumber Y_{\bm{\widehat{3}'},3}^{(-1)} &=& \dfrac{y^2}{2} - \dfrac{15 }{4 \pi^2}\left( \dfrac{\Gamma(2, 4 \pi y)}{ q}+ \dfrac{21 \Gamma(2, 8 \pi y)}{20 q^2}+ \dfrac{136 \Gamma(2, 12 \pi y)}{135 q^3}+ \dfrac{17 \Gamma(2, 16 \pi y)}{16 q^4} + \cdots  \right)  \\
&&\hskip-0.4in -\dfrac{\pi}{8} \dfrac{L(2,\chi_2)}{L(3,\chi_2)} - \dfrac{17 }{4 \pi^2}\left( q + \dfrac{65 q^2}{68}+ \dfrac{40 q^3}{51}+ \dfrac{257 q^4}{272}+ \dfrac{26 q^5}{25}+ \dfrac{14 q^6}{17} + \cdots \right)\,.
\end{eqnarray}

\item{$k_Y=0$}

At this weight, there is a $S'_4$ singlet $Y^{(0)}_{\bm{1}}(\tau)$, a doublet $Y^{(0)}_{\bm{2}}(\tau)$ and a triplet $Y_{\bm{3}}^{(0)}(\tau)$ of the polyharmonic Maa{\ss} forms.
\begin{eqnarray}
\nonumber Y^{(0)}_{\bm{1}} &=& 1\,,  \\
\nonumber Y^{(0)}_{\bm{2},1}&=& y - \dfrac{6\,e^{-4\pi y}}{\pi q}-\dfrac{3\,e^{-8\pi y}}{\pi q^2}-\dfrac{8\,e^{-12\pi y}}{\pi q^3}-\dfrac{3\,e^{-16\pi y}}{2\pi q^4}-\dfrac{36\,e^{-20\pi y}}{5\pi q^5} + \cdots \\
\nonumber&&\hskip-0.4in  - \dfrac{4\log 2}{\pi} -\dfrac{6q}{\pi}-\dfrac{3q^2}{\pi}-\dfrac{8q^3}{\pi}-\dfrac{3q^4}{2\pi}-\dfrac{36q^5}{5\pi}+\cdots \,, \\
\nonumber Y^{(0)}_{\bm{2},2}&=&4\sqrt{3}\,q^{1/2}\left( \dfrac{e^{-2\pi y}}{\pi q}+\dfrac{4\,e^{-6\pi y}}{3\pi q^2}+\dfrac{6\,e^{-10\pi y}}{5\pi q^3}+\dfrac{8\,e^{-14\pi y}}{7\pi q^4}+\dfrac{13\,e^{-18\pi y}}{9\pi q^5}+\cdots \right) \\
&&\hskip-0.4in+\dfrac{4\sqrt{3}\,q^{1/2}}{\pi}\left(1+\dfrac{4}{3}q+\dfrac{6}{5}q^2+\dfrac{8}{7}q^3+\dfrac{13}{9}q^4+\dfrac{12}{11}q^5+\cdots \right)\,,\\
\nonumber Y_{\bm{3},1}^{(0)}&=&y + \dfrac{2\,e^{-4\pi y}}{\pi q} - \dfrac{3\,e^{-8\pi y}}{\pi q^2} + \dfrac{8\, e^{-12\pi y}}{3\pi q^3} - \dfrac{3\,e^{-16\pi y}}{2\pi q^4} + \dfrac{12\,e^{-20\pi y}}{5\pi q^5} + \cdots \\
\nonumber &&\hskip-0.4in - \dfrac{2\log 2}{\pi} +\dfrac{2q}{\pi}-\dfrac{3q^2}{\pi}+\dfrac{8q^3}{3\pi}-\dfrac{3q^4}{2\pi}+\dfrac{12q^5}{5\pi}-\dfrac{4q^6}{\pi}+\cdots\,, \\
\nonumber Y_{\bm{3},2}^{(0)}&=&\dfrac{16\sqrt{2}q^{1/4}}{\pi}\left( \dfrac{e^{-3\pi y}}{3 q} + \dfrac{2\,e^{-7\pi y}}{7 q^2} + \dfrac{3\,e^{-11\pi y}}{11 q^3} + \dfrac{2\,e^{-15\pi y}}{5 q^4} + \dfrac{5\,e^{-19\pi y}}{19 q^5} + \cdots \right) \\
\nonumber &&\hskip-0.4in +\dfrac{4\sqrt{2}q^{1/4}}{\pi}\left( 1 + \dfrac{6q}{5} + \dfrac{13q^2}{9} + \dfrac{14q^3}{13} + \dfrac{18q^4}{17} + \dfrac{32q^5}{21} + \dfrac{31 q^6}{25} + \cdots \right) \,, \\
\nonumber Y_{\bm{3},3}^{(0)}&=&\dfrac{4\sqrt{2}q^{3/4}}{\pi} \left( \dfrac{e^{-\pi y}}{q} + \dfrac{6\,e^{-5\pi y}}{5q^2} + \dfrac{13\, e^{-9\pi y}}{9q^3} + \dfrac{14\,e^{-13\pi y}}{13q^4} + \dfrac{18\,e^{-17\pi y}}{14q^5} + \cdots \right) \,, \\
&&\hskip-0.4in +\dfrac{16\sqrt{2}q^{3/4}}{\pi} \left( \dfrac{1}{3} + \dfrac{2q}{7} + \dfrac{3q^2}{11} + \dfrac{2 q^3}{5} + \dfrac{5q^4}{19} + \dfrac{6q^5}{23} + \dfrac{10q^6}{27} + \cdots \right)\,.
\end{eqnarray}

\item{$k_Y=1$}

The weight $1$ non-holomorphic polyharmonic Maa{\ss} forms of level $4$ are linear combination of $E_1^{(1)}(\tau; i\infty)$, $E_1^{(1)}(\tau; 0)$, $E_1^{(1)}(\tau; 1)$, $E_1^{(1)}(\tau; 2)$, $E_1^{(1)}(\tau; 3)$, $E_1^{(1)}(\tau; 1/2)$, they transform as triplet $\bm{\widehat{3}}$ under $S'_4$, and the Fourier expansion is,
\begin{eqnarray}
\nonumber Y^{(1)}_{\bm{\widehat{3}},1} &=& 4\sqrt{2} q^{3/4} \left( \log 3 + q \log 7 + q^2 \log 11 + 2 q^3 \log 3 + q^4 \log 19 + q^5 \log 23 + \cdots \right)  \\
\nonumber && + 2\sqrt{2} q^{3/4} \left( \dfrac{\Gamma(0,\pi y)}{q} + \dfrac{2\Gamma(0,5\pi y)}{q^2} + \dfrac{\Gamma(0,9\pi y)}{q^3} + \dfrac{2\Gamma(0,13\pi y)}{y^4} + \cdots \right) \,, \\
\nonumber Y^{(1)}_{\bm{\widehat{3}},2} &=& a_0 - 4 \left( 2 q \log 2 + 3 q^2 \log 2 + 2 q^3 \log 3 + 4 q^4 \log 2 + 4 q^5 \log 2 + \cdots  \right)  \\
\nonumber&&  + \log y - 4 \left( \dfrac{\Gamma(0,4\pi y)}{q} + \dfrac{\Gamma(0,8\pi y)}{q^2} + \dfrac{\Gamma(0,16\pi y)}{q^4} + \dfrac{2\Gamma(0,20\pi y)}{q^5} + \cdots \right)\,,  \\
\nonumber Y^{(1)}_{\bm{\widehat{3}},3} &=& 4 q^{1/2} \left( \log 2 + 2 q \log 3 + 2 q^2 \log 2 + 2 q^3 \log 7 + q^4 \log 2 + 2 q^5 \log 11 + \cdots \right)  \\
&&  + \dfrac{1}{4} q^{1/2} \left( \dfrac{\Gamma(0,2\pi y)}{q} + \dfrac{2\Gamma(0,10\pi y)}{q^3} + \dfrac{\Gamma(0,18\pi y)}{q^5} + \dfrac{2\Gamma(26\pi y)}{q^7} + \cdots \right)\,,
\end{eqnarray}
where the constant term is given by
\begin{eqnarray}
a_0 = \gamma_E + \log 16\pi - 4 \log \Gamma\left( \dfrac{1}{4} \right) + 4\log  \Gamma\left( \dfrac{3}{4} \right) \simeq 0.1556\,.
\end{eqnarray}
As shown in Eq.~\eqref{eq:w1l4-3hatp}, the holomorphic weight 1 polyharmonic Maa{\ss} forms make up another triplet $\bm{\widehat{3}'}$, and it can be expressed in the terms of the Jacobi theta functions $\vartheta_1$ and $\vartheta_2$~\cite{Liu:2020msy},
\begin{equation}
Y_{\bm{\widehat{3}'}}^{(1)}(\tau)=\begin{pmatrix}
\sqrt{2}\vartheta_1\vartheta_2\\
-\vartheta_2^2\\
\vartheta_1^2\\
\end{pmatrix}\,.
\end{equation}
where
\begin{eqnarray}
\nonumber \vartheta_1(\tau) &=& \sum_{m\in \mathbb{Z}} e^{2\pi i \tau m^2} = 1 + 2 q + 2 q^4 + 2 q^9 + 2 q^{16} + \cdots \,, \\
\vartheta_2(\tau) &=& - \sum_{m\in \mathbb{Z}} e^{2\pi i \tau (m+1/2)^2} = - 2 q^{1/4} ( 1+ q^2 + q^6 + q^{12} + \cdots ) \,.
\end{eqnarray}

\item{$k_Y=2$}

\begin{eqnarray}
\nonumber && Y_{\bm{1}}^{(2)} = \widehat{E}_2(\tau)\,,  \\
\nonumber && Y_{\bm{2}}^{(2)} = \dfrac{1}{\sqrt{3}}\left( Y_{\bm{\widehat{3}'}}^{(1)} Y_{\bm{\widehat{3}'}}^{(1)} \right)_{\bm{2}} =\begin{pmatrix}
\vartheta_1^4+\vartheta_2^4\\
-2\sqrt{3}\vartheta_1^2\vartheta_2^2\\
\end{pmatrix}\,,\\
&&Y_{\bm{3}}^{(2)} = - \left( Y_{\bm{\widehat{3}'}}^{(1)} Y_{\bm{\widehat{3}'}}^{(1)} \right)_{\bm{3}} =\begin{pmatrix}
\vartheta_1^4-\vartheta_2^4\\
2\sqrt{2}\vartheta_1^3\vartheta_2\\
2\sqrt{2}\vartheta_1\vartheta_2^3\\
\end{pmatrix}\,.
\end{eqnarray}

\item{$k_Y=3$}

\begin{eqnarray}
\nonumber&&Y_{\bm{\widehat{1}'}}^{(3)} = \dfrac{1}{3\sqrt{2}} \left( Y_{\bm{\widehat{3}'}}^{(1)} Y_{\bm{3}}^{(2)} \right)_{\bm{\widehat{1}'}} =\vartheta_1\vartheta_2\left(\vartheta_1^4-\vartheta_2^4\right)\,, \\
\nonumber&&Y_{\bm{\widehat{3}}}^{(3)} = \dfrac{1}{\sqrt{3}} \left( Y_{\bm{\widehat{3}'}}^{(1)} Y_{\bm{2}}^{(2)} \right)_{\bm{\widehat{3}}} =\begin{pmatrix}
4\sqrt{2}\vartheta_1^3\vartheta_2^3\\
\vartheta_1^6+3\vartheta_1^2\vartheta_2^4\\
-\vartheta_2^2\left(3\vartheta_1^4+\vartheta_2^4\right)\\
\end{pmatrix}\,,\\
&&Y_{\bm{\widehat{3}'}}^{(3)} = \left( Y_{\bm{\widehat{3}'}}^{(1)} Y_{\bm{2}}^{(2)} \right)_{\bm{\widehat{3}'}} =
\begin{pmatrix}
2\sqrt{2}\vartheta_1\vartheta_2\left(\vartheta_1^4+\vartheta_2^4\right)\\
\vartheta_2^6-5\vartheta_1^4\vartheta_2^2\\
5\vartheta_1^2\vartheta_2^4-\vartheta_1^6\\
\end{pmatrix}\,.
\end{eqnarray}

\item{$k_Y=4$}

\begin{eqnarray}
\nonumber&&Y_{\bm{1}}^{(4)}= \left( Y_{\bm{\widehat{3}'}}^{(1)} Y_{\bm{\widehat{3}}}^{(3)} \right)_{\bm{1}} =
\vartheta_1^8+14\vartheta_1^4\vartheta_2^4+\vartheta_2^8\,,\\
\nonumber&&Y_{\bm{2}}^{(4)} = - \left( Y_{\bm{\widehat{3}'}}^{(1)} Y_{\bm{\widehat{3}}}^{(3)} \right)_{\bm{2}} =\begin{pmatrix}
\vartheta_1^8-10\vartheta_1^4\vartheta_2^4+\vartheta_2^8\\
4\sqrt{3}\vartheta_1^2\vartheta_2^2\left(\vartheta_1^4+\vartheta_2^4\right)\\
\end{pmatrix}\,,\\
\nonumber&&Y_{\bm{3}}^{(4)} = - \left( Y_{\bm{\widehat{3}'}}^{(1)} Y_{\bm{\widehat{3}}}^{(3)} \right)_{\bm{3}} =\begin{pmatrix}
\vartheta_2^8-\vartheta_1^8\\
\sqrt{2}\vartheta_2\left(\vartheta_1^7+7\vartheta_1^3\vartheta_2^4\right)\\
\sqrt{2}\vartheta_1\left(\vartheta_2^7+7\vartheta_1^4\vartheta_2^3\right)\\
\end{pmatrix}\,,\\
&&Y_{\bm{3'}}^{(4)} = \dfrac{1}{\sqrt{2}} \left( Y_{\bm{\widehat{3}'}}^{(1)} Y_{\bm{\widehat{3}}}^{(3)} \right)_{\bm{3}'} =\vartheta_1\vartheta_2\left(\vartheta_1^4-\vartheta_2^4\right)\begin{pmatrix}
\sqrt{2}\vartheta_1\vartheta_2\\
-\vartheta_2^2\\
\vartheta_1^2\\
\end{pmatrix}\,.
\end{eqnarray}

\item{$k_Y=5$}

\begin{eqnarray}
\nonumber&&Y_{\bm{\widehat{2}}}^{(5)} = \dfrac{1}{\sqrt{3}} \left( Y_{\bm{\widehat{3}'}}^{(1)} Y_{\bm{3'}}^{(4)} \right)_{\bm{\widehat{2}}} =\vartheta_1\vartheta_2\left(\vartheta_1^4-\vartheta_2^4\right)
\begin{pmatrix}
2\sqrt{3}\vartheta_1^2\vartheta_2^2\\
\vartheta_1^4+\vartheta_2^4\\
\end{pmatrix}\,, \\
\nonumber&&Y_{\bm{\widehat{3}}}^{(5)}= \left( Y_{\bm{\widehat{3}'}}^{(1)} Y_{\bm{3}}^{(4)} \right)_{\bm{\widehat{3}}} =
\begin{pmatrix}
-8\sqrt{2}\vartheta_1^3\vartheta_2^3\left(\vartheta_1^4+\vartheta_2^4\right)\\
\vartheta_1^2\left(\vartheta_1^8-14\vartheta_1^4\vartheta_2^4-3\vartheta_2^8\right)\\
\vartheta_2^2\left(3\vartheta_1^8+14\vartheta_1^4\vartheta_2^4-\vartheta_2^8\right)\\
\end{pmatrix}\,, \\
\nonumber&&Y_{\bm{\widehat{3}'}I}^{(5)}= \left( Y_{\bm{\widehat{3}'}}^{(1)} Y_{\bm{2}}^{(4)} \right)_{\bm{\widehat{3}'}} =
\begin{pmatrix}
2\sqrt{2}\vartheta_1\vartheta_2\left(\vartheta_1^8-10\vartheta_1^4\vartheta_2^4+\vartheta_2^8\right)\\
\vartheta_2^2\left(13\vartheta_1^8+2\vartheta_1^4\vartheta_2^4+\vartheta_2^8\right)\\
-\vartheta_1^2\left(\vartheta_1^8+2\vartheta_1^4\vartheta_2^4+13\vartheta_2^8\right)\\
\end{pmatrix}\,, \\
&&Y_{\bm{\widehat{3}'}II}^{(5)}= \left( Y_{\bm{\widehat{3}'}}^{(1)} Y_{\bm{1}}^{(4)} \right)_{\bm{\widehat{3}'}} =\left(\vartheta_1^8+14\vartheta_1^4\vartheta_2^4+\vartheta_2^8\right)
\begin{pmatrix}
\sqrt{2}\vartheta_1\vartheta_2\\
-\vartheta_2^2\\
\vartheta_1^2\\
\end{pmatrix}\,.
\end{eqnarray}

\item{$k_Y=6$}

\begin{eqnarray}
\nonumber&&Y_{\bm{1}}^{(6)}= \left( Y_{\bm{\widehat{3}'}}^{(1)} Y_{\bm{\widehat{3}}}^{(5)} \right)_{\bm{1}} =
\vartheta_1^{12}-33\vartheta_1^8\vartheta_2^4-33\vartheta_1^4\vartheta_2^8+\vartheta_2^{12}\,,\\
\nonumber&&Y_{\bm{1'}}^{(6)}= \dfrac{1}{18} \left( Y_{\bm{\widehat{3}'}}^{(1)} Y_{\bm{\widehat{3}'}I}^{(5)} \right)_{\bm{1}'} =
\vartheta_1^2\vartheta_2^2\left(\vartheta_1^4-\vartheta_2^4\right)^2 \,, \\
\nonumber&&Y_{\bm{2}}^{(6)}= \left( Y_{\bm{2}}^{(2)} Y_{\bm{1}}^{(4)} \right)_{\bm{2}} =\left(\vartheta_1^8+14\vartheta_1^4\vartheta_2^4+\vartheta_2^8\right)\begin{pmatrix}
\vartheta_1^4+\vartheta_2^4\\
-2\sqrt{3}\vartheta_1^2\vartheta_2^2\\
\end{pmatrix}\,,\\
\nonumber&&Y_{\bm{3}I}^{(6)}= \left( Y_{\bm{\widehat{3}'}}^{(1)} Y_{\bm{\widehat{3}'}I}^{(5)} \right)_{\bm{3}} =\begin{pmatrix}
\vartheta_1^{12}-11\vartheta_1^8\vartheta_2^4+11\vartheta_1^4\vartheta_2^8-\vartheta_2^{12}\\
-\sqrt{2}\vartheta_1^3\vartheta_2\left(\vartheta_1^8-22\vartheta_1^4\vartheta_2^4-11\vartheta_2^8\right)\\
\sqrt{2}\vartheta_1\vartheta_2^3\left(11\vartheta_1^8+22\vartheta_1^4\vartheta_2^4-\vartheta_2^8\right)\\
\end{pmatrix}\,,\\
\nonumber&&Y_{\bm{3}II}^{(6)}= \left( Y_{\bm{3}}^{(2)} Y_{\bm{1}}^{(4)} \right)_{\bm{3}} =\left(\vartheta_1^8+14\vartheta_2^4\vartheta_1^4+\vartheta_2^8\right)
\begin{pmatrix}
\vartheta_1^4-\vartheta_2^4\\
2\sqrt{2}\vartheta_1^3\vartheta_2\\
2\sqrt{2}\vartheta_1\vartheta_2^3\\
\end{pmatrix}\,,\\
&&Y_{\bm{3'}}^{(6)}= - \dfrac{1}{3\sqrt{2}} \left( Y_{\bm{\widehat{3}'}}^{(1)} Y_{\bm{\widehat{3}'}I}^{(5)} \right)_{\bm{3}'} =\vartheta_1\vartheta_2\left(\vartheta_1^4-\vartheta_2^4\right)
\begin{pmatrix}
2\sqrt{2}\vartheta_1\vartheta_2\left(\vartheta_1^4+\vartheta_2^4\right)\\
\vartheta_2^6-5\vartheta_1^4\vartheta_2^2\\
5\vartheta_1^2\vartheta_2^4-\vartheta_1^6\\
\end{pmatrix}\,.
\end{eqnarray}

\end{itemize}

\subsection{$N=5$\label{app:Polyharmonic_N_5}}

As shown in Eq.~\eqref{eq:cusps-N}, the principal congruence subgroup $\Gamma(5)$ has twelve cusps given by $\{0,1,2,3,4,1/2,3/2,5/2,7/2,9/2,2/5,i\infty\}=\{ \overline{0/1}, \overline{1/1}, \overline{2/1}, \overline{3/1}, \overline{4/1}, \overline{1/2}, \overline{3/2}, \overline{5/2}, \overline{7/2}, \overline{9/2}, \overline{2/5}, \overline{1/0}\}$. Therefore we have twelve weight $k$ non-holomorphic Eisenstein series $E_k(\tau; s; 0)$, $E_k(\tau; s; 1)$, $E_k(\tau; s; 2)$, $E_k(\tau; s; 3)$, $E_k(\tau; s; 4)$, $E_k(\tau; s; 1/2)$, $E_k(\tau; s; 3/2)$, $E_k(\tau; s; 5/2)$, $E_k(\tau; s; 7/2)$, $E_k(\tau; s; 9/2)$, $E_k(\tau; s; 2/5)$ and $E_k(\tau; s; i\infty)$ which are defined at the cusps of $\Gamma(5)$. Under the action of the modular generators $S$ and $T$, each of these functions is mapped to another up to the automorphic factor,
\begin{eqnarray}
\nonumber E_k(S\tau; s; i\infty) &=& (-\tau)^k E_k(\tau; s; 0) \,,  \\
\nonumber E_k(S\tau; s; 0) &=& (-1)^k (-\tau)^k E_k(\tau; s; i\infty) \,,   \\
\nonumber E_k(S\tau; s; 1) &=& (-\tau)^k E_k(\tau; s; 4) \,,  \\
\nonumber E_k(S\tau; s; 2) &=& (-\tau)^k E_k(\tau; s; 9/2) \,,  \\
\nonumber E_k(S\tau; s; 3) &=& (-1)^k (-\tau)^k E_k(\tau; s; 1/2) \,,  \\
\nonumber E_k(S\tau; s; 4) &=& (-1)^k (-\tau)^k E_k(\tau; s; 1) \,,  \\
\nonumber E_k(S\tau; s; 1/2) &=& (-\tau)^k E_k(\tau; s; 3) \,,  \\
\nonumber E_k(S\tau; s; 3/2) &=& (-1)^k (-\tau)^k E_k(\tau; s; 7/2) \,,  \\
\nonumber E_k(S\tau; s; 5/2) &=& (-1)^k (-\tau)^k E_k(\tau; s; 2/5)  \,, \\
\nonumber E_k(S\tau; s; 7/2) &=& (-\tau)^k E_k(\tau; s; 3/2) \,,  \\
\nonumber E_k(S\tau; s; 9/2) &=& (-1)^k (-\tau)^k E_k(\tau; s; 2) \,,  \\
 E_k(S\tau; s; 2/5) &=& (-\tau)^k E_k(\tau; s; 5/2) \,, \label{eq:Eien-series-N5-S}
\end{eqnarray}
and
\begin{eqnarray}
\nonumber E_k(T\tau; s; i\infty) &=& E_k(\tau; s; i\infty) \,,  \\
\nonumber E_k(T\tau; s; 0) &=& E_k(\tau; s; 4) \,,   \\
\nonumber E_k(T\tau; s; 1) &=& E_k(\tau; s; 0) \,,  \\
\nonumber E_k(T\tau; s; 2) &=& E_k(\tau; s; 1) \,,  \\
\nonumber E_k(T\tau; s; 3) &=& E_k(\tau; s; 2) \,,   \\
\nonumber E_k(T\tau; s; 4) &=& E_k(\tau; s; 3) \,,  \\
\nonumber E_k(T\tau; s; 1/2) &=& E_k(\tau; s; 9/2) \,,  \\
\nonumber E_k(T\tau; s; 3/2) &=& E_k(\tau; s; 1/2) \,,  \\
\nonumber E_k(T\tau; s; 5/2) &=& E_k(\tau; s; 3/2) \,,  \\
\nonumber E_k(T\tau; s; 7/2) &=& E_k(\tau; s; 5/2) \,,  \\
\nonumber E_k(T\tau; s; 9/2) &=& E_k(\tau; s; 7/2) \,,  \\
E_k(T\tau; s; 2/5) &=& E_k(\tau; s; 2/5)\,.\label{eq:Eien-series-N5-T}
\end{eqnarray}
These Eisenstein series can be arranged into irreducible representations of $A'_5$. For simplicity we introduce the following modular functions
\begin{eqnarray}
\nonumber u_k^{(n)}(\tau) &=& E_k(\tau; s; 0) + \omega_5^{n} E_k(\tau; s; 1) + \omega_5^{2n} E_k(\tau; s; 2) + \omega_5^{3n} E_k(\tau; s; 3) + \omega_5^{4n} E_k(\tau; s; 4)\,,  \\
\nonumber v_k^{(n)}(\tau) &=& E_k(\tau; s; 1/2) + \omega_5^{n} E_k(\tau; s; 3/2) + \omega_5^{2n} E_k(\tau; s; 5/2) + \omega_5^{3n} E_k(\tau; s; 7/2) + \omega_5^{4n} E_k(\tau; s; 9/2)\,,
\end{eqnarray}
where $\omega_5=e^{2\pi i/5}$ and the superscript $n=0, 1, 2, 3, 4$.
It can be seen that the modular generator $T$ maps $u_k^{(n)}$ and $v_k^{(n)}$ to themselves multiplying $\omega_5^n$, i.e
\begin{eqnarray}
u_k^{(n)} \stackrel{T}{\longrightarrow} \omega_5^n  u_k^{(n)}\,,~~~
v_k^{(n)} \stackrel{T}{\longrightarrow} \omega_5^n  v_k^{(n)}\,.
\end{eqnarray}
From Eqs.~(\ref{eq:Eien-series-N5-S}, \ref{eq:Eien-series-N5-T}), we know that the even weight Eisenstein series can be arranged into a singlet $Y^{(2k)}_{\bm{1}}(\tau)$, two triplets $Y^{(2k)}_{\bm{3}}(\tau)=(Y^{(2k)}_{\bm{3},1}, Y^{(2k)}_{\bm{3},2}, Y^{(2k)}_{\bm{3},3})^T$, $Y^{(2k)}_{\bm{3'}}(\tau)=(Y^{(2k)}_{\bm{3'},1},Y^{(2k)}_{\bm{3'},2},Y^{(2k)}_{\bm{3'},3})^T$, and a quintuplet $Y^{(2k)}_{\bm{5}}(\tau)=(Y^{(2k)}_{\bm{5},1}, Y^{(2k)}_{\bm{5},2}, Y^{(2k)}_{\bm{5},3},Y^{(2k)}_{\bm{5},4},Y^{(2k)}_{\bm{5},5})^T$ via the following linear combinations:

\begin{eqnarray}
\nonumber Y^{(2k)}_{\bm{1}}(\tau) &=& E_{2k}(\tau; s; i\infty) + E_{2k}(\tau; s; 2/5) + u_{2k}^{(0)}(\tau) + v_{2k}^{(0)}(\tau)\,,\\
\nonumber Y^{(2k)}_{\bm{3},1}(\tau) &=& E_{2k}(\tau; s; i\infty) - E_{2k}(\tau; s; 2/5) + \dfrac{1}{\sqrt{5}} u_{2k}^{(0)}(\tau) - \dfrac{1}{\sqrt{5}} v_{2k}^{(0)} \,,  \\
\nonumber Y^{(2k)}_{\bm{3},2}(\tau) &=& - \sqrt{\dfrac{2}{5}} u_{2k}^{(1)}(\tau) + \sqrt{\dfrac{2}{5}} \omega_5^3 v_{2k}^{(1)}(\tau) \,, \\
\nonumber Y^{(2k)}_{\bm{3},3}(\tau) &=& - \sqrt{\dfrac{2}{5}} u_{2k}^{(4)}(\tau) + \sqrt{\dfrac{2}{5}} \omega_5^2 v_{2k}^{(4)}(\tau)\,,\\
\nonumber Y^{(2k)}_{\bm{3'},1}(\tau) &=& E_{2k}(\tau; s; i\infty) - E_{2k}(\tau; s; 2/5) - \dfrac{1}{\sqrt{5}} u_{2k}^{(0)}(\tau) + \dfrac{1}{\sqrt{5}} v_{2k}^{(0)} \,, \\
\nonumber Y^{(2k)}_{\bm{3'},2}(\tau) &=& \sqrt{\dfrac{2}{5}} u_{2k}^{(2)}(\tau) - \sqrt{\dfrac{2}{5}} \omega_5 v_{2k}^{(2)}(\tau) \,, \\
\nonumber Y^{(2k)}_{\bm{3'},3}(\tau) &=& \sqrt{\dfrac{2}{5}} u_{2k}^{(3)}(\tau) - \sqrt{\dfrac{2}{5}} \omega_5^4 v_{2k}^{(3)}(\tau)
\,,\\
\nonumber Y^{(2k)}_{\bm{5},1}(\tau) &=& E_{2k}(\tau; s; i\infty) + E_{2k}(\tau; s; 2/5) - \dfrac{1}{5} u_{2k}^{(0)}(\tau) - \dfrac{1}{5} v_{2k}^{(0)}\,,  \\
\nonumber Y^{(2k)}_{\bm{5},2}(\tau) &=& \dfrac{\sqrt{6}}{5} u_{2k}^{(1)}(\tau) + \dfrac{\sqrt{6}}{5} \omega_5^3 v_{2k}^{(1)}(\tau) \,, \\
\nonumber Y^{(2k)}_{\bm{5},3}(\tau) &=& \dfrac{\sqrt{6}}{5} u_{2k}^{(2)}(\tau) + \dfrac{\sqrt{6}}{5} \omega_5 v_{2k}^{(2)}(\tau)\,,  \\
\nonumber Y^{(2k)}_{\bm{5},4}(\tau) &=& \dfrac{\sqrt{6}}{5} u_{2k}^{(3)}(\tau) + \dfrac{\sqrt{6}}{5} \omega_5^4 v_{2k}^{(3)}(\tau) \,, \\
\label{eq:N=5_Y1-Y3-Y3p-Y5} Y^{(2k)}_{\bm{5},5}(\tau) &=& \dfrac{\sqrt{6}}{5} u_{2k}^{(4)}(\tau) + \dfrac{\sqrt{6}}{5} \omega_5^2 v_{2k}^{(4)}(\tau)\,.
\end{eqnarray}
The odd weight Eisenstein series can be arranged into two independent sextets $Y^{(2k+1)}_{\bm{\widehat{6}}I}(\tau)=(Y^{(2k+1)}_{\bm{\widehat{6}}I, 1}, Y^{(2k+1)}_{\bm{\widehat{6}}I, 2}, Y^{(2k+1)}_{\bm{\widehat{6}}I, 3}, Y^{(2k+1)}_{\bm{\widehat{6}}I, 4}, Y^{(2k+1)}_{\bm{\widehat{6}}I, 5}, Y^{(2k+1)}_{\bm{\widehat{6}}I, 6})^T$ and $Y^{(2k+1)}_{\bm{\widehat{6}}II}(\tau)=(Y^{(2k+1)}_{\bm{\widehat{6}}II, 1}, Y^{(2k+1)}_{\bm{\widehat{6}}II, 2}, Y^{(2k+1)}_{\bm{\widehat{6}}II, 3}$, $Y^{(2k+1)}_{\bm{\widehat{6}}II, 4}, Y^{(2k+1)}_{\bm{\widehat{6}}II, 5}, Y^{(2k+1)}_{\bm{\widehat{6}}II, 6})^T$ in the representation $\bm{\widehat{6}}$ of $A'_5$,
\begin{eqnarray}
\nonumber Y^{(2k+1)}_{\bm{\widehat{6}}I,1}(\tau) &=& E_{2k+1}(\tau; s; i\infty) + i\sqrt{\dfrac{1}{5\sqrt{5}\phi}} u_{2k+1}^{(0)}(\tau) + i \sqrt{\dfrac{\phi}{5\sqrt{5}}} v_{2k+1}^{(0)}(\tau)\,,   \\
\nonumber Y^{(2k+1)}_{\bm{\widehat{6}}I,2}(\tau) &=& - E_{2k+1}(\tau; s; 2/5) - i \sqrt{\dfrac{\phi}{5\sqrt{5}}} u_{2k+1}^{(0)}(\tau) + i\sqrt{\dfrac{1}{5\sqrt{5}\phi}} v_{2k+1}^{(0)}(\tau) \,,\\
\nonumber Y^{(2k+1)}_{\bm{\widehat{6}}I,3}(\tau) &=& - i\sqrt{\dfrac{1}{5\sqrt{5}\phi}} \dfrac{1}{\phi} u_{2k+1}^{(1)}(\tau) + i\sqrt{\dfrac{1}{5\sqrt{5}\phi}} \phi^2 \omega_5^3 v_{2k+1}^{(1)}(\tau)\,, \\
\nonumber Y^{(2k+1)}_{\bm{\widehat{6}}I,4}(\tau) &=& - i\sqrt{\dfrac{1}{5\sqrt{5}\phi}} \sqrt{2}\phi u_{2k+1}^{(2)}(\tau) + i\sqrt{\dfrac{1}{5\sqrt{5}\phi}} \sqrt{2} \omega_5 v_{2k+1}^{(2)}(\tau)\,, \\
\nonumber Y^{(2k+1)}_{\bm{\widehat{6}}I,5}(\tau) &=& - i\sqrt{\dfrac{1}{5\sqrt{5}\phi}} \sqrt{2} u_{2k+1}^{(3)}(\tau) - i\sqrt{\dfrac{1}{5\sqrt{5}\phi}} \sqrt{2} \phi \omega_5^4 v_{2k+1}^{(3)}(\tau)\,, \\
\nonumber Y^{(2k+1)}_{\bm{\widehat{6}}I,6}(\tau) &=& - i\sqrt{\dfrac{1}{5\sqrt{5}\phi}} \phi^2 u_{2k+1}^{(4)}(\tau) - i\sqrt{\dfrac{1}{5\sqrt{5}\phi}} \dfrac{1}{\phi} \omega_5^2 v_{2k+1}^{(4)}(\tau)\,,\\
\nonumber Y^{(2k+1)}_{\bm{\widehat{6}}II,1}(\tau) &=& E_{2k+1}(\tau; s; 2/5) - i\sqrt{\dfrac{\phi}{5\sqrt{5}}} u_{2k+1}^{(0)}(\tau) + i\sqrt{\dfrac{1}{5\sqrt{5}\phi}} v_{2k+1}^{(0)}(\tau) \,,  \\
\nonumber Y^{(2k+1)}_{\bm{\widehat{6}}II,2}(\tau) &=& E_{2k+1}(\tau; s; i\infty) - i\sqrt{\dfrac{1}{5\sqrt{5}\phi}} u_{2k+1}^{(0)}(\tau) - i \sqrt{\dfrac{\phi}{5\sqrt{5}}} v_{2k+1}^{(0)}(\tau) \,, \\
\nonumber Y^{(2k+1)}_{\bm{\widehat{6}}II,3}(\tau) &=& - i\sqrt{\dfrac{1}{5\sqrt{5}\phi}} \phi^2 u_{2k+1}^{(1)}(\tau) - i\sqrt{\dfrac{1}{5\sqrt{5}\phi}} \dfrac{1}{\phi} \omega_5^3 v_{2k+1}^{(1)}(\tau) \,, \\
\nonumber Y^{(2k+1)}_{\bm{\widehat{6}}II,4}(\tau) &=& - i\sqrt{\dfrac{1}{5\sqrt{5}\phi}} \sqrt{2} u_{2k+1}^{(2)}(\tau) - i\sqrt{\dfrac{1}{5\sqrt{5}\phi}} \sqrt{2}\phi \omega_5 v_{2k+1}^{(2)}(\tau) \,, \\
\nonumber Y^{(2k+1)}_{\bm{\widehat{6}}II, 5}(\tau) &=& i\sqrt{\dfrac{1}{5\sqrt{5}\phi}} \sqrt{2}\phi u_{2k+1}^{(3)}(\tau) - i\sqrt{\dfrac{1}{5\sqrt{5}\phi}} \sqrt{2} \omega_5^4 v_{2k+1}^{(3)}(\tau)\,,  \\
\label{eq:N=5_Y6hI-Y6hII} Y^{(2k+1)}_{\bm{\widehat{6}}II,6}(\tau) &=& i\sqrt{\dfrac{1}{5\sqrt{5}\phi}} \dfrac{1}{\phi} u_{2k+1}^{(4)}(\tau) - i\sqrt{\dfrac{1}{5\sqrt{5}\phi}} \phi^2 \omega_5^2 v_{2k+1}^{(4)}(\tau)\,.
\end{eqnarray}
The linear space of the weight $k$ polyharmonic Maa{\ss} form of level $N=5$ is spanned by the above non-holomorphic Eisenstein series $E_{k}(\tau; s; \overline{A/C})$ for $s=1-k$. The Fourier expansion of the even weight $2k$ Eisenstein series for $k\leq 0$ reads as
\begin{footnotesize}
\begin{eqnarray}
\nonumber E_{2k}(\tau; 1-2k; i\infty) &=& y^{1-2k} + (-1)^k \dfrac{2^{2k+1} 5^{2k-1}}{1-5^{2k-2}}\dfrac{\zeta(1-2k)}{\zeta(2-2k)} \dfrac{\pi}{5}  \\
\nonumber &&\hskip-1.2in + (-1)^k 2^{2k-1} \left( \dfrac{1}{(1-5^{2k-2})\zeta(2-2k)} + \dfrac{1}{L(2-2k,\chi_3)} \right) \dfrac{\pi}{5} \sum_{n=1}^{+\infty} q^{\frac{n}{5}} \sum_{\substack{m|n,\, m\in\mathbb{Z} \\ \frac{n}{m}\equiv 0\,({\rm mod}\, 5)}} \left| \dfrac{n}{m} \right|^{2k-1}\, \omega_5^{4m} \\
\nonumber &&\hskip-1.2in + (-1)^k 2^{2k-1} \left( \dfrac{1}{(1-5^{2k-2})\zeta(2-2k)} - \dfrac{1}{L(2-2k,\chi_3)} \right) \dfrac{\pi}{5} \sum_{n=1}^{+\infty} q^{\frac{n}{5}} \sum_{\substack{m|n,\, m\in\mathbb{Z} \\ \frac{n}{m}\equiv 0\,({\rm mod}\, 5)}} \left| \dfrac{n}{m} \right|^{2k-1}\, \omega_5^{3m} \\
\nonumber &&\hskip-1.2in + \dfrac{(-1)^k 2^{2k-1}}{(-2k)!}  \left( \dfrac{1}{(1-5^{2k-2})\zeta(2-2k)} + \dfrac{1}{L(2-2k,\chi_3)} \right) \dfrac{\pi}{5} \sum_{n=-1}^{-\infty} q^{\frac{n}{5}} \Gamma\left(1-2k,-\dfrac{4\pi n y}{5} \right) \sum_{\substack{m|n,\, m\in\mathbb{Z} \\ \frac{n}{m}\equiv 0\,({\rm mod}\, 5)}} \left| \dfrac{n}{m} \right|^{2k-1}\, \omega_5^{4m} \\
\nonumber &&\hskip-1.2in + \dfrac{(-1)^k 2^{2k-1}}{(-2k)!}  \left( \dfrac{1}{(1-5^{2k-2})\zeta(2-2k)} - \dfrac{1}{L(2-2k,\chi_3)} \right) \dfrac{\pi}{5} \sum_{n=-1}^{-\infty} q^{\frac{n}{5}} \Gamma\left(1-2k,-\dfrac{4\pi n y}{5} \right) \sum_{\substack{m|n,\, m\in\mathbb{Z} \\ \frac{n}{m}\equiv 0\,({\rm mod}\, 5)}} \left| \dfrac{n}{m} \right|^{2k-1}\, \omega_5^{3m} \,, \\
\nonumber E_{2k}(\tau; 1-2k; j) &=& (-1)^k 2^{2k-1} \left( \dfrac{1-5^{2k-1}}{1-5^{2k-2}}\dfrac{\zeta(1-2k)}{\zeta(2-2k)} + \dfrac{L(2-2k, \chi_3)}{L(1-2k, \chi_3)} \right) \dfrac{\pi}{5} \\
\nonumber &&\hskip-1.2in + (-1)^k 2^{2k-1} \left( \dfrac{1}{(1-5^{2k-2})\zeta(2-2k)} + \dfrac{1}{L(2-2k,\chi_3)} \right) \dfrac{\pi}{5} \sum_{n=1}^{+\infty} q^{\frac{n}{5}} \sum_{\substack{m|n,\, m\in\mathbb{Z} \\ \frac{n}{m}\equiv 1\,({\rm mod}\, 5)}} \left| \dfrac{n}{m} \right|^{2k-1}\, \omega_5^{4jm} \\
\nonumber &&\hskip-1.2in + (-1)^k 2^{2k-1} \left( \dfrac{1}{(1-5^{2k-2})\zeta(2-2k)} - \dfrac{1}{L(2-2k,\chi_3)} \right) \dfrac{\pi}{5} \sum_{n=1}^{+\infty} q^{\frac{n}{5}} \sum_{\substack{m|n,\, m\in\mathbb{Z} \\ \frac{n}{m}\equiv 2\,({\rm mod}\, 5)}} \left| \dfrac{n}{m} \right|^{2k-1}\, \omega_5^{3jm} \\
\nonumber &&\hskip-1.2in + \dfrac{(-1)^k 2^{2k-1}}{(-2k)!}  \left( \dfrac{1}{(1-5^{2k-2})\zeta(2-2k)} + \dfrac{1}{L(2-2k,\chi_3)} \right) \dfrac{\pi}{5} \sum_{n=-1}^{-\infty} q^{\frac{n}{5}} \Gamma\left(1-2k,-\dfrac{4\pi n y}{5} \right) \sum_{\substack{m|n,\, m\in\mathbb{Z} \\ \frac{n}{m}\equiv 1\,({\rm mod}\, 5)}} \left| \dfrac{n}{m} \right|^{2k-1}\, \omega_5^{4jm} \\
\nonumber &&\hskip-1.2in + \dfrac{(-1)^k 2^{2k-1}}{(-2k)!}  \left( \dfrac{1}{(1-5^{2k-2})\zeta(2-2k)} - \dfrac{1}{L(2-2k,\chi_3)} \right) \dfrac{\pi}{5} \sum_{n=-1}^{-\infty} q^{\frac{n}{5}} \Gamma\left(1-2k,-\dfrac{4\pi n y}{5} \right) \sum_{\substack{m|n,\, m\in\mathbb{Z} \\ \frac{n}{m}\equiv 2\,({\rm mod}\, 5)}} \left| \dfrac{n}{m} \right|^{2k-1}\, \omega_5^{3jm}\,, \\
\nonumber E_{2k}(\tau; 1-2k; l/2) &=& (-1)^k 2^{2k-1} \left( \dfrac{1-5^{2k-1}}{1-5^{2k-2}}\dfrac{\zeta(1-2k)}{\zeta(2-2k)} - \dfrac{L(2-2k, \chi_3)}{L(1-2k, \chi_3)} \right) \dfrac{\pi}{5} \\
\nonumber &&\hskip-1.2in + (-1)^k 2^{2k-1} \left( \dfrac{1}{(1-5^{2k-2})\zeta(2-2k)} + \dfrac{1}{L(2-2k,\chi_3)} \right) \dfrac{\pi}{5} \sum_{n=1}^{+\infty} q^{n/5} \sum_{\substack{m|n,\, m\in\mathbb{Z} \\ \frac{n}{m}\equiv 2\,({\rm mod}\, 5)}} \left| \dfrac{n}{m} \right|^{2k-1}\, \omega_5^{4lm} \\
\nonumber &&\hskip-1.2in + (-1)^k 2^{2k-1} \left( \dfrac{1}{(1-5^{2k-2})\zeta(2-2k)} - \dfrac{1}{L(2-2k,\chi_3)} \right) \dfrac{\pi}{5} \sum_{n=1}^{+\infty} q^{n/5} \sum_{\substack{m|n,\, m\in\mathbb{Z} \\ \frac{n}{m}\equiv 4\,({\rm mod}\, 5)}} \left| \dfrac{n}{m} \right|^{2k-1}\, \omega_5^{3lm} \\
\nonumber &&\hskip-1.2in + \dfrac{(-1)^k 2^{2k-1}}{(-2k)!}  \left( \dfrac{1}{(1-5^{2k-2})\zeta(2-2k)} + \dfrac{1}{L(2-2k,\chi_3)} \right) \dfrac{\pi}{5} \sum_{n=-1}^{-\infty} q^{\frac{n}{5}} \Gamma\left(1-2k,-\dfrac{4\pi n y}{5} \right) \sum_{\substack{m|n,\, m\in\mathbb{Z} \\ \frac{n}{m}\equiv 2\,({\rm mod}\, 5)}} \left| \dfrac{n}{m} \right|^{2k-1}\, \omega_5^{4lm} \\
\nonumber &&\hskip-1.2in + \dfrac{(-1)^k 2^{2k-1}}{(-2k)!}  \left( \dfrac{1}{(1-5^{2k-2})\zeta(2-2k)} - \dfrac{1}{L(2-2k,\chi_3)} \right) \dfrac{\pi}{5} \sum_{n=-1}^{-\infty} q^{\frac{n}{5}} \Gamma\left(1-2k,-\dfrac{4\pi n y}{5} \right) \sum_{\substack{m|n,\, m\in\mathbb{Z} \\ \frac{n}{m}\equiv 4\,({\rm mod}\, 5)}} \left| \dfrac{n}{m} \right|^{2k-1}\, \omega_5^{3lm} \,,\\
\nonumber E_{2k}(\tau; 1-2k; 2/5) &=& (-1)^k 2^{2k+1} \dfrac{5^{2k-1}}{1-5^{2k-2}}\dfrac{\zeta(1-2k)}{\zeta(2-2k)} \dfrac{\pi}{5}  \\
\nonumber &&\hskip-1.2in + (-1)^k 2^{2k-1} \left( \dfrac{1}{(1-5^{2k-2})\zeta(2-2k)} + \dfrac{1}{L(2-2k,\chi_3)} \right) \dfrac{\pi}{5} \sum_{n=1}^{+\infty} q^{\frac{n}{5}} \sum_{\substack{m|n,\, m\in\mathbb{Z} \\ \frac{n}{m}\equiv 0\,({\rm mod}\, 5)}} \left| \dfrac{n}{m} \right|^{2k-1}\, \omega_5^{3m} \\
\nonumber &&\hskip-1.2in + (-1)^k 2^{2k-1} \left( \dfrac{1}{(1-5^{2k-2})\zeta(2-2k)} - \dfrac{1}{L(2-2k,\chi_3)} \right) \dfrac{\pi}{5} \sum_{n=1}^{+\infty} q^{\frac{n}{5}} \sum_{\substack{m|n,\, m\in\mathbb{Z} \\ \frac{n}{m}\equiv 0\,({\rm mod}\, 5)}} \left| \dfrac{n}{m} \right|^{2k-1}\, \omega_5^{m} \\
\nonumber &&\hskip-1.2in + \dfrac{(-1)^k 2^{2k-1}}{(-2k)!}  \left( \dfrac{1}{(1-5^{2k-2})\zeta(2-2k)} + \dfrac{1}{L(2-2k,\chi_3)} \right) \dfrac{\pi}{5} \sum_{n=-1}^{-\infty} q^{\frac{n}{5}} \Gamma\left(1-2k,-\dfrac{4\pi n y}{5} \right) \sum_{\substack{m|n,\, m\in\mathbb{Z} \\ \frac{n}{m}\equiv 0\,({\rm mod}\, 5)}} \left| \dfrac{n}{m} \right|^{2k-1}\, \omega_5^{3m} \\
\nonumber &&\hskip-1.2in + \dfrac{(-1)^k 2^{2k-1}}{(-2k)!}  \left( \dfrac{1}{(1-5^{2k-2})\zeta(2-2k)} - \dfrac{1}{L(2-2k,\chi_3)} \right) \dfrac{\pi}{5} \sum_{n=-1}^{-\infty} q^{\frac{n}{5}} \Gamma\left(1-2k,-\dfrac{4\pi n y}{5} \right) \sum_{\substack{m|n,\, m\in\mathbb{Z} \\ \frac{n}{m}\equiv 0\,({\rm mod}\, 5)}} \left| \dfrac{n}{m} \right|^{2k-1}\, \omega_5^{m}\,,
\end{eqnarray}
\end{footnotesize}
where $j$ are the integer cusps $\{0,1,2,3,4\}$ and $l/2$ are the half integer cusps $\{1/2,3/2,5/2,7/2,9/2\}$.

Following the same procedure as the case of $N=3$ and $N=4$, we insert the above expressions of the even weight non-holomorphic Eisenstein series into Eq.~\eqref{eq:N=5_Y1-Y3-Y3p-Y5}, then the $q$-expansion of the polyharmonic Maa{\ss} form multiplets of level $N=5$ follows,
\begin{eqnarray}
\nonumber Y_{\bm{1}}^{(2k)} &=& (-1)^k \dfrac{2^{2k}\, \pi}{1-2k}\dfrac{\zeta(1-2k)}{\zeta(2-2k)} + \dfrac{y^{1-2k}}{1-2k}  + \dfrac{(-1)^k\, 2^{2k}\,\pi}{(1-2k)\zeta(2-2k)} \sum_{n=1}^{+\infty} \sigma_{2k-1}(n) q^n \\
\nonumber&&\hskip-0.4in + \dfrac{(-1)^k\, 2^{2k}\,\pi}{(1-2k)!\zeta(2-2k)} \sum_{n=1}^{+\infty} \sigma_{2k-1}(n) q^{-n}\Gamma(1-2k,4\pi n y)\,,\\
\nonumber Y_{\bm{3},1}^{(2k)} &=& \dfrac{(-1)^k\, 2^{2k}\,\pi}{\sqrt{5} (1-2k)} \dfrac{L(1-2k,\chi_3)}{L(2-2k,\chi_3)} + \dfrac{y^{1-2k}}{1-2k}  \\
\nonumber && \hskip-0.4in + \dfrac{(-1)^k\, 2^{2k}\,\pi}{\sqrt{5}(1-2k)L(2-2k,\chi_3)} \sum_{n=1}^{+\infty} \Big( \widehat{\sigma}^{\chi_3}_{2k-1}(n) + 5^{2k-1} \widetilde{\sigma}^{\chi_3}_{2k-1}(n) \Big) q^n  \\
\nonumber &&\hskip-0.4in  + \dfrac{(-1)^k\, 2^{2k}\,\pi}{\sqrt{5}(1-2k)!L(2-2k,\chi_3)} \sum_{n=1}^{+\infty} \Big( \widehat{\sigma}^{\chi_3}_{2k-1}(n) + 5^{2k-1} \widetilde{\sigma}^{\chi_3}_{2k-1}(n) \Big) q^{-n}\Gamma(1-2k,4\pi n y) \,, \\
\nonumber Y_{\bm{3},2}^{(2k)} &=& - \dfrac{(-1)^k\, 2^{2k}\sqrt{2}\,\pi}{\sqrt{5}(1-2k)L(2-2k,\chi_3)} \sum_{n=0}^{+\infty} \widehat{\sigma}^{\chi_3}_{2k-1}(5n+1)  q^{(5n+1)/5}  \\
\nonumber &&\hskip-0.4in  - \dfrac{(-1)^k\, 2^{2k}\sqrt{2}\,\pi}{\sqrt{5}(1-2k)!L(2-2k,\chi_3)} \sum_{n=0}^{+\infty}  \widehat{\sigma}^{\chi_3}_{2k-1}(5n+4) q^{-(5n+4)/5}\Gamma(1-2k,4\pi (5n+4) y/5) \,, \\
\nonumber Y_{\bm{3},3}^{(2k)} &=& - \dfrac{(-1)^k\, 2^{2k}\sqrt{2}\,\pi}{\sqrt{5}(1-2k)L(2-2k,\chi_3)} \sum_{n=0}^{+\infty} \widehat{\sigma}^{\chi_3}_{2k-1}(5n+4)  q^{(5n+4)/5}  \\
\nonumber &&\hskip-0.4in   - \dfrac{(-1)^k\, 2^{2k}\sqrt{2}\,\pi}{\sqrt{5}(1-2k)!L(2-2k,\chi_3)} \sum_{n=0}^{+\infty}  \widehat{\sigma}^{\chi_3}_{2k-1}(5n+1) q^{-(5n+1)/5}\Gamma(1-2k,4\pi (5n+1) y/5)\,,\\
\nonumber Y_{\bm{3'},1}^{(2k)} &=& -\dfrac{(-1)^k\, 2^{2k}\,\pi}{\sqrt{5} (1-2k)} \dfrac{L(1-2k,\chi_3)}{L(2-2k,\chi_3)} + \dfrac{y^{1-2k}}{1-2k}  \\
\nonumber &&\hskip-0.4in - \dfrac{(-1)^k\, 2^{2k}\,\pi}{\sqrt{5}(1-2k)L(2-2k,\chi_3)} \sum_{n=1}^{+\infty} \Big( \widehat{\sigma}^{\chi_3}_{2k-1}(n) - 5^{2k-1} \widetilde{\sigma}^{\chi_3}_{2k-1}(n) \Big) q^n  \\
\nonumber &&\hskip-0.4in - \dfrac{(-1)^k\, 2^{2k}\,\pi}{\sqrt{5}(1-2k)!L(2-2k,\chi_3)} \sum_{n=1}^{+\infty} \Big( \widehat{\sigma}^{\chi_3}_{2k-1}(n) - 5^{2k-1} \widetilde{\sigma}^{\chi_3}_{2k-1}(n) \Big) q^{-n}\Gamma(1-2k,4\pi n y) \,, \\
\nonumber Y_{\bm{3'},2}^{(2k)} &=& \dfrac{(-1)^k\, 2^{2k}\sqrt{2}\,\pi}{\sqrt{5}(1-2k)L(2-2k,\chi_3)} \sum_{n=0}^{+\infty} \widehat{\sigma}^{\chi_3}_{2k-1}(5n+2)  q^{(5n+2)/5}  \\
\nonumber &&\hskip-0.4in  + \dfrac{(-1)^k\, 2^{2k}\sqrt{2}\,\pi}{\sqrt{5}(1-2k)!L(2-2k,\chi_3)} \sum_{n=0}^{+\infty}  \widehat{\sigma}^{\chi_3}_{2k-1}(5n+3) q^{-(5n+3)/5}\Gamma(1-2k,4\pi (5n+3) y/5) \,, \\
\nonumber Y_{\bm{3'},3}^{(2k)} &=& \dfrac{(-1)^k\, 2^{2k}\sqrt{2}\,\pi}{\sqrt{5}(1-2k)L(2-2k,\chi_3)} \sum_{n=0}^{+\infty} \widehat{\sigma}^{\chi_3}_{2k-1}(5n+3)  q^{(5n+3)/5}  \\
\nonumber &&\hskip-0.4in  + \dfrac{(-1)^k\, 2^{2k}\sqrt{2}\,\pi}{\sqrt{5}(1-2k)!L(2-2k,\chi_3)} \sum_{n=0}^{+\infty}  \widehat{\sigma}^{\chi_3}_{2k-1}(5n+2) q^{-(5n+2)/5}\Gamma(1-2k,4\pi (5n+2) y/5)\,,\\
\nonumber Y_{\bm{5},1}^{(2k)} &=& \dfrac{(-1)^k\,(1-5^{2k})2^{2k}\, \pi}{(1-2k)(5^{2k-1}-5)} \dfrac{\zeta(1-2k)}{\zeta(2-2k)} + \dfrac{y^{1-2k}}{1-2k}  \\
\nonumber &&\hskip-0.4in  + \dfrac{(-1)^k\, 2^{2k}\,\pi}{(1-2k)(5-5^{2k-1})\zeta(2-2k)} \sum_{n=1}^{+\infty} \Big( (5+ 5^{2k})\sigma_{2k-1}(n) - 6\sigma_{2k-1}(5n)  \Big) q^n \\
\nonumber &&\hskip-0.4in  + \dfrac{(-1)^k\, 2^{2k}\,\pi}{(1-2k)!(5-5^{2k-1})\zeta(2-2k)} \sum_{n=1}^{+\infty} \Big( (5+ 5^{2k})\sigma_{2k-1}(n) - 6\sigma_{2k-1}(5n)  \Big) q^{-n}\Gamma(1-2k,4\pi n y) \,,  \\
\nonumber Y_{\bm{5},2}^{(2k)} &=&  \dfrac{(-1)^k\, 2^{2k} \sqrt{6}\,\pi}{(1-2k)(5-5^{2k-1})\zeta(2-2k)} \sum_{n=1}^{+\infty} \sigma_{2k-1}(5n+1) q^{(5n+1)/5} \\
\nonumber &&\hskip-0.4in  + \dfrac{(-1)^k\, 2^{2k} \sqrt{6}\,\pi}{(1-2k)!(5-5^{2k-1})\zeta(2-2k)} \sum_{n=1}^{+\infty} \sigma_{2k-1}(5n+4) q^{-(5n+4)/5} \Gamma(1-2k,4\pi(5n+4)y/5)\,, \\
\nonumber Y_{\bm{5},3}^{(2k)} &=&  \dfrac{(-1)^k\, 2^{2k} \sqrt{6}\,\pi}{(1-2k)(5-5^{2k-1})\zeta(2-2k)} \sum_{n=1}^{+\infty} \sigma_{2k-1}(5n+2) q^{(5n+2)/5} \\
\nonumber &&\hskip-0.4in  + \dfrac{(-1)^k\, 2^{2k} \sqrt{6}\,\pi}{(1-2k)!(5-5^{2k-1})\zeta(2-2k)} \sum_{n=1}^{+\infty} \sigma_{2k-1}(5n+4) q^{-(5n+3)/5} \Gamma(1-2k,4\pi(5n+3)y/5) \,,\\
\nonumber Y_{\bm{5},4}^{(2k)} &=&  \dfrac{(-1)^k\, 2^{2k} \sqrt{6}\,\pi}{(1-2k)(5-5^{2k-1})\zeta(2-2k)} \sum_{n=1}^{+\infty} \sigma_{2k-1}(5n+3) q^{(5n+3)/5} \\
\nonumber &&\hskip-0.4in  + \dfrac{(-1)^k\, 2^{2k} \sqrt{6}\,\pi}{(1-2k)!(5-5^{2k-1})\zeta(2-2k)} \sum_{n=1}^{+\infty} \sigma_{2k-1}(5n+2) q^{-(5n+2)/5} \Gamma(1-2k,4\pi(5n+2)y/5)\,, \\
\nonumber Y_{\bm{5},5}^{(2k)} &=&  \dfrac{(-1)^k\, 2^{2k} \sqrt{6}\,\pi}{(1-2k)(5-5^{2k-1})\zeta(2-2k)} \sum_{n=1}^{+\infty} \sigma_{2k-1}(5n+4) q^{(5n+4)/5} \\
 &&\hskip-0.4in  + \dfrac{(-1)^k\, 2^{2k} \sqrt{6}\,\pi}{(1-2k)!(5-5^{2k-1})\zeta(2-2k)} \sum_{n=1}^{+\infty} \sigma_{2k-1}(5n+1) q^{-(5n+1)/5} \Gamma(1-2k,4\pi(5n+1)y/5)\,.~~~
\end{eqnarray}
Considering $\lim_{k\rightarrow0}(1-5^{-2k}) \zeta(1-2k) = - \log 5$ and the values $L(1,\chi_3) = \dfrac{2}{\sqrt{5}} \log \phi$ and $L(2,\chi_3) = \dfrac{4\pi^2}{25\sqrt{5}}$ of the $L$-functions, the weight $0$ polyharmonic Maa{\ss} forms can be recast from the above expressions as previously stated. Likewise the Fourier expansion of the negative odd weight $2k+1$ Eisenstein series for $k<0$ are given by
\begin{footnotesize}
\begin{eqnarray}
\nonumber E_{2k+1}(\tau; -2k; i\infty) &=& y^{-2k} - (-1)^k 2^{2k} \left( \dfrac{1}{ L(1-2k,\chi_2)} + \dfrac{1}{L(1-2k,\chi_4)}\right) \dfrac{i\pi}{5} \sum_{n=1}^{+\infty} q^{\frac{n}{5}} \sum_{\substack{m|n,\, m\in\mathbb{Z} \\ \frac{n}{m}\equiv 0\,({\rm mod}\, 5)}} {\rm sign}(m) \left| \dfrac{n}{m} \right|^{2k}\, \omega_5^{4m}   \\
\nonumber &&\hskip-1.2in - (-1)^k 2^{2k} \left( \dfrac{1}{ L(1-2k,\chi_2)} - \dfrac{1}{L(1-2k,\chi_4)}\right) \dfrac{i\pi}{5} \sum_{n=1}^{+\infty} q^{\frac{n}{5}} \sum_{\substack{m|n,\, m\in\mathbb{Z} \\ \frac{n}{m}\equiv 0\,({\rm mod}\, 5)}} {\rm sign}(m) \left| \dfrac{n}{m} \right|^{2k}\, \omega_5^{3m} \\
\nonumber &&\hskip-1.2in - \dfrac{(-1)^k 2^{2k}}{(-2k-1)!} \left( \dfrac{1}{ L(1-2k,\chi_2)} + \dfrac{1}{L(1-2k,\chi_4)}\right) \dfrac{i\pi}{5} \sum_{n=-1}^{-\infty} q^{\frac{n}{5}} \Gamma\left(-2k, - \dfrac{4\pi n y}{5} \right) \sum_{\substack{m|n,\, m\in\mathbb{Z} \\ \frac{n}{m}\equiv 0\,({\rm mod}\, 5)}} {\rm sign}(m) \left| \dfrac{n}{m} \right|^{2k}\, \omega_5^{4m}   \\
\nonumber &&\hskip-1.2in - \dfrac{(-1)^k 2^{2k}}{(-2k-1)!} \left( \dfrac{1}{ L(1-2k,\chi_2)} - \dfrac{1}{L(1-2k,\chi_4)}\right) \dfrac{i\pi}{5} \sum_{n=-1}^{-\infty} q^{\frac{n}{5}} \Gamma\left(-2k, - \dfrac{4\pi n y}{5} \right) \sum_{\substack{m|n,\, m\in\mathbb{Z} \\ \frac{n}{m}\equiv 0\,({\rm mod}\, 5)}} {\rm sign}(m) \left| \dfrac{n}{m} \right|^{2k}\, \omega_5^{3m}\,,   \\
\nonumber
E_{2k+1}(\tau; -2k; j) &=& (-1)^k 2^{2k+1} \left( \dfrac{L(-2k,\chi_4)}{L(1-2k,\chi_4)} + \dfrac{L(-2k,\chi_2)}{L(1-2k,\chi_2)} \right) \dfrac{i\pi}{5} \\
\nonumber &&\hskip-1.2in - (-1)^k 2^{2k} \left( \dfrac{1}{ L(1-2k,\chi_2)} + \dfrac{1}{L(1-2k,\chi_4)}\right) \dfrac{i\pi}{5} \sum_{n=1}^{+\infty} q^{\frac{n}{5}} \sum_{\substack{m|n,\, m\in\mathbb{Z} \\ \frac{n}{m}\equiv 1\,({\rm mod}\, 5)}} {\rm sign}(m) \left| \dfrac{n}{m} \right|^{2k}\, \omega_5^{4jm}   \\
\nonumber &&\hskip-1.2in - (-1)^k 2^{2k} \left( \dfrac{1}{ L(1-2k,\chi_2)} - \dfrac{1}{L(1-2k,\chi_4)}\right) \dfrac{i\pi}{5} \sum_{n=1}^{+\infty} q^{\frac{n}{5}} \sum_{\substack{m|n,\, m\in\mathbb{Z} \\ \frac{n}{m}\equiv 2\,({\rm mod}\, 5)}} {\rm sign}(m) \left| \dfrac{n}{m} \right|^{2k}\, \omega_5^{3jm} \\
\nonumber &&\hskip-1.2in - \dfrac{(-1)^k 2^{2k}}{(-2k-1)!} \left( \dfrac{1}{ L(1-2k,\chi_2)} + \dfrac{1}{L(1-2k,\chi_4)}\right) \dfrac{i\pi}{5} \sum_{n=-1}^{-\infty} q^{\frac{n}{5}} \Gamma\left(-2k, - \dfrac{4\pi n y}{5} \right) \sum_{\substack{m|n,\, m\in\mathbb{Z} \\ \frac{n}{m}\equiv 1\,({\rm mod}\, 5)}} {\rm sign}(m) \left| \dfrac{n}{m} \right|^{2k}\, \omega_5^{4jm}   \\
\nonumber &&\hskip-1.2in - \dfrac{(-1)^k 2^{2k}}{(-2k-1)!} \left( \dfrac{1}{ L(1-2k,\chi_2)} - \dfrac{1}{L(1-2k,\chi_4)}\right) \dfrac{i\pi}{5} \sum_{n=-1}^{-\infty} q^{\frac{n}{5}} \Gamma\left(-2k, - \dfrac{4\pi n y}{5} \right) \sum_{\substack{m|n,\, m\in\mathbb{Z} \\ \frac{n}{m}\equiv 2\,({\rm mod}\, 5)}} {\rm sign}(m) \left| \dfrac{n}{m} \right|^{2k}\, \omega_5^{3jm}\,,   \\
E_{2k+1}(\tau; -2k; l/2) &=& (-1)^k 2^{2k+1} \left( -i \dfrac{L(-2k,\chi_4)}{L(1-2k,\chi_4)} + i\dfrac{L(-2k,\chi_2)}{L(1-2k,\chi_2)} \right) \dfrac{i\pi}{5} \\
\nonumber &&\hskip-1.2in - (-1)^k 2^{2k} \left( \dfrac{1}{ L(1-2k,\chi_2)} + \dfrac{1}{L(1-2k,\chi_4)}\right) \dfrac{i\pi}{5} \sum_{n=1}^{+\infty} q^{\frac{n}{5}} \sum_{\substack{m|n,\, m\in\mathbb{Z} \\ \frac{n}{m}\equiv 2\,({\rm mod}\, 5)}} {\rm sign}(m) \left| \dfrac{n}{m} \right|^{2k}\, \omega_5^{4lm}   \\
\nonumber &&\hskip-1.2in - (-1)^k 2^{2k} \left( \dfrac{1}{ L(1-2k,\chi_2)} - \dfrac{1}{L(1-2k,\chi_4)}\right) \dfrac{i\pi}{5} \sum_{n=1}^{+\infty} q^{\frac{n}{5}} \sum_{\substack{m|n,\, m\in\mathbb{Z} \\ \frac{n}{m}\equiv 4\,({\rm mod}\, 5)}} {\rm sign}(m) \left| \dfrac{n}{m} \right|^{2k}\, \omega_5^{3lm} \\
\nonumber &&\hskip-1.2in - \dfrac{(-1)^k 2^{2k}}{(-2k-1)!} \left( \dfrac{1}{ L(1-2k,\chi_2)} + \dfrac{1}{L(1-2k,\chi_4)}\right) \dfrac{i\pi}{5} \sum_{n=-1}^{-\infty} q^{\frac{n}{5}} \Gamma\left(-2k, - \dfrac{4\pi n y}{5} \right) \sum_{\substack{m|n,\, m\in\mathbb{Z} \\ \frac{n}{m}\equiv 2\,({\rm mod}\, 5)}} {\rm sign}(m) \left| \dfrac{n}{m} \right|^{2k}\, \omega_5^{4lm}   \\
\nonumber &&\hskip-1.2in - \dfrac{(-1)^k 2^{2k}}{(-2k-1)!} \left( \dfrac{1}{ L(1-2k,\chi_2)} - \dfrac{1}{L(1-2k,\chi_4)}\right) \dfrac{i\pi}{5} \sum_{n=-1}^{-\infty} q^{\frac{n}{5}} \Gamma\left(-2k, - \dfrac{4\pi n y}{5} \right) \sum_{\substack{m|n,\, m\in\mathbb{Z} \\ \frac{n}{m}\equiv 4\,({\rm mod}\, 5)}} {\rm sign}(m) \left| \dfrac{n}{m} \right|^{2k}\, \omega_5^{3lm} \,,  \\
\nonumber E_{2k+1}(\tau; -2k; 2/5) &=& - (-1)^k 2^{2k} \left( \dfrac{1}{ L(1-2k,\chi_2)} + \dfrac{1}{L(1-2k,\chi_4)}\right) \dfrac{i\pi}{5} \sum_{n=1}^{+\infty} q^{\frac{n}{5}} \sum_{\substack{m|n,\, m\in\mathbb{Z} \\ \frac{n}{m}\equiv 0\,({\rm mod}\, 5)}} {\rm sign}(m) \left| \dfrac{n}{m} \right|^{2k}\, \omega_5^{3m}   \\
\nonumber &&\hskip-1.2in - (-1)^k 2^{2k} \left( \dfrac{1}{ L(1-2k,\chi_2)} - \dfrac{1}{L(1-2k,\chi_4)}\right) \dfrac{i\pi}{5} \sum_{n=1}^{+\infty} q^{\frac{n}{5}} \sum_{\substack{m|n,\, m\in\mathbb{Z} \\ \frac{n}{m}\equiv 0\,({\rm mod}\, 5)}} {\rm sign}(m) \left| \dfrac{n}{m} \right|^{2k}\, \omega_5^{m} \\
\nonumber &&\hskip-1.2in - \dfrac{(-1)^k 2^{2k}}{(-2k-1)!} \left( \dfrac{1}{ L(1-2k,\chi_2)} + \dfrac{1}{L(1-2k,\chi_4)}\right) \dfrac{i\pi}{5} \sum_{n=-1}^{-\infty} q^{\frac{n}{5}} \Gamma\left(-2k, - \dfrac{4\pi n y}{5} \right) \sum_{\substack{m|n,\, m\in\mathbb{Z} \\ \frac{n}{m}\equiv 0\,({\rm mod}\, 5)}} {\rm sign}(m) \left| \dfrac{n}{m} \right|^{2k}\, \omega_5^{3m}   \\
 &&\hskip-1.2in - \dfrac{(-1)^k 2^{2k}}{(-2k-1)!} \left( \dfrac{1}{ L(1-2k,\chi_2)} - \dfrac{1}{L(1-2k,\chi_4)}\right) \dfrac{i\pi}{5} \sum_{n=-1}^{-\infty} q^{\frac{n}{5}} \Gamma\left(-2k, - \dfrac{4\pi n y}{5} \right) \sum_{\substack{m|n,\, m\in\mathbb{Z} \\ \frac{n}{m}\equiv 0\,({\rm mod}\, 5)}} {\rm sign}(m) \left| \dfrac{n}{m} \right|^{2k}\, \omega_5^{m}\,.
\end{eqnarray}
\end{footnotesize}
They form two linearly independent sextuplets polyharmonic Maa{\ss} forms. By plugging the above expressions into Eq.~\eqref{eq:N=5_Y6hI-Y6hII} with $s=-2k$, we can obtain the $q$-expansion of the negative odd weight polyharmonic Maa{\ss} form multiplets as follow,
\begin{small}
\begin{eqnarray}
\nonumber Y_{\bm{6}I,1}^{(2k+1)} &=&  (-1)^k \dfrac{2^{2k}\,\pi}{2k} \left( \sqrt{\dfrac{-1+2i}{5\sqrt{5}}} \dfrac{L(-2k,\chi_4)}{L(1-2k,\chi_4)} + \sqrt{\dfrac{-1-2i}{5\sqrt{5}}} \dfrac{L(-2k,\chi_2)}{L(1-2k,\chi_2)} \right) - \dfrac{y^{-2k}}{2k} \\
\nonumber &&\hskip-0.4in - \dfrac{(-1)^k\, 2^{2k}\,\pi i}{5\cdot(-2k)} \sum_{n=1}^{+\infty} \left( \dfrac{\sqrt{\sqrt{5}(1+2i)}}{L(1-2k,\chi_2)}(i 5^{2k}\widetilde{\sigma}^{\chi_2}_{2k}(n) - \widehat{\sigma}^{\chi_2}_{2k}(n)) + \dfrac{\sqrt{\sqrt{5}(1-2i)}}{L(1-2k,\chi_4)}(i 5^{2k}\widetilde{\sigma}^{\chi_4}_{2k}(n) + \widehat{\sigma}^{\chi_4}_{2k}(n)) \right) q^n  \\
\nonumber &&\hskip-0.4in + \dfrac{(-1)^k\, 2^{2k}\,\pi i}{5\cdot(-2k)!} \sum_{n=1}^{+\infty} \left( \dfrac{\sqrt{\sqrt{5}(1+2i)}}{L(1-2k,\chi_2)}(i 5^{2k}\widetilde{\sigma}^{\chi_2}_{2k}(n) + \widehat{\sigma}^{\chi_2}_{2k}(n)) + \dfrac{\sqrt{\sqrt{5}(1-2i)}}{L(1-2k,\chi_4)}(i 5^{2k}\widetilde{\sigma}^{\chi_4}_{2k}(n) - \widehat{\sigma}^{\chi_4}_{2k}(n)) \right)  \\
\nonumber &&\hskip-0.4in \times q^{-n}\Gamma(-2k,4\pi n y)\,,  \\
\nonumber Y_{\bm{6}I,2}^{(2k+1)} &=& - (-1)^k \dfrac{2^{2k}\,\pi}{2k} \left( \sqrt{\dfrac{1-2i}{5\sqrt{5}}} \dfrac{L(-2k,\chi_4)}{L(1-2k,\chi_4)} + \sqrt{\dfrac{1+2i}{5\sqrt{5}}} \dfrac{L(-2k,\chi_2)}{L(1-2k,\chi_2)} \right) \\
\nonumber &&\hskip-0.4in - \dfrac{(-1)^k\, 2^{2k}\,\pi i}{5\cdot(-2k)} \sum_{n=1}^{+\infty} \left( \dfrac{\sqrt{\sqrt{5}(1+2i)}}{L(1-2k,\chi_2)}(-5^{2k}\widetilde{\sigma}^{\chi_2}_{2k}(n) + i \widehat{\sigma}^{\chi_2}_{2k}(n)) + \dfrac{\sqrt{\sqrt{5}(1-2i)}}{L(1-2k,\chi_4)}(5^{2k}\widetilde{\sigma}^{\chi_4}_{2k}(n) + i\widehat{\sigma}^{\chi_4}_{2k}(n)) \right) q^n  \\
\nonumber &&\hskip-0.4in + \dfrac{(-1)^k\, 2^{2k}\,\pi i}{5\cdot(-2k)!} \sum_{n=1}^{+\infty} \left( \dfrac{\sqrt{\sqrt{5}(1+2i)}}{L(1-2k,\chi_2)}(-5^{2k}\widetilde{\sigma}^{\chi_2}_{2k}(n) - i \widehat{\sigma}^{\chi_2}_{2k}(n)) + \dfrac{\sqrt{\sqrt{5}(1-2i)}}{L(1-2k,\chi_4)}(5^{2k}\widetilde{\sigma}^{\chi_4}_{2k}(n) - i \widehat{\sigma}^{\chi_4}_{2k}(n)) \right)  \\
\nonumber &&\hskip-0.4in \times q^{-n}\Gamma(-2k,4\pi n y)\,,  \\
\nonumber Y_{\bm{6}I,3}^{(2k+1)} &=& - \dfrac{(-1)^k\, 2^{2k}\,\pi i}{5\cdot(-2k)} \sum_{n=1}^{+\infty} \left( - \dfrac{ \sqrt{2\sqrt{5}(2-i)}}{L(1-2k,\chi_2)} \widehat{\sigma}^{\chi_2}_{2k}(5n + 1) + \dfrac{\sqrt{2\sqrt{5}(2+i)}}{L(1-2k,\chi_4)} \widehat{\sigma}^{\chi_4}_{2k}(5n+1) \right) q^{(5n+1)/5}  \\
\nonumber &&\hskip-0.4in - \dfrac{(-1)^k\, 2^{2k}\,\pi i}{5\cdot(-2k)!} \sum_{n=1}^{+\infty} \left( - \dfrac{ \sqrt{2\sqrt{5}(2-i)}}{L(1-2k,\chi_2)} \widehat{\sigma}^{\chi_2}_{2k}(5n + 4) + \dfrac{\sqrt{2\sqrt{5}(2+i)}}{L(1-2k,\chi_4)} \widehat{\sigma}^{\chi_4}_{2k}(5n+4) \right)  \\
\nonumber &&\hskip-0.4in \times q^{-(5n+4)/5} \Gamma(-2k,4\pi(5n+4)y/5)\,,  \\
\nonumber Y_{\bm{6}I,4}^{(2k+1)} &=& - \dfrac{(-1)^k\, 2^{2k}\,\pi i}{5\cdot(-2k)} \sum_{n=1}^{+\infty} \left( - \dfrac{ \sqrt{-2\sqrt{5}(1+2i)}}{L(1-2k,\chi_2)} \widehat{\sigma}^{\chi_2}_{2k}(5n + 2) + \dfrac{\sqrt{-2\sqrt{5}(1-2i)}}{L(1-2k,\chi_4)} \widehat{\sigma}^{\chi_4}_{2k}(5n+2) \right) q^{(5n+2)/5}  \\
\nonumber &&\hskip-0.4in - \dfrac{(-1)^k\, 2^{2k}\,\pi i}{5\cdot(-2k)!} \sum_{n=1}^{+\infty} \left( - \dfrac{ \sqrt{-2\sqrt{5}(1+2i)}}{L(1-2k,\chi_2)} \widehat{\sigma}^{\chi_2}_{2k}(5n + 3) + \dfrac{\sqrt{-2\sqrt{5}(1-2i)}}{L(1-2k,\chi_4)} \widehat{\sigma}^{\chi_4}_{2k}(5n+3) \right)  \\
\nonumber &&\hskip-0.4in \times q^{-(5n+3)/5} \Gamma(-2k,4\pi(5n+3)y/5) \,, \\
\nonumber Y_{\bm{6}I,5}^{(2k+1)} &=& -\dfrac{(-1)^k\, 2^{2k}\,\pi i}{5\cdot(-2k)} \sum_{n=1}^{+\infty} \left( \dfrac{ \sqrt{2\sqrt{5}(1+2i)}}{L(1-2k,\chi_2)} \widehat{\sigma}^{\chi_2}_{2k}(5n + 3) - \dfrac{\sqrt{2\sqrt{5}(1-2i)}}{L(1-2k,\chi_4)} \widehat{\sigma}^{\chi_4}_{2k}(5n+3) \right) q^{(5n+3)/5}  \\
\nonumber &&\hskip-0.4in - \dfrac{(-1)^k\, 2^{2k}\,\pi i}{5\cdot(-2k)!} \sum_{n=1}^{+\infty} \left( \dfrac{ \sqrt{2\sqrt{5}(1+2i)}}{L(1-2k,\chi_2)} \widehat{\sigma}^{\chi_2}_{2k}(5n + 2) - \dfrac{\sqrt{2\sqrt{5}(1-2i)}}{L(1-2k,\chi_4)} \widehat{\sigma}^{\chi_4}_{2k}(5n+2) \right)  \\
\nonumber &&\hskip-0.4in \times q^{-(5n+2)/5} \Gamma(-2k,4\pi(5n+2)y/5) \,, \\
\nonumber Y_{\bm{6}I,6}^{(2k+1)} &=& -\dfrac{(-1)^k\, 2^{2k}\,\pi i}{5\cdot(-2k)} \sum_{n=1}^{+\infty} \left( \dfrac{ \sqrt{-2\sqrt{5}(2-i)}}{L(1-2k,\chi_2)} \widehat{\sigma}^{\chi_2}_{2k}(5n + 4) - \dfrac{\sqrt{-2\sqrt{5}(2+i)}}{L(1-2k,\chi_4)} \widehat{\sigma}^{\chi_4}_{2k}(5n+4) \right) q^{(5n+4)/5}  \\
\nonumber &&\hskip-0.4in - \dfrac{(-1)^k\, 2^{2k}\,\pi i}{5\cdot(-2k)!} \sum_{n=1}^{+\infty} \left( \dfrac{ \sqrt{-2\sqrt{5}(2-i)}}{L(1-2k,\chi_2)} \widehat{\sigma}^{\chi_2}_{2k}(5n + 1) - \dfrac{\sqrt{-2\sqrt{5}(2+i)}}{L(1-2k,\chi_4)} \widehat{\sigma}^{\chi_4}_{2k}(5n+1) \right)  \\
\nonumber &&\hskip-0.4in \times q^{-(5n+1)/5} \Gamma(-2k,4\pi(5n+1)y/5) \,,\\
\nonumber Y_{\bm{6}II,1}^{(2k+1)} &=& - (-1)^k \dfrac{2^{2k}\,\pi}{2k} \left( \sqrt{\dfrac{1-2i}{5\sqrt{5}}} \dfrac{L(-2k,\chi_4)}{L(1-2k,\chi_4)} + \sqrt{\dfrac{1+2i}{5\sqrt{5}}} \dfrac{L(-2k,\chi_2)}{L(1-2k,\chi_2)} \right)  \\
\nonumber &&\hskip-0.4in - \dfrac{(-1)^k\, 2^{2k}\,\pi i}{5\cdot(-2k)} \sum_{n=1}^{+\infty} \left( \dfrac{\sqrt{\sqrt{5}(1+2i)}}{L(1-2k,\chi_2)}(5^{2k}\widetilde{\sigma}^{\chi_2}_{2k}(n) + i \widehat{\sigma}^{\chi_2}_{2k}(n)) + \dfrac{\sqrt{\sqrt{5}(1-2i)}}{L(1-2k,\chi_4)}(- 5^{2k}\widetilde{\sigma}^{\chi_4}_{2k}(n) + i \widehat{\sigma}^{\chi_4}_{2k}(n)) \right) q^n  \\
\nonumber &&\hskip-0.4in + \dfrac{(-1)^k\, 2^{2k}\,\pi i}{5\cdot(-2k)!} \sum_{n=1}^{+\infty} \left( \dfrac{\sqrt{\sqrt{5}(1+2i)}}{L(1-2k,\chi_2)}(5^{2k}\widetilde{\sigma}^{\chi_2}_{2k}(n) -i \widehat{\sigma}^{\chi_2}_{2k}(n)) + \dfrac{\sqrt{\sqrt{5}(-1+2i)}}{L(1-2k,\chi_4)}(- 5^{2k}\widetilde{\sigma}^{\chi_4}_{2k}(n) - i \widehat{\sigma}^{\chi_4}_{2k}(n)) \right)  \\
\nonumber &&\hskip-0.4in \times q^{-n}\Gamma(-2k,4\pi n y) \,, \\
\nonumber Y_{\bm{6}II,2}^{(2k+1)} &=& - (-1)^k \dfrac{2^{2k}\,\pi}{2k} \left( \sqrt{\dfrac{-1+2i}{5\sqrt{5}}} \dfrac{L(-2k,\chi_4)}{L(1-2k,\chi_4)} + \sqrt{\dfrac{-1-2i}{5\sqrt{5}}} \dfrac{L(-2k,\chi_2)}{L(1-2k,\chi_2)} \right) - \dfrac{y^{-2k}}{2k} \\
\nonumber &&\hskip-0.4in - \dfrac{(-1)^k\, 2^{2k}\,\pi i}{5\cdot(-2k)} \sum_{n=1}^{+\infty} \left( \dfrac{\sqrt{\sqrt{5}(1+2i)}}{L(1-2k,\chi_2)}(i 5^{2k}\widetilde{\sigma}^{\chi_2}_{2k}(n) + \widehat{\sigma}^{\chi_2}_{2k}(n)) + \dfrac{\sqrt{\sqrt{5}(1-2i)}}{L(1-2k,\chi_4)}(i 5^{2k}\widetilde{\sigma}^{\chi_4}_{2k}(n) - \widehat{\sigma}^{\chi_4}_{2k}(n)) \right) q^n  \\
\nonumber &&\hskip-0.4in + \dfrac{(-1)^k\, 2^{2k}\,\pi i}{5\cdot(-2k)!} \sum_{n=1}^{+\infty} \left( \dfrac{\sqrt{\sqrt{5}(1+2i)}}{L(1-2k,\chi_2)}(i 5^{2k}\widetilde{\sigma}^{\chi_2}_{2k}(n) - \widehat{\sigma}^{\chi_2}_{2k}(n)) + \dfrac{\sqrt{\sqrt{5}(1-2i)}}{L(1-2k,\chi_4)}(i 5^{2k}\widetilde{\sigma}^{\chi_4}_{2k}(n) + \widehat{\sigma}^{\chi_4}_{2k}(n)) \right)  \\
\nonumber &&\hskip-0.4in \times q^{-n}\Gamma(-2k,4\pi n y) \,,  \\
\nonumber Y_{\bm{6}II,3}^{(2k+1)} &=& -\dfrac{(-1)^k\, 2^{2k}\,\pi i}{5\cdot(-2k)} \sum_{n=1}^{+\infty} \left( \dfrac{ \sqrt{-2\sqrt{5}(2-i)}}{L(1-2k,\chi_2)} \widehat{\sigma}^{\chi_2}_{2k}(5n + 1) - \dfrac{\sqrt{-2\sqrt{5}(2+i)}}{L(1-2k,\chi_4)} \widehat{\sigma}^{\chi_4}_{2k}(5n+1) \right) q^{(5n+1)/5}  \\
\nonumber &&\hskip-0.4in - \dfrac{(-1)^k\, 2^{2k}\,\pi i}{5\cdot(-2k)!} \sum_{n=1}^{+\infty} \left( \dfrac{ \sqrt{-2\sqrt{5}(2-i)}}{L(1-2k,\chi_2)} \widehat{\sigma}^{\chi_2}_{2k}(5n + 4) - \dfrac{\sqrt{-2\sqrt{5}(2+i)}}{L(1-2k,\chi_4)} \widehat{\sigma}^{\chi_4}_{2k}(5n+4) \right)  \\
\nonumber &&\hskip-0.4in \times q^{-(5n+4)/5} \Gamma(-2k,4\pi(5n+4)y/5)\,,  \\
\nonumber Y_{\bm{6}II,4}^{(2k+1)} &=& -\dfrac{(-1)^k\, 2^{2k}\,\pi i}{5\cdot(-2k)} \sum_{n=1}^{+\infty} \left( \dfrac{ \sqrt{2\sqrt{5}(1+2i)}}{L(1-2k,\chi_2)} \widehat{\sigma}^{\chi_2}_{2k}(5n + 2) - \dfrac{\sqrt{2\sqrt{5}(1-2i)}}{L(1-2k,\chi_4)} \widehat{\sigma}^{\chi_4}_{2k}(5n+2) \right) q^{(5n+2)/5}  \\
\nonumber &&\hskip-0.4in - \dfrac{(-1)^k\, 2^{2k}\,\pi i}{5\cdot(-2k)!} \sum_{n=1}^{+\infty} \left( \dfrac{ \sqrt{2\sqrt{5}(1+2i)}}{L(1-2k,\chi_2)} \widehat{\sigma}^{\chi_2}_{2k}(5n + 3) - \dfrac{\sqrt{2\sqrt{5}(1-2i)}}{L(1-2k,\chi_4)} \widehat{\sigma}^{\chi_4}_{2k}(5n+3) \right)  \\
\nonumber &&\hskip-0.4in \times q^{-(5n+3)/5} \Gamma(-2k,4\pi(5n+3)y/5) \,, \\
\nonumber Y_{\bm{6}II,5}^{(2k+1)} &=& -\dfrac{(-1)^k\, 2^{2k}\,\pi i}{5\cdot(-2k)} \sum_{n=1}^{+\infty} \left( \dfrac{ \sqrt{-2\sqrt{5}(1+2i)}}{L(1-2k,\chi_2)} \widehat{\sigma}^{\chi_2}_{2k}(5n + 3) - \dfrac{\sqrt{-2\sqrt{5}(1-2i)}}{L(1-2k,\chi_4)} \widehat{\sigma}^{\chi_4}_{2k}(5n+3) \right) q^{(5n+3)/5}  \\
\nonumber &&\hskip-0.4in - \dfrac{(-1)^k\, 2^{2k}\,\pi i}{5\cdot(-2k)!} \sum_{n=1}^{+\infty} \left( \dfrac{ \sqrt{-2\sqrt{5}(1+2i)}}{L(1-2k,\chi_2)} \widehat{\sigma}^{\chi_2}_{2k}(5n + 2) - \dfrac{\sqrt{2\sqrt{5}(1-2i)}}{L(1-2k,\chi_4)} \widehat{\sigma}^{\chi_4}_{2k}(5n+2) \right)  \\
\nonumber &&\hskip-0.4in \times q^{-(5n+2)/5} \Gamma(-2k,4\pi(5n+2)y/5) \,, \\
\nonumber Y_{\bm{6}II,6}^{(2k+1)} &=& -\dfrac{(-1)^k\, 2^{2k}\,\pi i}{5\cdot(-2k)} \sum_{n=1}^{+\infty} \left( \dfrac{ \sqrt{2\sqrt{5}(2-i)}}{L(1-2k,\chi_2)} \widehat{\sigma}^{\chi_2}_{2k}(5n + 4) - \dfrac{\sqrt{2\sqrt{5}(2+i)}}{L(1-2k,\chi_4)} \widehat{\sigma}^{\chi_4}_{2k}(5n+4) \right) q^{(5n+4)/5}  \\
\nonumber &&\hskip-0.4in - \dfrac{(-1)^k\, 2^{2k}\,\pi i}{5\cdot(-2k)!} \sum_{n=1}^{+\infty} \left( \dfrac{ \sqrt{2\sqrt{5}(2-i)}}{L(1-2k,\chi_2)} \widehat{\sigma}^{\chi_2}_{2k}(5n + 1) - \dfrac{\sqrt{2\sqrt{5}(2+i)}}{L(1-2k,\chi_4)} \widehat{\sigma}^{\chi_4}_{2k}(5n+1) \right)  \\
 &&\hskip-0.4in \times q^{-(5n+1)/5} \Gamma(-2k,4\pi(5n+1)y/5) \,.
\end{eqnarray}
\end{small}
Thus we have derived the analytic expressions for all negative weight polyharmonic Maa{\ss} forms of level $N=5$. Then we proceed to discuss the weight one non-holomorphic Eisenstein series, which can be used to construct the weight one polyharmonic Maa{\ss} forms.
The weight one Eisenstein series $E_1(\tau; 0; \overline{A/C})$ for $s=0$, are given by
\begin{eqnarray}
\nonumber E_1(\tau; 0; \overline{A/C})&=& \dfrac{1}{2} \delta\left(\dfrac{C}{5} \right) \left( \chi^{*}_2(A) + \chi^{*}_4(A) \right) + \dfrac{i\pi}{10} \left(\chi^{*}_2(C)\dfrac{L(0,\chi_2)}{L(1,\chi_2)} + \chi^{*}_4(C) \dfrac{L(0,\chi_4)}{L(1,\chi_4)} \right)  \\
\nonumber && + i\sqrt{1+\dfrac{2}{\sqrt{5}}} \sum_{n=1}^{+\infty} \sum_{\substack{m|n\,, m\in\mathbb{Z} \\ \frac{n}{m}\equiv C\, ({\rm mod}\, 5)}} {\rm sign}(m) \omega_5^{4Am} q^{n/5} \\
&& + i\sqrt{1-\dfrac{2}{\sqrt{5}}} \sum_{n=1}^{+\infty} \sum_{\substack{m|n\,, m\in\mathbb{Z} \\ \frac{n}{m}\equiv 2C\, ({\rm mod}\, 5)}} {\rm sign}(m) \omega_5^{3Am} q^{n/5}  \,,
\end{eqnarray}
where $\overline{A/C}$ runs through every cusp of $\Gamma(5)$. Note that $E_1(\tau; 0; \overline{A/C})$ is a holomorphic function of $\tau$, their linear combination as Eq.~\eqref{eq:N=5_Y6hI-Y6hII} can give rise to the weight one holomorphic modular form sextuplet of level $N=5$~\cite{Yao:2020zml}. In a similar manner, we can obtain analytical formulas for the Fourier expansion of the first derivative $E_1^{(1)}(\tau ; \overline{A/C})$ of the weight one non-holomorphic Eisenstein series at $s=0$. However, it is too lengthy to provide some useful insight, consequently we don’t show this cumbersome expression here. Taking appropriate linear combinations of $E_1^{(1)}(\tau ; \overline{A/C})$, we can obtain two polyharmonic Maa{\ss} form sextuplets, one of which is the holomorphic modular form.

In the following we present the explicit $q$-expansions of the polyharmonic Maa{\ss} form multiplets from weight $-4$ to weight $6$
\begin{itemize}[labelindent=-0.8em,leftmargin=0.3em]

\item{$k_Y = -4$}

The weight $k_Y=-4$ polyharmonic Maa{\ss} forms of level $5$ can be arranged to a singlet $\bm{1}$, two triplets $\bm{3}$, $\bm{3'}$ and a quintuplet $\bm{5}$ of $A'_5$, each component of these multiplets has the following $q$-expansion,
\begin{eqnarray}
\nonumber Y_{\bm{1}}^{(-4)}&=& \dfrac{y^5}{5} + \dfrac{63\Gamma(5,4\pi y)}{128\pi^5 q} + \dfrac{2079\Gamma(5,8\pi y)}{4096\pi^5 q^2} + \dfrac{427\Gamma(5,12\pi y)}{864\pi^5 q^3} + \dfrac{66591\Gamma(5,16\pi y)}{131072 \pi^5 q^4} + \cdots   \\
\nonumber &&\hskip-0.4in +\dfrac{\pi}{80}\dfrac{\zeta(5)}{\zeta(6)} + \dfrac{189 q}{16\pi^5} + \dfrac{6237 q^2}{512 \pi^5} + \dfrac{427 q^3}{36 \pi^5} + \dfrac{199773 q^4}{16384 \pi^5} + \cdots \,, \\
\nonumber  Y_{\bm{3},1}^{(-4)} &=& \dfrac{y^5}{5} + \dfrac{7815 }{34304 \pi^5}\left( \dfrac{\Gamma(5, 4 \pi y)}{ q}+ \dfrac{24211 \Gamma(5, 8 \pi y)}{25008 q^2}+ \dfrac{378004 \Gamma(5, 12 \pi y)}{379809 q^3}+ \dfrac{993 \Gamma(5, 16 \pi y)}{1024 q^4} + \cdots  \right) \\
\nonumber &&\hskip-0.4in + \dfrac{\pi}{80\sqrt{5}} \dfrac{L(5,\chi_3)}{L(6,\chi_3)} + \dfrac{23445 }{4288 \pi^5}\left( q + \dfrac{24211 q^2}{25008}+ \dfrac{378004 q^3}{379809}+ \dfrac{993 q^4}{1024}+ \dfrac{4882813 q^5}{4884375} + \cdots \right) \,,  \\
\nonumber  Y_{\bm{3},2}^{(-4)} &=& - \dfrac{15515625 q^{1/5}}{35127296 \sqrt{2} \pi^5}\left( \dfrac{\Gamma(5, 16 \pi y/5)}{ q}+ \dfrac{60218368 \Gamma(5, 36 \pi y/5)}{58635657 q^2}+ \dfrac{5557184 \Gamma(5, 56 \pi y/5)}{5563117 q^3} + \cdots  \right) \\
\nonumber &&\hskip-0.4in - \dfrac{46875 q^{1/5}}{4288 \sqrt{2} \pi^5}\left( 1 + \dfrac{3751 q}{3888}+ \dfrac{161052 q^2}{161051}+ \dfrac{1016801 q^3}{1048576}+ \dfrac{1355684 q^4}{1361367} + \cdots \right) \,, \\
\nonumber  Y_{\bm{3},3}^{(-4)} &=& - \dfrac{15625 q^{4/5}}{34304 \sqrt{2} \pi^5}\left( \dfrac{\Gamma(5, 4 \pi y/5)}{ q}+ \dfrac{3751 \Gamma(5, 24 \pi y/5)}{3888 q^2}+ \dfrac{161052 \Gamma(5, 44 \pi y/5)}{161051 q^3} + \cdots  \right) \\
\nonumber &&\hskip-0.4in - \dfrac{46546875 q^{4/5}}{4390912 \sqrt{2} \pi^5}\left( 1 + \dfrac{60218368 q}{58635657}+ \dfrac{5557184 q^2}{5563117}+ \dfrac{2535526400 q^3}{2458766307}+ \dfrac{3844775 q^4}{3860784} + \cdots \right) \,, \\
\nonumber  Y_{\bm{3}',1}^{(-4)} &=& \dfrac{y^5}{5} - \dfrac{3905 }{17152 \pi^5}\left( \dfrac{\Gamma(5, 4 \pi y)}{ q}+ \dfrac{48453 \Gamma(5, 8 \pi y)}{49984 q^2}+ \dfrac{5731 \Gamma(5, 12 \pi y)}{5751 q^3}+ \dfrac{993 \Gamma(5, 16 \pi y)}{1024 q^4} + \cdots  \right) \\
\nonumber &&\hskip-0.4in - \dfrac{\pi}{80\sqrt{5}} \dfrac{L(5,\chi_3)}{L(6,\chi_3)} - \dfrac{11715 }{2144 \pi^5}\left( q + \dfrac{48453 q^2}{49984}+ \dfrac{5731 q^3}{5751}+ \dfrac{993 q^4}{1024}+ \dfrac{3126 q^5}{3125} + \cdots \right) \,, \\
\nonumber  Y_{\bm{3}',2}^{(-4)} &=& \dfrac{1890625 q^{2/5}}{4167936 \sqrt{2} \pi^5}\left( \dfrac{\Gamma(5, 12 \pi y/5)}{ q}+ \dfrac{7721325 \Gamma(5, 32 \pi y/5)}{7929856 q^2}+ \dfrac{45111978 \Gamma(5, 52 \pi y/5)}{44926453 q^3} + \cdots  \right) \\
\nonumber &&\hskip-0.4in + \dfrac{1453125 q^{2/5}}{137216 \sqrt{2} \pi^5}\left( 1 + \dfrac{537792 q}{521017}+ \dfrac{40051 q^2}{40176}+ \dfrac{45435392 q^3}{44015567}+ \dfrac{161052 q^4}{161051} + \cdots \right)\,, \\
\nonumber  Y_{\bm{3}',3}^{(-4)} &=& \dfrac{484375 q^{3/5}}{1097728 \sqrt{2} \pi^5}\left( \dfrac{\Gamma(5, 8 \pi y/5)}{ q}+ \dfrac{537792 \Gamma(5, 28 \pi y/5)}{521017 q^2}+ \dfrac{40051 \Gamma(5, 48 \pi y/5)}{40176 q^3} + \cdots  \right) \\
\nonumber &&\hskip-0.4in + \dfrac{1890625 q^{3/5}}{173664 \sqrt{2} \pi^5}\left( 1 + \dfrac{7721325 q}{7929856}+ \dfrac{45111978 q^2}{44926453}+ \dfrac{1823017 q^3}{1881792}+ \dfrac{71092323 q^4}{70799773} + \cdots \right) \,, \\
\nonumber  Y_{\bm{5},1}^{(-4)} &=& \dfrac{y^5}{5} - \dfrac{1563 }{15872 \pi^5}\left( \dfrac{\Gamma(5, 4 \pi y)}{ q}+ \dfrac{33 \Gamma(5, 8 \pi y)}{32 q^2}+ \dfrac{244 \Gamma(5, 12 \pi y)}{243 q^3}+ \dfrac{1057 \Gamma(5, 16 \pi y)}{1024 q^4} + \cdots  \right) \\
\nonumber &&\hskip-0.4in - \dfrac{13\pi}{5208} \dfrac{\zeta(5)}{\zeta(6)} - \dfrac{4689 }{1984 \pi^5}\left( q + \dfrac{33 q^2}{32}+ \dfrac{244 q^3}{243}+ \dfrac{1057 q^4}{1024}+ \dfrac{1625521 q^5}{1628125} + \cdots \right)\,, \\
\nonumber  Y_{\bm{5},2}^{(-4)} &=& \dfrac{3303125 \sqrt{3} q^{1/5}}{16252928 \sqrt{2} \pi^5}\left( \dfrac{\Gamma(5, 16 \pi y/5)}{ q}+ \dfrac{60716032 \Gamma(5, 36 \pi y/5)}{62414793 q^2}+ \dfrac{17749248 \Gamma(5, 56 \pi y/5)}{17764999 q^3} + \cdots  \right) \\
\nonumber &&\hskip-0.4in + \dfrac{9375 \sqrt{3} q^{1/5}}{1984 \sqrt{2} \pi^5}\left( 1 + \dfrac{671 q}{648}+ \dfrac{161052 q^2}{161051}+ \dfrac{1082401 q^3}{1048576}+ \dfrac{4101152 q^4}{4084101} + \cdots \right) \,, \\
\nonumber  Y_{\bm{5},3}^{(-4)} &=& \dfrac{190625 q^{2/5}}{321408 \sqrt{6} \pi^5}\left( \dfrac{\Gamma(5, 12 \pi y/5)}{ q}+ \dfrac{8219475 \Gamma(5, 32 \pi y/5)}{7995392 q^2}+ \dfrac{45112221 \Gamma(5, 52 \pi y/5)}{45297746 q^3} + \cdots  \right) \\
\nonumber &&\hskip-0.4in + \dfrac{309375 \sqrt{3} q^{2/5}}{63488 \sqrt{2} \pi^5}\left( 1 + \dfrac{48896 q}{50421}+ \dfrac{64477 q^2}{64152}+ \dfrac{1376832 q^3}{1419857}+ \dfrac{161052 q^4}{161051} + \cdots \right)\,, \\
\nonumber  Y_{\bm{5},4}^{(-4)} &=& \dfrac{103125 \sqrt{3} q^{3/5}}{507904 \sqrt{2} \pi^5}\left( \dfrac{\Gamma(5, 8 \pi y/5)}{ q}+ \dfrac{48896 \Gamma(5, 28 \pi y/5)}{50421 q^2}+ \dfrac{64477 \Gamma(5, 48 \pi y/5)}{64152 q^3} + \cdots  \right) \\
\nonumber &&\hskip-0.4in + \dfrac{190625 q^{3/5}}{13392 \sqrt{6} \pi^5}\left( 1 + \dfrac{8219475 q}{7995392}+ \dfrac{45112221 q^2}{45297746}+ \dfrac{652223 q^3}{632448}+ \dfrac{391007898 q^4}{392616923} + \cdots \right) \,, \\
\nonumber  Y_{\bm{5},5}^{(-4)} &=& \dfrac{3125 \sqrt{3} q^{4/5}}{15872 \sqrt{2} \pi^5}\left( \dfrac{\Gamma(5, 4 \pi y/5)}{ q}+ \dfrac{671 \Gamma(5, 24 \pi y/5)}{648 q^2}+ \dfrac{161052 \Gamma(5, 44 \pi y/5)}{161051 q^3} + \cdots  \right) \\
&&\hskip-0.4in + \dfrac{9909375 \sqrt{3} q^{4/5}}{2031616 \sqrt{2} \pi^5}\left( 1 + \dfrac{60716032 q}{62414793}+ \dfrac{17749248 q^2}{17764999}+ \dfrac{2535526400 q^3}{2617236643}+ \dfrac{687775 q^4}{684936} + \cdots \right)\,.
\end{eqnarray}

\item{$k_Y = -3$}

One can organize the weight $k_Y=-3$ polyharmonic Maa{\ss} forms of level $5$ into two sextuplets $\bm{\widehat{6}}$ of $A'_5$ as follow,
\begin{footnotesize}
\begin{eqnarray}
\nonumber Y_{\bm{\widehat{6}}I,1}^{(-3)} &=& \dfrac{y^4}{4} - \dfrac{26949 }{75008 \pi^4}\left( \dfrac{\Gamma(4, 4 \pi y)}{ q}+ \dfrac{39751 \Gamma(4, 8 \pi y)}{35932 q^2}+ \dfrac{54544 \Gamma(4, 12 \pi y)}{55971 q^3}+ \dfrac{2524981 \Gamma(4, 16 \pi y)}{2299648 q^4} + \cdots  \right)  \\
\nonumber &&\hskip-0.4in + c_{\bm{\widehat{6}}I,1}^{(-3)} - \dfrac{80403 }{37504 \pi^4}\left( q + \dfrac{29718 q^2}{26801}+ \dfrac{2134034 q^3}{2170881}+ \dfrac{7611307 q^4}{6861056}+ \dfrac{16796801 q^5}{16750625}+ \dfrac{18884971 q^6}{17367048} + \cdots \right) \,,  \\
\nonumber Y_{\bm{\widehat{6}}I,2}^{(-3)} &=& \dfrac{46293 }{75008 \pi^4}\left( \dfrac{\Gamma(4, 4 \pi y)}{ q}+ \dfrac{1141 \Gamma(4, 8 \pi y)}{1187 q^2}+ \dfrac{3779162 \Gamma(4, 12 \pi y)}{3749733 q^3}+ \dfrac{3783867 \Gamma(4, 16 \pi y)}{3950336 q^4} + \cdots  \right)    \\
\nonumber &&\hskip-0.4in + c_{\bm{\widehat{6}}I,2}^{(-3)} + \dfrac{138621 }{37504 \pi^4}\left( q + \dfrac{51019 q^2}{52808}+ \dfrac{1255696 q^3}{1247589}+ \dfrac{11375899 q^4}{11828992}+ \dfrac{28906207 q^5}{28879375}+ \dfrac{29089397 q^6}{29942136} + \cdots \right) \,,  \\
\nonumber Y_{\bm{\widehat{6}}I,3}^{(-3)} &=&  \dfrac{99432789375 q^{1/5}}{83961254912}\left( \dfrac{562760251 \
\Gamma(4, 16 \pi y/5)}{3393972544 \pi^4 q}+ \dfrac{238242075668 \Gamma(4, 36 \pi y/5)}{1043805649743 \pi^4 q^2}+ \dfrac{21176407993 \Gamma(4, 56 \pi y/5)}{127327001221 \pi^4 q^3} + \cdots  \right)  \\
\nonumber &&\hskip-0.4in + \dfrac{58125 q^{1/5}}{37504 \pi^4}\left( 1 + \dfrac{16301 q}{20088}+ \dfrac{14642 q^2}{14641}+ \dfrac{1546351 q^3}{2031616}+ \dfrac{6300382 q^4}{6028911}+ \dfrac{5413261 q^5}{7083128}+ \dfrac{923522 q^6}{923521} + \cdots \right) \,, \\
\nonumber Y_{\bm{\widehat{6}}I,4}^{(-3)} &=& \dfrac{3773125 q^{2/5}}{3037824 \sqrt{2} \pi^4}\left( \dfrac{\Gamma(4, 12 \pi y/5)}{ q}+ \dfrac{23567355 \Gamma(4, 32 \pi y/5)}{24727552 q^2}+ \dfrac{171198117 \Gamma(4, 52 \pi y/5)}{172422757 q^3}+ \dfrac{3773341 \Gamma(4, 72 \pi y/5)}{3911976 q^4} + \cdots  \right)   \\
\nonumber && \hskip-0.4in + \dfrac{2139375 q^{2/5}}{300032 \sqrt{2} \pi^4}\left( 1 + \dfrac{2842096 q}{2739541}+ \dfrac{1484891 q^2}{1478736}+ \dfrac{98888176 q^3}{95297461}+ \dfrac{14642 q^4}{14641}+ \dfrac{633643520 q^5}{606374181}+ \dfrac{65281 q^6}{65536} + \cdots \right)\,,   \\
\nonumber Y_{\bm{\widehat{6}}I,5}^{(-3)} &=& \dfrac{238125 q^{3/5}}{300032 \sqrt{2} \pi^4}\left( \dfrac{\Gamma(4, 8 \pi y/5)}{ q}+ \dfrac{275512 \Gamma(4, 28 \pi y/5)}{304927 q^2}+ \dfrac{965887 \Gamma(4, 48 \pi y/5)}{987552 q^3}+ \dfrac{9577272 \Gamma(4, 68 \pi y/5)}{10607167 q^4} + \cdots  \right)   \\
\nonumber &&\hskip-0.4in + \dfrac{2130625 q^{3/5}}{506304 \sqrt{2} \pi^4}\left( 1 + \dfrac{7869555 q}{6981632}+ \dfrac{99471969 q^2}{97364449}+ \dfrac{1635433 q^3}{1472688}+ \dfrac{974680209 q^4}{953977969}+ \dfrac{2364222411 q^5}{2095362304}+ \dfrac{14642 q^6}{14641} + \cdots \right) \,,  \\
\nonumber Y_{\bm{\widehat{6}}I,6}^{(-3)} &=& \dfrac{73125 q^{4/5}}{75008 \pi^4}\left( \dfrac{\Gamma(4, 4 \pi y/5)}{ q}+ \dfrac{2957 \Gamma(4, 24 \pi y/5)}{2916 q^2}+ \dfrac{14642 \Gamma(4, 44 \pi y/5)}{14641 q^3}+ \dfrac{2588119 \Gamma(4, 64 \pi y/5)}{2555904 q^4} + \cdots  \right)   \\
&&\hskip-0.4in + \dfrac{56870625 q^{4/5}}{9601024 \pi^4}\left( 1 + \dfrac{3108608 q}{3158757}+ \dfrac{73111712 q^2}{72824731}+ \dfrac{3903344640 q^3}{3952766251}+ \dfrac{3267485 q^4}{3275748}+ \dfrac{3026350080 q^5}{3064648573} + \cdots \right)\,,
\end{eqnarray}
\end{footnotesize}
and
\begin{footnotesize}
\begin{eqnarray}
\nonumber Y_{\bm{\widehat{6}}II,1}^{(-3)} &=& \dfrac{46207 }{75008 \pi^4}\left( \dfrac{\Gamma(4, 4 \pi y)}{ q}+ \dfrac{51019 \Gamma(4, 8 \pi y)}{52808 q^2}+ \dfrac{1255696 \Gamma(4, 12 \pi y)}{1247589 q^3}+ \dfrac{11375899 \Gamma(4, 16 \pi y)}{11828992 q^4} + \cdots  \right)   \\
\nonumber &&\hskip-0.4in + c_{\bm{\widehat{6}}II,1}^{(-3)} + \dfrac{138879 }{37504 \pi^4}\left( q + \dfrac{1141 q^2}{1187}+ \dfrac{3779162 q^3}{3749733}+ \dfrac{3783867 q^4}{3950336}+ \dfrac{741187 q^5}{741875}+ \dfrac{1121153 q^6}{1153764} + \cdots \right)\,,  \\
\nonumber Y_{\bm{\widehat{6}}II,2}^{(-3)} &=& \dfrac{y^4}{4} + \dfrac{26801 }{75008 \pi^4}\left( \dfrac{\Gamma(4, 4 \pi y)}{ q}+ \dfrac{29718 \Gamma(4, 8 \pi y)}{26801 q^2}+ \dfrac{2134034 \Gamma(4, 12 \pi y)}{2170881 q^3}+ \dfrac{7611307 \Gamma(4, 16 \pi y)}{6861056 q^4} + \cdots  \right)  \\
\nonumber &&\hskip-0.4in + c_{\bm{\widehat{6}}II,2}^{(-3)} + \dfrac{80847 }{37504 \pi^4}\left( q + \dfrac{39751 q^2}{35932}+ \dfrac{54544 q^3}{55971}+ \dfrac{2524981 q^4}{2299648}+ \dfrac{430691 q^5}{431875}+ \dfrac{729929 q^6}{671652} + \cdots \right)  \\
\nonumber Y_{\bm{\widehat{6}}II,3}^{(-3)} &=& \dfrac{18956875 q^{1/5}}{19202048 \pi^4}\left( \dfrac{\Gamma(4, 16 \pi y/5)}{ q}+ \dfrac{3108608 \Gamma(4, 36 \pi y/5)}{3158757 q^2}+ \dfrac{73111712 \Gamma(4, 56 \pi y/5)}{72824731 q^3}+ \dfrac{3903344640 \Gamma(4, 76 \pi y/5)}{3952766251 q^4} + \cdots  \right)   \\
\nonumber &&\hskip-0.4in + \dfrac{219375 q^{1/5}}{37504 \pi^4}\left( 1 + \dfrac{2957 q}{2916}+ \dfrac{14642 q^2}{14641}+ \dfrac{2588119 q^3}{2555904}+ \dfrac{22682474 q^4}{22754277}+ \dfrac{4529267 q^5}{4455516} + \cdots \right) \,, \\
\nonumber Y_{\bm{\widehat{6}}II,4}^{(-3)} &=& \dfrac{2130625 q^{2/5}}{3037824 \sqrt{2} \pi^4}\left( \dfrac{\Gamma(4, 12 \pi y/5)}{ q}+ \dfrac{7869555 \Gamma(4, 32 \pi y/5)}{6981632 q^2}+ \dfrac{99471969 \Gamma(4, 52 \pi y/5)}{97364449 q^3}+ \dfrac{1635433 \Gamma(4, 72 \pi y/5)}{1472688 q^4} + \cdots  \right)   \\
\nonumber &&\hskip-0.4in + \dfrac{714375 q^{2/5}}{150016 \sqrt{2} \pi^4}\left( 1 + \dfrac{275512 q}{304927}+ \dfrac{965887 q^2}{987552}+ \dfrac{9577272 q^3}{10607167}+ \dfrac{14642 q^4}{14641}+ \dfrac{178904320 q^5}{202479021} + \cdots \right)\,,  \\
\nonumber Y_{\bm{\widehat{6}}II,5}^{(-3)} &=& - \dfrac{713125 q^{3/5}}{600064 \sqrt{2} \pi^4}\left( \dfrac{\Gamma(4, 8 \pi y/5)}{ q}+ \dfrac{2842096 \Gamma(4, 28 \pi y/5)}{2739541 q^2}+ \dfrac{1484891 \Gamma(4, 48 \pi y/5)}{1478736 q^3}+ \dfrac{98888176 \Gamma(4, 68 \pi y/5)}{95297461 q^4} + \cdots  \right)   \\
\nonumber &&\hskip-0.4in - \dfrac{3773125 q^{3/5}}{506304 \sqrt{2} \pi^4}\left( 1 + \dfrac{23567355 q}{24727552}+ \dfrac{171198117 q^2}{172422757}+ \dfrac{3773341 q^3}{3911976}+ \dfrac{1677370437 q^4}{1689400117}+ \dfrac{3535069473 q^5}{3710678272} + \cdots \right) \,, \\
\nonumber Y_{\bm{\widehat{6}}II,6}^{(-1)} &=& - \dfrac{19375 q^{4/5}}{75008 \pi^4}\left( \dfrac{\Gamma(4, 4 \pi y/5)}{ q}+ \dfrac{16301 \Gamma(4, 24 \pi y/5)}{20088 q^2}+ \dfrac{14642 \Gamma(4, 44 \pi y/5)}{14641 q^3}+ \dfrac{1546351 \Gamma(4, 64 \pi y/5)}{2031616 q^4} + \cdots  \right)   \\
&&\hskip-0.4in - \dfrac{11311875 q^{4/5}}{9601024 \pi^4}\left( 1 + \dfrac{54486272 q}{39582513}+ \dfrac{4843072 q^2}{4828411}+ \dfrac{344739840 q^3}{262075531}+ \dfrac{1385585 q^4}{1303128}+ \dfrac{1870991360 q^5}{1422342091} + \cdots \right)\,,
\end{eqnarray}
\end{footnotesize}
where the constant terms are
\begin{small}
\begin{eqnarray}
\nonumber c_{\bm{\widehat{6}}I,1}^{(-3)} &=& - c_{\bm{\widehat{6}}II,2}^{(-3)}
= -\dfrac{\pi}{64}\left( \sqrt{\dfrac{1}{5\sqrt{5}\phi}} + i \sqrt{\dfrac{\phi}{5\sqrt{5}}} \right) \dfrac{L(4,\chi_4)}{L(5,\chi_4)} - \dfrac{\pi}{64} \left( \sqrt{\dfrac{1}{5\sqrt{5}\phi}} - i \sqrt{\dfrac{\phi}{5\sqrt{5}}} \right) \dfrac{L(4,\chi_2)}{L(5,\chi_2)} \,,  \\
c_{\bm{\widehat{6}}I,2}^{(-3)} &=& c_{\bm{\widehat{6}}II,1}^{(-3)}
= \dfrac{\pi}{64}\left( \sqrt{\dfrac{\phi}{5\sqrt{5}}} - i \sqrt{\dfrac{1}{5\sqrt{5}\phi}} \right) \dfrac{L(4,\chi_4)}{L(5,\chi_4)} + \dfrac{\pi}{64} \left( \sqrt{\dfrac{\phi}{5\sqrt{5}}} + i \sqrt{\dfrac{1}{5\sqrt{5}\phi}} \right) \dfrac{L(4,\chi_2)}{L(5,\chi_2)}\,.
\end{eqnarray}
\end{small}

\item{$k_Y = -2$}

The weight $k_Y=-2$ polyharmonic Maa{\ss} forms of level $5$ can be arranged into a singlet $\bm{1}$, two triplets $\bm{3}$, $\bm{3'}$ and a quintuplet $\bm{5}$ of $A'_5$ with
\begin{small}
\begin{eqnarray}
\nonumber Y_{\bm{1}}^{(-2)}&=& \dfrac{y^3}{3} - \dfrac{15\Gamma(3,4\pi y)}{4\pi^3 q} - \dfrac{135\Gamma(3,8\pi y)}{32\pi^3 q^2} - \dfrac{35\Gamma(3,12\pi y)}{9\pi^3 q^3} + \cdots \\
\nonumber&&\hskip-0.4in -\dfrac{\pi}{12}\dfrac{\zeta(3)}{\zeta(4)} - \dfrac{15 q}{2\pi^3} - \dfrac{135 q^2}{16\pi^3} - \dfrac{70 q^3}{q\pi^3} - \dfrac{1095 q^4}{128\pi^3} - \dfrac{189 q^5}{25\pi^3} - \dfrac{35 q^6}{4\pi^3} + \cdots \,, \\
\nonumber  Y_{\bm{3},1}^{(-2)} &=& \dfrac{y^3}{3} - \dfrac{63 }{32 \pi^3}\left( \dfrac{\Gamma(3, 4 \pi y)}{ q}+ \dfrac{31 \Gamma(3, 8 \pi y)}{36 q^2}+ \dfrac{1612 \Gamma(3, 12 \pi y)}{1701 q^3}+ \dfrac{57 \Gamma(3, 16 \pi y)}{64 q^4} + \cdots  \right) \\
\nonumber &&\hskip-0.4in - \dfrac{\pi}{12\sqrt{5}} \dfrac{L(3,\chi_3)}{L(4,\chi_3)} - \dfrac{63 }{16 \pi^3}\left( q + \dfrac{31 q^2}{36}+ \dfrac{1612 q^3}{1701}+ \dfrac{57 q^4}{64}+ \dfrac{7813 q^5}{7875} + \cdots \right) \,, \\
\nonumber  Y_{\bm{3},2}^{(-2)} &=& \dfrac{7125 q^{1/5}}{2048 \sqrt{2} \pi^3}\left( \dfrac{\Gamma(3, 16 \pi y/5)}{ q}+ \dfrac{2368 \Gamma(3, 36 \pi y/5)}{2187 q^2}+ \dfrac{48 \Gamma(3, 56 \pi y/5)}{49 q^3}+ \dfrac{439040 \Gamma(3, 76 \pi y/5)}{390963 q^4} + \cdots  \right) \\
\nonumber&&\hskip-0.4in + \dfrac{125 q^{1/5}}{16 \sqrt{2} \pi^3}\left( 1 + \dfrac{91 q}{108}+ \dfrac{1332 q^2}{1331}+ \dfrac{3641 q^3}{4096}+ \dfrac{988 q^4}{1029}+ \dfrac{3843 q^5}{4394} + \cdots \right) \,, \\
\nonumber  Y_{\bm{3},3}^{(-2)} &=& \dfrac{125 q^{4/5}}{32 \sqrt{2} \pi^3}\left( \dfrac{\Gamma(3, 4 \pi y/5)}{ q}+ \dfrac{91 \Gamma(3, 24 \pi y/5)}{108 q^2}+ \dfrac{1332 \Gamma(3, 44 \pi y/5)}{1331 q^3}+ \dfrac{3641 \Gamma(3, 64 \pi y/5)}{4096 q^4} + \cdots  \right) \\
\nonumber&&\hskip-0.4in + \dfrac{7125 q^{4/5}}{1024 \sqrt{2} \pi^3}\left( 1 + \dfrac{2368 q}{2187}+ \dfrac{48 q^2}{49}+ \dfrac{439040 q^3}{390963}+ \dfrac{5915 q^4}{6156}+ \dfrac{520320 q^5}{463391} + \cdots \right) \,, \\
\nonumber  Y_{\bm{3}',1}^{(-2)} &=& \dfrac{y^3}{3} + \dfrac{31 }{16 \pi^3}\left( \dfrac{\Gamma(3, 4 \pi y)}{ q}+ \dfrac{441 \Gamma(3, 8 \pi y)}{496 q^2}+ \dfrac{91 \Gamma(3, 12 \pi y)}{93 q^3}+ \dfrac{57 \Gamma(3, 16 \pi y)}{64 q^4} + \cdots  \right) \\
\nonumber &&\hskip-0.4in + \dfrac{\pi}{12\sqrt{5}} \dfrac{L(3,\chi_3)}{L(4,\chi_3)} + \dfrac{31 }{8 \pi^3}\left( q + \dfrac{441 q^2}{496}+ \dfrac{91 q^3}{93}+ \dfrac{57 q^4}{64}+ \dfrac{126 q^5}{125} + \cdots \right)  \,, \\
\nonumber  Y_{\bm{3}',2}^{(-2)} &=& - \dfrac{1625 q^{2/5}}{432 \sqrt{2} \pi^3}\left( \dfrac{\Gamma(3, 12 \pi y/5)}{ q}+ \dfrac{945 \Gamma(3, 32 \pi y/5)}{1024 q^2}+ \dfrac{29646 \Gamma(3, 52 \pi y/5)}{28561 q^3}+ \dfrac{4921 \Gamma(3, 72 \pi y/5)}{5616 q^4} + \cdots  \right) \\
\nonumber &&\hskip-0.4in - \dfrac{875 q^{2/5}}{128 \sqrt{2} \pi^3}\left( 1 + \dfrac{2736 q}{2401}+ \dfrac{247 q^2}{252}+ \dfrac{39296 q^3}{34391}+ \dfrac{1332 q^4}{1331}+ \dfrac{151840 q^5}{137781} + \cdots \right) \,, \\
\nonumber  Y_{\bm{3}',3}^{(-2)} &=& - \dfrac{875 q^{3/5}}{256 \sqrt{2} \pi^3}\left( \dfrac{\Gamma(3, 8 \pi y/5)}{ q}+ \dfrac{2736 \Gamma(3, 28 \pi y/5)}{2401 q^2}+ \dfrac{247 \Gamma(3, 48 \pi y/5)}{252 q^3}+ \dfrac{39296 \Gamma(3, 68 \pi y/5)}{34391 q^4} + \cdots  \right) \\
\nonumber &&\hskip-0.4in - \dfrac{1625 q^{3/5}}{216 \sqrt{2} \pi^3}\left( 1 + \dfrac{945 q}{1024}+ \dfrac{29646 q^2}{28561}+ \dfrac{4921 q^3}{5616}+ \dfrac{164241 q^4}{158171}+ \dfrac{263169 q^5}{285376} + \cdots \right) \,, \\
\nonumber  Y_{\bm{5},1}^{(-2)} &=& \dfrac{y^3}{3} + \dfrac{315 }{416 \pi^3}\left( \dfrac{\Gamma(3, 4 \pi y)}{ q}+ \dfrac{9 \Gamma(3, 8 \pi y)}{8 q^2}+ \dfrac{28 \Gamma(3, 12 \pi y)}{27 q^3}+ \dfrac{73 \Gamma(3, 16 \pi y)}{64 q^4} + \cdots  \right) \\
\nonumber&&\hskip-0.4in + \dfrac{5\pi}{312} \dfrac{\zeta(3)}{\zeta(4)} + \dfrac{315 }{208 \pi^3}\left( q + \dfrac{9 q^2}{8}+ \dfrac{28 q^3}{27}+ \dfrac{73 q^4}{64}+ \dfrac{2521 q^5}{2625} + \cdots \right) \,, \\
\nonumber  Y_{\bm{5},2}^{(-2)} &=& - \dfrac{45625 \sqrt{3} q^{1/5}}{26624 \sqrt{2} \pi^3}\bigg( \dfrac{\Gamma(3, 16 \pi y/5)}{ q}+ \dfrac{48448 \Gamma(3, 36 \pi y/5)}{53217 q^2}+ \dfrac{24768 \Gamma(3, 56 \pi y/5)}{25039 q^3} \\
\nonumber &&\hskip-0.4in +\dfrac{439040 \Gamma(3, 76 \pi y/5)}{500707 q^4}+ \cdots\bigg)  - \dfrac{625 \sqrt{3} q^{1/5}}{208 \sqrt{2} \pi^3}\left( 1 + \dfrac{7 q}{6}+ \dfrac{1332 q^2}{1331}+ \dfrac{4681 q^3}{4096}+ \dfrac{1376 q^4}{1323}+ \dfrac{9891 q^5}{8788} + \cdots \right) \,,\\
\nonumber  Y_{\bm{5},3}^{(-2)} &=& - \dfrac{4375 q^{2/5}}{936 \sqrt{6} \pi^3}\left( \dfrac{\Gamma(3, 12 \pi y/5)}{ q}+ \dfrac{15795 \Gamma(3, 32 \pi y/5)}{14336 q^2}+ \dfrac{4239 \Gamma(3, 52 \pi y/5)}{4394 q^3}+ \dfrac{757 \Gamma(3, 72 \pi y/5)}{672 q^4} + \cdots  \right) \\
\nonumber&&\hskip-0.4in - \dfrac{5625 \sqrt{3} q^{2/5}}{1664 \sqrt{2} \pi^3}\left( 1 + \dfrac{2752 q}{3087}+ \dfrac{511 q^2}{486}+ \dfrac{4368 q^3}{4913}+ \dfrac{1332 q^4}{1331}+ \dfrac{163520 q^5}{177147} + \cdots \right) \,, \\
\nonumber  Y_{\bm{5},4}^{(-2)} &=& - \dfrac{5625 \sqrt{3} q^{3/5}}{3328 \sqrt{2} \pi^3}\left( \dfrac{\Gamma(3, 8 \pi y/5)}{ q}+ \dfrac{2752 \Gamma(3, 28 \pi y/5)}{3087 q^2}+ \dfrac{511 \Gamma(3, 48 \pi y/5)}{486 q^3}+ \dfrac{4368 \Gamma(3, 68 \pi y/5)}{4913 q^4} + \cdots  \right) \\
\nonumber &&\hskip-0.4in - \dfrac{4375 q^{3/5}}{468 \sqrt{6} \pi^3}\left( 1 + \dfrac{15795 q}{14336}+ \dfrac{4239 q^2}{4394}+ \dfrac{757 q^3}{672}+ \dfrac{82134 q^4}{85169}+ \dfrac{84753 q^5}{76832} + \cdots \right) \,, \\
\nonumber  Y_{\bm{5},5}^{(-2)} &=& - \dfrac{625 \sqrt{3} q^{4/5}}{416 \sqrt{2} \pi^3}\left( \dfrac{\Gamma(3, 4 \pi y/5)}{ q}+ \dfrac{7 \Gamma(3, 24 \pi y/5)}{6 q^2}+ \dfrac{1332 \Gamma(3, 44 \pi y/5)}{1331 q^3}+ \dfrac{4681 \Gamma(3, 64 \pi y/5)}{4096 q^4} + \cdots  \right) \\
&&\hskip-0.4in  - \dfrac{45625 \sqrt{3} q^{4/5}}{13312 \sqrt{2} \pi^3}\left( 1 + \dfrac{48448 q}{53217}+ \dfrac{24768 q^2}{25039}+ \dfrac{439040 q^3}{500707}+ \dfrac{455 q^4}{438}+ \dfrac{1560960 q^5}{1780397} + \cdots \right)\,.
\end{eqnarray}
\end{small}

\item{$k_Y = -1$}

The weight $k_Y=-1$ polyharmonic Maa{\ss} forms of level $5$ can be arranged into two linearly independent sextuplets $\bm{\widehat{6}}$ of $A'_5$ as follow,
\begin{eqnarray}
\nonumber Y_{\bm{\widehat{6}}I,1}^{(-1)} &=& \dfrac{y^2}{2} +  \dfrac{27 }{16 \pi^2}\left( \dfrac{\Gamma(2, 4 \pi y)}{ q}+ \dfrac{13 \Gamma(2, 8 \pi y)}{9 q^2}+ \dfrac{56 \Gamma(2, 12 \pi y)}{81 q^3}+ \dfrac{61 \Gamma(2, 16 \pi y)}{48 q^4} + \cdots  \right)  \\
\nonumber &&\hskip-0.4in + c_{\bm{\widehat{6}}I,1}^{(-1)} + \dfrac{23 }{16 \pi^2}\left( q + \dfrac{36 q^2}{23}+ \dfrac{182 q^3}{207}+ \dfrac{601 q^4}{368}+ \dfrac{623 q^5}{575}+ \dfrac{553 q^6}{414} + \cdots \right) \,,  \\
\nonumber Y_{\bm{\widehat{6}}I,2}^{(-1)} &=& - \dfrac{51 }{16 \pi^2}\left( \dfrac{\Gamma(2, 4 \pi y)}{ q}+ \dfrac{14 \Gamma(2, 8 \pi y)}{17 q^2}+ \dfrac{494 \Gamma(2, 12 \pi y)}{459 q^3}+ \dfrac{209 \Gamma(2, 16 \pi y)}{272 q^4} + \cdots  \right)   \\
\nonumber &&\hskip-0.4in + c_{\bm{\widehat{6}}I,2}^{(-1)} - \dfrac{49 }{16 \pi^2}\left( q + \dfrac{13 q^2}{14}+ \dfrac{152 q^3}{147}+ \dfrac{673 q^4}{784}+ \dfrac{1249 q^5}{1225}+ \dfrac{839 q^6}{882} + \cdots \right) \,,\\
\nonumber Y_{\bm{\widehat{6}}I,3}^{(-1)} &=& - \dfrac{75 q^{1/5}}{256 \pi^2}\left( \dfrac{\Gamma(2, 16 \pi y/5)}{ q}+ \dfrac{1712 \Gamma(2, 36 \pi y/5)}{243 q^2}+ \dfrac{48 \Gamma(2, 56 \pi y/5)}{49 q^3}+ \dfrac{1920 \Gamma(2, 76 \pi y/5)}{361 q^4} + \cdots  \right)   \\
\nonumber &&\hskip-0.4in - \dfrac{25 q^{1/5}}{16 \pi^2}\left( 1 + \dfrac{11 q}{18}+ \dfrac{122 q^2}{121}+ \dfrac{61 q^3}{256}+ \dfrac{562 q^4}{441}+ \dfrac{7 q^5}{26}+ \dfrac{962 q^6}{961} + \cdots \right) \,, \\
\nonumber Y_{\bm{\widehat{6}}I,4}^{(-1)} &=& - \dfrac{475 q^{2/5}}{72 \sqrt{2} \pi^2}\left( \dfrac{\Gamma(2, 12 \pi y/5)}{ q}+ \dfrac{945 \Gamma(2, 32 \pi y/5)}{1216 q^2}+ \dfrac{3051 \Gamma(2, 52 \pi y/5)}{3211 q^3}+ \dfrac{307 \Gamma(2, 72 \pi y/5)}{342 q^4} + \cdots  \right)   \\
\nonumber &&\hskip-0.4in - \dfrac{175 q^{2/5}}{32 \sqrt{2} \pi^2}\left( 1 + \dfrac{388 q}{343}+ \dfrac{257 q^2}{252}+ \dfrac{2308 q^3}{2023}+ \dfrac{122 q^4}{121}+ \dfrac{6080 q^5}{5103}+ \dfrac{241 q^6}{256} + \cdots \right) \,, \\
\nonumber Y_{\bm{\widehat{6}}I,5}^{(-1)} &=& - \dfrac{75 q^{3/5}}{16 \sqrt{2} \pi^2}\left( \dfrac{\Gamma(2, 8 \pi y/5)}{ q}+ \dfrac{34 \Gamma(2, 28 \pi y/5)}{49 q^2}+ \dfrac{181 \Gamma(2, 48 \pi y/5)}{216 q^3}+ \dfrac{194 \Gamma(2, 68 \pi y/5)}{289 q^4} + \cdots  \right)   \\
\nonumber &&\hskip-0.4in - \dfrac{175 q^{3/5}}{72 \sqrt{2} \pi^2}\left( 1 + \dfrac{405 q}{224}+ \dfrac{1503 q^2}{1183}+ \dfrac{139 q^3}{84}+ \dfrac{4743 q^4}{3703}+ \dfrac{10377 q^5}{5488}+ \dfrac{122 q^6}{121} + \cdots \right) \,, \\
\nonumber Y_{\bm{\widehat{6}}I,6}^{(-1)} &=& - \dfrac{75 q^{4/5}}{16 \pi^2}\left( \dfrac{\Gamma(2, 4 \pi y/5)}{ q}+ \dfrac{29 \Gamma(2, 24 \pi y/5)}{27 q^2}+ \dfrac{122 \Gamma(2, 44 \pi y/5)}{121 q^3}+ \dfrac{261 \Gamma(2, 64 \pi y/5)}{256 q^4} + \cdots  \right)   \\
\nonumber &&\hskip-0.4in  - \dfrac{1225 q^{4/5}}{256 \pi^2}\left( 1 + \dfrac{176 q}{189}+ \dfrac{2552 q^2}{2401}+ \dfrac{17280 q^3}{17689}+ \dfrac{145 q^4}{147}+ \dfrac{5760 q^5}{5887}+ \dfrac{15032 q^6}{14161} + \cdots \right)\,, \\
\nonumber Y_{\bm{\widehat{6}}II,1}^{(-1)} &=& - \dfrac{49 }{16 \pi^2}\left( \dfrac{\Gamma(2, 4 \pi y)}{ q}+ \dfrac{13 \Gamma(2, 8 \pi y)}{14 q^2}+ \dfrac{152 \Gamma(2, 12 \pi y)}{147 q^3}+ \dfrac{673 \Gamma(2, 16 \pi y)}{784 q^4} + \cdots  \right)   \\
\nonumber &&\hskip-0.4in + c_{\bm{\widehat{6}}II,1}^{(-1)} - \dfrac{51 }{16 \pi^2}\left(+ \dfrac{ q}{1}+ \dfrac{14 q^2}{17}+ \dfrac{494 q^3}{459}+ \dfrac{209 q^4}{272}+ \dfrac{417 q^5}{425}+ \dfrac{443 q^6}{459} + \cdots \right)\,,  \\
\nonumber Y_{\bm{\widehat{6}}II,2}^{(-1)} &=& \dfrac{y^2}{2} - \dfrac{23 }{16 \pi^2}\left( \dfrac{\Gamma(2, 4 \pi y)}{ q}+ \dfrac{36 \Gamma(2, 8 \pi y)}{23 q^2}+ \dfrac{182 \Gamma(2, 12 \pi y)}{207 q^3}+ \dfrac{601 \Gamma(2, 16 \pi y)}{368 q^4} + \cdots  \right)  \\
\nonumber &&\hskip-0.4in + c_{\bm{\widehat{6}}II,2}^{(-1)} - \dfrac{27 }{16 \pi^2}\left(+ \dfrac{ q}{1}+ \dfrac{13 q^2}{9}+ \dfrac{56 q^3}{81}+ \dfrac{61 q^4}{48}+ \dfrac{209 q^5}{225}+ \dfrac{311 q^6}{243} + \cdots \right) \,, \\
\nonumber Y_{\bm{\widehat{6}}II,3}^{(-1)} &=& - \dfrac{1225 q^{1/5}}{256 \pi^2}\bigg( \dfrac{\Gamma(2, 16 \pi y/5)}{ q}+ \dfrac{176 \Gamma(2, 36 \pi y/5)}{189 q^2}+ \dfrac{2552 \Gamma(2, 56 \pi y/5)}{2401 q^3}   \\
\nonumber &&\hskip-0.4in+ \dfrac{17280 \Gamma(2, 76 \pi y/5)}{17689 q^4} + \cdots  \bigg) - \dfrac{75 q^{1/5}}{16 \pi^2}\left( 1 + \dfrac{29 q}{27}+ \dfrac{122 q^2}{121}+ \dfrac{261 q^3}{256}+ \dfrac{1286 q^4}{1323}+ \dfrac{183 q^5}{169} + \cdots \right)\,,  \\
\nonumber Y_{\bm{\widehat{6}}II,4}^{(-1)} &=& - \dfrac{175 q^{2/5}}{72 \sqrt{2} \pi^2}\left( \dfrac{\Gamma(2, 12 \pi y/5)}{ q}+ \dfrac{405 \Gamma(2, 32 \pi y/5)}{224 q^2}+ \dfrac{1503 \Gamma(2, 52 \pi y/5)}{1183 q^3}+ \dfrac{139 \Gamma(2, 72 \pi y/5)}{84 q^4} + \cdots  \right)   \\
\nonumber &&\hskip-0.4in - \dfrac{75 q^{2/5}}{16 \sqrt{2} \pi^2}\left( 1 + \dfrac{34 q}{49}+ \dfrac{181 q^2}{216}+ \dfrac{194 q^3}{289}+ \dfrac{122 q^4}{121}+ \dfrac{1120 q^5}{2187} + \cdots \right) \,, \\
\nonumber Y_{\bm{\widehat{6}}II,5}^{(-1)} &=& \dfrac{175 q^{3/5}}{32 \sqrt{2} \pi^2}\left( \dfrac{\Gamma(2, 8 \pi y/5)}{ q}+ \dfrac{388 \Gamma(2, 28 \pi y/5)}{343 q^2}+ \dfrac{257 \Gamma(2, 48 \pi y/5)}{252 q^3}+ \dfrac{2308 \Gamma(2, 68 \pi y/5)}{2023 q^4} + \cdots  \right)   \\
\nonumber &&\hskip-0.4in + \dfrac{475 q^{3/5}}{72 \sqrt{2} \pi^2}\left( 1 + \dfrac{945 q}{1216}+ \dfrac{3051 q^2}{3211}+ \dfrac{307 q^3}{342}+ \dfrac{9531 q^4}{10051}+ \dfrac{11259 q^5}{14896} + \cdots \right)\,,  \\
\nonumber Y_{\bm{\widehat{6}}II,6}^{(-1)} &=& \dfrac{25 q^{4/5}}{16 \pi^2}\left( \dfrac{\Gamma(2, 4 \pi y/5)}{ q}+ \dfrac{11 \Gamma(2, 24 \pi y/5)}{18 q^2}+ \dfrac{122 \Gamma(2, 44 \pi y/5)}{121 q^3}+ \dfrac{61 \Gamma(2, 64 \pi y/5)}{256 q^4} + \cdots  \right)   \\
&&\hskip-0.4in + \dfrac{75 q^{4/5}}{256 \pi^2}\left( 1 + \dfrac{1712 q}{243}+ \dfrac{48 q^2}{49}+ \dfrac{1920 q^3}{361}+ \dfrac{55 q^4}{18}+ \dfrac{4480 q^5}{841}+ \dfrac{368 q^6}{289} + \cdots \right)\,,
\end{eqnarray}
where the constant terms are
\begin{small}
\begin{eqnarray}
\nonumber c_{\bm{\widehat{6}}I,1}^{(-1)} &=& - c_{\bm{\widehat{6}}II,2}^{(-1)}
= \dfrac{\pi}{8}\left( \sqrt{\dfrac{1}{5\sqrt{5}\phi}} + i \sqrt{\dfrac{\phi}{5\sqrt{5}}} \right) \dfrac{L(2,\chi_4)}{L(3,\chi_4)} + \dfrac{\pi}{8} \left( \sqrt{\dfrac{1}{5\sqrt{5}\phi}} - i \sqrt{\dfrac{\phi}{5\sqrt{5}}} \right) \dfrac{L(2,\chi_2)}{L(3,\chi_2)}\,,  \\
c_{\bm{\widehat{6}}I,2}^{(-1)} &=& c_{\bm{\widehat{6}}II,1}^{(-1)}
= - \dfrac{\pi}{8}\left( \sqrt{\dfrac{\phi}{5\sqrt{5}}} - i \sqrt{\dfrac{1}{5\sqrt{5}\phi}} \right) \dfrac{L(2,\chi_4)}{L(3,\chi_4)} - \dfrac{\pi}{8} \left( \sqrt{\dfrac{\phi}{5\sqrt{5}}} + i \sqrt{\dfrac{1}{5\sqrt{5}\phi}} \right) \dfrac{L(2,\chi_2)}{L(3,\chi_2)}\,.~~~~~~
\end{eqnarray}
\end{small}

\item{$k_Y = 0$}

At weight $k_Y=0$, the non-holomorphic polyharmonic Maa{\ss} forms of level $5$ can be arranged two triplets $\bm{3}$, $\bm{3'}$ and a quintuplet $\bm{5}$ of $A'_5$. The holomorphic singlet is a constant which can be taken to be 1 without loss of generality.
\begin{eqnarray}
\nonumber Y_{\bm{1}}^{(0)} &=& 1 \,, \\
\nonumber Y_{\bm{3},1}^{(0)} &=& y + \dfrac{15 }{2 \pi}\left( \dfrac{e^{-4 \pi y}}{ q}+ \dfrac{e^{-8 \pi \
y}}{3 q^2}+ \dfrac{4\, e^{-12 \pi y}}{9 q^3}+ \dfrac{3\, e^{-16 \pi y}}{4 q^4} + \cdots  \right)  \\
\nonumber &&\hskip-0.4in + \dfrac{5\sqrt{5}\log \phi}{2\pi} + \dfrac{15 }{2 \pi}\left( q + \dfrac{ q^2}{3}+ \dfrac{4 q^3}{9}+ \dfrac{3 q^4}{4}+ \dfrac{13 q^5}{15} + \cdots \right) \,,   \\
\nonumber Y_{\bm{3},2}^{(0)} &=& - \dfrac{75 q^{1/5}}{8 \sqrt{2} \pi}\left( \dfrac{e^{-16 \pi y/5}}{ q}+ \dfrac{28\, e^{-36 \pi y/5}}{27 q^2}+ \dfrac{4\, e^{-56 \pi y/5}}{7 q^3}+ \dfrac{80\, e^{-76 \pi y/5}}{57 q^4} + \cdots  \right)   \\
\nonumber &&\hskip-0.4in - \dfrac{25 q^{1/5}}{2 \sqrt{2} \pi}\left( 1 + \dfrac{ q}{3}+ \dfrac{12 q^2}{11}+ \dfrac{11 q^3}{16}+ \dfrac{4 q^4}{7}+ \dfrac{6 q^5}{13} + \cdots \right) \,,   \\
\nonumber Y_{\bm{3},3}^{(0)} &=& - \dfrac{25 q^{4/5}}{2 \sqrt{2} \pi}\left( \dfrac{e^{-4 \pi y/5}}{ q}+ \dfrac{e^{-24 \pi y/5}}{3 q^2}+ \dfrac{12\, e^{-44 \pi y/5}}{11 q^3}+ \dfrac{11\, e^{-64 \pi y/5}}{16 q^4} + \cdots  \right)  \\
\nonumber &&\hskip-0.4in - \dfrac{75 q^{4/5}}{8 \sqrt{2} \pi}\left( 1 + \dfrac{28 q}{27}+ \dfrac{4 q^2}{7}+ \dfrac{80 q^3}{57}+ \dfrac{5 q^4}{9}+ \dfrac{40 q^5}{29} + \cdots \right)\,,  \\
\nonumber Y_{\bm{3}',1}^{(0)} &=& y - \dfrac{5 }{\pi}\left( \dfrac{e^{-4 \pi y}}{ q}+ \dfrac{3\, e^{-8 \pi y}}{4 q^2}+ \dfrac{e^{-12 \pi y}}{ q^3}+ \dfrac{3\, e^{-16 \pi y}}{4 q^4} + \cdots  \right)   \\
\nonumber &&\hskip-0.4in - \dfrac{5\sqrt{5}\log \phi}{2\pi} - \dfrac{5 }{\pi}\left( q + \dfrac{3 q^2}{4}+ \dfrac{q^3}{1}+ \dfrac{3 q^4}{4}+ \dfrac{6 q^5}{5} + \cdots \right)  \,, \\
\nonumber Y_{\bm{3}',2}^{(0)} &=& \dfrac{25 q^{2/5}}{3 \sqrt{2} \pi}\left( \dfrac{e^{-12 \pi y/5}}{ q}+ \dfrac{15\, e^{-32 \pi y/5}}{16 q^2}+ \dfrac{18\, e^{-52 \pi y/5}}{13 q^3}+ \dfrac{7\, e^{-72 \pi y/5}}{12 q^4} + \cdots  \right)   \\
\nonumber &&\hskip-0.4in + \dfrac{25 q^{2/5}}{4 \sqrt{2} \pi}\left( 1 + \dfrac{12 q}{7}+ \dfrac{ q^2}{1}+ \dfrac{32 q^3}{17}+ \dfrac{12 q^4}{11}+ \dfrac{40 q^5}{27} + \cdots \right) \,, \\
\nonumber Y_{\bm{3}',3}^{(0)} &=& \dfrac{25 q^{3/5}}{4 \sqrt{2} \pi}\left( \dfrac{e^{-8 \pi y/5}}{ q}+ \dfrac{12\, e^{-28 \pi y/5}}{7 q^2}+ \dfrac{e^{-48 \pi y/5}}{ q^3}+ \dfrac{32\, e^{-68 \pi y/5}}{17 q^4} + \cdots  \right)  \\
\nonumber &&\hskip-0.4in + \dfrac{25 q^{3/5}}{3 \sqrt{2} \pi}\left( 1 + \dfrac{15 q}{16}+ \dfrac{18 q^2}{13}+ \dfrac{7 q^3}{12}+ \dfrac{33 q^4}{23}+ \dfrac{27 q^5}{28} + \cdots \right)\,,  \\
\nonumber Y_{\bm{5},1}^{(0)} &=& y  - \dfrac{3 }{2 \pi}\left( \dfrac{e^{-4 \pi y}}{ q}+ \dfrac{3\, e^{-8 \pi y}}{2 q^2}+ \dfrac{4\, e^{-12 \pi y}}{3 q^3}+ \dfrac{7\, e^{-16 \pi y}}{4 q^4} + \cdots  \right) \\
\nonumber &&\hskip-0.4in - \dfrac{5\log 5}{4\pi} - \dfrac{3 }{2 \pi}\left( q + \dfrac{3 q^2}{2}+ \dfrac{4 q^3}{3}+ \dfrac{7 q^4}{4}+ \dfrac{ q^5}{5} + \cdots \right)\,,   \\
\nonumber Y_{\bm{5},2}^{(0)} &=& \dfrac{35 \sqrt{3} q^{1/5}}{8 \sqrt{2} \pi}\left( \dfrac{e^{-16 \pi y/5}}{ q}+ \dfrac{52\, e^{-36 \pi y/5}}{63 q^2}+ \dfrac{48\, e^{-56 \pi y/5}}{49 q^3}+ \dfrac{80\, e^{-76 \pi y/5}}{133 q^4} + \cdots  \right)   \\
\nonumber &&\hskip-0.4in + \dfrac{5 \sqrt{3} q^{1/5}}{2 \sqrt{2} \pi}\left( 1 + 2 q + \dfrac{12 q^2}{11}+ \dfrac{31 q^3}{16}+ \dfrac{32 q^4}{21}+ \dfrac{21 q^5}{13} + \cdots \right) \,, \\
\nonumber Y_{\bm{5},3}^{(0)} &=& \dfrac{5 \sqrt{2} q^{2/5}}{\sqrt{3} \pi}\left( \dfrac{e^{-12 \pi y/5}}{ q}+ \dfrac{45\, e^{-32 \pi y/5}}{32 q^2}+ \dfrac{21\, e^{-52 \pi y/5}}{26 q^3}+ \dfrac{13\, e^{-72 \pi y/5}}{8 q^4} + \cdots  \right)  \\
\nonumber &&\hskip-0.4in + \dfrac{15 \sqrt{3} q^{2/5}}{4 \sqrt{2} \pi}\left( 1 + \dfrac{16 q}{21}+ \dfrac{14 q^2}{9}+ \dfrac{12 q^3}{17}+ \dfrac{12 q^4}{11}+ \dfrac{80 q^5}{81} + \cdots \right)\,,  \\
\nonumber Y_{\bm{5},4}^{(0)} &=& \dfrac{15 \sqrt{3} q^{3/5}}{4 \sqrt{2} \pi}\left( \dfrac{e^{-8 \pi y/5}}{ q}+ \dfrac{16\, e^{-28 \pi y/5}}{21 q^2}+ \dfrac{14\, e^{-48 \pi y/5}}{9 q^3}+ \dfrac{12\, e^{-68 \pi y/5}}{17 q^4} + \cdots  \right)  \\
\nonumber&&\hskip-0.4in + \dfrac{5 \sqrt{2} q^{3/5}}{\sqrt{3} \pi}\left( 1 + \dfrac{45 q}{32}+ \dfrac{21 q^2}{26}+ \dfrac{13 q^3}{8}+ \dfrac{18 q^4}{23}+ \dfrac{3 q^5}{2} + \cdots \right) \,, \\
\nonumber Y_{\bm{5},5}^{(0)} &=& \dfrac{5 \sqrt{3} q^{4/5}}{2 \sqrt{2} \pi}\left( \dfrac{e^{-4 \pi y/5}}{ q}+ \dfrac{2\, e^{-24 \pi y/5}}{ q^2}+ \dfrac{12\, e^{-44 \pi y/5}}{11 q^3}+ \dfrac{31\, e^{-64 \pi y/5}}{16 q^4} + \cdots  \right)  \\
&&\hskip-0.4in + \dfrac{35 \sqrt{3} q^{4/5}}{8 \sqrt{2} \pi}\left( 1 + \dfrac{52 q}{63}+ \dfrac{48 q^2}{49}+ \dfrac{80 q^3}{133}+ \dfrac{10 q^4}{7}+ \dfrac{120 q^5}{203} + \cdots \right)\,,
\end{eqnarray}

\item{$k_Y = 1$}

The weight one polyharmonic Maa{\ss} forms of level $N=5$ consist of two sextuplets $Y^{(1)}_{\bm{\widehat{6}}I}(\tau)$ and $Y^{(1)}_{\bm{\widehat{6}}II}(\tau)$, $Y^{(1)}_{\bm{\widehat{6}}I}(\tau)$ is non-holomorphic and its $q$-expansion is given by
\begin{eqnarray}
\nonumber Y^{(1)}_{\bm{\widehat{6}}I,1} &=& a_1 - 5 \left( q \log 5 + 2 q^2 \log 5 + 2 q^3 \log 3 + q^4 \log 80 + 2 q^5 \log 5 + 2 q^6 \log \dfrac{15}{2} + \cdots  \right)  \\
\nonumber &&\hskip-0.4in + 2 \log y - \dfrac{5 \Gamma(0,4\pi y)}{q} - \dfrac{10 \Gamma(0,8\pi y)}{q^2} - \dfrac{5 \Gamma(0,16\pi y)}{q^4} - \dfrac{5 \Gamma(0,20\pi y)}{q^5} + \cdots \,, \\
\nonumber Y^{(1)}_{\bm{\widehat{6}}I,2} &=& a_2 + 5 \left( q \log 5 + 2 q^2 \log 2  + 2 q^3 \log 5 + q^4 \log \dfrac{16}{5} + 2 q^5 \log 5 + 2 q^6 \log \dfrac{10}{3} + \cdots \right) \\
\nonumber &&\hskip-0.4in - \log y + \dfrac{5 \Gamma(0,4\pi y)}{q} + \dfrac{10 \Gamma(0,12\pi y)}{q^3} - \dfrac{5 \Gamma(0,16\pi y)}{q^4} + \dfrac{5\Gamma(0,20\pi y)}{q^5} + \cdots \,, \\
\nonumber Y^{(1)}_{\bm{\widehat{6}}I,3} &=& - 10 q^{1/5} \left( q \log \dfrac{3}{2} + 2 q^3 \log 2 - q^4 \log \dfrac{7}{3} + q^5 \log \dfrac{13}{2} + 2 q^7 \log \dfrac{3}{2} + q^9 \log \dfrac{23}{2} + \cdots \right) \\
\nonumber &&\hskip-0.4in - 5 q^{1/5} \left( \dfrac{\Gamma(0,16\pi y/5)}{q} - \dfrac{\Gamma(0,36\pi y/5)}{q^2} + \dfrac{2\Gamma(0,56\pi y/5)}{q^3} + \dfrac{2\Gamma(0,136\pi y/5)}{q^7} + \cdots \right) \,, \\
\nonumber Y^{(1)}_{\bm{\widehat{6}}I,4} &=& 5\sqrt{2} q^{2/5} \left( \log 2 + q \log        7 + q^2 \log \dfrac{16}{3} + q^3 \log 17 + 2 q^4 \log 2  + 4 q^5 \log 81 + q^6 \log 2 + \cdots  \right) \\
\nonumber &&\hskip-0.4in + 5\sqrt{2}q^{2/5} \left( \dfrac{\Gamma(0,12\pi y/5)}{q} + \dfrac{\Gamma(0,52\pi y/5)}{q^3} + \dfrac{\Gamma(0,72\pi y/5)}{q^4} + \dfrac{\Gamma(0,92\pi y/5)}{q^5} + \cdots \right)\,,  \\
\nonumber Y^{(1)}_{\bm{\widehat{6}}I,5} &=& 5\sqrt{2} q^{3/5} \left( \log 3 + 4 q \log 2 + q^2 \log 13 + q^3 \log \dfrac{81}{2} + q^4 \log 23 + q^5 \log 112 + 2 q^6 \log 3 + \cdots \right) \\
\nonumber &&\hskip-0.4in + 5\sqrt{2} q^{3/5} \left( \dfrac{\Gamma(0,8\pi y/5)}{q} + \dfrac{\Gamma(0,28\pi y/5)}{q^2} + \dfrac{\Gamma(0,48\pi y/5)}{q^3} + \dfrac{\Gamma(0,68\pi y/5)}{q^4} + \cdots \right) \,,  \\
\nonumber Y^{(1)}_{\bm{\widehat{6}}I,6} &=& 10 q^{4/5} \left( 2 \log 2 + 2 q \log 3 + q^2 \log 14 + q^3 \log 19 + 4 q^4 \log 2 + q^5 \log 29 + q^6 \log 34 + \cdots \right) \\
&&\hskip-0.4in + 5 q^{4/5} \left( \dfrac{\Gamma(0,4\pi y/5)}{q} + \dfrac{2 \Gamma(0,24\pi y/5)}{q^2} + \dfrac{2\Gamma(0,44\pi y/5)}{q^3} + \dfrac{\Gamma(0,64\pi y/5)}{q^4} + \cdots \right)\,,
\end{eqnarray}
where the constants $a_1$ and $a_2$ are
\begin{eqnarray}
\nonumber a_1 &=& 2 \gamma_E + 2\log 20\pi - 5 \left[\log \Gamma\left( \dfrac{1}{5} \right) + \log \Gamma\left( \dfrac{2}{5} \right) - \log \Gamma\left( \dfrac{3}{5} \right) - \log \Gamma\left( \dfrac{4}{5} \right) \right] \simeq 0.5831\,, \\
a_2 &=& - \gamma_E - \log 20 \pi + 5 \left[ \log \Gamma\left( \dfrac{1}{5} \right) - \log \Gamma\left( \dfrac{2}{5} \right) + \log \Gamma\left( \dfrac{3}{5} \right) - \log \Gamma\left( \dfrac{4}{5} \right) \right] \simeq 0.1501 \,.~~~~\quad~~
\end{eqnarray}
Another sextuplet $Y^{(1)}_{\bm{\widehat{6}}II}(\tau)$ can be taken to be holomorphic function with the following $q$-expansion,
\begin{eqnarray}
\nonumber && Y^{(1)}_{\bm{\widehat{6}}II,1}=1+5q+10q^3-5q^4+5q^5+10q^6+5q^9+\ldots\equiv Y_1(\tau)\,, \\
\nonumber && Y^{(1)}_{\bm{\widehat{6}}II,2}=2+5 q+10 q^2+5 q^4+5 q^5+10 q^6+10q^7-5q^9+\ldots\equiv Y_2(\tau)\,, \\
\nonumber && Y^{(1)}_{\bm{\widehat{6}}II,3}=5q^{1/5}\left(1+2q+2q^2+q^3+2q^4+2q^5+2q^6+q^7+2q^8+2q^9+\ldots\right)\equiv Y_3(\tau)\,,\\
\nonumber && Y^{(1)}_{\bm{\widehat{6}}II,4}=5\sqrt{2}q^{2/5}\left(1+q+q^2+q^3+2q^4+q^6+q^7+2 q^8+q^9+\ldots\right)\equiv Y_4(\tau)\,,\\
\nonumber && Y^{(1)}_{\bm{\widehat{6}}II,5}=-5\sqrt{2}q^{3/5}\left(1+q^2+q^3+q^4-q^5+2 q^6+q^8+q^9+\ldots\right)\equiv Y_5(\tau)\,,\\
\label{eq:q-series-wt1-N6}&&Y^{(1)}_{\bm{\widehat{6}}II,6}=5q^{4/5}\left(1-q+2q^2+2q^6-2q^7+2q^8+q^9+\ldots\right)\equiv Y_6(\tau)\,,
\end{eqnarray}
which coincides with the weight one and level $N=5$ modular forms constructed from eta function and the Klein form ~\cite{Yao:2020zml}. The positive integer weight modular forms at level 5 can be generated from the tensor products of $Y^{(1)}_{\bm{\widehat{6}}II}(\tau)$.

\item{$k_Y=2$}

\begin{eqnarray}
\nonumber Y_{\bm{3}}^{(2)}&=&(Y_{\bm{\widehat{6}}II}^{(1)}Y_{\bm{\widehat{6}}II}^{(1)})_{\bm{3}_{1, s}}=
\begin{pmatrix}
 -2 \left(Y_1 Y_2+Y_4 Y_5-Y_3 Y_6\right) \\
\sqrt{2} \left(Y_5^2-2 Y_2 Y_3\right) \\
 -\sqrt{2} \left(Y_4^2+2 Y_1 Y_6\right) \\
\end{pmatrix}\,,\\
\nonumber Y_{\bm{3'}}^{(2)}&=&(Y_{\bm{\widehat{6}}II}^{(1)}Y_{\bm{\widehat{6}}II}^{(1)})_{\bm{3'}_{1, s}}=
\begin{pmatrix}
 2 \left(Y_1 Y_2+Y_3 Y_6\right) \\
 -2 (Y_2 Y_4-Y_5 Y_6) \\
 -2 \left(Y_1 Y_5+Y_3 Y_4\right) \\
\end{pmatrix}\,,\\
Y_{\bm{5}}^{(2)}&=&(Y_{\bm{\widehat{6}}}^{(1)}Y_{\bm{\widehat{6}}}^{(1)})_{\bm{5}_{1, s}}=
\begin{pmatrix}
 -\sqrt{6} \left(Y_1^2+Y_2^2\right) \\
 2 \left(Y_5^2+Y_1 Y_3+Y_2 Y_3+\sqrt{2} Y_4 Y_6\right) \\
 2 \left(Y_3^2+\sqrt{2} Y_2 Y_4+\sqrt{2}Y_5 Y_6\right) \\
 2 \left(Y_6^2-\sqrt{2} Y_1 Y_5+\sqrt{2} Y_3 Y_4\right) \\
 2 \left(Y_4^2-\sqrt{2} Y_3 Y_5+\left(Y_2-Y_1\right) Y_6\right) \\
\end{pmatrix}\,.
\end{eqnarray}

\item{$k_Y = 3$}

\begin{eqnarray}
\nonumber Y_{\bm{\widehat{4}'}}^{(3)}&=&(Y_{\bm{\widehat{6}}II}^{(1)}Y_{\bm{3'}}^{(2)})_{\bm{\widehat{4}'}}=
\begin{pmatrix}
 -\sqrt{6} Y_3 Y_{\bm{3}',1}^{(2)}-\sqrt{3} Y_6 Y_{\bm{3}',2}^{(2)}+\sqrt{6} Y_5 Y_{\bm{3}',3}^{(2)} \\
 -2 Y_4 Y_{\bm{3}',1}^{(2)}+Y_1 Y_{\bm{3}',2}^{(2)}-3 Y_2 Y_{\bm{3}',2}^{(2)}+Y_6 Y_{\bm{3}',3}^{(2)} \\
 -2 Y_5 Y_{\bm{3}',1}^{(2)}-Y_3 Y_{\bm{3}',2}^{(2)}+\left(3 Y_1+Y_2\right) Y_{\bm{3}',3}^{(2)} \\
 -\sqrt{6} Y_6 Y_{\bm{3}',1}^{(2)}-\sqrt{6} Y_4 Y_{\bm{3}',2}^{(2)}+\sqrt{3} Y_3 Y_{\bm{3}',3}^{(2)} \\
\end{pmatrix}\,, \\
\nonumber Y_{\bm{\widehat{6}}I}^{(3)}&=&(Y_{\bm{\widehat{6}}II}^{(1)}Y_{\bm{3}}^{(2)})_{\bm{\widehat{6}}_{1}}=
\begin{pmatrix}
 -Y_1 Y_{\bm{3},1}^{(2)}-\sqrt{2} Y_3 Y_{\bm{3},3}^{(2)} \\
 Y_2 Y_{\bm{3},1}^{(2)}+\sqrt{2} Y_6 Y_{\bm{3},2}^{(2)} \\
 Y_3 Y_{\bm{3},1}^{(2)}-\sqrt{2} Y_1 Y_{\bm{3},2}^{(2)} \\
\sqrt{2} Y_5 Y_{\bm{3},3}^{(2)}-Y_4 Y_{\bm{3},1}^{(2)} \\
 Y_5 Y_{\bm{3},1}^{(2)}+\sqrt{2} Y_4 Y_{\bm{3},2}^{(2)} \\
\sqrt{2} Y_2 Y_{\bm{3},3}^{(2)}-Y_6 Y_{\bm{3},1}^{(2)} \\
\end{pmatrix} \,, \\
Y_{\bm{\widehat{6}}II}^{(3)}&=&(Y_{\bm{\widehat{6}}II}^{(1)}Y_{\bm{5}}^{(2)})_{\bm{\widehat{6}}_{3}}=
\begin{pmatrix}
\sqrt{2} Y_1 Y_{\bm{5},1}^{(2)}+\sqrt{3} \left(Y_6 Y_{\bm{5},2}^{(2)}-\sqrt{2} Y_5 Y_{\bm{5},3}^{(2)}-\sqrt{2} Y_4 Y_{\bm{5},4}^{(2)}+Y_3 Y_{\bm{5},5}^{(2)}\right) \\
\sqrt{2} Y_2 Y_{\bm{5},1}^{(2)}+\sqrt{3} \left(Y_6 Y_{\bm{5},2}^{(2)}-\sqrt{2} Y_5 Y_{\bm{5},3}^{(2)}+\sqrt{2} Y_4 Y_{\bm{5},4}^{(2)}-Y_3 Y_{\bm{5},5}^{(2)}\right) \\
\sqrt{3} \left(Y_1 Y_{\bm{5},2}^{(2)}-Y_2 Y_{\bm{5},2}^{(2)}+\sqrt{2} Y_4 Y_{\bm{5},5}^{(2)}\right)-2\sqrt{2} Y_3 Y_{\bm{5},1}^{(2)} \\
\sqrt{2} Y_4 Y_{\bm{5},1}^{(2)}+\sqrt{6} \left(Y_3 Y_{\bm{5},2}^{(2)}+\left(Y_2-Y_1\right) Y_{\bm{5},3}^{(2)}\right) \\
\sqrt{2} Y_5 Y_{\bm{5},1}^{(2)}-\sqrt{6} \left(Y_1 Y_{\bm{5},4}^{(2)}+Y_2 Y_{\bm{5},4}^{(2)}-Y_6 Y_{\bm{5},5}^{(2)}\right) \\
\sqrt{3} \left(\sqrt{2} Y_5 Y_{\bm{5},2}^{(2)}+\left(Y_1+Y_2\right) Y_{\bm{5},5}^{(2)}\right)-2\sqrt{2} Y_6 Y_{\bm{5},1}^{(2)} \\
\end{pmatrix}\,.~~~~~~~~~
\end{eqnarray}

\item{$k_Y=4$}

\begin{eqnarray}
\nonumber Y_{\bm{1}}^{(4)}&=&(Y_{\bm{\widehat{6}}II}^{(1)}Y_{\bm{\widehat{6}}I}^{(3)})_{\bm{1}_{a}}=Y_2 Y_{\bm{6}II,1}^{(3)}-Y_1 Y_{\bm{6}II,2}^{(3)}+Y_6 Y_{\bm{6}II,3}^{(3)}-Y_3 Y_{\bm{6}II,6}^{(3)}+Y_5 Y_{\bm{6}II,4}^{(3)}-Y_4 Y_{\bm{6}II,5}^{(3)}\,, \\
\nonumber Y_{\bm{3}}^{(4)}&=&(Y_{\bm{\widehat{6}}II}^{(1)}Y_{\bm{\widehat{6}}II}^{(3)})_{\bm{3}_{1,s}}=
\begin{pmatrix}
-Y_1Y_{\bm{6}II,2}^{(3)}-Y_2Y_{\bm{6}II,1}^{(3)}+Y_3Y_{\bm{6}II,6}^{(3)}-Y_4Y_{\bm{6}II,5}^{(3)}-Y_5Y_{\bm{6}II,4}^{(3)}+Y_6Y_{\bm{6}II,3}^{(3)}\\
-\sqrt{2}Y_2Y_{\bm{6}II,3}^{(3)}-\sqrt{2}Y_3Y_{\bm{6}II,2}^{(3)}+\sqrt{2}Y_5Y_{\bm{6}II,5}^{(3)}\\
-\sqrt{2}Y_1Y_{\bm{6}II,6}^{(3)}-\sqrt{2}Y_4Y_{\bm{6}II,4}^{(3)}-\sqrt{2}Y_6Y_{\bm{6}II,1}^{(3)}
\end{pmatrix}\,,\\
\nonumber Y_{\bm{3'}}^{(4)}&=&(Y_{\bm{\widehat{6}}II}^{(1)}Y_{\bm{\widehat{6}}I}^{(3)})_{\bm{3'}_{2, s}}=\begin{pmatrix}
-Y_1 Y_{\bm{6}I,1}^{(3)}+Y_2 Y_{\bm{6}I,2}^{(3)}-Y_6 Y_{\bm{6}I,3}^{(3)}-Y_5 Y_{\bm{6}I,4}^{(3)}-Y_4 Y_{\bm{6}I,5}^{(3)}-Y_3 Y_{\bm{6}I,6}^{(3)} \\
Y_4 \left(Y_{\bm{6}I,2}^{(3)}-Y_{\bm{6}I,1}^{(3)}\right)-\sqrt{2} Y_3 Y_{\bm{6}I,3}^{(3)}+\left(Y_2-Y_1\right) Y_{\bm{6}I,4}^{(3)} \\
Y_5 \left(Y_{\bm{6}I,1}^{(3)}+Y_{\bm{6}I,2}^{(3)}\right)+(Y_1+Y_2) Y_{\bm{6}I,5}^{(3)}+\sqrt{2} Y_6 Y_{\bm{6}I,6}^{(3)} \\
\end{pmatrix} \,, \\
\nonumber Y_{\bm{4}}^{(4)}&=&(Y_{\bm{\widehat{6}}II}^{(1)}Y_{\bm{\widehat{6}}I}^{(3)})_{\bm{4}_{a}}=
\begin{pmatrix}
-\sqrt{2} Y_3 Y_{\bm{6}I,2}^{(3)}+\sqrt{2} Y_2 Y_{\bm{6}I,3}^{(3)}-Y_6 Y_{\bm{6}I,4}^{(3)}+Y_4 Y_{\bm{6}I,6}^{(3)} \\
Y_4 \left(Y_{\bm{6}I,1}^{(3)}-Y_{\bm{6}I,2}^{(3)}\right)-Y_1 Y_{\bm{6}I,4}^{(3)}+Y_2 Y_{\bm{6}I,4}^{(3)}+Y_6 Y_{\bm{6}I,5}^{(3)}-Y_5 Y_{\bm{6}I,6}^{(3)} \\
Y_5 \left(Y_{\bm{6}I,1}^{(3)}+Y_{\bm{6}I,2}^{(3)}\right)-Y_4 Y_{\bm{6}I,3}^{(3)}+Y_3 Y_{\bm{6}I,4}^{(3)}-Y_1 Y_{\bm{6}I,5}^{(3)}-Y_2 Y_{\bm{6}I,5}^{(3)} \\
\sqrt{2} Y_6 Y_{\bm{6}I,1}^{(3)}-Y_5 Y_{\bm{6}I,3}^{(3)}+Y_3 Y_{\bm{6}I,5}^{(3)}-\sqrt{2} Y_1 Y_{\bm{6}I,6}^{(3)} \\
\end{pmatrix}  \,, \\
\nonumber Y_{\bm{5}I}^{(4)}&=&(Y_{\bm{\widehat{6}}II}^{(1)}Y_{\bm{\widehat{6}}I}^{(3)})_{\bm{5}_{1, s}}=
\begin{pmatrix}
-\sqrt{6} \left(Y_1 Y_{\bm{6}I,1}^{(3)}+Y_2 Y_{\bm{6}I,2}^{(3)}\right) \\
Y_3 \left(Y_{\bm{6}I,1}^{(3)}+Y_{\bm{6}I,2}^{(3)}\right)+Y_1 Y_{\bm{6}I,3}^{(3)}+Y_2 Y_{\bm{6}I,3}^{(3)}+\sqrt{2} Y_6 Y_{\bm{6}I,4}^{(3)}+2 Y_5 Y_{\bm{6}I,5}^{(3)}+\sqrt{2} Y_4 Y_{\bm{6}I,6}^{(3)} \\
\sqrt{2} Y_4 Y_{\bm{6}I,2}^{(3)}+2 Y_3 Y_{\bm{6}I,3}^{(3)}+\sqrt{2} \left(Y_2 Y_{\bm{6}I,4}^{(3)}+Y_6 Y_{\bm{6}I,5}^{(3)}+Y_5 Y_{\bm{6}I,6}^{(3)}\right) \\
-\sqrt{2} Y_5 Y_{\bm{6}I,1}^{(3)}+\sqrt{2} Y_4 Y_{\bm{6}I,3}^{(3)}+\sqrt{2} Y_3 Y_{\bm{6}I,4}^{(3)}-\sqrt{2} Y_1 Y_{\bm{6}I,5}^{(3)}+2 Y_6 Y_{\bm{6}I,6}^{(3)} \\
Y_6 \left(Y_{\bm{6}I,2}^{(3)}-Y_{\bm{6}I,1}^{(3)}\right)-\sqrt{2} Y_5 Y_{\bm{6}I,3}^{(3)}+2 Y_4 Y_{\bm{6}I,4}^{(3)}-\sqrt{2} Y_3 Y_{\bm{6}I,5}^{(3)}-Y_1 Y_{\bm{6}I,6}^{(3)}+Y_2 Y_{\bm{6}I,6}^{(3)} \\
\end{pmatrix} \,, \\
Y_{\bm{5}II}^{(4)}&=&(Y_{\bm{\widehat{6}}II}^{(1)}Y_{\bm{\widehat{6}}I}^{(3)})_{\bm{5}_{2, a}}=
\begin{pmatrix}
-Y_2 Y_{\bm{6}I,1}^{(3)}+Y_1 Y_{\bm{6}I,2}^{(3)}-Y_6 Y_{\bm{6}I,3}^{(3)}+2 Y_5 Y_{\bm{6}I,4}^{(3)}-2 Y_4 Y_{\bm{6}I,5}^{(3)}+Y_3 Y_{\bm{6}I,6}^{(3)} \\
\sqrt{6} \left(Y_1 Y_{\bm{6}I,3}^{(3)}-Y_3 Y_{\bm{6}I,1}^{(3)}\right) \\
\sqrt{3} \left(Y_4 Y_{\bm{6}I,2}^{(3)}-Y_2 Y_{\bm{6}I,4}^{(3)}+Y_6 Y_{\bm{6}I,5}^{(3)}-Y_5 Y_{\bm{6}I,6}^{(3)}\right) \\
\sqrt{3} \left(-Y_5 Y_{\bm{6}I,1}^{(3)}-Y_4 Y_{\bm{6}I,3}^{(3)}+Y_3 Y_{\bm{6}I,4}^{(3)}+Y_1 Y_{\bm{6}I,5}^{(3)}\right) \\
\sqrt{6} \left(Y_2 Y_{\bm{6}I,6}^{(3)}-Y_6 Y_{\bm{6}I,2}^{(3)}\right)
\end{pmatrix}\,.
\end{eqnarray}

\item{$k_Y = 5$}

\begin{eqnarray}
\nonumber Y_{\bm{\widehat{2}}}^{(5)}&=&(Y_{\bm{\widehat{6}}II}^{(1)}Y_{\bm{5}II}^{(4)})_{\bm{2}}=
\begin{pmatrix}
\sqrt{3} Y_4 Y_{\bm{5}II,1}^{(4)}-2 Y_3 Y_{\bm{5}II,2}^{(4)}+Y_1 Y_{\bm{5}II,3}^{(4)}+2 Y_2 Y_{\bm{5}II,3}^{(4)}+Y_6 Y_{\bm{5}II,4}^{(4)}-\sqrt{2} Y_5 Y_{\bm{5}II,5}^{(4)} \\
 -\sqrt{3} Y_5 Y_{\bm{5}II,1}^{(4)}-\sqrt{2} Y_4 Y_{\bm{5}II,2}^{(4)}+Y_3 Y_{\bm{5}II,3}^{(4)}+2 Y_1 Y_{\bm{5}II,4}^{(4)}-Y_2 Y_{\bm{5}II,4}^{(4)}+2 Y_6 Y_{\bm{5}II,5}^{(4)} \\
\end{pmatrix} \,,\\
\nonumber Y_{\bm{\widehat{2}'}}^{(5)}&=&(Y_{\bm{\widehat{6}}II}^{(1)}Y_{\bm{5}II}^{(4)})_{\bm{2'}}=
\begin{pmatrix}
\sqrt{6} Y_3 Y_{\bm{5}II,1}^{(4)}+3 Y_1 Y_{\bm{5}II,2}^{(4)}+Y_2 Y_{\bm{5}II,2}^{(4)}-2 Y_6 Y_{\bm{5}II,3}^{(4)}+2\sqrt{2} Y_5 Y_{\bm{5}II,4}^{(4)}+\sqrt{2} Y_4 Y_{\bm{5}II,5}^{(4)} \\
\sqrt{6} Y_6 Y_{\bm{5}II,1}^{(4)}+\sqrt{2} Y_5 Y_{\bm{5}II,2}^{(4)}-2\sqrt{2} Y_4 Y_{\bm{5}II,3}^{(4)}+2 Y_3 Y_{\bm{5}II,4}^{(4)}-Y_1 Y_{\bm{5}II,5}^{(4)}+3 Y_2 Y_{\bm{5}II,5}^{(4)} \\
\end{pmatrix}\,, \\
\nonumber Y_{\bm{\widehat{4}'}}^{(5)}&=&(Y_{\bm{\widehat{6}}II}^{(1)}Y_{\bm{5}II}^{(4)})_{\bm{4'}_{1}}=
\begin{pmatrix}
\sqrt{6} Y_3 Y_{\bm{5}II,1}^{(4)}-2 Y_2 Y_{\bm{5}II,2}^{(4)}+Y_6 Y_{\bm{5}II,3}^{(4)}-\sqrt{2} Y_5 Y_{\bm{5}II,4}^{(4)}+\sqrt{2} Y_4 Y_{\bm{5}II,5}^{(4)} \\
\sqrt{3} \left(Y_1 Y_{\bm{5}II,3}^{(4)}+Y_2 Y_{\bm{5}II,3}^{(4)}-Y_6 Y_{\bm{5}II,4}^{(4)}+\sqrt{2} Y_5 Y_{\bm{5}II,5}^{(4)}\right) \\
\sqrt{3} \left(\sqrt{2} Y_4 Y_{\bm{5}II,2}^{(4)}-Y_3 Y_{\bm{5}II,3}^{(4)}+\left(Y_1-Y_2\right) Y_{\bm{5}II,4}^{(4)}\right) \\
-\sqrt{6} Y_6 Y_{\bm{5}II,1}^{(4)}-\sqrt{2} Y_5 Y_{\bm{5}II,2}^{(4)}-\sqrt{2} Y_4 Y_{\bm{5}II,3}^{(4)}+Y_3 Y_{\bm{5}II,4}^{(4)}-2 Y_1 Y_{\bm{5}II,5}^{(4)} \\
\end{pmatrix}\,, \\
\nonumber Y_{\bm{\widehat{6}}I}^{(5)}&=&(Y_{\bm{\widehat{6}}II}^{(1)}Y_{\bm{1}}^{(4)})_{\bm{6}}=
\begin{pmatrix}
Y_1 Y_{\bm{1},1}^{(4)} \\
Y_2 Y_{\bm{1},1}^{(4)} \\
Y_3 Y_{\bm{1},1}^{(4)} \\
Y_4 Y_{\bm{1},1}^{(4)} \\
Y_5 Y_{\bm{1},1}^{(4)} \\
Y_6 Y_{\bm{1},1}^{(4)}
\end{pmatrix}\,, ~~~~~
Y_{\bm{\widehat{6}}II}^{(5)}=(Y_{\bm{\widehat{6}}II}^{(1)}Y_{\bm{3'}}^{(4)})_{\bm{6}_{1}}=
\begin{pmatrix}
Y_1 Y_{\bm{3}',1}^{(4)}-Y_4 Y_{\bm{3}',3}^{(4)} \\
Y_5 Y_{\bm{3}',2}^{(4)}-Y_2 Y_{\bm{3}',1}^{(4)} \\
Y_3 Y_{\bm{3}',1}^{(4)}+Y_5 Y_{\bm{3}',3}^{(4)} \\
Y_6 Y_{\bm{3}',3}^{(4)}-Y_1 Y_{\bm{3}',2}^{(4)} \\
Y_3 Y_{\bm{3}',2}^{(4)}+Y_2 Y_{\bm{3}',3}^{(4)} \\
Y_4 Y_{\bm{3}',2}^{(4)}-Y_6 Y_{\bm{3}',1}^{(4)} \\
\end{pmatrix} \,, \\
Y_{\bm{\widehat{6}}III}^{(5)}&=&(Y_{\bm{\widehat{6}}II}^{(1)}Y_{\bm{5}II}^{(4)})_{\bm{6}_{2}}=
\begin{pmatrix}
\sqrt{3} Y_1 Y_{\bm{5}II,1}^{(4)}-2 Y_5 Y_{\bm{5}II,3}^{(4)}-Y_4 Y_{\bm{5}II,4}^{(4)}+\sqrt{2} Y_3 Y_{\bm{5}II,5}^{(4)} \\
\sqrt{3} Y_2 Y_{\bm{5}II,1}^{(4)}+\sqrt{2} Y_6 Y_{\bm{5}II,2}^{(4)}-Y_5 Y_{\bm{5}II,3}^{(4)}+2 Y_4 Y_{\bm{5}II,4}^{(4)} \\
-\sqrt{3} Y_3 Y_{\bm{5}II,1}^{(4)}+\sqrt{2} Y_1 Y_{\bm{5}II,2}^{(4)}-Y_5 Y_{\bm{5}II,4}^{(4)}+2 Y_4 Y_{\bm{5}II,5}^{(4)} \\
2 Y_3 Y_{\bm{5}II,2}^{(4)}-Y_1 Y_{\bm{5}II,3}^{(4)}+2 Y_2 Y_{\bm{5}II,3}^{(4)}+Y_6 Y_{\bm{5}II,4}^{(4)} \\
-Y_3 Y_{\bm{5}II,3}^{(4)}-2 Y_1 Y_{\bm{5}II,4}^{(4)}-Y_2 Y_{\bm{5}II,4}^{(4)}+2 Y_6 Y_{\bm{5}II,5}^{(4)} \\
-\sqrt{3} Y_6 Y_{\bm{5}II,1}^{(4)}+2 Y_5 Y_{\bm{5}II,2}^{(4)}+Y_4 Y_{\bm{5}II,3}^{(4)}+\sqrt{2} Y_2 Y_{\bm{5}II,5}^{(4)}
\end{pmatrix}\,.
\end{eqnarray}

\item{$k_Y = 6$}

\begin{eqnarray}
\nonumber Y_{\bm{1}}^{(6)}&=&(Y_{\bm{\widehat{6}}II}^{(1)}Y_{\bm{\widehat{6}}II}^{(5)})_{\bm{1}_{a}}=
 Y_2 Y_{\bm{6}II,1}^{(5)}-Y_1 Y_{\bm{6}II,2}^{(5)}+Y_6 Y_{\bm{6}II,3}^{(5)}+Y_5 Y_{\bm{6}II,4}^{(5)}-Y_4 Y_{\bm{6}II,5}^{(5)}-Y_3 Y_{\bm{6}II,6}^{(5)}\,, \\
\nonumber Y_{\bm{3}I}^{(6)}&=&(Y_{\bm{\widehat{6}}II}^{(1)}Y_{\bm{\widehat{6}}I}^{(5)})_{\bm{3}_{1, s}}=
\begin{pmatrix}
 -Y_2 Y_{\bm{6}I,1}^{(5)}-Y_1 Y_{\bm{6}I,2}^{(5)}+Y_6 Y_{\bm{6}I,3}^{(5)}-Y_5 Y_{\bm{6}I,4}^{(5)}-Y_4 Y_{\bm{6}I,5}^{(5)}+Y_3 Y_{\bm{6}I,6}^{(5)} \\
 -\sqrt{2} \left(Y_3 Y_{\bm{6}I,2}^{(5)}+Y_2 Y_{\bm{6}I,3}^{(5)}-Y_5 Y_{\bm{6}I,5}^{(5)}\right) \\
 -\sqrt{2} \left(Y_6 Y_{\bm{6}I,1}^{(5)}+Y_4 Y_{\bm{6}I,4}^{(5)}+Y_1 Y_{\bm{6}I,6}^{(5)}\right) \\
\end{pmatrix}\,,\\
\nonumber Y_{\bm{3}II}^{(6)}&=&(Y_{\bm{\widehat{6}}II}^{(1)}Y_{\bm{\widehat{6}}III}^{(5)})_{\bm{3}_{1, s}}=
\begin{pmatrix}
-Y_2 Y_{\bm{6}III,1}^{(5)}-Y_1 Y_{\bm{6}III,2}^{(5)}+Y_6 Y_{\bm{6}III,3}^{(5)}-Y_5 Y_{\bm{6}III,4}^{(5)}-Y_4 Y_{\bm{6}III,5}^{(5)}+Y_3 Y_{\bm{6}III,6}^{(5)} \\
-\sqrt{2} \left(Y_3 Y_{\bm{6}III,2}^{(5)}+Y_2 Y_{\bm{6}III,3}^{(5)}-Y_5 Y_{\bm{6}III,5}^{(5)}\right) \\
-\sqrt{2} \left(Y_6 Y_{\bm{6}III,1}^{(5)}+Y_4 Y_{\bm{6}III,4}^{(5)}+Y_1 Y_{\bm{6}III,6}^{(5)}\right) \\
\end{pmatrix}\,,\\
\nonumber Y_{\bm{3'}I}^{(6)}&=&(Y_{\bm{\widehat{6}}II}^{(1)}Y_{\bm{\widehat{6}}I}^{(5)})_{\bm{3'}_{1, s}}=\begin{pmatrix}
Y_2 Y_{\bm{6}I,1}^{(5)}+Y_1 Y_{\bm{6}I,2}^{(5)}+Y_6 Y_{\bm{6}I,3}^{(5)}+Y_3 Y_{\bm{6}I,6}^{(5)} \\
-Y_4 Y_{\bm{6}I,2}^{(5)}-Y_2 Y_{\bm{6}I,4}^{(5)}+Y_6 Y_{\bm{6}I,5}^{(5)}+Y_5 Y_{\bm{6}I,6}^{(5)} \\
-Y_5 Y_{\bm{6}I,1}^{(5)}-Y_4 Y_{\bm{6}I,3}^{(5)}-Y_3 Y_{\bm{6}I,4}^{(5)}-Y_1 Y_{\bm{6}I,5}^{(5)} \\
\end{pmatrix}\,,\\
\nonumber Y_{\bm{3'}II}^{(6)}&=&(Y_{\bm{\widehat{6}}II}^{(1)}Y_{\bm{\widehat{6}}III}^{(5)})_{\bm{3'}_{1, s}}=\begin{pmatrix}
Y_2 Y_{\bm{6}III,1}^{(5)}+Y_1 Y_{\bm{6}III,2}^{(5)}+Y_6 Y_{\bm{6}III,3}^{(5)}+Y_3 Y_{\bm{6}III,6}^{(5)} \\
-Y_4 Y_{\bm{6}III,2}^{(5)}-Y_2 Y_{\bm{6}III,4}^{(5)}+Y_6 Y_{\bm{6}III,5}^{(5)}+Y_5 Y_{\bm{6}III,6}^{(5)} \\
-Y_5 Y_{\bm{6}III,1}^{(5)}-Y_4 Y_{\bm{6}III,3}^{(5)}-Y_3 Y_{\bm{6}III,4}^{(5)}-Y_1 Y_{\bm{6}III,5}^{(5)} \\
\end{pmatrix}\,, \\
\nonumber Y_{\bm{4}I}^{(6)}&=&(Y_{\bm{\widehat{6}}II}^{(1)}Y_{\bm{\widehat{2}'}}^{(5)})_{\bm{4}}=
\begin{pmatrix}
Y_4 Y_{\bm{2}',2}^{(5)}-\sqrt{2} Y_1 Y_{\bm{2}',1}^{(5)} \\
\sqrt{2} Y_3 Y_{\bm{2}',1}^{(5)}+Y_5 Y_{\bm{2}',2}^{(5)} \\
Y_4 Y_{\bm{2}',1}^{(5)}+\sqrt{2} Y_6 Y_{\bm{2}',2}^{(5)} \\
-Y_5 Y_{\bm{2}',1}^{(5)}-\sqrt{2} Y_2 Y_{\bm{2}',2}^{(5)} \\
\end{pmatrix}\,, \\
\nonumber Y_{\bm{4}II}^{(6)}&=&(Y_{\bm{\widehat{6}}II}^{(1)}Y_{\bm{\widehat{4}'}}^{(5)})_{\bm{4}_{1}}=
\begin{pmatrix}
Y_1 Y_{\bm{4}',1}^{(5)}-\sqrt{3} Y_6 Y_{\bm{4}',2}^{(5)}-\sqrt{2} Y_4 Y_{\bm{4}',4}^{(5)} \\
-Y_3 Y_{\bm{4}',1}^{(5)}-\sqrt{3} Y_2 Y_{\bm{4}',2}^{(5)}-\sqrt{2} Y_5 Y_{\bm{4}',4}^{(5)} \\
\sqrt{2} Y_4 Y_{\bm{4}',1}^{(5)}-\sqrt{3} Y_1 Y_{\bm{4}',3}^{(5)}+Y_6 Y_{\bm{4}',4}^{(5)} \\
-\sqrt{2} Y_5 Y_{\bm{4}',1}^{(5)}-\sqrt{3} Y_3 Y_{\bm{4}',3}^{(5)}-Y_2 Y_{\bm{4}',4}^{(5)} \\
\end{pmatrix}\,, \\
\nonumber Y_{\bm{5}I}^{(6)}&=&(Y_{\bm{\widehat{6}}II}^{(1)}Y_{\bm{\widehat{4}'}}^{(5)})_{\bm{5}_{1}}=
\begin{pmatrix}
\sqrt{6} \left(Y_5 Y_{\bm{4}',2}^{(5)}+Y_4 Y_{\bm{4}',3}^{(5)}\right) \\
\sqrt{6} Y_2 Y_{\bm{4}',1}^{(5)}+\sqrt{2} Y_6 Y_{\bm{4}',2}^{(5)}+Y_5 Y_{\bm{4}',3}^{(5)}-\sqrt{3} Y_4 Y_{\bm{4}',4}^{(5)} \\
\sqrt{2} \left(-\sqrt{3} Y_3 Y_{\bm{4}',1}^{(5)}+Y_1 Y_{\bm{4}',2}^{(5)}+Y_2 Y_{\bm{4}',2}^{(5)}+Y_6 Y_{\bm{4}',3}^{(5)}\right) \\
\sqrt{2} \left(-Y_3 Y_{\bm{4}',2}^{(5)}+Y_1 Y_{\bm{4}',3}^{(5)}-Y_2 Y_{\bm{4}',3}^{(5)}+\sqrt{3} Y_6 Y_{\bm{4}',4}^{(5)}\right) \\
 -\sqrt{3} Y_5 Y_{\bm{4}',1}^{(5)}-Y_4 Y_{\bm{4}',2}^{(5)}+\sqrt{2} Y_3 Y_{\bm{4}',3}^{(5)}+\sqrt{6} Y_1 Y_{\bm{4}',4}^{(5)} \\
\end{pmatrix}\,, \\
Y_{\bm{5}II}^{(6)}&=&(Y_{\bm{\widehat{6}}II}^{(1)}Y_{\bm{\widehat{6}}II}^{(5)})_{\bm{5}_{2, a}}=\begin{pmatrix}
-Y_2 Y_{\bm{6}II,1}^{(5)}+Y_1 Y_{\bm{6}II,2}^{(5)}-Y_6 Y_{\bm{6}II,3}^{(5)}+2 Y_5 Y_{\bm{6}II,4}^{(5)}-2 Y_4 Y_{\bm{6}II,5}^{(5)}+Y_3 Y_{\bm{6}II,6}^{(5)} \\
\sqrt{6} \left(Y_1 Y_{\bm{6}II,3}^{(5)}-Y_3 Y_{\bm{6}II,1}^{(5)}\right) \\
\sqrt{3} \left(Y_4 Y_{\bm{6}II,2}^{(5)}-Y_2 Y_{\bm{6}II,4}^{(5)}+Y_6 Y_{\bm{6}II,5}^{(5)}-Y_5 Y_{\bm{6}II,6}^{(5)}\right) \\
\sqrt{3} \left(-Y_5 Y_{\bm{6}II,1}^{(5)}-Y_4 Y_{\bm{6}II,3}^{(5)}+Y_3 Y_{\bm{6}II,4}^{(5)}+Y_1 Y_{\bm{6}II,5}^{(5)}\right) \\
\sqrt{6} \left(Y_2 Y_{\bm{6}II,6}^{(5)}-Y_6 Y_{\bm{6}II,2}^{(5)}\right)
\end{pmatrix}\,.~~~~~~~~
\end{eqnarray}

\end{itemize}

\end{appendix}

\providecommand{\href}[2]{#2}\begingroup\raggedright\endgroup

\end{document}